\documentclass[a4paper]{article}

\usepackage{latexsym}
\usepackage{graphics}
\usepackage{epsfig,hangcaption}
\usepackage{multirow}
\usepackage{amssymb,amsmath}
\usepackage{times}

\newcommand{\shat}{\mbox{\ensuremath{\hat{s}}}}

\newcommand{\Mfunction}[1]{\mathbf{#1}}

\DeclareOldFontCommand{\tt}{\normalfont\sf\small}{\mathtt}

\newcommand{\authorrunning}[1]{\edef\authorrun{#1}}
\newcommand{\titlerunning}[1]{\edef\titlerun{#1}}
\usepackage{fancyhdr}
\begin{document}

\title{The Monte Carlo Event Generator \\
 AcerMC versions 2.0 to 3.8
with interfaces to \\ PYTHIA 6.4, HERWIG 6.5 and ARIADNE 4.1}

\titlerunning{AcerMC Event Generator}
\authorrunning{{B. P. Kersevan,E. Richter-Was}}

\author{
Borut Paul Kersevan \\
\small Faculty of Mathematics and Physics, University of Ljubljana,
 Jadranska 19, SI-1000 Ljubljana, Slovenia. \\ \small and  \\
\small  Jozef Stefan Institute, Jamova 39, SI-1000 Ljubljana, Slovenia.\\[10pt]
Elzbieta Richter-W\c{a}s\thanks{ \scriptsize 
Partly supported by Marie Curie Host Fellowship for the Transfer of 
  Knowledge Contract No. MTKD-CT-2004-510126  and  
  by the  EC FP5 Centre of Excellence ``COPIRA'' under the contract 
  No. IST-2001-37259.}\\
\small Institute of Physics, Jagellonian University,\\
\small 30-059 Krakow, ul. Reymonta 4, Poland.\\
}

\maketitle

\begin{picture}(0,0)
\put(350,370){\rm   TPJU-6/2004}\\
\put(350,360){\rm   hep-ph/0405247}\\
\end{picture}

\abstract{\it 
The {\bf AcerMC} Monte Carlo Event Generator is dedicated for the
generation of Standard Model background processes at $pp$ LHC
collisions. The program itself provides a library of the massive
matrix elements and phase space modules for generation of selected
processes: $gg, q \bar q \to t \bar t b \bar b$; $ q \bar q W(\to \ell
\nu) b \bar b$; $q \bar q W(\to \ell \nu) t \bar t$; $gg, q \bar q \to
Z/\gamma^*(\to \ell \ell) b \bar b$; $gg, q \bar q \to Z/\gamma^*(\to
\ell \ell, \nu \nu, b \bar b) t \bar t$; complete electroweak $gg, q
\bar q \to (Z/W/\gamma^* \to) b \bar b t \bar t$; $gg, q \bar q \to t
\bar t t \bar t$; $gg, q \bar q \to (t \bar t \to) f \bar fb f \bar f
\bar b $; $ gg, q \bar q \to (W W b b~ \to) f \bar f f \bar f b \bar
b$; $ gg, q \bar q \to (W W b b~ \to) f \bar f f \bar f b \bar b$,
single top production, $Z^0 b$ and $Z^{0\prime} \to t \bar t$ processes.  The
hard process event, generated with one of these modules, can be
completed by the initial and final state radiation, hadronisation and
decays, simulated with either {\tt PYTHIA}, {\tt ARIADNE} or {\tt
HERWIG} Monte Carlo event generator and (optionally) with {\tt TAUOLA}
and {\tt PHOTOS}. Interfaces to all these packages are provided in the
distribution version. The matrix element codes have been derived with
the help of the {\tt MADGRAPH} package.  The phase-space generation is
based on the multi-channel self-optimising approach using the modified
Kajantie-Byckling formalism for phase space construction and further
smoothing of the phase space was obtained by using a modified {\tt
ac-VEGAS} algorithm.
\bigskip \hrule}

\begin{center}

\end{center}
\vfill
\clearpage
 
\tableofcontents

\newpage

{\scriptsize
\begin{verbatim}      
  -----------------------------------------------------------------------------
                                   ._                           
                                 .j%3]:,                        
                               ~!%%%%%%% ,._.                   
                            _|xx%xxxx%%%%%+`                    
                             :~]%xxxx]xx%x_,+_x_%`              
                      -__||x||xx]+]]]+]]]]x|xxx]`               
                       -+%%xxxx]]]]+]]++]+]x]|>- .;..;.:_/`     
                          -+x]]]]|+]+]+]++]++]]+|+|]+|]]+-      
              ,.   .  ., |x]+]+||=++]+]=++++++=|++]=+]|~-       
              -|%x]]];x]]||=++++++++|];|++++++++++=+]=];   ..   ..  :..,; 
                -/]|]+]|]++++++++||||==|:;:=|==++==+;|,   :;;. :;=;;===|` 
                 _|]|+>+++]+]]|+|||=|=;|;;:|;=:===|==;::.;;:;,;:;;;;;;:-  
                     -++]+++:+|x+::||=||:=:::::;;::::::::;:;:;:;:;;;=::   
                   .|x+|+|]++:,-..:::|=;=:-.:.:-::.:.:.::-::::::::;:-     
         .,  .:||_   --||;:|:::.-.:.-:|;||::.:...-.-.-....::::-:::;::.    
     __._;++;;;|=|;. -:::::::---:.--;;|==|:;:.:-:-:.-.-.:::.--:::;:;--    
     -+++=+=======;==:::::-:.::...:.|+=;;===:-..:.:.:::::-...--:  -       
     :|:-:|===;:;:;::::::--::.:-::::.|||===:::.:.:::::.-......--:: .      
        -;|===;;:;;;;::;::.:-:: -:=;;|+||=;==|=::::::::...-.:..-.::..     
         ---|;:,::::.::::=;=:=::  -++===|======:--:   ...-...-:-:.-       
             ---;:::::;:--:===;;;:|-|+=+|+|==;:`    ...-....-..           
             .::;;:::;:;;:--- -- --.|=||=+|=:=,      ..... .              
            .;;=;;:-:::;:::.        -+;,|]| ;;=`                          
              - -:- --:;;:- -         - :|; -                             
                       -                 :`                               
                                          :                               
                                          :                               
                                          .                               
                                                                  
   40000L,                                           |0000i   j000&  .a00000L#0
   --?##0L        .aaaa aa     .aaaa;     aaaa, _aaa, -000A  _0001- _d0!`  -400
     d0 40,     _W0#V9N0#&    d0#V*N#0,   0##0LW0@4#@' 00j#; J0|01  d0'      40
    J0l -#W     #0'    ?#W   ##~    -#0;    j##9       00 4#|01|01  00     
   _00yyyW0L   :0f      ^-  :000###00001    j#1        00 ?#0@`|01  #0      
   ##!!!!!#0;  -0A       _  -0A             j#1        00  HH< |01  j0L       _
 ad0La,  aj0Aa  4#Aaa_aj#0`  ?0Laa_aaa0L  aaJ0Laaa,  _a00aa  _aj0La  *0Aaa_aad
 HHHRHl  HHHHH   `9##009!     `9NW00@!!`  HHHHRHHHl  :HHHHH  ?HHHRH   ?!##00P!`
      
  -----------------------------------------------------------------------------
      
          AcerMC 3.8 (May 2011),  B. P. Kersevan, E. Richter-Was
      
  -----------------------------------------------------------------------------
\end{verbatim}
}
\vspace{0.5cm}
\begin{flushleft}
{\it Available from web page: {\bf http://cern.ch/Borut.Kersevan} }\\
\vspace{0.10cm}
{\it Contact author email: {\bf Borut.Kersevan@cern.ch}}
\end{flushleft}

\newpage
\boldmath
\section{PROGRAM SUMMARY}
\unboldmath

\noindent
{\it Title of the program:} {\bf AcerMC version 3.8}\\
{\it Operating system:} Linux\\
{\it Programming language:} FORTRAN 77 with popular extensions (g77, gfortran).\\
{\it External libraries:} CERNLIB, LHAPDF.\\
{\it Size of the compressed distribution directory:} about 57 MB. The distribution includes modified
versions of {\tt PYTHIA~6.4}, {\tt HERWIG 6.5}, {\tt ARIADNE} and {\tt HELAS} libraries, {\tt TAUOLA} and 
{\tt PHOTOS} packages.\\
{\it Key words:} Standard Model backgrounds at LHC, massive matrix elements, Monte Carlo generator,
heavy flavor production, multi-channel phase-space generation.\\
{\it Does the new version supersede the previous version?:} Yes.\\
{\it Reasons for the new version:} Implementation of several new processes and methods.\\
{\it Summary of revisions:} Each version added new processes or functionalities, a detailed list is given in the section
'Changes since AcerMC 1.0'\\
{\it Nature of physical problem:}
Despite a large repertoire of processes implemented for generation in event
generators like {\tt PYTHIA} [1] or {\tt HERWIG} [2] a number of background processes, 
crucial for studying the expected physics of the LHC experiments, is missing.
For some of these processes the matrix element expressions are rather lengthly and/or
to achieve a reasonable generation efficiency it is necessary to tailor the
phase-space selection procedure to the dynamics of the process. 
That is why it is not practical to imagine that any of the above general
purpose generators will contain {\it every}, or even only {\it observable}, processes which
will occur at LHC collisions. A more practical solution
can be found in  a library of dedicated  matrix-element-based generators,
with the standardised interfaces  like that proposed in [3],
 to the more universal one which is used to complete the event generation.\\
{\it Method of solution:}
The {\bf AcerMC} Event Generator provides itself library of the matrix-element-based 
generators for several 
processes. The initial- and final- state showers, beam remnants and underlying events,
fragmentation and remaining 
decays are supposed to be performed by the other universal generator
to which this one is interfaced. We will call it {\it supervising
generator}. 
The interfaces to {\tt PYTHIA~6.4}, {\tt ARIADNE 4.1} and {\tt HERWIG 6.5}, as such  generators, 
are provided. Provided is also interface to {\tt TAUOLA} [4] and {\tt PHOTOS} [5] packages
for $\tau$-lepton decays (including spin correlations treatement) and QED radiations
in decays of particles.
At present, the following  matrix-element-based processes have
 been implemented: 
$gg, q \bar q \to t \bar t b \bar b$, 
$q \bar q \to W (\to \ell \nu) b \bar b$;
$q \bar q \to W (\to \ell \nu) t \bar t$;
$gg, q \bar q \to Z/\gamma^*(\to \ell \ell) b \bar b$;
$gg, q \bar q \to Z/\gamma^*(\to \ell \ell, \nu \nu, b \bar b) t \bar t$;
complete EW $gg, q \bar q  \to (Z/W/\gamma^* \to) t \bar t b \bar b$;
$gg, q \bar q  \to t \bar t t \bar t$;
 $gg, q \bar q  \to   (t \bar t \to) f \bar f b f \bar f \bar b$; 
$ gg, q \bar q  \to  (W W b b~ \to) f \bar f f \bar f b \bar b$.
Both interfaces allow the use of the {\tt LHAPDF/LHAGLUE} library of  parton density functions.
Provided is also set of {\it control processes}: $q \bar q \to W \to \ell \nu$;
 $q \bar q \to Z/\gamma^* \to \ell \ell$; 
$gg, q \bar q  \to t \bar t$ and  $gg \to (t \bar t \to )Wb  W \bar b$; \\
{\it Restriction on the complexity of the problem:}
The package is optimized for the 14 TeV $pp$ collision simulated in the LHC environment and also
works at the achieved LHC energies of 7 TeV and 8 TeV. 
The consistency between results of the complete generation using {\tt PYTHIA~6.4} 
or {\tt HERWIG~6.5} interfaces is technically limited by the different
approaches taken in both these generators for evaluating $\alpha_{QCD}$ 
and $\alpha_{QED}$ couplings and by the different models
for fragmentation/hadronisation.
For the consistency check, in the {\bf AcerMC} library  contains native coded 
definitions of the $\alpha_{QCD}$ and  $\alpha_{QED}$. Using these native
definitions leads to the same total cross-sections both with 
 {\tt PYTHIA~6.4} or {\tt HERWIG~6.5} interfaces. \\
{\it Typical running time:} On an PIII 800 MHz PC it amounts to $\sim 0.05 \to 1.1$ events/sec, 
depending on the choice of process. 
  
[1]. T. Sjostrand et al., {\it High energy physics generation with PYTHIA~6.2},
eprint hep-ph/0108264, LU-TP  01-21, August 2001.

[2]. G. Julyesini et al., Comp. Phys. Commun. {\bf 67} (1992) 465,
G. Corcella et al., JHEP {\bf 0101} (2001) 010.

[3]. E. Boos at al., {\it Generic user process interface for event generators}, hep-ph/0109068.

[4]. S. Jadach, J. H. Kuhn, Z. Was, Comput. Phys. Commun. {\bf 64} (1990) 275;
M. Jezabek, Z. Was, S. Jadach, J. H. Kuhn, Comput. Phys. Commun. {\bf 70}
(1992) 69; R. Decker, S. Jadach, J. H. Kuhn, Z. Was, Comput. Phys. Commun. {\bf 76} 
(1993) 361.  

[5]. E. Barberio and Z. Was, Comp. Phys. Commun. {\bf 79} (1994) 291.

\newpage
\boldmath
\section{Changes since AcerMC 1.0 [Comput.\ Phys.\ Commun.\  {\bf 149} (2003) 142]}
\unboldmath

\begin{itemize}
\item {\bf AcerMC version 1.1 (11. 7. 2002)}: The changes include transition to {\tt HERWIG 6.4}, 
updated scale choices for processes {\bf 5-8} (c.f. Section \ref{s:scdef}) and the
inclusion of control processes 91-94 for consistent process evaluation 
(c.f. Section \ref{s:cont_chan}). Also, a  possibility of an event dump 
according to the Les Houches standard was added (see Section \ref{s:leshdump}).
\item  {\bf AcerMC version 1.2 (20. 9. 2002)}:  A bug fix in {\tt HERWIG 6.4}, affecting the shower 
evolution, was made. It carries no immediate impact on the AcerMC processes but was
discovered by the AcerMC authors and added for the convenience of the users. This bug
fix will be included in future versions of {\tt HERWIG}. Also, the {\tt PYTHIA} version
was upgraded to {\tt PYTHIA 6.208} and the the implementation of storing/reading back 
of hard process events according to the Les Hauches standard was simplified with respect
to {\bf AcerMC 1.1} (see Section \ref{s:leshdump}). Also, for the convenience of the users
the {\tt Pythia} code was modified so that the top decay products from {\bf AcerMC} 
processes are now stored in the history part of the event record (status code 21) and
have the correct pointers to the top quark they originate from. This feature will 
be added to the future versions of {\tt PYTHIA}.  
\item  {\bf AcerMC version 1.3 (10. 2. 2003)}: A transition to the {\tt HERWIG 6.5} was made, which now 
supports the Les Houches standard for handling the external processes. Consequently,
the {\bf AcerMC} interface to {\tt HERWIG} was completely rewritten (c.f. Section 
\ref{s:hwintf}). As a direct consequence, the same event record can freely be 
swapped (i.e. read back) to either {\tt PYTHIA} or {\tt HERWIG} for fragmentation and 
hadronisation treatment. In addition, the {\tt PYTHIA} version was upgraded to 
{\tt PYTHIA 6.214}. Also, the build procedure of the libraries and executables was
greatly simplified (see Section \ref{s:insprod}).
\item  {\bf AcerMC version 1.4 (10. 5. 2003)}: The interfaces to external {\tt TAUOLA} and {\tt PHOTOS}
libraries were added. The necessary modifications in the interface routines and the
native {\tt PHOTOS} code were made to enable the user to process the events with  
{\tt TAUOLA} and/or {\tt PHOTOS} using {\tt PYTHIA} or {\tt HERWIG} as the supervising 
generators.
\item  {\bf AcerMC version 2.0 (25. 5. 2004)}: 
{\bf New algorithm for phase space generation implemented and optimised.}\\
New processes were added: $ q \bar q  \to (Z/W/\gamma^* \to) t \bar t b \bar b$; 
$gg, q \bar q  \to t \bar t t \bar t$;  $gg, q \bar q  \to   (t \bar t \to) f \bar f b f \bar f \bar b$; 
$gg, q \bar q  \to t \bar t$ and  $gg \to (t \bar t \to )Wb  W \bar b$. 
New control channel added:  $gg \to (t \bar t \to )Wb  W \bar b$.
\item  {\bf AcerMC version 2.1 (23. 6. 2004)}: Interface to {\tt ARIADNE} implemented (current version 4.12). 
\item  {\bf AcerMC version 2.2 (27. 9. 2004)}: Certain minor bug fixes implemented.
\item  {\bf AcerMC version 2.3 (31. 10. 2004)}: Another switch for fully leptonic boson pair decay (ACSET(13)=17) added.
\item  {\bf AcerMC version 2.4 (21. 3. 2005)}: Interface to  {\tt LHAPDF/LHAGLUE} implemented (current version 3). 
\item  {\bf AcerMC version 3.1 (18. 2. 2006)}: Interface to  {\tt LHAPDF/LHAGLUE} implemented (current version 4.2), added single top and
$Z^0 + b$ production processes as well as the $Z^{0}` \to t \bar t$ process. Interfaced to {\tt PYTHIA 6.3xx}. 
\item  {\bf AcerMC version 3.2 (23. 6. 2006)}: New Z-prime coupling options and mass choice of 0.5 TeV added 
as well as the new tt~ combined process code and branching option for inclusive semi-leptonic and leptonic mode. 
\item  {\bf AcerMC version 3.3 (20. 7. 2006)}: General code cleaning and minor bug fixes.
\item  {\bf AcerMC version 3.4 (11. 9. 2006)}: Added the combined processes 24 ( 5+6 ) and 25 (7+8) for convenience.
\item {\bf AcerMC version 3.5 (15. 4. 2008)}: Added massive corrections to the
  splitting kernels in ME+PS matching. Also, the settings for any
  possible $V-A$ and $V+A$ mixture in top quark pair decays. One can
  separately define the couplings for hadronically and leptonically
  decaying top quarks. \\
  A new PS+ME matched process $b \bar b \oplus b g \oplus g g
    \to Z/\gamma^*(\to f \bar f) \oplus  b \oplus \bar b$ is added.\\
  Latest versions of  {\tt PYTHIA 6.416} and  {\tt PYTHIA 6.510} used.
\item {\bf AcerMC version 3.6 (14. 12. 2008)}: A bug fix for top pair processes is made because the angular distributions in W+ decays for quark initial state (process 12) were reversed. The bug was introduced in AcerMC 3.2.
\item {\bf AcerMC version 3.7 (22.6. 2009)}: An improvement for top pair processes is made because due to numerical accuracy the branching ratios in top pair decays were off by a few percent when using the non-full-hadronic branching mode (ACSET13=6) which is now corrected.
\item {\bf AcerMC version 3.8 (2. 5. 2011)}: New ATLAS default parameters (top quark mass etc.) are added as the configuration options.
\end{itemize}

\newpage
\boldmath
\section{Introduction}
\unboldmath

Despite a large repertoire of processes implemented for generation in the
universal generators like {\tt PYTHIA} \cite{Pythia62} or {\tt HERWIG}
\cite{Herwig6.3} a number of Standard Model background processes for studying
expected physics potential of the LHC experiments were found missing at the start of the AcerMC project.  
For some of theseprocesses the matrix element expressions are rather lengthy and/or to achieve a reasonable
generation efficiency it is necessary to tailor the phase-space selection procedure to the
dynamics of the process. In the last years huge progress was made in developing automated Monte--Carlo systems
generating the matrix elements and phase space sampling on-the-fly, such as Sherpa \cite{sherpa} or Madgraph5 \cite{madgraph5} and also including next-to-leading order (real and virtual) corrections, e.g. MCFM \cite{mcfm},  MC@NLO \cite{mcatnlo} and Powheg \cite{powheg} and even combining the two features, such as aMC@NLO \cite{amcatnlo}, Powheg-Box \cite{powhegbox} and Sherpa implementations \cite{sherpanlo}. Nevertheless, in complex automated setups it is sometimes hard to achieve optimal phase space sampling and the user interfaces are necessarily generic, thus for now dedicated matrix-element-based generators like AcerMC, with standardised interfaces (defined e.g. in \cite{Boo01}), still can play a visible role.

The {\bf AcerMC} Monte Carlo Event Generator follows up on this idea.
It is dedicated for the simulation of the specific Standard Model
background and other processes at LHC collisions: the $gg, q \bar q \to t \bar t b \bar b$, 
$q \bar q \to W (\to \ell \nu) b \bar b$;
$q \bar q \to W (\to \ell \nu) t \bar t$;
$gg, q \bar q \to Z/\gamma^*(\to \ell \ell) b \bar b$;
$gg, q \bar q \to Z/\gamma^*(\to \ell \ell, \nu \nu, b \bar b) t \bar t$;
complete EW $gg, q \bar q \to (Z/W/\gamma^* \to) t \bar t b \bar b$;
4 top-quark production  $gg, q \bar q \to t \bar t t \bar t$; extended treatement
of the 2 top-quark production  $gg, q \bar q  \to  ( t \bar t \to)f \bar f b f \bar f \bar b $  and  
$ gg, q \bar q  \to  (W W b b~ \to) f \bar f f \bar f b \bar b$.
They are characterised by the
presence of the heavy flavour jets and multiple isolated leptons in
the final state. For the Higgs boson searches,  the $t \bar t H$, $ZH, WH$ 
with $H \to b \bar b$, the $gg \to H$ with $H\to ZZ^* \to 4 \ell$,
the $b\bar b h/H/A$ with $h/H/A \to \tau \tau, \mu \mu $ are the most obvious
examples of signals where the implemented processes would contribute
to the dominant irreducible backgrounds. The same background processes
should also be considered for e.g. estimating the observability of SUSY events
with a signature of multi-b-jet and multi-lepton production.

The program itself provides library of the massive matrix elements 
and phase space modules for the generation of the implemented processes.
The hard process event, generated with these modules, can be completed by the 
initial and final state radiation, hadronisation and decays, simulated with either 
{\tt PYTHIA~6.4}, {\tt ARIADNE 4.1} \cite{ariadne41} or {\tt HERWIG 6.5} Monte Carlo Event Generators. 
These will subsequently be called the
{\it Supervising Generators}. Interfaces of {\bf AcerMC} to
 {\tt PYTHIA~6.4}, {\tt ARIADNE 4.1}  and {\tt HERWIG 6.5}
generators, are provided in the distribution version. Provided is also the interface 
to {\tt TAUOLA} \cite{TAUOLA} and {\tt PHOTOS} \cite{PHOTOS} packages, 
for the more correct treatement of the 
$\tau$-lepton decays and photon radiation, than what available in the 
{\it Supervising Generators}.
The {\bf AcerMC} also uses
several other external libraries: {\tt CERNLIB}, {\tt HELAS} \cite{HELAS}, 
{\tt VEGAS} \cite{vegas}.
The matrix element codes have been derived with the help of {\tt MADGRAPH} 
\cite{Madgraph} package.
The achieved typical efficiency for the generation of unweighted events 
is of {\bf 20\%~-~30\%}, rather high given a complicated topology of the 
implemented processes.
 
This paper superseeds the first version of the manual, published in  \cite{AcerMC-CPC}. 
The outline of this paper is as follows. In Section 3, we describe physics
motivation for implementing each of the above processes and we collect some
numerical results (plots, tables) which can be used as benchmarks.  In Section 4
we describe the overall Monte Carlo algorithm. Section 5 gives details on the
structure of the program.  Section 6 collects information on how to use this
program and existing interfaces to {\tt PYTHIA~6.4}, {\tt ARIADNE 4.1} and {\tt HERWIG
6.5}, {\tt TAUOLA} and {\tt PHOTOS}. 
Summary, Section 7, closes the paper. Appendix A documents sets of Feynman
diagrams used for calculation of the matrix element for each subprocess,
Appendices B and C give examples of the input/output of the program.


\boldmath
\section{Physics content}
\unboldmath

The physics programme of the general purpose LHC experiments, ATLAS \cite{ATL-PHYS-TDR}
and CMS \cite{CMS-Documents}, focuses on the searches for the {\it New Physics} with the
distinctive signatures indicating production of the Higgs boson, SUSY particles, exotic
particles, etc.  The expected environment will in most cases be very difficult, with the
signal to background ratio being quite low, on the level of a few
percent after final selection in the signal window \cite{cscbook}.

Efficient and reliable Monte Carlo generators, which allow one to understand and predict
background contributions, are becoming the key point to the discovery. As the
cross-section for signal events is rather low, even rare Standard Model processes might
become the overwhelming background in such searches. In several cases, generation of such
	a process is not implemented in the general purpose Monte Carlo generators, when the
complicated phase space behaviour requires dedicated (and often rather complex)
	pre-sampling, whilst the general purpose Monte Carlo generators due to a large number of
	implemented processes tend to use simpler (albeit more generic) phase space sampling
	algorithms.  In addition, the matrix element for these processes is often lengthy and thus
	requiring complicated calculations.  Only recently, with the appearance of modern
	techniques for automatic computations, their availability {\it on demand} became feasible
	for the tree-type processes (and more recently even for next-to-leading order corrections, see e.g. \cite{amcatnlo}).
         With the computation power becoming more and more easily
	available, even very complicated formulas can now be calculated within a reasonable time
	frame.

	The physics processes implemented in {\bf  AcerMC} library represent such a set of cases.
	They are all being key background processes for the  discovery in the channels
	characterised by the presence of the heavy flavour jets and/or 
	multiple isolated leptons. For the Higgs boson searches,  the $t \bar t H$, $ZH, WH$ 
	with $H \to b \bar b$, the $gg \to H$ with $H\to ZZ^* \to 4 \ell$,
	the $b\bar b h/H/A$ with $h/H/A \to \tau \tau, \mu \mu $ are the most obvious
	examples of such channels. 

	It is not always the case that the matrix element calculations in the lowest
	order for a given topology represent the total expected background of a given
	type. This particularly concerns the heavy flavour content of the event.  The
	heavy flavour in a given event might occur in the hard process of a much simpler
	topology, as the effect of including higher order QCD corrections (eg. in the
			shower mechanism). This is the case for the b-quarks present in the inclusive
	Z-boson or W-boson production, which has a total cross-section orders of
	magnitude higher than the discussed matrix-element-based $Wb \bar b$ or $Zb \bar
	b$ production.  Nevertheless, the matrix-element-based calculation is a very
	good reference point to compare with parton shower approaches in different
	fragmentation/hadronisation models.  It also helps to study matching
	procedures between calculations in a fixed $\alpha_{QCD}$ order and parton
	shower approaches.  For very exclusive hard topologies matrix-element-based
	calculations represent a much more conservative approximation than the parton
	shower ones
	\cite{ATLCOMP032}.

	Let us shortly discuss the motivation for these few Standard Model background 
	processes which are implemented in the {\bf AcerMC} library. 

	\boldmath {\bf The $t \bar t b \bar b$ production  } \unboldmath
	at LHC is a dominant irreducible background for the Standard Model (SM) and
	Minimal Supersymmetric Standard Model (MSSM) Higgs boson search in the
	associated production, $ t \bar t H$, followed by the decay $H \to b \bar b$.
	The potential for the observability of this channel has been carefully studied
	and documented in \cite{ATL-PHYS-TDR} and \cite{ATL-PHYS-98-132}.  Proposed
	analysis requires identifying four b-jets, reconstruction of both top-quarks in
	the hadronic and leptonic mode and visibility of the peak in the invariant mass
	distribution of the remaining b-jets.  The irreducible $t \bar t b \bar b$
	background contributes about 60-70\% of the total background from the $t \bar t$
	events ($t \bar t b \bar b$, $t \bar t b j$, $t \bar t j j$).

	\boldmath {\bf The $W b \bar b$ production} \unboldmath at LHC is
recognised as a substantial irreducible background for the Standard Model (SM)
	and Minimal Supersymmetric Standard Model (MSSM) Higgs boson search in the
	associated production, $WH$, followed by the decay $H \to b \bar b$. The {\bf AcerMC}
	library discussed here includes even more efficient implementation of the
	algorithm presented in \cite{ATLCOMP013}.

	\boldmath {\bf The $W t \bar t$ production } \unboldmath 
	at LHC has to our knowledge been the first implementation in the publicly available code\footnote{ We thank M. L. Mangano for bringing this process to our
		attention and for providing benchmark numbers for verifying the total
			cross-section.}. It is of interest because it contributes an overwhelming
			background \cite{ATL-PHYS-2002-029} for the measurement of the Standard Model Higgs
			self-couplings at LHC in the most promising channel $pp \to HH \to WWWW$. More recently other implementations at next-to-leading order are now also available (see Ref. \cite{amcatnlo}).

			\boldmath {\bf The  $Z/\gamma^*(\to f \bar{f})  b \bar b$ production} \unboldmath at LHC 
			has since several years been recognised as one of the most substantial
			irreducible (or reducible) backgrounds for the several Standard Model (SM) and
			Minimal Supersymmetric Standard Model (MSSM) Higgs boson decay modes as well as
			for observability of the SUSY particles.  There is a rather wide spectrum of
{\it regions of interest} for this background.  In all cases the leptonic
$Z/\gamma^*$ decay is asked for, but events with di-lepton invariant mass around
the mass of the Z-boson mass or with the masses above or below the resonance
peak could be of interest.  The presented process enters an analysis either by
the accompanying b-quarks being tagged as b-jets, or by the presence of leptons
from the b-quark semi-leptonic decays in these events, in both cases thus
contributing to the respective backgrounds.

Good understanding of this background, and having a credible Monte Carlo
generator available, which allows studying of expected acceptances for different final states
topologies, is crucial for several analyses at LHC. 
The {\bf AcerMC} library discussed here includes  more efficient implementation of
the algorithm presented in \cite{ATLCOMP014}.

The new $\mathbf{ b \bar b \oplus b g \oplus g g \to Z/\gamma^*(\to f \bar f) \oplus  b
	\oplus \bar b}$ implementation takes the advantage of the developed
	parton shower and matrix element massive matching technique as
	described in \cite{acot_zbb} and aims to give an improved description
	covering the full phase space. 

	\boldmath {\bf The  $Z/\gamma^*(\to f \bar{f}, \nu \nu, b \bar b)  t \bar t$ production}
	\unboldmath at LHC is an irreducible background to the Higgs search in the
	invisible decay mode (case of $Z \to \nu \nu)$ in the production with association
	to the top-quark pair \cite{Gunion94}. With the  $Z/\gamma^*(\to b \bar b)$ it is also
	an irreducible resonant background to the Higgs search in the $t \bar t H$  production channel
	but with the Higgs boson decaying to the b-quark pair \cite{ATL-PHYS-98-132}. 

	The complete {\bf EW production}  of the
	\boldmath {\bf $gg, q \bar q \to (Z/W/\gamma^* \to)  b \bar b t \bar t$} \unboldmath  final
	state is also provided. It can be considered as a benchmark for the previous
	process, where only the diagrams with resonant $gg, q \bar q \to (Z/\gamma^* \to) b \bar b
	t \bar t$ are included. It thus allows the verification of the question, whether
	the EW resonant contribution is sufficient in case of studying the $t \bar t b
	\bar b$ background away from the Z-boson peak, like for the $t \bar t H$ with
	Higgs-boson mass of 120~GeV.

	\boldmath {\bf The  $gg, q \bar q \to   t \bar t t \bar t$ production} \unboldmath, interesting process per se,
	is a background to the possible Higgs self-coupling measurement in the $gg \to HH \to WWWW$ decay,
	\cite{ATL-PHYS-2002-029}.

	\boldmath {\bf The $ gg, q \bar q  \to  (W W b b~ \to) f \bar f f \bar f b \bar b$ and  
		$gg, q \bar q \to ( t \bar t \to) f \bar f b f \bar f \bar b $ }\unboldmath
		processes give posiblity to study spin correlations in the top-quark pair
		production and decays as well as the effect from the off-shell
		production. Those are important for the selection optimisation eg. in the $gg
		\to H \to WW$ channel, see the discussion in \cite{Krauer2002}

		\boldmath {\bf $ b b \oplus b g \to Z^0 \oplus b \to f \bar f \oplus b$} associated $Z*0$ and b-quark production at
		the LHC, important for e.g. b-quark PDF determination and background to Higgs searches. 

		\boldmath {\bf The single top processes $ g b \to t W \to b f \bar{f} f \bar f$, $ q q \to t b  \to b f \bar{f} b $ and 
$ q b \oplus q g \to q t \oplus b \to  q  b f \bar{f} \oplus b $} which are of relevance for single top production
searches at the LHC and top quark polarisation studies. 

\boldmath {\bf $ q q \to Z^{0\prime} \to t \bar t \to b \bar b f \bar f f \bar f$} as the channel for new boson searches at
the LHC including full spin correlations between the decay products.

A set of {\bf control channels}, i.e. the \boldmath {\bf  $q \bar q \to Z/\gamma^* \to f \bar{f}$,
$gg, q \bar q \to t \bar t$} \unboldmath, \boldmath {\bf $q \bar q \to W \to  f \bar{f}$} and
$gg  \to (t \bar t \to) W b W \bar b$ \unboldmath processes, have been added
to {\bf AcerMC} in order to provide a means of consistency and cross-check studies.

This completes the list of the native {\bf AcerMC} processes implemented so far.
Having all these different production processes implemented in the consistent framework, 
which can also be directly used for generating standard subprocesses implemented in either
{\tt PYTHIA} or {\tt HERWIG} Monte Carlo, represents a very convenient environment 
for several phenomenological studies dedicated to the LHC physics.

For the cases, where radiative photon emission from final state leptons is important 
the package {\tt PHOTOS} \cite{PHOTOS} can be used in the chain of event generation.
In similar way also package {\tt TAUOLA} \cite{TAUOLA} can be interfaced directly to
the generation chain and used for events generation in cases where
more detailed treatment of the tau-lepton decay and including spin
correlations effects is relevant.

At this point it also needs to be acknowledged that several other recent Monte--Carlo tools provide implementations 
which in some cases surpass the complexity and accuracy of the processes as implemented in AcerMC, by including also next-to-leading order corrections which are
not present in AcerMC, like for example the $W t \bar t$ process in aMC@NLO \cite{amcatnlo} or choose a somewhat different approach to real QCD next-to-leading order corrections in processes like single top production, as for example MC@NLO \cite{mcatnlo} or Powheg \cite{powheg}. 

In the following subsections we discuss in more detail implementation of each
subprocess. We also give benchmark Tables with the total
cross-sections obtained with {\tt AcerMC} processes but  different 
implementations and setting of 
$\alpha_{\rm QCD}(Q_{QCD})$: the native {\tt AcerMC}, {\tt PYTHIA} and
{\tt HERWIG} ones. For a more detailed discussion on this
topic the reader is referred to Section \ref{s:alphas}. If the  native
{\bf AcerMC} definition is used, the same cross-section is obtained
either  with {\tt PYTHIA} or {\tt HERWIG} generation chains.

\clearpage
\newpage

\begin{table}[h]
\newcommand{\lstrut}{{$\strut\atop\strut$}}
  \isucaption {\em All AcerMC processes implemented so far with the corresponding process code. 
\label{T:process}}
\vspace{2mm}
\begin{center}
\begin{tabular}{|c|c|} \hline \hline
Process   & Description  \\
\hline \hline
 [1] & $gg \to t \bar t b \bar b$           \\
\hline 
 [2] & $q \bar q \to t \bar t b \bar b$           \\
\hline 
 [3] & $q \bar q \to W(\to f \bar f ) b \bar b$   \\
\hline 
 [4] &$q \bar q \to W(\to f \bar f ) t \bar t$   \\
\hline 
 [5] &$gg \to Z/\gamma^*(\to f \bar f) b \bar b$ \\
\hline 
 [6] &$q \bar q \to Z/\gamma^*(\to f \bar f ) b \bar b$ \\
\hline 
 [7] &$gg \to Z/\gamma^*(\to f \bar f, \nu \nu) t \bar t$ \\
\hline 
 [8] &$q \bar q \to Z/\gamma^*(\to f \bar f, \nu \nu) t \bar t$ \\
\hline
 [9] & $gg  \to (Z/W/\gamma^* \to) t \bar t b \bar b $ \\
\hline 
 [10] & $q \bar q  \to (Z/W/\gamma^* \to) t \bar t b \bar b $ \\
\hline
 [11] & $gg  \to (t \bar t \to) f \bar f b f \bar f b$ \\
\hline
 [12] & $q \bar q \to (t \bar t \to) f \bar f b f \bar f b$ \\
\hline
 [13] & $gg  \to (WW b \bar b \to) f \bar f  f \bar f b \bar b$ \\
\hline
 [14] & $q \bar q \to (WW b \bar b \to) f \bar f bf \bar f b \bar b$ \\
\hline 
 [15] & $gg  \to t \bar t t \bar t $ \\
\hline
 [16] & $q \bar q \to t \bar t t \bar t$ \\
\hline
 [17] & $ q b \oplus q g   \to q t \oplus b \to  q  b f \bar{f} \oplus b $ (100+101) \\
\hline
 [18] & $ b b \oplus b g \to Z^0 \oplus b \to f \bar f \oplus b$ (96+97) \\
\hline
 [19] & $ q q \to t b  \to b f \bar{f} b $ \\
\hline
 [20] & $gb \oplus gg  \to (WW b \bar \oplus b \to) f \bar f  f \bar f \bar b \oplus b$ (13+105)\\
\hline
 [21] & $ g b \to t W \to b f \bar{f} f \bar f$ \\
\hline
 [22] & $ q \bar q \to Z^{0\prime} \to t \bar t \to b \bar b f \bar f f \bar f$ \\
\hline
 [23] & $gg,q \bar q  \to (t \bar t \to) f \bar f b f \bar f b$ (11+12) \\
\hline 
 [24] &$gg,q \bar q \to Z/\gamma^*(\to f \bar f) b \bar b$ (5+6)\\
\hline 
 [25] &$gg,q \bar q \to Z/\gamma^*(\to f \bar f) t \bar t$ (7+8) \\
\hline 
 [26] &$ b \bar b \oplus b g \oplus g g \to Z/\gamma^*(\to f \bar f) \oplus  b
 \oplus \bar b$  (5+96+97)\\
\hline 
 [27] &$gg,q \bar q \to Z/\gamma^*(\to f \bar f) b \bar b$ (26+6) \\
\hline \hline
     &   Control processes  \\
\hline 
 [91] &$q \bar q \to Z/\gamma^* \to f \bar f $ \\
\hline 
 [92] & $gg \to t \bar t$         \\
\hline 
 [93] & $q \bar q \to t \bar t$           \\
\hline 
 [94] &$q \bar q \to W \to f \bar f $   \\
\hline 
 [95] & $gg \to (t \bar t \to) Wb W \bar b$         \\
\hline
 [96] &  $ b b \to Z^0  \to f \bar f $ \\
\hline
 [97] &  $ b g \to Z^0 b \to f \bar f b$ \\
\hline
 [98] & $q b   \to q t $ \\
\hline
 [99] & $q g   \to q t b $ \\
\hline
 [100] & $ q b \to q t  \to  q  b f \bar{f}  $ \\
\hline
 [101] & $ q g   \to q t b \to  q  b f \bar{f} b $ \\
\hline
 [102] & $ q b \to q t  \to  q  b W  $ \\
\hline
 [103] &  $qb \oplus qg  \to q t \oplus b$  (98+99) \\
\hline
 [104] &  $ g b \to t W \to t f \bar{f} $ \\
\hline
 [105] & $ g b \to t W \to b f \bar{f} f \bar f$ (equal to 21)  \\
\hline
 [106] & $ g g \to ( t W b  \to)  t f \bar{f} b$  \\
\hline
 [107] &  $ g g \to ( t W b  \to) f \bar f  f \bar f \bar b \oplus b $ \\
\hline \hline
\end{tabular}
\end{center}
\end{table}

\clearpage

\boldmath
\subsection{The  $g g, q \bar q \to t \bar t b \bar b$ processes}
\unboldmath

In the implementation discussed here, the matrix element was derived using the
{\tt MADGRAPH} package \cite{Madgraph}. These matrix elements are not covering
the decay of the top-quarks, the latter are considered as massive final states
of the process. The top-quark decays is than performed by the supervising
generator.  Rather important spin effects (spin correlations) in the top decays
are therefore not yet included. The similar solution, like for tau decay in
the Z-boson production process discussed in \cite{Acta2001}, is planned to be
implemented here in the near future.

As a benchmark, the processes $g g, q \bar q \to t \bar t b \bar b$ have been
simulated for pp collisions with 14~TeV centre-of-mass energy and CTEQ5L \cite{cteq5l}
parton density functions, using event generation with massive $2 \to 4$ matrix
element implemented as an external process to {\tt PYTHIA 6.2} (see Section 4
and 5).  The decays of the top-quarks have been left under control of {\tt
PYTHIA 6.2} generator.  The $q \bar q \to t \bar t b \bar b$ subprocess
contributes less than 10\% of the total cross-section.

The total cross-section is very sensitive to the choice of the QCD energy scale
used for calculation of that process, thus indicating potentially large
contributions from higher order corrections. The same definition for the
factorisation and renormalisation scale is used. The example values of the total
cross-section for implemented choices of the QCD energy scale are given in
Table~\ref{T1:1}\footnote{Numbers obtained with {\tt HERWIG} generator (third collumn) are slightly different 
than what published in \cite{AcerMC-CPC}. This is related to the 
internal changes in {\tt HERWIG} between version 6.3 and 6.5.}.

As a cross-check, the processes $gg, q \bar q \to t \bar t b \bar b$ have been
coded independently using the {\tt COMPHEP} package \cite{COMPHEP}.  The same
set of diagrams was selected and only the integrating part of the package was
used to calculate total cross-section.  The choices for the QCD energy scale
were kept consistent.  A very good agreement between the cross-sections obtained
with two independent calculation streams prepared for this study has been
achieved \cite{ATLCOMP025}.

\begin{table}
\newcommand{\lstrut}{{$\strut\atop\strut$}}
  \isucaption {\em {\bf AcerMC} cross-sections  for the $gg, q \bar q \to t \bar t b \bar b$
production at different choices of the QCD energy scale and $\alpha_{QCD}$ implementations. 
The 14 TeV centre-of-mass
energy and CTEQ5L parton density functions were used for the simulation with interfaces
to {\tt PYTHIA 6.2} and {\tt HERWIG 6.5}.  The $m_H~=~120$~GeV and $m_t~=~175$~GeV 
were used for calculating the $Q^2_{QCD}$ in the last row of this table. The default 
settings of $\alpha_{QCD}$ as implemented in {\bf AcerMC}, {\tt PYTHIA 6.2} and {\tt HERWIG 6.5}
were used. \label{T1:1}}\vspace{0.5cm}
\begin{center}
\scriptsize
\begin{tabular}{|c|c|c|c|} \hline \hline
Factorisation scale  &  $\alpha_{QCD}$ (1L)  &  $\alpha_{QCD}$ (1L) &  $\alpha_{QCD}$ (2L)  \\ 
       & native AcerMC & as in {\tt PYTHIA 6.2} & as in {\tt HERWIG 6.5}  \\
\hline \hline
\cline{2-4} \multicolumn{1}{|c|}{}& \multicolumn{3}{|c|}{$gg \to t \bar t b \bar b$} \\
\hline 
$Q^2_{QCD}~=~\hat{s}$ &  4.2 [pb]     &  3.9 [pb]     & 3.0 [pb]     \\
\hline
$Q^2_{QCD}~=~\sum{({p^i_T}^2 + m_i^2)}/4$ &  10.3 [pb]      & 10.2 [pb]    & 7.2 [pb]    \\
\hline
$Q^2_{QCD}~=~\sum{({p^i_T}^2)}/4$ &  17.0 [pb]      &  16.9 [pb]   & 11.5 [pb]    \\
\hline
$Q^2_{QCD}~=~(m_t + m_H/2)^2$ & 8.2  [pb]     & 8.1 [pb]   & 5.8 [pb]      \\
\hline \hline
\cline{2-3} \multicolumn{1}{|c|}{}& \multicolumn{3}{|c|}{$q \bar q \to t \bar t b \bar b$} \\
\hline 
$Q^2_{QCD}~=~\hat{s}$ &  0.30 [pb]    &  0.29 [pb]     & 0.22 [pb]    \\
\hline
$Q^2_{QCD}~=~\sum{({p^i_T}^2 + m_i^2)}/4$ & 0.61  [pb]   & 0.60 [pb]    &  0.43 [pb]    \\
\hline
$Q^2_{QCD}~=~\sum{({p^i_T}^2)}/4$ &  0.91 [pb]       &  0.90 [pb]     & 0.62 [pb]   \\
\hline
$Q^2_{QCD}~=~(m_t + m_H/2)^2$ & 0.52  [pb]     &  0.51  [pb]     & 0.37 [pb]     \\
\hline \hline 
\end{tabular}
\end{center}
\end{table}

One can observe a very strong scale dependence of the cross-section for the $gg, q \bar q
\to t \bar t b \bar b$ process (c.f. Table \ref{T1:1}).  Factor four (!!) can be expected
on the predicted cross-section when changing from the scale $Q^2_{QCD}~=~\hat{s}$ to the
scale $Q^2_{QCD}~=~<p_T^2>$.  This very strong dependence on the energy scale is also
observed in the case of the $t \bar t H$ production, for recent discussion see
\cite{NLOttH}. There, the recommended {\it central} factorisation and renormalisation
energy scale is $\mu_0~=~(m_t + m_H/2)$.  Having in mind that the primary interest of
evaluating this background is the Higgs search in the $t \bar t H$ production, i.e.  with
the b-quark system being produced with the invariant mass of the expected Higgs boson, we
have also introduced this {\it central} energy scale, with $m_H~=~120$~GeV as one of the
possible choices.

\begin{figure}[ht]
\vspace{-0.5cm}
\begin{center}
\mbox{
     \hspace{-0.2cm}
     \epsfxsize=5.0cm
     \epsffile{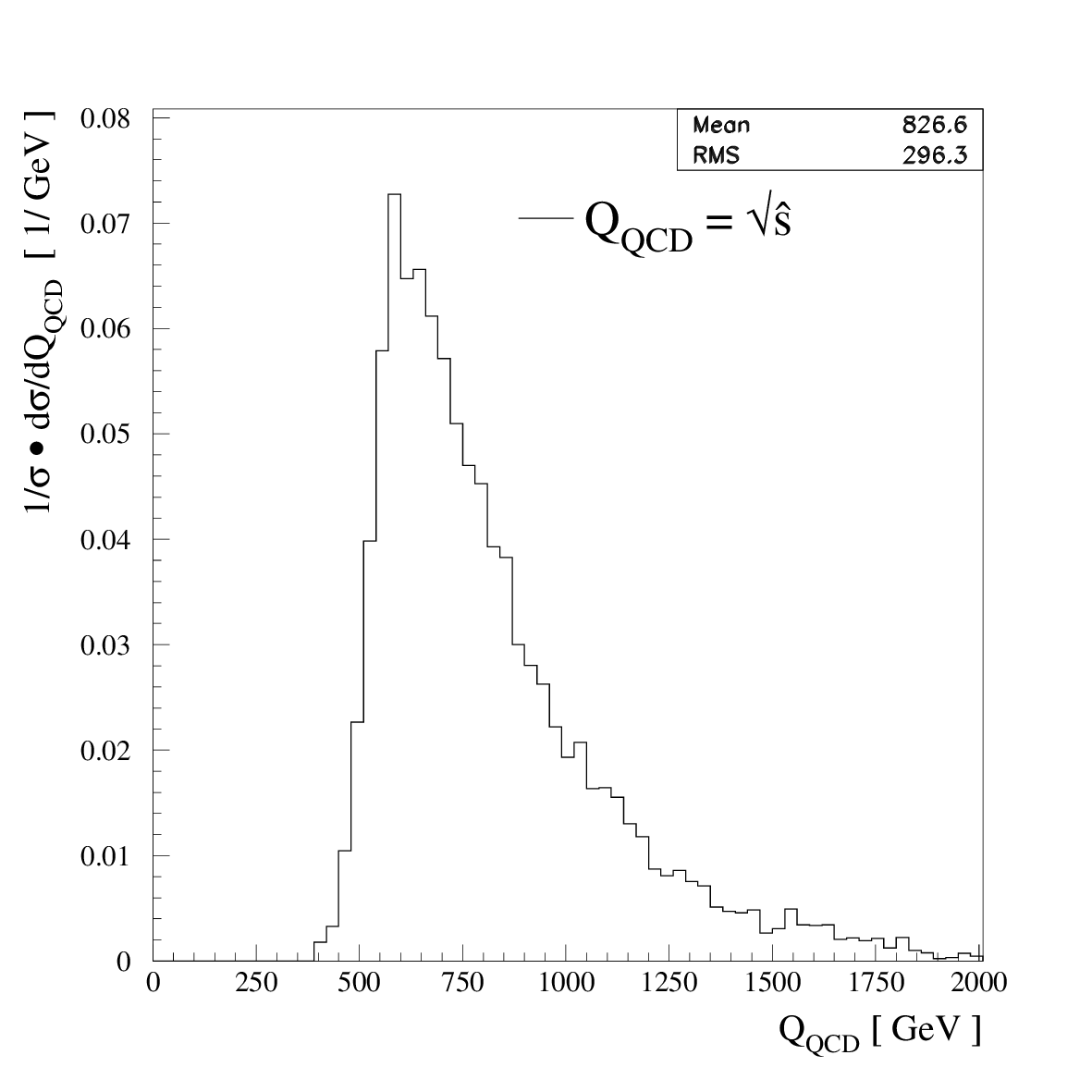} 
     \hspace{-0.7cm}
     \epsfxsize=5.0cm
     \epsffile{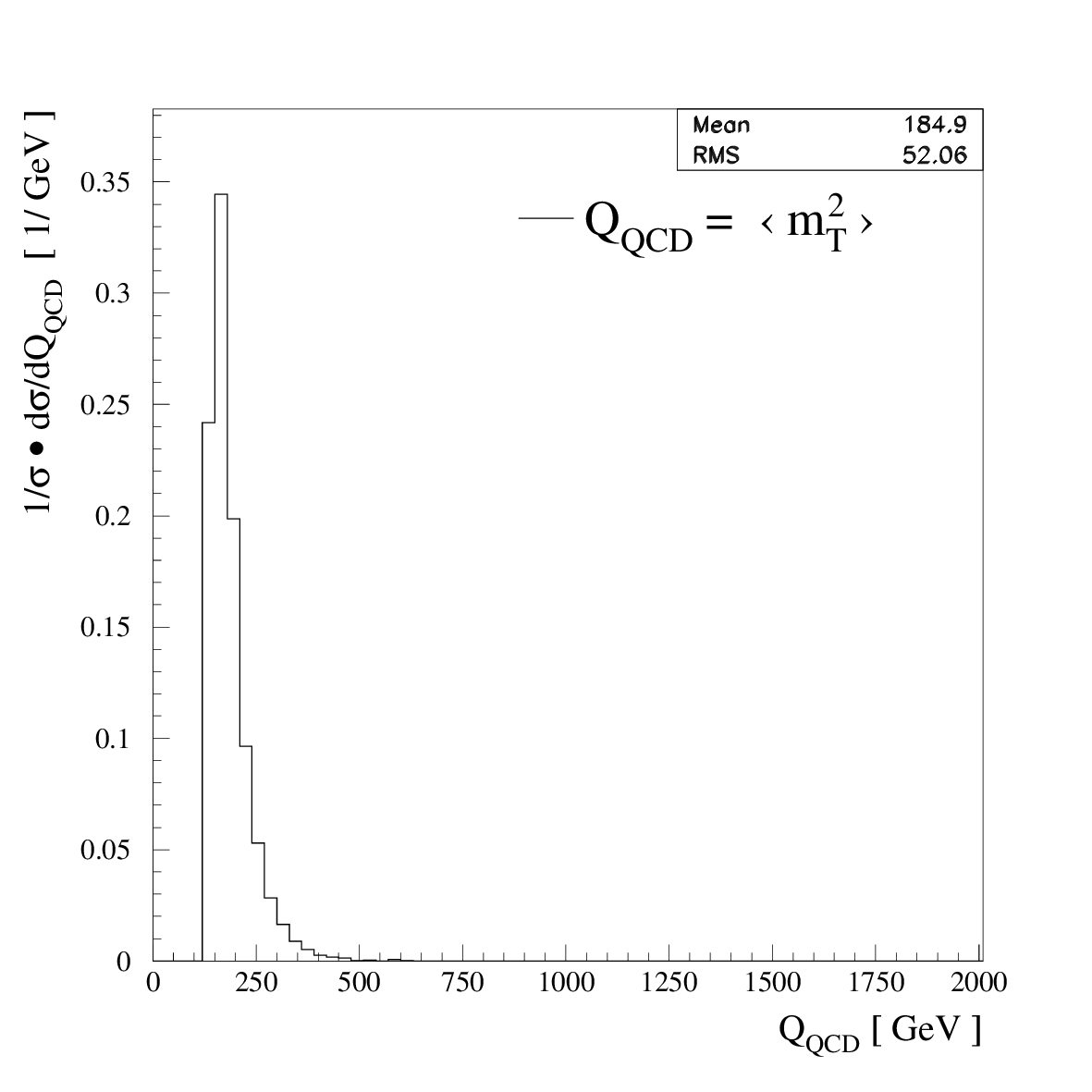}
     \hspace{-0.7cm}
     \epsfxsize=5.0cm
     \epsffile{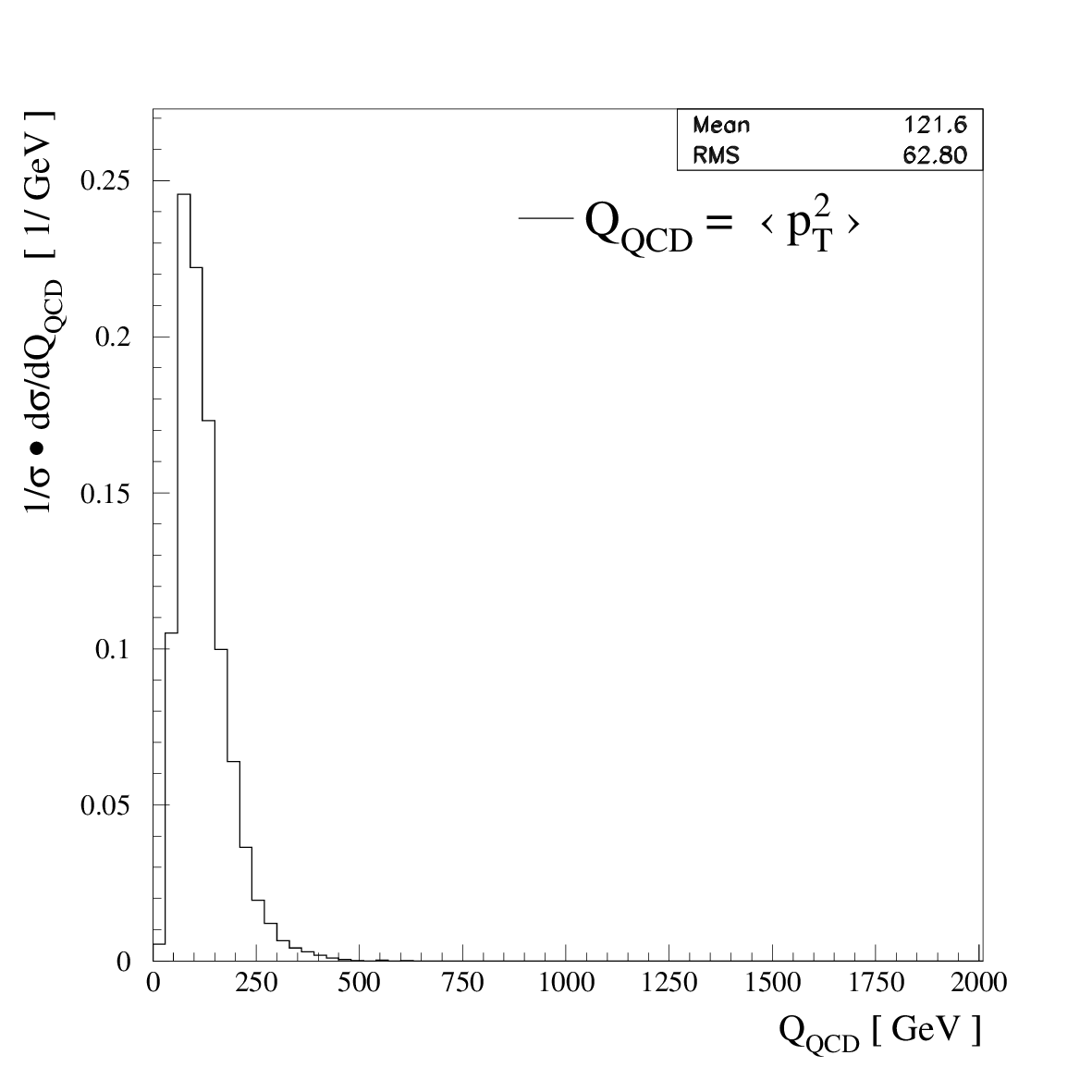}
}
\vglue -0.3cm
\mbox{
     \epsfxsize=5.0cm
     \epsffile{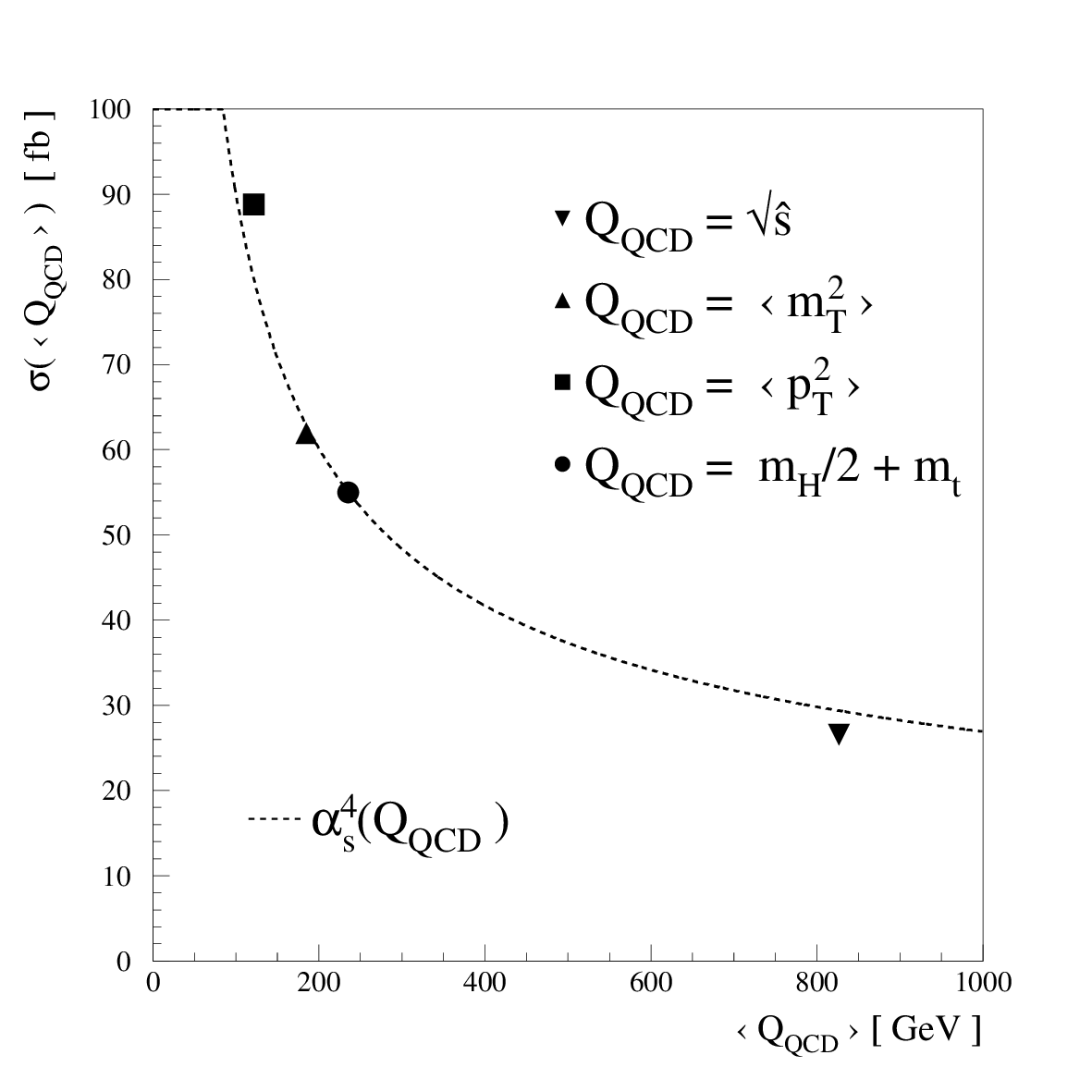}
}
\end{center}
\isucaption{\em
Top: the $Q^2_{QCD}$ distributions for $t \bar t b \bar b$ events with 
the invariant mass of the b-jets system $m_{bb-jets}=120~\pm~30$~GeV.
Bottom: the total cross-section of accepted
events as a function of the averaged $Q^2_{QCD}$ (for these events). 
\label{FS2a:a}}
\end{figure}

Fig.~\ref{FS2a:a} shows the distributions of the $Q_{QCD}=\sqrt{Q^2_{QCD}}$ (distributions
have been normalised to one) for the $t \bar t b \bar b$ events with the invariant mass
of the b-jets system, calculated using the default {\tt PYTHIA} (LO) $\alpha_{\rm QCD}$ 
implementation\footnote{ This would makes distributions directly relevant for the $t
\bar tH$ analysis. For details on the jet reconstruction see \cite{ATL-PHYS-98-131}.}.
$m_{bb-jets}~=~120~\pm~30$~GeV. The distribution is well collimated around the
average value when $Q^2_{QCD}$ is defined as $<m_T^2>$ or $<p_T^2>$ while it is
much broader when $Q^2_{QCD}$ is defined as $\hat{s}$. The kinematic
distributions are very similar in shape for separate $gg \to t \bar t b \bar b$
and $q \bar q \to t \bar t b \bar b$ contributions.  The total cross-section for
accepted events as a function of the averaged $Q^2_{QCD}$ (for these events) is
shown in the bottom plot. It can be noted that the cross-section decreases
rather fast with the  increasing value of the average $<Q^2_{QCD}>$.  Also shown is
the $\alpha_s^4(Q_{QCD})$ dependence scaled to match the cross-section at
$Q_{QCD}=(m_H/2 + m_t)$ with $ m_H=120\;$ GeV, it being the only calculated
cross-section point with a fixed scale. The other cross-sections are shown to
follow the expected $\alpha_s^4(Q_{QCD})$ dependence rather well, while the
deviations are induced by the parton density function dependence on the $Q^2_{QCD}$
scale, most notably at $Q^2_{QCD}=\hat{s}$ value. The deviations induced by
the parton density functions dependence on the $Q^2_{QCD}$ scale are different for the
$gg$ and $q \bar q$ contributions, as can be concluded from results given in
Table~\ref{T1:1}.

The series of plots illustrating the most relevant differential distributions for the
top-quarks and b-quarks can be found in \cite{ATLCOMP025}.


\boldmath
\subsection{The  $q \bar q \to W (\to  f \bar{f'}) g^*(\to b \bar b)$ 
process}
\unboldmath

The matrix element for the implemented process was  coded by using the {\tt
MADGRAPH} package \cite{Madgraph}.  This process is represented by only two
Feynman diagrams, with quark exchange in the t-channel, leading to the
production of the $W$-boson and virtual gluon splitting into $ b \bar b$
pair. Only the $u, d, s, c$ quarks were considered in this implementation, the
possibility of the b-quark in the initial state was omitted as expected to be
negligible numerically (e.g.. $|V_{bc}/V_{ud}|^2 \sim 0.002$) but leading to
several additional diagrams which would have to be included.  The massive matrix
element takes into account spin correlations in the W-boson decay and angular
correlations between leptons and quarks.  Due to the massive treatment of the
final state fermions the amplitude has no singularities; the total cross-section
is well defined.  The effect from the $W$-boson natural width and the $W$-boson
propagator are also properly included.

\vspace{0.2cm}
\begin{table}[hb]
\newcommand{\lstrut}{{$\strut\atop\strut$}}
  \isucaption {\em {\bf AcerMC} production cross-sections for the $q \bar q \to W b \bar b$ with
  $W \to e \nu~$ decay (single flavour). The 14 TeV centre-of-mass energy and
  CTEQ5L parton density functions were used with different definitions of
  $\alpha_{QED}$, $\alpha_{QCD}$ ( as in default {\tt PYTHIA 6.2} and {\tt
  HERWIG 6.5}) and several choices of the factorisation scale, $\alpha_{QED}$ and $\alpha_{QCD}$ 
  implementations.
\label{T2:1c}}
\vspace{0.5cm}
\begin{center}
\scriptsize
\begin{tabular}{|c|c|c|c|} \hline \hline
Factorisation scale  &  $\alpha_{QED}$, $\alpha_{QCD}$ (1L)  & 
       $\alpha_{QED}$, $\alpha_{QCD}$ (1L) &  $\alpha_{QED}$, $\alpha_{QCD}$ (2L)  \\ 
       & native AcerMC & as in {\tt PYTHIA 6.2} & as in {\tt HERWIG 6.5}  \\
\hline 
 $Q^2 = M_W^2$   &  36.5 [pb]   &  36.4 [pb]   & 30.6 [pb]  \\
\hline 
 $Q^2 = s^*_{b \bar b}$   &  44.1 [pb]   &  44.0 [pb]   & 36.0 [pb]  \\
\hline 
 $Q^2 = M_W^2+pT_W^2$   &  36.0 [pb]   &  36.0 [pb]   & 29.8 [pb]  \\
\hline 
 $Q^2 = (s^*_W + s^*_{b \bar b})/2 +pT_W^2$     &  37.2 [pb]   &  37.1 [pb]   & 30.4 [pb]  \\
\hline \hline
\end{tabular} 
\end{center}
\vspace{0.5cm}
\end{table}

As a benchmark, the process $ q \bar q \to W (\to \ell \nu) g^*(\to b \bar b)$ has
been simulated for pp
collision with 14~TeV centre-of-mass energy.  The total cross-section, including branching
ratio for $W \to \ell \nu$ (single flavour) is 36.5~pb (CTEQ5L parton density functions, $Q^2 =
M_W^2$, {\tt PYTHIA 6.2} interface)\footnote{This can be compared with the matrix element implementation to {\tt HERWIG 5.6},
used in \cite{ATL-PHYS-94-043},\cite{ActaB31}, where originally this cross-section was estimated
to 19.8~pb (CTEQ2L parton density functions) but, when implementing CTEQ5L parton density functions
and setting kinematic parameters to be in approximate accordance with {\tt PYTHIA} defaults, rises to
36.0~pb, which is consistent with the AcerMC implementation by taking into account the remaining 
differences in the two calculations (e.g. the former implementation uses an on-shell W boson in the ME
calculation).}.

The dependence on the choice of the factorisation scale is rather modest (c.f. Table
\ref{T2:1c}) and does not exceed 20\% for the choices implemented in {\bf AcerMC}
library. The variation of the cross-section due to different $\alpha_{QED}$ and
$\alpha_{QCD}$ implementations and default settings is again evident; as one can expect the two-loop 
$\alpha_{QCD}$ implementation given in {\tt HERWIG} gives a $\sim$20 \% lower cross-section when compared to
the cases when native {\bf AcerMC} and {\tt PYTHIA} one-loop $\alpha_{QCD}$ were used\footnote{While 
performing further comparisons of native {\bf AcerMC} and {\tt PYTHIA} processes we discovered a 
misinterpretation of our CKM matrix implementation. This correction efectively changes the cross-section for 
$q \bar q \to W b \bar b$ and $q \bar q \to W t \bar t$ processes by $\sim$10\% compared to the draft 
versions of this paper, which is nevertheless  still well within the physics precision of the program. The 
affected tables in this paper are already updated.}.

The differential distributions of the $q \bar q \to Wb \bar b$ events turn out to be interesting
when compared to the corresponding ones of the $q \bar q \to Z b \bar b$ and $g
g \to Z b \bar b$ events (generated with pure Z-boson exchange). Such comparison
is well documented in
\cite{ATLCOMP014}.

\boldmath
\subsection{The  $q \bar q \to W (\to  f \bar{f'}) g^*(\to t \bar t)$ 
process}
\unboldmath

The $2 \to 4$ matrix elements, coded by the {\tt MADGRAPH} package \cite{Madgraph}, are not covering
the decay of the top-quarks; the latter are considered as massive final states of the
process. The top decay is than performed by the supervising generator. As in the case of
$gg, q \bar q \to t \bar t b \bar b$ process spin effects in the top decays are therefore
not yet included. This process, although rare, contributes an overwhelming irreducible
background to possible measurement of the Higgs-boson self-coupling in the $HH \to WWWW$
decay mode \cite{ATL-PHYS-2002-029}.
\vspace{0.2cm} 
\begin{table}[htb]
\newcommand{\lstrut}{{$\strut\atop\strut$}}
  \isucaption {\em {\bf AcerMC} production cross-sections for the $q \bar q \to W t \bar t$ with primary
  $W \to e \nu~$ decay (single flavour). The 14 TeV centre-of-mass energy, CTEQ5L parton
  density functions with different factorisation scales and different definitions of the
  $\alpha_{QED}$ and $\alpha_{QCD}$ were used in the matrix element calculations.
\label{T2:1d}}
\vspace{2mm} 
\begin{center}
\scriptsize
\begin{tabular}{|c|c|c|c|} \hline \hline
Factorisation scale  &  $\alpha_{QED}$, $\alpha_{QCD}$ (1L)  & 
       $\alpha_{QED}$, $\alpha_{QCD}$ (1L) &  $\alpha_{QED}$, $\alpha_{QCD}$ (2L)  \\ 
       & native AcerMC & as in {\tt PYTHIA 6.2} & as in {\tt HERWIG 6.5}  \\
\hline 
$Q^2_{QCD}~=~ M_W^2$     &  69.3 [fb]   &  69.1 [fb]   &  57.4 [fb] \\
\hline
$Q^2_{QCD}~=~ s^*_{t \bar t}$ & 40.9 [fb]    & 39.9 [fb]   & 34.7  [fb]     \\
\hline
$Q^2_{QCD}~=~ M_W^2+pT_W^2$ & 59.7 [fb]    &  59.5  [fb]    &  49.6 [fb]   \\
\hline
$Q^2_{QCD}~=~(s^*_W + s^*_{t \bar t})/2 +pT_W^2$  & 43.7[fb]   &  42.8 [fb]    &  36.9 [fb]     \\
\hline \hline 
\end{tabular} 
\end{center}
\end{table}

Table~\ref{T2:1d} shows the expected {\bf AcerMC} cross-sections for different choices of the energy
scale and coupling ($\alpha_{QED}$, $\alpha_{QCD}$) definitions. One should notice the
effect of almost a factor two from different choices of the energy scale.

\newpage

\boldmath
\subsection{The  $gg, q \bar q \to  Z/\gamma^* (\to f \bar{f}) b \bar b$ 
processes}
\unboldmath 

The matrix elements, derived using the {\tt MADGRAPH} package \cite{Madgraph},
properly take into account spin correlations in the Z-boson decay and angular
correlations between leptons and quarks.  Thank to keeping non-zero b-quark
masses the amplitude has no singularities; the total cross-section is well
defined.

The full $Z/\gamma^*$ exchange proves to be important: For events well below the
Z-boson resonance the contribution from $\gamma^*$ becomes dominant; the
$\gamma^*$ contribution is also sizeable in the high mass tail and increases
proportionally with the effective mass of the di-lepton system.

As a benchmark result, the process has been simulated for pp collisions at
14~TeV centre-of-mass energy.  The total cross-sections, including the branching
ratio for $Z/\gamma^* \to f \bar{f}$ (single flavour) are given in
Table~\ref{T2:1a} for different definitions of  $\alpha_{QED}$, $\alpha_{QCD}$ couplings.

Several differential benchmark distributions for leptons and b-quarks
originating from the hard process has been collected and discussed in
\cite{ATLCOMP013}.

\vspace{0.2cm} 
\begin{table}[ht]
\newcommand{\lstrut}{{$\strut\atop\strut$}}
  \isucaption {\em {\bf AcerMC} production cross-sections for the $gg, q \bar q \to Z/\gamma^* b
  \bar b$ with $Z/\gamma^* \to e e$ decay (single flavour). The 14 TeV
  centre-of-mass energy, CTEQ5L parton density functions and different
  definitions for the $\alpha_{QED}$, $\alpha_{QCD}$ (as in default {\tt PYTHIA
  6.2} and {\tt HERWIG 6.5})  were used in the matrix element calculations.  The threshold
  $m_{f \bar{f}}~\geq~10$~GeV was used in the event generation.
\label{T2:1a}}
\vspace{0.2cm} 
\begin{center}
\scriptsize
\begin{tabular}{|c|c|c|c|} \hline \hline
Factorisation scale  &  $\alpha_{QED}$, $\alpha_{QCD} (1L)$ & 
 $\alpha_{QED}$, $\alpha_{QCD} (1L)$ &  $\alpha_{QED}$, $\alpha_{QCD} (2L)$  \\ 
 & native AcerMC     & as in {\tt PYTHIA 6.2} & as in {\tt HERWIG 6.5}  \\
\hline \hline 
\cline{2-3} \multicolumn{1}{|c|}{}& \multicolumn{3}{|c|}{$gg \to Z/\gamma^* b \bar b$} \\
\hline 
 $Q^2 = m_Z^2$                      &  49.5 [pb]   & 45.8  [pb]   & 38.0 [pb]  \\
\hline 
 $Q^2 = s^*_{b \bar b}$             &  53.8 [pb]   & 53.9  [pb]   & 44.0 [pb]  \\
\hline 
 $Q^2 = s^*_Z$                      &  54.7 [pb]   & 54.6  [pb]   & 44.4 [pb]  \\
\hline 
 $Q^2 = (pT_Z^2 + s^*_{b \bar b})/2$ &  49.7 [pb]   & 49.7  [pb]   & 40.5 [pb]  \\
\hline 
\cline{2-3} \multicolumn{1}{|c|}{}& \multicolumn{3}{|c|}{$q \bar q \to Z/\gamma^* b \bar b$} \\
\hline 
 $Q^2 = m_Z^2 $                     &  6.7 [pb]   &  6.7 [pb]  &  5.6 [pb]   \\
\hline 
 $Q^2 = s^*_{b \bar b}$             &  8.0 [pb]   & 8.0  [pb]   & 6.4 [pb]  \\
\hline 
 $Q^2 = s^*_Z$                      &  7.0  [pb]   & 7.0  [pb]   & 5.7 [pb]  \\
\hline 
 $Q^2 = (pT_Z^2 + s^*_{b \bar b})/2$ & 6.9  [pb]   & 6.9  [pb]   & 5.7 [pb]  \\
\hline \hline
\end{tabular}
\end{center}
\end{table}

The new 'heavy' associated Drell-Yan process is the combined production of $ b b \to
Z^0$ and $g b \to Z^0 b$ and the above $g g \to Z^0 b \bar{b}$
processes 
while removing the double counting between the initial state shower
(ISR) $ g \to b \bar b$ splitting and the higher-order $\alpha_S$
processes using a procedure described in \cite{acot} and specifically
in \cite{acot_zbb} for this process. The method incorporates part of
the NLO corrections to the process by removing the collinear singularities. The
process is important for e.g. b-quark PDF determination and background to Higgs
searches. The boson decays into any relevant final state (quarks or
leptons). Note that a fraction of events due to this procedure now has
negative weights \emph{equal to -1}, i.e. the events are weighted with
$\pm 1$ weights.

\begin{figure}[htb]
\begin{center}
     \includegraphics[width=4cm]{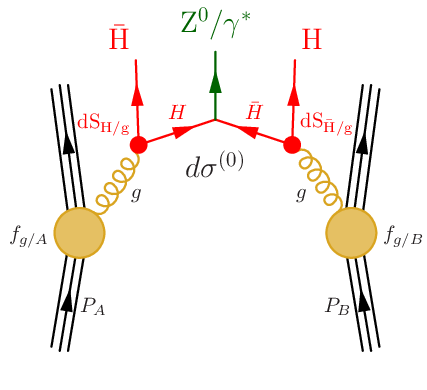}
     \includegraphics[width=4cm]{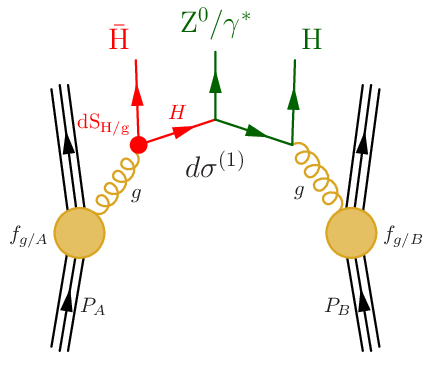}
     \includegraphics[width=4cm]{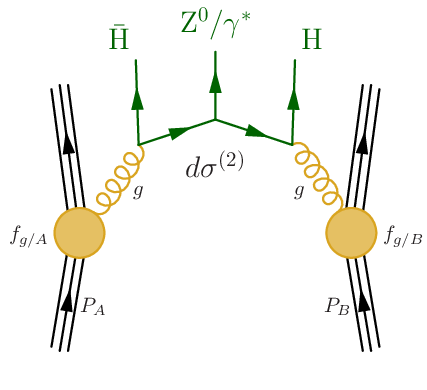}
\end{center}
\isucaption{\small Schematic representation of contributions resulting in
exclusive $Z^0 H \bar{H}$ final state:  two fully evolved heavy (H=b)
quarks entering `pure' Drell-Yan at order $\alpha_s^0$ in combination
with double initial state parton shower (left), one heavy quark and
one gluon entering the hard process at order $\alpha_s^1$ in combination with one parton
shower (middle) and fully perturbative calculation involving two
incoming gluons in a hard process of order $\alpha_s^2$ (right). These
three processes need to be combined with appropriate overlap removal
as detailed in the paper \cite{acot_zbb}. \label{f:zbbchoice}}
\vspace*{-0.2cm}
\end{figure}

The cross-section values are omitted, the user can quickly and easily extract them by running
the {\bf AcerMC} with a couple of thousand events and the relevant final state.

\boldmath
\subsection{The  $gg, q \bar q \to  Z/\gamma^* (\to f \bar{f}, \nu \nu, b \bar b) t \bar t$ 
processes}
\unboldmath 
 
This process, in spite of having a very small cross-section at LHC energies,
contributes as irreducible background to the $t \bar t H$ production at low
masses. In case the Higgs boson is searched within the $H \to b \bar b$ mode, this
contribution becomes less and less important with the Higgs boson mass moving
away from the Z-boson mass. In case of the Higgs-boson search in the invisible
decaying mode, the $Z \to \nu \nu$ might be more relevant also for the
higher masses, as the mass peak cannot be reconstructed for signal events.  The
$Z/\gamma^* \to \ell \ell$ decay is of less interest, as the expected observability
 at LHC is very low (Table \ref{T2:1b}). 
\vspace{-0.5cm}
\begin{table}[htb]
\newcommand{\lstrut}{{$\strut\atop\strut$}}
  \isucaption {\em {\bf AcerMC} production cross-sections for the $gg, q \bar q \to Z t
  \bar t$ with $Z \to \nu_e \nu_e $ decay (3 flavours). The 14 TeV
  centre-of-mass energy, CTEQ5L parton density functions and different
  definitions for the $\alpha_{QED}$, $\alpha_{QCD}$ (as in native {\bf AcerMC},
  default {\tt PYTHIA 6.2} and {\tt HERWIG 6.5}) were used.\vspace{0.5cm}
\label{T2:1b}}
\vspace{-0.3cm} 
\begin{center}
\scriptsize
\begin{tabular}{|c|c|c|c|} \hline \hline
Factorisation scale  &  $\alpha_{QED}$, $\alpha_{QCD}(1L)$ &  $\alpha_{QED}$, $\alpha_{QCD}(1L)$ &  $\alpha_{QED}$, $\alpha_{QCD}(2L)$  \\ 
   & native AcerMC    & as in {\tt PYTHIA 6.2} & as in {\tt HERWIG 6.5}  \\
\hline \hline
\cline{2-4} \multicolumn{1}{|c|}{}& \multicolumn{3}{|c|}{$gg \to Z (\to \nu_e \nu_e) t \bar t$} \\
\hline 
 $Q^2 = m_Z^2$                      &  126.0  [fb]  & 125.8  [fb]   & 104. [fb]     \\
\hline 
 $Q^2 = s^*_{t \bar t}$             &  61.9  [fb]  & 60.3  [fb]   & 52.4 [fb]  \\
\hline 
 $Q^2 = s^*_Z$                      &  126.2 [fb]   &  126.1 [fb]   & 105. [fb]  \\
\hline 
 $Q^2 = pT_Z^2 + s^*_{t \bar t})/2$ &  67.6 [fb]   &  66.3 [fb]   & 57.0 [fb]  \\
\hline \hline
\cline{2-4} \multicolumn{1}{|c|}{}& \multicolumn{3}{|c|}{$q \bar q \to Z (\to \nu_e \nu_e) t \bar t$} \\
\hline  
 $Q^2 = m_Z^2 $                     & 64.7 [fb]   &  64.6 [fb]   & 53.7 [fb]     \\
\hline 
 $Q^2 = s^*_{t \bar t}$             & 39.2  [fb]  & 38.2  [fb]   & 33.3 [fb]  \\
\hline 
 $Q^2 = s^*_Z$                      & 64.8  [fb]  & 64.7  [fb]   & 53.5 [fb]  \\
\hline 
 $Q^2 = (pT_Z^2 + s^*_{t \bar t})/2$ & 41.7  [fb]  & 41.0  [fb]   & 35.2 [fb]  \\
\hline \hline
\end{tabular}
\end{center}
\end{table}

\clearpage

\boldmath
\subsection{The electroweak  $gg, q \bar q \to  (Z/W/\gamma^* \to) b \bar b t \bar t$ 
process}
\unboldmath

One should be well aware, that the $gg, q \bar q \to  Z/\gamma^* t \bar t$ with 
$Z/\gamma^* \to b \bar b$ does not represent a complete electroweak production of the 
$t \bar t b \bar b$ final state. Consequently, a separate implementation for generation
of the complete set of such diagrams (including as well W-boson exchange) was addressed.
In fact this final state leads to complicated pattern of the 72 Feynman diagrams 
(in case of the gg initial state).

The contribution from all non-resonant channels is a dominant one for the
inclusive cross-section, see Table~\ref{T2:1e}.  An almost factor 10 higher
cross-section is calculated with the full electroweak $gg \to (Z/W/\gamma^* \to)
b \bar b t \bar t$ with respect to calculated with the $gg \to (Z/\gamma^* \to b \bar b) t \bar
t$ process only.  
One should also note that the electroweak $gg \to (Z/W/\gamma^* \to) b \bar b t
\bar t$ inclusive cross-section is on the level of 10\% of the QCD $gg \to b
\bar b t \bar t$ cross-section, see Table~\ref{T1:1}, for the same choice of the
energy scale. But in the mass range around 120~GeV it is on the level of 50\% of 
the QCD contribution, \cite{AcerMC-CPC}.
\vspace{-0.3cm}

\begin{table}[htb]
\newcommand{\lstrut}{{$\strut\atop\strut$}}
  \isucaption {\em {\bf AcerMC} production cross-sections for the electroweak $gg, q \bar q \to
  (Z/W/\gamma^* \to) b \bar b t \bar t$.  The 14 TeV centre-of-mass energy and
  CTEQ5L parton density functions were used along with different definitions for the
  $\alpha_{QED}$, $\alpha_{QCD}$ (as in native {\bf AcerMC}, default {\tt PYTHIA 6.2}
  and {\tt HERWIG 6.5}).  The $m_H~=~120$~GeV was used for calculation
  of the energy scale.
\label{T2:1e}}
\vspace{-0.3cm} 
\begin{center}
\scriptsize
\begin{tabular}{|c|c|c|c|} \hline \hline
Factorisation scale  &  $\alpha_{QED}$, $\alpha_{QCD}(1L)$ &  $\alpha_{QED}$, $\alpha_{QCD}(1L)$ &  $\alpha_{QED}$, $\alpha_{QCD}(2L)$  \\ 
   & native AcerMC    &  as in {\tt PYTHIA 6.2} & as in {\tt HERWIG 6.5}  \\
\hline \hline
\cline{2-4} \multicolumn{1}{|c|}{}& \multicolumn{3}{|c|}{$gg \to (Z/W/\gamma^* \to) b \bar b t \bar t$} \\
\hline 
$Q^2_{QCD}~=~\hat{s}$                     &  0.58 [pb]     &  0.56 [pb]  & 0.50 [pb]     \\
\hline
$Q^2_{QCD}~=~\sum{({p^i_T}^2 + m_i^2)}/4$ &  1.10 [pb]     &  1.05 [pb]  & 0.84 [pb]    \\
\hline
$Q^2_{QCD}~=~\sum{({p^i_T}^2)}/4$         &  1.50 [pb]     & 1.50  [pb]  & 1.16 [pb]    \\
\hline
$Q^2_{QCD}~=~(m_t + m_H/2)^2$             &  0.90 [pb]     & 0.89 [pb]   & 0.71 [pb]     \\
\hline \hline
\cline{2-4} \multicolumn{1}{|c|}{}& \multicolumn{3}{|c|}{$q \bar q \to (Z/W/\gamma^* \to) b \bar b t \bar t$} \\
\hline 
$Q^2_{QCD}~=~\hat{s}$                  & 0.029 [pb]    & 0.029  [pb]    & 0.025 [pb]    \\
\hline 
 $Q^2~=~\sum{({p^i_T}^2 + m_i^2)}/4$               & 0.043 [pb]    & 0.042  [pb]    & 0.036 [pb]  \\
\hline 
 $Q^2~= ~\sum{({p^i_T}^2)}/4$                       & 0.049 [pb]    & 0.048  [pb]    & 0.040 [pb]  \\
\hline 
 $Q^2~= ~(m_t + m_H/2)^2$ & 0.041 [pb]    & 0.041  [pb]    & 0.035 [pb]  \\
\hline \hline 
\end{tabular}
\end{center}
\end{table} 
 
\clearpage

\boldmath
\subsection{The  $gg, q \bar q \to  (W W b \bar b \to) f \bar f f \bar f b \bar b$;
$gg, q \bar q \to  ( t \bar t \to)  f \bar f f \bar f b \bar b$ processes}
\unboldmath 

The implemented $2 \to 6$ matrix elements for the resonant $gg, q \bar q \to (t \bar t \to)  f \bar f f \bar f b \bar b$ and
complete  $gg, q \bar q \to (W W b \bar b \to) f \bar f f \bar f b \bar b$ processes give possibility to study background
from top-quark production in more details, than with resonant on-shell $t \bar t$ production only ( as implemented in
{\tt PYTHIA} and {\tt HERWIG}).
In particular, for the Higgs boson search in the $H \to WW \to \ell \nu \ell \nu$ decay channel,
the analysis foresees strong suppression against $t \bar t$ background using topological 
feaures of events (jet veto, lepton angluar correlations), but does not foresees implicit
top-quarks reconstruction. To reliably predict such backgrounds,
availability of the complete $2 \to 6$ matrix element in Monte Carlo is mandatory, see eg. 
discussion in \cite{Krauer2002}. The total cross-sections are given in
Table~\ref{T2:1ee} for different definitions of  $\alpha_{QED}$, $\alpha_{QCD}$ couplings.

\begin{table}[hb]
\newcommand{\lstrut}{{$\strut\atop\strut$}}
  \isucaption {\em {\bf AcerMC} production cross-sections for the 
 $gg, q \bar q \to  (W W b b~ \to) f \bar f f \bar f b \bar b$ process.  The 14 TeV centre-of-mass energy and
  CTEQ5L parton density functions were used along with different definitions for the
  $\alpha_{QED}$, $\alpha_{QCD}$ (as in native {\bf AcerMC}, default {\tt PYTHIA 6.2}
  and {\tt HERWIG 6.5}).
\label{T2:1ee}}
\vspace{0.5cm} 
\begin{center}
\scriptsize
\begin{tabular}{|c|c|c|c|} \hline \hline
Factorisation scale  &  $\alpha_{QED}$, $\alpha_{QCD}(1L)$ &  $\alpha_{QED}$, $\alpha_{QCD}(1L)$ &  $\alpha_{QED}$, $\alpha_{QCD}(2L)$  \\ 
   & native AcerMC    &  as in {\tt PYTHIA 6.2} & as in {\tt HERWIG 6.5}  \\
\hline \hline
\cline{2-4} \multicolumn{1}{|c|}{}& \multicolumn{3}{|c|}{$gg \to  (W W b b~ \to) f \bar f f \bar f b \bar b $} \\
\hline 
$Q^2_{QCD}~=~(2 \cdot m_t^2)  $         &  400. [pb]     & 390.  [pb]  & 330. [pb]     \\
\hline
$Q^2_{QCD}~=~\sum{({p^i_T}^2 + m_i^2)}/4 $         &  450. [pb]     & 450.  [pb]  & 380. [pb]    \\
\hline
$Q^2_{QCD}~=~\sum{({p^i_T}^2)}/2 $         &  550. [pb]     & 550.  [pb]  & 460. [pb]    \\
\hline
$Q^2_{QCD}~=~\hat{s} $         &  355. [pb]     & 350.  [pb]   & 300. [pb]     \\
\hline \hline
\cline{2-4} \multicolumn{1}{|c|}{}& \multicolumn{3}{|c|}{$q \bar q \to  (W W b b~ \to) f \bar f f \bar f b \bar b$} \\
\hline 
$Q^2_{QCD}~=~(2 \cdot m_t^2)$          & 63. [pb]    &  62. [pb]    &  53. [pb]    \\
\hline 
 $Q^2_{QCD}~=~\sum{({p^i_T}^2 + m_i^2)}/2$         & 69. [pb]    &  69. [pb]    &  58. [pb]  \\
\hline 
 $Q^2_{QCD}~= ~\sum{({p^i_T}^2)}/2$        & 78. [pb]    &  78. [pb]    &  65. [pb]  \\
\hline 
 $Q^2_{QCD}~= ~\hat{s}$        & 59. [pb]    &  58. [pb]    &  50. [pb]  \\
\hline \hline 
\end{tabular}
\end{center}
\end{table}

\begin{table}[hb]
\newcommand{\lstrut}{{$\strut\atop\strut$}}
  \isucaption {\em {\bf AcerMC} production cross-sections for the 
 $gg, q \bar q \to  (t \bar t \to) f \bar f b f \bar f \bar b)$.  The 14 TeV centre-of-mass energy and
  CTEQ5L parton density functions were used along with different definitions for the
  $\alpha_{QED}$, $\alpha_{QCD}$ (as in native {\bf AcerMC}, default {\tt PYTHIA 6.2}
  and {\tt HERWIG 6.5}).
\label{T2:1g}}
\vspace{0.5cm} 
\begin{center}
\scriptsize
\begin{tabular}{|c|c|c|c|} \hline \hline
Factorisation scale  &  $\alpha_{QED}$, $\alpha_{QCD}(1L)$ &  $\alpha_{QED}$, $\alpha_{QCD}(1L)$ &  $\alpha_{QED}$, $\alpha_{QCD}(2L)$  \\ 
   & native AcerMC    &  as in {\tt PYTHIA 6.2} & as in {\tt HERWIG 6.5}  \\
\hline \hline
\cline{2-4} \multicolumn{1}{|c|}{}& \multicolumn{3}{|c|}{$gg \to  (t \bar t \to) f \bar f b f \bar f \bar b$} \\
\hline 
$Q^2_{QCD}~=~(2 \cdot m_t^2) $         & 370.  [pb]     & 365.  [pb]  & 310. [pb]     \\
\hline
$Q^2_{QCD}~=~\sum{({p^i_T}^2 + m_i^2)}/2 $         & 425.  [pb]     & 420.  [pb]  & 355. [pb]    \\
\hline
$Q^2_{QCD}~=~\sum{({p^i_T}^2)}/2 $         & 512.  [pb]     & 510.  [pb]  & 425. [pb]    \\
\hline
$Q^2_{QCD}~=~\hat{s} $         & 330.  [pb]     & 320.  [pb]   & 280. [pb]     \\
\hline \hline
\cline{2-4} \multicolumn{1}{|c|}{}& \multicolumn{3}{|c|}{$q \bar q \to t  \bar t \to) f \bar f b f \bar f \bar b$} \\
\hline 
$Q^2_{QCD}~=~(2 \cdot m_t^2)$          & 63. [pb]       & 62.  [pb]    & 53. [pb]    \\
\hline 
 $Q^2_{QCD}~=~\sum{({p^i_T}^2 + m_i^2)}/2$         & 69. [pb]       & 68.  [pb]    & 58. [pb]  \\
\hline  
 $Q^2_{QCD}~= ~\sum{({p^i_T}^2)}/2$        & 78. [pb]       & 78.  [pb]    & 65. [pb]  \\
\hline 
 $Q^2_{QCD}~= ~\hat{s}$        & 59. [pb]       & 57.  [pb]    & 50. [pb]  \\
\hline \hline 
\end{tabular}
\end{center}
\end{table} 

\begin{figure}[htb]
\begin{center}
\mbox{
     \epsfxsize=7.0cm
     \epsffile{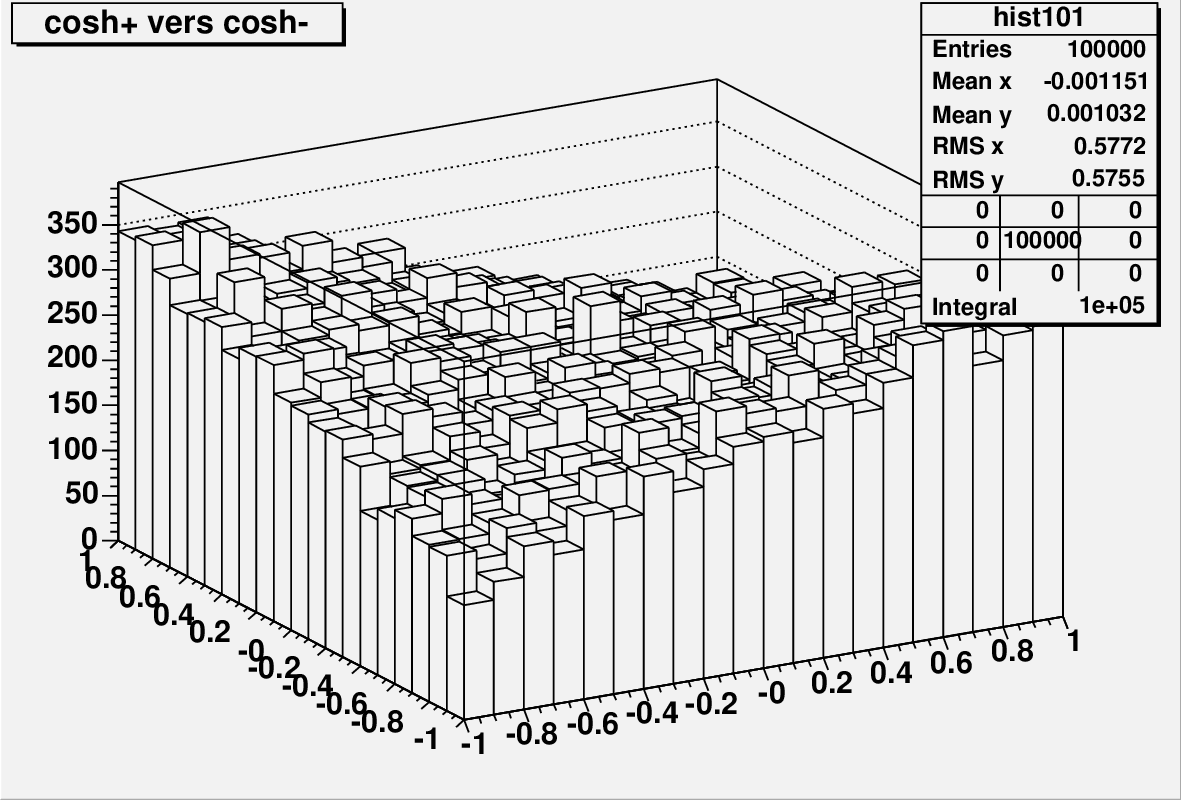}
     \epsfxsize=7.0cm
     \epsffile{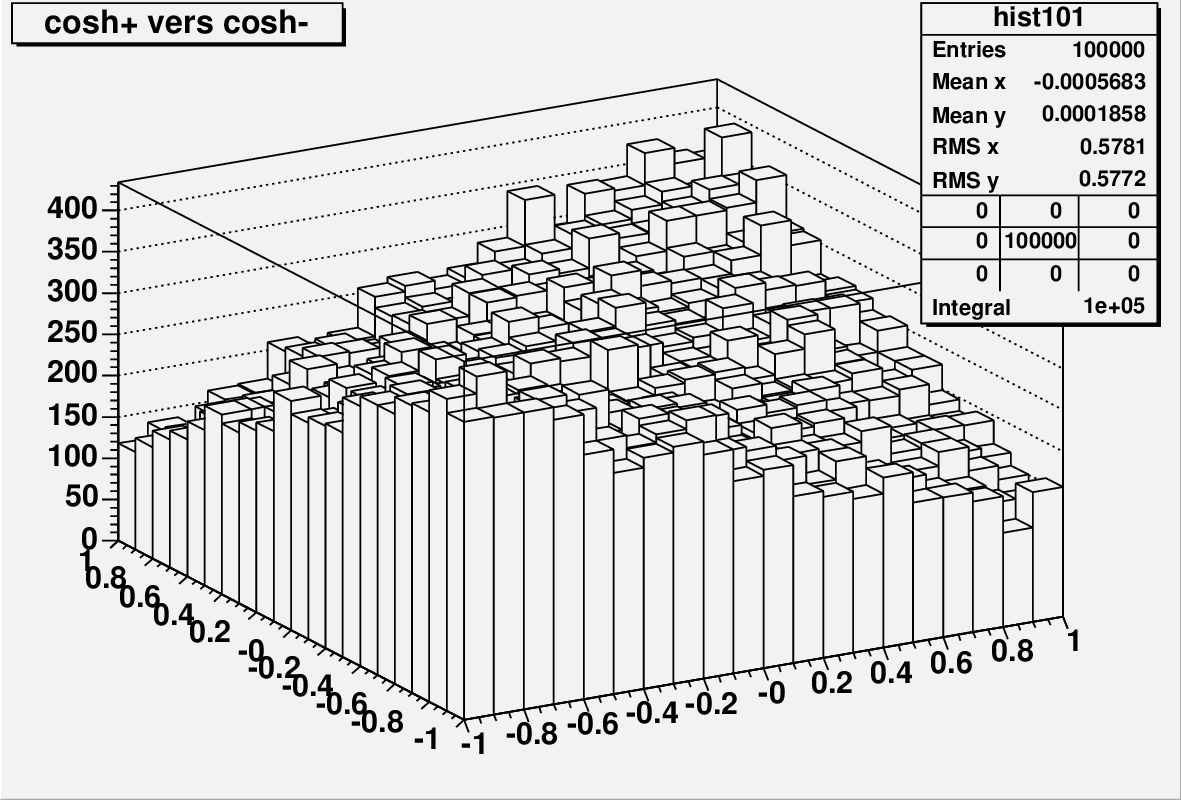}
}
\end{center}
\isucaption{\em
The correlations between $\cos \Theta$ (azimutal angle) of lepton and antylepton from 
$t \bar t \to \ell \bar \nu b \bar \ell \nu \bar b$ decays
measured in the rest frame of the top-quark with respect to the anty-top quark direction.
Left plot is for $gg\to  (W W b \bar b \to) f \bar f f \bar f b \bar b$ process,
right plot for  $q \bar q \to  (W W b \bar b \to) f \bar f f \bar f b \bar b$ process.
\label{f:ttasym}}
\end{figure}

As an example, Fig.~\ref{f:ttasym} illustrates
spin correlation effects in the top-pair production and decays, namely asymmetry in the correlations
between lepton and antylepton direction in the rest frame of top-quark, for events generated with $2 \to 6$
matrix element. 
Such correlation is absent if only $2 \to 2$ matrix element is used for events generation, followed by the
independent decays of each top-quark.


\boldmath
\subsection{The  $gg, q \bar q \to   t \bar t  t \bar t$ 
process}
\unboldmath

This process, in spite of having a very small cross-section at LHC energies,
contributes as reducible background to the $H H \to WWWW$ production at low
masses, \cite{ATL-PHYS-2002-029}. Availability of the complete Monte Carlo generator 
is mandatory to give reliable predictions of theis background and to optimise selection 
criteria.

\begin{table}[hb]
\newcommand{\lstrut}{{$\strut\atop\strut$}}
  \isucaption {\em {\bf AcerMC} production cross-sections for the 
 $gg, q \bar q \to   t \bar t  t \bar t$ process.  The 14 TeV centre-of-mass energy and
  CTEQ5L parton density functions were used along with different definitions for the
  $\alpha_{QED}$, $\alpha_{QCD}$ (as in native {\bf AcerMC}, default {\tt PYTHIA 6.2}
  and {\tt HERWIG 6.5}). The $m_H~=~120$~GeV and $m_t~=~175$~GeV 
were used for calculating the $Q^2_{QCD}$.
\label{T2:1h}}
\begin{center}
\scriptsize
\begin{tabular}{|c|c|c|c|} \hline \hline
Factorisation scale  &  $\alpha_{QED}$, $\alpha_{QCD}(1L)$ &  $\alpha_{QED}$, $\alpha_{QCD}(1L)$ &  $\alpha_{QED}$, $\alpha_{QCD}(2L)$  \\ 
   & native AcerMC    &  as in {\tt PYTHIA 6.2} & as in {\tt HERWIG 6.5}  \\
\hline \hline
\cline{2-4} \multicolumn{1}{|c|}{}& \multicolumn{3}{|c|}{$gg \to  t \bar t t \bar t$} \\
\hline 
$Q^2_{QCD}~=~\hat{s} $         &  2.65 [fb]     &  2.44 [fb]  & 1.93 [fb]     \\
\hline
$Q^2_{QCD}~=~\sum{({p^i_T}^2 + m_i^2)}/4 $         &  7.57 [fb]     &  7.38 [fb]  & 5.34 [fb]    \\
\hline
$Q^2_{QCD}~=~\sum{({p^i_T}^2)}/4 $         &  9.47 [fb]     &  9.32 [fb]  & 6.62 [fb]    \\
\hline
$Q^2_{QCD}~=~(m_t + m_H/2)^2 $         &  8.95 [fb]     &  8.78 [fb]  & 6.29 [fb]     \\
\hline \hline
\cline{2-4} \multicolumn{1}{|c|}{}& \multicolumn{3}{|c|}{$q \bar q \to t  \bar t t \bar t$} \\
\hline 
$Q^2_{QCD}~=~\hat{s}$          &    0.5 [fb]    &  0.5 [fb]    & 0.4 [fb]    \\
\hline 
 $Q^2_{QCD}~=~\sum{({p^i_T}^2 + m_i^2)}/4$         &    1.2 [fb]    &  1.2 [fb]    & 0.9 [fb]  \\
\hline 
 $Q^2_{QCD}~= ~\sum{({p^i_T}^2)}/4$        &    1.5 [fb]    &  1.5 [fb]    & 1.0 [fb]  \\
\hline 
 $Q^2_{QCD}~= ~(m_t + m_H/2)^2$        &    1.5 [fb]    &  1.4 [fb]    & 1.0 [fb]  \\
\hline \hline 
\end{tabular}
\end{center}
\end{table} 

\boldmath
\subsection{The  $ b b \oplus b g \to Z^0 \oplus b \to f \bar f \oplus b$
process}
\unboldmath

The 'heavy' associated Drell-Yan process is the combined production of $ b b \to
Z^0$ and $g b \to Z^0 b$ processes while removing the double counting between
the ISR $ g \to b \bar b$ splitting and the next-order $\alpha_S$ process $g b
\to Z^0$ using a procedure described in \cite{acot}, which incorporates part of
the NLO corrections to the process by removing the collinear singularities. The
process is important for e.g. b-quark PDF determination and background to Higgs
searches. The boson decays into any relevant final state (quarks or leptons). In itself
this process also 'double counts' the {\bf AcerMC} process $gg \to Z  b \bar{b} \to f \bar f b \bar{b}$
but is of relevance when only one distinct b-jet (high transverse momentum) is required in the
event selection.  

\begin{figure}[htb]
\begin{center}
     \epsfig{file=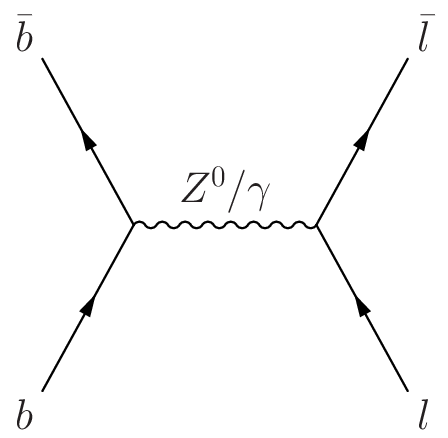,width=3.6cm}\parbox{0.3cm}{\vskip -3.2cm \Large$\oplus$}
     \epsfig{file=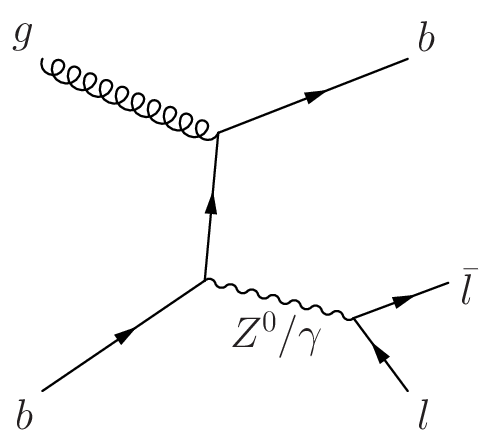,width=3.6cm}\parbox{0.3cm}{\vskip -3.2cm \Large$\ominus$}
     \epsfig{file=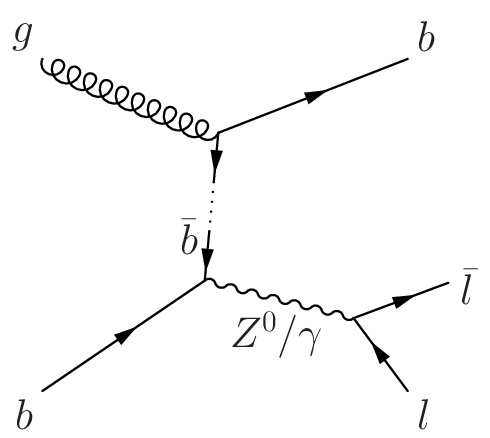,width=3.6cm}
\end{center}
\isucaption{\small Representative Feynman diagrams for the Drell-Yan with
associated b-quark production process for (from left to right): Order $\rm
\alpha_s^{(0)}$, order $\rm \alpha_s^{(1)}$ and order $\rm \alpha_s^{(1)}$ subtraction
term.\label{f:dy}}
\vspace*{-0.2cm}
\end{figure}

The cross-section values are omitted, the user can quickly and easily extract them by running
the {\bf AcerMC} with a couple of thousand events and the relevant final state.
Note that a fraction of events due to this procedure now has
negative weights \emph{equal to -1}, i.e. the events are weighted with
$\pm 1$ weights.

\boldmath
\subsection{The  single top production processes}
\unboldmath 
 
The single top production processes are now implemented in the {\bf AcerMC}:
\begin{itemize}
\item the associated Wt production process  $ g b \to t W \to b f \bar{f} f \bar f$,
\item the s-channel production process $ q q \to t b  \to b f \bar{f} b $ and 
\item the t-channel production process $ q b \oplus q g   \to q t \oplus b \to  q  b f \bar{f} \oplus b $,
\end{itemize}
which are of relevance for single top production searches at the LHC and top quark polarisation studies. 
The t-channel process is the combined production of the $q b \to q t$ and $ q g \to q t b $ W-exchange processes 
while removing the double counting between
the ISR $ g \to b \bar b$ splitting and the next-order $\alpha_S$ process $q g 
\to q t b $ using a procedure described in \cite{acot}, which incorporates part of
the NLO corrections to the process by removing the collinear singularities.
\begin{figure}[htb]
\begin{center}
     \epsfig{file=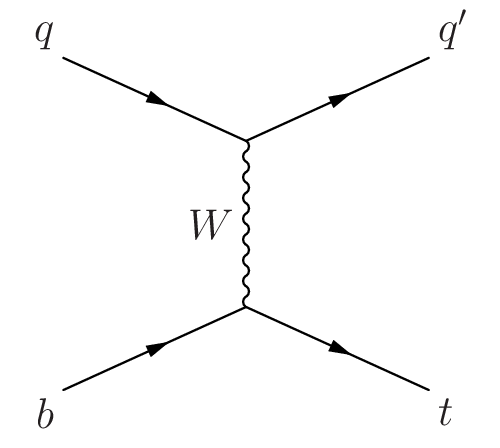,width=3.6cm}\parbox{0.3cm}{\vskip -3.2cm \Large$\oplus$}
     \epsfig{file=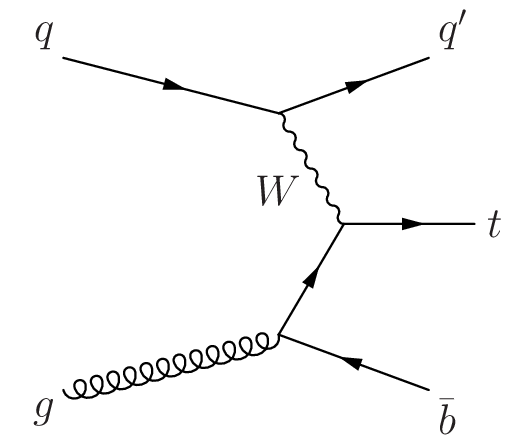,width=3.6cm}\parbox{0.3cm}{\vskip -3.2cm \Large$\ominus$}
     \epsfig{file=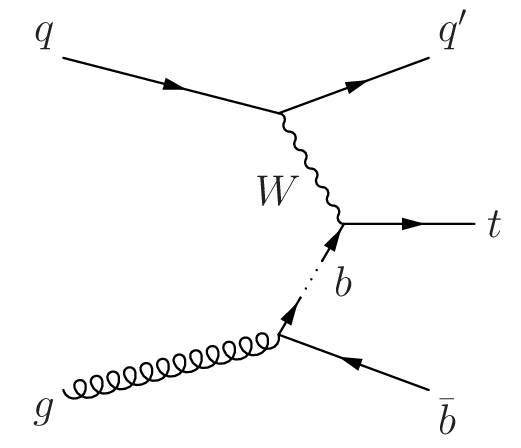,width=3.6cm}
\end{center}
\isucaption{\small Representative Feynman diagrams for the single top production process for (from
left to right): Order $\rm \alpha_s^{(0)}$, order $\rm \alpha_s^{(1)}$ and order $\rm \alpha_s^{(1)}$
subtraction term.\label{f:onet}}
\vspace*{-0.2cm}
\end{figure}

In addition, the associated Wt process is combined  with the process  $gg, q \bar q \to  (W W b b~ \to) f \bar f f \bar f b \bar b$, which 
contains the same Wt diagrams as when  $ g b \to t W \to b f \bar{f} f \bar f$ is showered to  $ g g \to t W \to b f \bar{f} f \bar f \oplus \bar b$. The same \cite{acot} formalism is used.

The cross-section values are omitted, the user can quickly and easily extract them by running
the {\bf AcerMC} with a couple of thousand events and the relevant final state.
Note that a fraction of events due to this procedure now has
negative weights \emph{equal to -1}, i.e. the events are weighted with
$\pm 1$ weights.


\boldmath
\subsection{The  $ q q \to Z^{0\prime} \to t \bar t \to b \bar b f \bar f f \bar f$
process}
\unboldmath 

The $Z^{0\prime} \to t \bar t $ (Z-prime) production process is of relevance to the new
boson searches beyond the Standard Model at the LHC. The process is a $ 2 \to 6$
process including full $\gamma/Z^0/Z^{0\prime}$ interference terms in the matrix
element and full spin correlations between the decay processes. The matrix
element was obtained by modifying the QCD top pair production process matrix
element produced by Madgraph/HELAS by hand to add beyond-the-SM $Z^{0\prime}$
production and the full $\gamma/Z^0/Z^{0\prime}$ which is not in the version of
Madgraph used for {\bf AcerMC} matrix element calculations. The cross-section
predictions agree well with the corresponding {\tt PYTHIA} process in terms of validating the code.

The cross-section values are omitted, the user can quickly and easily extract them by running
the {\bf AcerMC} with a couple of thousand events and the relevant final state.

\vspace{-0.2cm}

\subsection{The {\it control channel} processes \label{s:cont_chan}}

The set of simple $2 \to 2$ control channel processes was added to {\bf AcerMC} in order
to provide a means of consistency and cross-check studies.  Although these processes are
already implemented in {\tt PYTHIA} and/or {\tt HERWIG} (except the $gg \to WbWb$ one), 
the availability of the native
implementations is supposed to offer a more consistent control of generation parameters
when performing e.g. the comparison of parton shower {\tt PYTHIA/HERWIG} produced
additional pair of heavy quarks with the exact leading-order matrix elements implemented
in the core group of the {\bf AcerMC} $2 \to 4$ processes.

To the list of  control channel processes we have added also  $2 \to 4$ process, the
$gg \to WbWb$ process, as a control channel for the $2 \to 6$ process $gg \to f \bar f b f \bar f \bar b$. 
The  $2 \to 4$ process  we consider as very usefull for studying in more detail the resonant 
and complete  $WbWb$ production at LHC.

The benchmark results, given in  Table~\ref{T2:1f}, are obtained for simulated pp collisions at
14~TeV centre-of-mass energy.  The total cross-sections are listed for 
different definitions of  $\alpha_{QED}$, $\alpha_{QCD}$ couplings and different definitions 
of the energy scale $Q^2$.
\vspace{-0.6cm}
\begin{table}[ht]
\newcommand{\lstrut}{{$\strut\atop\strut$}}
\isucaption {\em {\bf AcerMC} production cross-sections for the 
$q \bar q \to Z/\gamma^* \to f \bar{f}$, $gg, q \bar q \to t \bar t$ and $q \bar q \to W
\to \ell \nu$ with single flavour $Z/\gamma^* \to f \bar{f}$ and/or $W \to \ell \nu$ decays.
The 14 TeV centre-of-mass energy, CTEQ5L parton density functions and different
definitions for the $\alpha_{QED}$, $\alpha_{QCD}$ (as in native {\bf AcerMC}, default
{\tt PYTHIA 6.2} and {\tt HERWIG 6.5}) were used.  The threshold $m_{f \bar{f}}~\geq~60$~GeV 
was used in the event generation. In {\tt PYTHIA 6.2} the setting {\tt MSTU(115)=2} was used to set
the lower $Q^2$ limit in $\alpha_{QCD}$ evolution to $\rm 4~GeV^2$ as done in the native {\bf AcerMC}
implementation. 
\vspace{-0.2cm}
\label{T2:1f}}
\begin{center}
\scriptsize
\begin{tabular}{|c|c|c|c|} \hline \hline
Factorisation scale  &  $\alpha_{QED}$, $\alpha_{QCD}(1L)$ &  $\alpha_{QED}$, $\alpha_{QCD}(1L)$ &  $\alpha_{QED}$, $\alpha_{QCD}(2L)$  \\ 
   & native AcerMC    & as in {\tt PYTHIA 6.2} & as in {\tt HERWIG 6.5}  \\
\hline \hline
\cline{2-4} \multicolumn{1}{|c|}{}& \multicolumn{3}{|c|}{$q \bar q \to Z/\gamma^* \to f \bar{f}$} \\
\hline 
 $Q^2~=~\hat{s}$                         & 1620  [pb]   &  1630 [pb]   & 1630 [pb]     \\
\hline 
 $Q^2~=~(\sum{({p^i_T}^2)} + M_Z^2)/2$   & 1550  [pb]   &  1560 [pb]   & 1560 [pb]     \\
\hline 
 $Q^2~=~\sum{({p^i_T}^2)}/2$             & 1260  [pb]   &  1260 [pb]   & 1260 [pb]     \\
\hline 
 $Q^2~=~M_Z^2$                           & 1630  [pb]   &  1630 [pb]   & 1640 [pb]     \\
\hline \hline
\cline{2-4} \multicolumn{1}{|c|}{}& \multicolumn{3}{|c|}{$gg \to t \bar t$} \\
\hline 
 $Q^2~=~\hat{s}$                         & 365  [pb]   &  360 [pb]   & 310 [pb]     \\
\hline 
 $Q^2~=~\sum{({p^i_T}^2 + m_i^2)}/2$     & 430  [pb]   &  420 [pb]   & 355 [pb]     \\
\hline 
 $Q^2~=~\sum{({p^i_T}^2)}/2$             & 595  [pb]   &  590  [pb]  & 490 [pb]     \\
\hline 
 $Q^2~=~(2 m_t)^2$                       & 320  [pb]   &  315 [pb]   & 270 [pb]     \\
\hline \hline
\cline{2-4} \multicolumn{1}{|c|}{}& \multicolumn{3}{|c|}{$q \bar q \to t \bar t$} \\
\hline 
 $Q^2~=~\hat{s}$                         &  62. [pb]   & 61. [pb]   & 52. [pb]     \\
\hline 
 $Q^2~=~\sum{({p^i_T}^2 + m_i^2)}/2$     &  69. [pb]   & 68. [pb]   & 58. [pb]     \\
\hline 
 $Q^2~=~\sum{({p^i_T}^2)}/2$             &  86. [pb]   & 85. [pb]  &  71. [pb]     \\
\hline 
 $Q^2~=~(2 m_t)^2$                       &  57. [pb]   & 56. [pb]   & 48. [pb]     \\
\hline \hline
\cline{2-4} \multicolumn{1}{|c|}{}& \multicolumn{3}{|c|}{$q \bar q \to W \to \ell \nu$} \\
\hline 
 $Q^2~=~\hat{s}$                         & 17200  [pb]   &  17230 [pb]   & 17310 [pb]     \\
\hline 
 $Q^2~=~(\sum{({p^i_T}^2)} + M_W^2)/2$   & 16480  [pb]   &  16490 [pb]   & 16460 [pb]     \\
\hline 
 $Q^2~=~\sum{({p^i_T}^2)}/2$             & 12920  [pb]   &  12920 [pb]   & 13020 [pb]     \\
\hline 
 $Q^2~=~M_W^2$                           & 17360  [pb]   &  17380 [pb]   & 17300 [pb]     \\
\hline \hline
\cline{2-4} \multicolumn{1}{|c|}{}& \multicolumn{3}{|c|}{$gg \to WbWb$} \\
\hline 
 $Q^2~=~\hat{s}$                 & 370  [pb]   &  365 [pb]   &  310 [pb]     \\
\hline 
 $Q^2~=~\sum{({p^i_T}^2 + m_i^2)}/2$  & 430  [pb]   &  425 [pb]   &  355 [pb]     \\
\hline 
 $Q^2~=~\sum{({p^i_T}^2)}/2$       & 525  [pb]   &  520  [pb]  &  435 [pb]     \\
\hline 
 $Q^2~=~(2 m_t)^2$                 & 330  [pb]   &  320 [pb]   &  275 [pb]     \\
\hline \hline
\end{tabular}
\vspace{-0.7cm}
\end{center}
\end{table}

\boldmath 
\section{Monte Carlo algorithm} 
\unboldmath

The conceptual motivation leading to the present implementation of {\bf AcerMC}
was to exploit the possibility of dedicated matrix-element-based generation
interfaced to a more general event generator, called {\it supervising} event
generator, which is subsequently used to complete the event generation procedure.

The goal of the dedicated matrix-element-based part is to efficiently generate
complicated event topologies using native (multi-channel based) phase space
generation procedures.  The strategy is based on the understanding that a
case-by-case optimisation is in complex cases of phase space topologies
preferable to an universal algorithm.  Given that phase-space is optimised on a
case-by-case basis, an user-defined pre-selection for the generated regions of
the phase-space is not implemented. Due to the fact that 
the $2 \to 4$ and $2 \to 6$ matrix
elements, provided by the {\tt MADGRAPH/HELAS} \cite{Madgraph} package, contain
full massive treatment of the final state particles, there are no explicit
divergences present for implemented processes and {\bf AcerMC} can indeed cover
the {\it full} (kinematically allowed) phase space of the processes at hand.

The matrix-element-based part uses $\alpha_{QCD}(Q^2)$ and $\alpha_{QED}(Q^2)$
couplings and mass spectra, as calculated by the supervising event generator, to
insure the full internal consistency in treatment of the event itself.
Optionally, the native $\alpha_{QCD}(Q^2)$ and $\alpha_{QED}(Q^2)$ definitions
can also be invoked.

The generation chain is built from the following steps:

\begin{itemize}
\item
The {\tt PYTHIA 6.2} or {\tt HERWIG 6.5} interfaces to the library of the structure
functions {\tt LHAPDF} \cite{PDF} are used to calculate convolution of the partonic density.
\item
{\bf AcerMC} modules produce unweighted hard-process events with colour flow information and
pass them to the supervising  generator {\tt PYTHIA 6.2} or {\tt HERWIG 6.5} as an external
event.
\item 
The generated events are then further treated within {\tt PYTHIA 6.2} or
{\tt HERWIG 6.5} event generators, where the fragmentation and hadronisation procedures,
as well as the initial and final state radiation are added and final unweighted events are
produced.
\end{itemize}

The {\bf AcerMC} efficiency\footnote{ Note that efficiency is energy scale
dependent and phase-space optimisation is done individually for each choice.  So
it might vary for the same process but different choices of the energy scale
definition.}  for generating unweighted events, using the implementation of the
phase-space generation discussed below, is summarised in Table~\ref{T3:1}. A
certain (very small) fraction of events is further rejected in the
showering/fragmentation procedures of the supervising generators.

\begin{table}[h]
\newcommand{\lstrut}{{$\strut\atop\strut$}}
  \isucaption {\em Efficiency  for the generation of unweighted events with the default
definition of the energy scale, {\tt ACSET2=1} (see Section \ref{s:scdef} for details).
 For generation of the $q \bar q, gg \to Z/\gamma^*(\to \ell \ell) b \bar b$  and 
 $q \bar q, gg \to Z/\gamma^*(\to \ell \ell) t \bar t$ events 
 threshold  $m_{\ell \ell}~\geq~60$~GeV has been used. The $f=e,\mu,\tau,q, b$. 
\label{T3:1}}
\vspace{2mm}
\begin{center}
\begin{tabular}{|c|c|c|} \hline \hline
Process   & Description & Internal AcerMC  efficiency \\
\hline \hline
     &   $2 \to 4$  & \\
\hline \hline
 [1] & $gg \to t \bar t b \bar b$           &   36.3 \%    \\
\hline 
 [2] & $q \bar q \to t \bar t b \bar b$           &  29.7 \%    \\
\hline 
 [3] & $q \bar q \to W(\to \nu \ell) b \bar b$   &  35.2 \%    \\
\hline 
 [4] &$q \bar q \to W(\to \nu \ell) t \bar t$   & 30.0 \%    \\
\hline 
 [5] &$gg \to Z/\gamma^*(\to \ell \ell) b \bar b$ &  42.8 \%     \\
\hline 
 [6] &$q \bar q \to Z/\gamma^*(\to \ell \ell) b \bar b$ & 35.1 \%   \\
\hline 
 [7] &$gg \to Z/\gamma^*(\to f \bar f, \nu \nu) t \bar t$ & 47.0 \%     \\
\hline 
 [8] &$q \bar q \to Z/\gamma^*(\to f \bar f, \nu \nu) t \bar t$ &  42.6 \%   \\
\hline
 [9] & $gg  \to (Z/W/\gamma^* \to) t \bar t b \bar b $ & 9.3 \% \\
\hline 
 [10] & $q \bar q  \to (Z/W/\gamma^* \to) t \bar t b \bar b $ & 32.4  \% \\
\hline
 [15] & $gg  \to t \bar t t \bar t $ & 48.0 \% \\
\hline
 [16] & $q \bar q \to t \bar t t \bar t$ & 50.2 \% \\
\hline \hline
     &   $2 \to 6$  & \\
\hline \hline
 [11] & $gg  \to (t \bar t \to) f \bar f b f \bar f b$ & 14.2 \% \\
\hline
 [12] & $q \bar q \to (t \bar t \to) f \bar f b f \bar f b$ & 12.0 \% \\
\hline
 [13] & $gg  \to (WW b \bar b \to) f \bar f  f \bar f b \bar b$ & 18.2 \% \\
\hline
 [14] & $q \bar q \to (WW b \bar b \to) f \bar f bf \bar f b \bar b$ & 4.2 \%  \\
\hline \hline
     &   Control processes  & \\
\hline 
 [91] &$q \bar q \to Z/\gamma^* \to \ell \ell $ &  68.4 \%   \\
\hline 
 [92] & $gg \to t \bar t$         &  65.6 \%    \\
\hline 
 [93] & $q \bar q \to t \bar t$           &  62.1 \%    \\
\hline 
 [94] &$q \bar q \to W \to \nu \ell$   &  69.4 \%    \\
\hline 
 [95] & $gg \to (t \bar t \to) Wb W \bar b$         &  40.2 \%    \\
\hline \hline
\end{tabular}
\end{center}
\end{table}

In the following we will briefly describe the key points of the implemented {\bf
AcerMC} modules and developed algorithms: matrix element calculations,
n-fermion phase-space generation based on the modified Kajantie-Byckling methods\cite{KB}, 
the issue of the s-dependent width and mass threshold effects 
for resonances  and finally, the modification of the {\tt VEGAS} algorithm.

\boldmath
\subsection{The Matrix Element Calculation}
\unboldmath

The squared matrix elements of the processes were obtained by using the {\tt
MADGRAPH/HELAS} \cite{Madgraph} package. They take properly into account the
masses and helicity contributions of final states particles, incoming quarks are
considered as massless. The particle masses, charges and coupling values that
are passed to the code derived with the {\tt MADGRAPH} package are calculated
from functions consistent with the ones used in supervising generators ({\tt
PYTHIA/HERWIG}). This allows to preserve the internal consistency of the event
generation procedure.  In particular, the (constant) coupling values of
$\alpha_s$ and $\alpha_{\rm QED}$ were replaced with the appropriate running
functions that were either taken from the interfaced generators or provided by
the {\bf AcerMC} code according to user settings.  Slightly modified {\tt
MADGRAPH/HELAS} allowed for obtaining colour flow information of the implemented
processes.

The sets of the  {\tt MADGRAPH/HELAS} coded diagrams, for each of the  implemented 
processes, are collected in Appendix A.
\vspace{3cm}

\boldmath
\subsection{The Phase Space Generation Procedure}
\unboldmath

The general objective in simulation of physics processes for the LHC environment
is to improve the integration of the differential cross-section using
Monte-Carlo sampling methods\footnote{For a nice discussion on the topic see
e.g. \cite{jadach,was}\ldots}. The sampling method used should aim to minimise
the variance of the integral as well as maximise the sampling efficiency given a
certain number of iterations and the construction of the sampling method itself
should aim to be sufficiently general and/or modular to be applicable to a wide
range of processes. Writing down a (process) cross-section integral for LHC type
(hadron-hadron) collisions:
\begin{equation}
\sigma = \int \sum_{a,b} f_a(x_1,Q^2) f_b(x_2,Q^2) \frac{|{\mathcal{M}_n}|^2}{(2 \pi)^{3n-4} 
(2\hat{s})}\, dx_1\, dx_2\, d\Phi_n,
\end{equation}
where $\rm f_{a,b}(x,Q^2)$ represent the gluon or (anti)quark parton density
functions, $\rm |{\mathcal{M}_n}|^2$ the squared n-particle matrix element divided
by the flux factor $\rm [(2 \pi)^{3n-4} 2\hat{s}]$ and $d\Phi_n$ denotes the
n-particle phase space differential. The quantity $\rm \hat{s} = x_1\, x_2\, s$ is
the effective centre-of-mass energy, and the sum $\rm \sum_{a,b}$ runs in case of
quark-antiquark incident partons over all possible quark-antiquark combinations
($\rm a,b = u,d,s,c,\bar{u},\bar{d},\bar{s},\bar{c}$). In case of $\rm g g$ initial
state the sum has only one term with $\rm a=b=g$.

It is often  convenient to re-write the differential cross-section in the
form:
\begin{equation}
\sigma = \int \sum_{a,b} x_1 f_a(x_1,Q^2) \; x_2 f_b(x_2,Q^2) \frac{|{\mathcal{M}_{n}}|^2}
{(2 \pi)^{3n-4} (2 \hat{s}^2)}\, dy\, d\hat{s}\, d\Phi_{n},
\label{e:dsig}
\end{equation}

with the new (rapidity) variable given by $\rm y = 0.5 \log(x_1/x_2)$. 
The n-body phase-space differential $\rm d\Phi_n$ and its integral $\Phi_n$
depend only on $\hat{s}$ and particle masses $m_i$  due to Lorentz invariance:
\begin{equation}
\Phi_n(\hat{s},m_1,m_2,\ldots,m_n) = \int d \Phi_n(\hat{s},m_1,m_2,\ldots,m_n) = 
\int \delta^4\left((p_a + p_b) - \sum_{i=1}^n p_i\right) \prod_{i=1}^n d^4p_i \delta (p_i^2 - m_i^2) \Theta(p_i^0),
\label{e:phins}
\end{equation}
with $\rm a$ and $\rm b$ denoting the incident particles and $\rm i$ running
over all outgoing particles $i=1,\ldots,n$. What one would like to do is to
split the n-body phase parameterised by 3n-4 essential (i.e. non-trivial)
independent variables into manageable subsets (modules) to be handled by
techniques which reduce the variance of the result and/or the sampling
efficiency (e.g. importance sampling\cite{Kleiss1994} or adaptive integration 
like VEGAS\cite{vegas} or FOAM\cite{foam}). Stating this in formal terms, 
the above Equation \ref{e:dsig} should be transformed into an expression like:
\begin{equation}
\sigma = \left(\prod_{i=1}^n \int\limits_{s_{i}^{-}}^{s_{i}^{+}} ds_i \right) 
\left(\prod_{j=1}^m \int\limits_{t_{j}^{-}}^{t_{j}^{+}} dt_j \right) 
\left(\prod_{k=1}^l \int\limits_{\Omega_{k}^{-}}^{\Omega_{k}^{+}} d\Omega_k \right) 
\left|\mathcal{J}_n\right| \int \sum_{a,b} x_1 f_a(x_1,Q^2) \; x_2 f_b(x_2,Q^2) \frac{|{\mathcal{M}_{n}}|^2}
{(2 \pi)^{3n-4} (2 \hat{s}^2)}\, dy\, d\hat{s}\,  
\label{e:dsigmod}
\end{equation}
where one integrates over Mandelstam type (Lorentz invariant) momenta transfers
$\rm s_i,t_j$ and space angles $\Omega_k \equiv (\cos \vartheta_k, \phi_k) $
 within the kinematically allowed limits (3n-4 variables in total) with the term $\rm
|\mathcal{J}_n|$ denoting the Jacobian of the transformation.  If one would then
decide to introduce importance sampling functions in order to reduce the peaking
behavior of the integrand \cite{Kleiss1994}, the integrals would take the form:
\begin{equation}
 \int\limits_{s_{i}^{-}}^{s_{i}^{+}} ds_i = 
\int\limits_{s_{i}^{-}}^{s_{i}^{+}} \frac{g_i(s_i)}{g_i(s_i)} ds_i,
\end{equation}
where the importance sampling function $\rm g_i$ is  probability density function
normalised in the integration region $\rm [s_{i}^{-},s_{i}^{+}]$:
\begin{equation}
\int\limits_{s_{i}^{-}}^{s_{i}^{+}}{g_i(s_i)} ds_i =1,
\end{equation}
which exhibits a similar peaking behavior as the integrand. Formally, one then inserts the 
identity:
\begin{equation}
  1 = \int\limits_0^1 \delta\left(r_i - \int\limits_{s_i^{-}}^{s_{i}} g_i(s_i) ds_i \right)  dr_i
\end{equation}
into the integral and then derives the \emph{unitary} sampling prescription: 
\begin{equation}
  \int\limits_0^1 dr_i \int\limits_{s_{i}^{-}}^{s_{i}^{+}} 
\delta\left(r_i - \int\limits_{s_i^{-}}^{s_i} g_i(s_i) ds_i \right)  \frac{g_i(s_i)}{g_i(s_i)} ds_i = 
\int\limits_0^1 dr_i \int\limits_{s_{i}^{-}}^{s_{i}^{+}} 
\delta\left(s_i - G^{-1}(r_i)\right)  \frac{1}{g_i(s_i)} ds_i = \int\limits_0^1
\frac{dr_i}{g_i(G^{-1}(r_i))},
\label{e:usampling}
\end{equation}
which formally means that the $\rm s_i$ values are sampled from the interval
according to the $\rm g_i(s_i)$ distribution by using \mbox{(pseudo-)random}
variable $\rm r_i$ together with the $\rm g_i(s_i)$ cumulant
$G(s_i)=\int_{s_i^{-}}^{s_{i}} g_i(s_i) ds_i$ with its inverse $G^{-1}$. The
unitarity of the algorithm states that each trial ( $\rm r_i$ value) produces a
result (i.e. a corresponding $\rm s_i$ value distributed according to $\rm g_i(s_i)$). 

Performing such substitutions on all integration parameters would give as the cross-section
expression;
\begin{equation}
\sigma = \prod_{i=1}^{3n-4} \int\limits_0^1 dr_i \frac{f(r_1,r_2\ldots)}{g(r_1,r_2\ldots)}
\label{e:gsampling}
\end{equation}
where the integrand would (hopefully) have as low variation as possible at least for a subset
of contributing Feynman diagrams\footnote{The 'modularisation' can be performed for several 
topologies at the same time and multi-channel techniques can be applied.}. To improve the sampling 
method further, the $\rm r_i$ (pseudo-)random variables can be sampled from adaptive algorithms of the VEGAStype \cite{vegas}. 

\begin{figure}[ht]
\begin{center}
     \epsfig{file=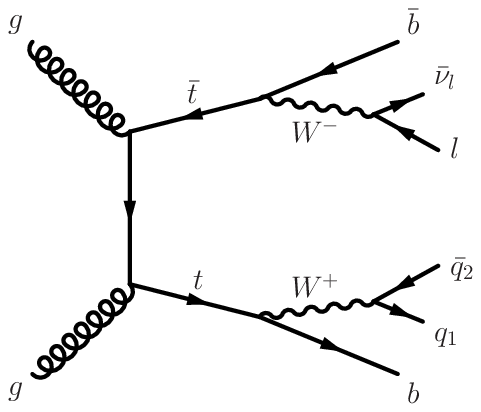,width=4.0cm}
     \epsfig{file=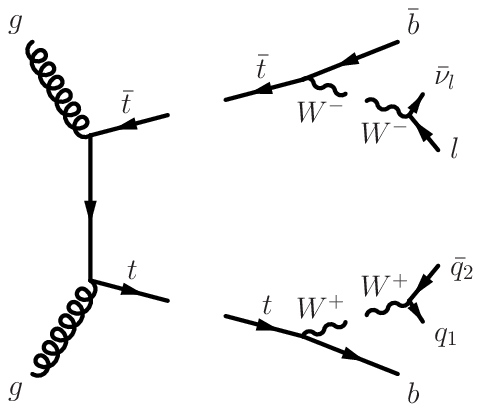,width=4.0cm}
     \epsfig{file=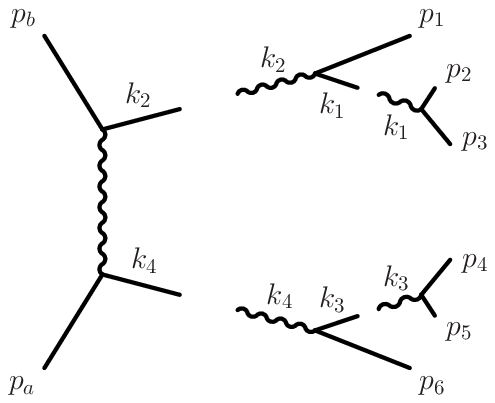,width=4.0cm}
\end{center}
\isucaption{
\small A representative Feynman diagram describing a $2 \to 6$ process $gg \to t\bar{t} \to b \bar{b} W^+ W^- \to b \bar{b} \ell \bar{\nu}_\ell q_1 \bar{q}_2$ and its decomposition into a set of $2 \to 2$ t-channel and s-channel sub-processes.
\label{f:ttbar}} 
\end{figure}

A representative Feynman diagram describing a $2 \to 6$ process is shown in
Figure \ref{f:ttbar}. As one can see, the process can be split in
several consecutive branchings, this approximation is often used in
matrix element (probability amplitude) calculations. It seems rather
obvious that any Feynman diagram can be split in a series of
horizontal and vertical branchings that one can denote as s-type and
t-type(u-type) using the analogy with the Mandelstam variables. What
one would like to do is thus to modularise the phase space in the form
of sequential s- and t-type splits.

The s-splitting of phase space is relatively easy to do and has as such been
used in many instances of Monte--Carlo generation (e.g. {\tt FermiSV}  \cite{fermisv}, 
{\tt Excalibur} \cite{excal}, {\tt Tauola} \cite{TAUOLA} etc..); the t-type
branchings (often tagged as multi-(peri)pheral topologies) have in contrast
generally been calculated only for specific cases (e.g. for 3 or 4 particles in
the final state \cite{fermisv,excal}). As it turns out, the problem of several
massive particles in the final state has already appeared more than 30 years ago
when several hadrons (e.g. pions) have been produced in (comparatively low
energy) nuclear interactions. At that time Kajantie and Byckling \cite{KB} have
derived the formulae for simulating any sequence of s- and t- type branchings
which, with some modifications, can also be applied to the EW and QCD processes
involving heavy quarks and/or massive bosons at LHC.

\clearpage 

\subsection{Modified Kajantie-Byckling Formalism\label{s:kb}}

\subsubsection{The s-type Branching Algorithms}
\begin{figure}[htb]
\begin{center}
     \epsfig{file=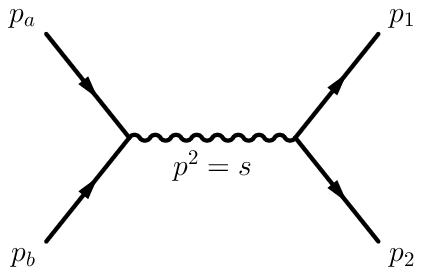,width=3.5cm}
\end{center}
\isucaption{
\small A diagram of a generic $\rm 2 \to 2$ s-channel process.
\label{f:schan}}
\end{figure}

The s-splits are the simplest method in the KB formalism. For the sake
of completeness one should start with the definition of the two-body
phase space integral (c.f. Fig \ref{f:schan}):
\begin{equation}
\Phi_2(s,m_1,m_2)=\int d^4p_1 d^4p_2 \delta (p_1^2 - m_1^2) \delta
(p_2^2 - m_2^2) \delta^4 (p - p_1 -p_2) \Theta(p_1^0) \Theta(p_2^0)  ,
\end{equation}
with the incoming momentum sum $\rm p = (p_a + p_b), p^2 = s$ and outgoing
momenta $\rm p_{1,2}, p_{1,2}^2 = m_{1,2}^2 $. The phase space integral is
Lorentz invariant (as one can observe in the above Equation where it is written
in a manifestly Lorentz invariant form). Subsequently, due to Lorentz
invariance, the integral is necessarily a function of the Lorentz scalars $\rm
s, m_1$ and $\rm m_2$ only. The step function product $\rm \Theta(p_1^0)
\Theta(p_2^0)$ is the explicit requirement of the positiveness of the energy
terms in $\rm p_{1,2}$ while the delta functions represent the on-shell
conditions on $\rm p_{1,2}$ and the total momentum conservation.

The integral can be transformed into a more compact form by
integrating out the spurious variables; one thus first integrates over
$\rm d^4 p_2$ and chooses the centre-of-mass system (CMS) as the
integration system of reference with $\rm p = (\sqrt{s},0,0,0)$ and
then evaluates the integrals over $\rm p_1^0$ and $\rm E_1^*$:
\begin{eqnarray}
\Phi_2(s,m_1,m_2) &=& \int  d^4p_1 \delta (p_1^2 - m_1^2) \delta
((p-p_1)^2 - m_2^2)  \Theta(p_1^0) \\ \notag
&=& \int \frac{d^3p_1^*}{2 E_1^*} \delta (s + m_1^2 - 2\sqrt{s} E_1^* - m_2^2)\\ \notag
&=& \frac{1}{4 \sqrt{s}} \int p^1_* dE_1^* d\Omega_1^* \delta \left( E_1^* - \frac{s +
m_1^2 -m_2^2}{2 \sqrt{s}} \right)\\ \notag
&=&\frac{p_1^*(s,m_1,m_2)}{4 \sqrt{s}} \int d\Omega_1^*,
\label{e:phi2s}
\end{eqnarray}
with the stars explicitly denoting the values in the centre-of mass system. 
The first integration simply sets $\rm p_1^0 = \sqrt{ (p_1^*)^2 + m_1^2 } =
E_1^*$ and the second integral leads to the well known relations for the energy:
\begin{equation}
E_1^*= \frac{s + m_1^2 -m_2^2}{2 \sqrt{s}}, ~~~~ E_2^* = \sqrt{s}- E_1^* = 
\frac{s + m_2^2 -m_1^2}{2 \sqrt{s}},
\label{e:cmse}
\end{equation}
and momenta sizes: 
\begin{equation}
p_1^*= |\vec{p}_1^*|=\frac{\sqrt{\lambda(s,m_1^2,m_2^2)}}{2\sqrt{s}}, ~~~~ 
p_2^*= p_1^*
\label{e:cmsp}
\end{equation}
of two particle production. The $\rm \lambda(s,m_1^2,m_2^2)$ denotes the Lorentz
invariant function:
\begin{equation}
\lambda(s,m_1^2,m_2^2)=(s-(m_1+m_2)^2)(s-(m_1-m_2)^2)
\label{e:lambda}
\end{equation}
and thus explicitly contains the phase space cutoff, i.e. the requirement that
the available CMS energy $\rm \sqrt{s}$ should be bigger than the mass sum $\rm
\sqrt{s} \geq (m_1+m_2)$. Note that the integration was so far done only over
the spurious parameters, leaving the polar and azimuthal angle of the $\rm p_1$
particle as the two independent parameters $\rm d\Omega^* = d\cos\theta^*
d\varphi^*$.  The integral becomes trivial to sample in case the outgoing
particles can be approximated as massless (the 'boost' factor lambda transforms
to unity). As already claimed, the latter approximation is however often
unjustified when studying processes representative for the LHC environment.

\begin{figure}[ht]
\begin{center}
     \epsfig{file=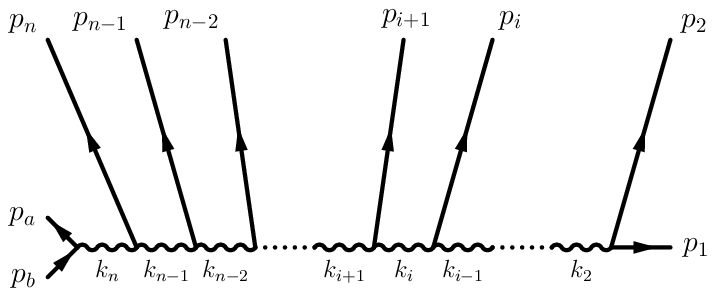,width=8.0cm}
\end{center}
\isucaption{
\small The diagrammatic representation of consecutive s-splits.
\label{f:ssplits}} 
\end{figure}

Kajantie and Byckling \cite{KB} introduced the \emph{recursion} and \emph{splitting} relations
for the n-particle phase space $\rm \Phi_n(s)$ given by Eq. \ref{e:phins}. The
recursion relation can be derived by defining the momentum sum:
\begin{equation}
k_i = \sum_{j=1}^{i} p_j = (k_i^0,\vec{k_i}) ;~~~ M_i^2 = k_i^2.
\label{e:kidef}
\end{equation}
Subsequently one can interpret $\rm p = k_n$ and $ s = M_n^2$ from Eq. \ref{e:phins}.
One continues by introducing the identities:
\begin{equation}
1 = \int dM_{n-1}^2 \delta(k_{n-1}^2 - M_{n-1}^2) \Theta(k_{n-1}^0)
\label{e:rec1}
\end{equation}
and
\begin{equation}
1 = \int d^4k_{n-1} \delta^4(p - k_{n-1} - p_n)
\label{e:rec2}
\end{equation}
into the integral of Equation \ref{e:phins}; separating out the arguments
containing $\rm k_{n-1}$ and $p_n$ terms one obtains:
\begin{eqnarray}
\Phi_n(M_n^2,m_1,m_2,\ldots,m_n) &=& \int dM_{n-1}^2 \times\\ \notag
&\times&\left\{ \int d^4k_{n-1} d^4p_n \delta(k_{n-1}^2 - M_{n-1}^2) \delta (p_n^2 - m_n^2)
\delta^4(p - k_{n-1} - p_n)\Theta(k_{n-1}^0)\Theta(p_{n}^0) \right\}\times\\ \notag
&\times& \Phi_{n-1}(M_{n-1}^2,m_1,m_2,\ldots,m_{n-1}),
\end{eqnarray}
where the remaining $p_i$ terms form the (n-1)-particle phase space integral
$\rm \Phi_{n-1}(M_{n-1}^2,m_1,m_2,\ldots,m_{n-1})$ and the terms in curly
brackets give a two particle phase space term (c.f. Eq. \ref{e:phi2s}):
\begin{eqnarray}
\Phi_n(M_n^2,m_1,m_2,\ldots,m_n) &=& \int dM_{n-1}^2 \Phi_2(M_n^2,M_{n-1},m_n) 
 \Phi_{n-1}(M_{n-1}^2,m_1,m_2,\ldots,m_{n-1}) \label{e:recur}
\\ \notag
&=&  \int dM_{n-1}^2 \frac{p_n^*}{4 M_n}
\Phi_{n-1}(M_{n-1}^2,m_1,m_2,\ldots,m_{n-1}) \\ \notag
&=& \int\limits_{(\sum_{i=1}^{n-1}m_i)^2}^{(M_n - m_n)^2} dM_{n-1}^2 
\frac{\sqrt{\lambda(M_n^2,M_{n-1}^2,m_n^2)}}{8 M_n^2} \int d\Omega_n^*
 \Phi_{n-1}(M_{n-1}^2,m_1,m_2,\ldots,m_{n-1}),
\end{eqnarray}
with the integration limits on $\rm M_{n-1}^2$ following from its definition in
Eq. \ref{e:kidef}. It has to be emphasized that the angles in $d\Omega_i^*$ are
each time calculated in the centre-of-mass system of $\rm k_i$ with the
invariant mass $\rm M_i$. The resulting recursion relation is clearly of
advantage when describing cascade decays of particles $k_n \to k_{n-1} p_n \to
k_{n-2}, p_n, p_{n-1} \to \ldots$; it also proves that the n-particle phase space
of Eq. \ref{e:phins} can be reduced into a sequence of two-particle phase space
terms, as shown in Figure \ref{f:ssplits}.
 
It can further prove of advantage to loosen up the splitting terms of
Eqns. \ref{e:rec1},\ref{e:rec2} so that instead of summing to n-1 one groups an
arbitrary set of $\rm \ell$ particles:
\begin{eqnarray}
1 &=& \int dM_{l}^2 \delta(k_{l}^2 - M_{l}^2) \Theta(k_{l}^0), \\ 
1 &=& \int d^4k_{l} \delta^4(p - k_{l} - \sum_{j=l+1}^{n} p_j),
\label{e:split}
\end{eqnarray}
which, when repeating the procedure in recursion relation of Eq. \ref{e:recur},
results in an expression:
\begin{equation}
\Phi_n(M_n^2,m_1,m_2,\ldots,m_n) = 
\int\limits_{(\sum_{i=1}^{l} m_i)^2}^{(M_{l+1} - m_{l+1})^2} dM_{l}^2 
\Phi_{n-l+1}(M_n^2,M_l,m_{l+1},\ldots,m_n) \Phi_{l}(M_l^2,m_1,m_2,\ldots,m_l),
\end{equation}
and thus effectively splits the phase space into two subsets, equivalent to
introducing an intermediate(virtual) particle with momentum $k_l$. 

The number of splitting relations and the number of particles in each group as
given in Eq. \ref{e:split} can be chosen in any possible sequence, thus meaning
that the grouping sequence is arbitrary and can be adjusted to fit the topology
in question.\footnote{Suggestions of \cite{KB} on how to pick random number
sequences will not be used since one might like to couple this method with an
adaptive algorithm to improve the sampling efficiencies.}

At this point some modifications were introduced to the algorithm in order to
adapt it to the specifics of the processes expected at the  LHC.
Kajantie and Byckling namely assumed that the generation sequence would be 'down' the
cascade (i.e. by sampling first a $\rm M_n$ value, then $M_{n-1}$ value
etc\ldots~ as is indeed most often done in Monte-Carlo Generators). This might
however not be optimal in the LHC environment since the available centre-of-mass
energy for the hard process (\shat)\; can vary in a wide range of values
(c.f. Equation \ref{e:dsig}) and has to be sampled from a distribution
itself. The shape of the distribution function for \shat\; is expected to behave
as a convolution of the peaking behavior of all participating invariant masses
times the parton density functions (c.f. Eq. \ref{e:dsig}); it subsequently
seems to be more natural (and efficient) first to sample the individual
propagator peaks and then their subsequent convolutions. Furthermore, by
generating the invariant masses 'up' the cascade (i.e. first $\rm M_2$, 
$\rm M_3$ \ldots $\rm M_n$ and finally $\rm \hat{s}$) the kinematic limits on
the branchings occur in a more efficient way (bound on the $\sqrt{\lambda}$
values, see Equations \ref{e:lambda} and \ref{e:revlim}), which is very
convenient since in the LHC environment no stringent generation cuts should be
made on the inherently non-measurable $\rm \hat{s}$  as it cannot be
accounted for by an analogous cut in a physics analysis.

A necessary modification of the algorithm would thus be to reverse the
generation steps by starting with the last pair(s) of particles. In terms of
integration (i.e. sampling) limits this translates into changing the limits of
Eq. \ref{e:recur}:
\begin{eqnarray}
\Phi_{n}(M_n^2,m_1,m_2,\ldots,m_n) &=& \\  \notag
&=& \int\limits_{(\sum_{i=1}^{n-1} m_i)^2}^{(M_n - m_n)^2} dM_{n-1}^2 
\frac{\sqrt{\lambda(M_n^2,M_{n-1}^2,m_n^2)}}{8 M_n^2} \int d\Omega_n^*\\ \notag &\times&
 \int\limits_{(\sum_{i=1}^{n-2} m_i)^2}^{(M_{n-1} - m_{n-1})^2} dM_{2-1}^2 
\frac{\sqrt{\lambda(M_{n-1}^2,M_{n-2}^2,m_{n-1}^2)}}{8 M_{n-1}^2} \int d\Omega_{n-1}^*
\\ \notag &\times& \ldots  
 \int\limits_{(\sum_{j=1}^{i-1} m_j)^2}^{(M_{i} - m_{i})^2} dM_{i}^2 
\frac{\sqrt{\lambda(M_{i}^2,M_{i-1}^2,m_{i}^2)}}{8 M_{i}^2} \int d\Omega_{i}^*
\ldots  \\ \notag &\times&
\int\limits_{(m_1 +m_2)^2}^{(M_3 - m_{3})^2} dM_{2}^2 
\frac{\sqrt{\lambda(M_{2}^2,m_1^2,m_2^2)}}{8 M_{2}^2} \int d\Omega_{2}^*,
\end{eqnarray}
which accommodates the mass generation sequence: $k_n \to k_{n-1} + p_n \to
\ldots$ (i.e. first sample $M_{n-1}^2$, then $M_{n-2}$ etc\ldots), into 
\begin{eqnarray}
\Phi_{n}(M_n^2,m_1,m_2,\ldots,m_n) &=& \label{e:revlim} \int\limits_{(M_{n-2}+m_{n-1})^2}^{(M_n - m_n)^2} dM_{n-1}^2 
\frac{\sqrt{\lambda(M_n^2,M_{n-1}^2,m_n^2)}}{8 M_n^2} \int d\Omega_n^* \\ \notag &\times&
 \int\limits_{(M_{n-3}+m_{n-2})^2}^{(M_n - m_n - m_{n-1})^2} dM_{n-2}^2 
\frac{\sqrt{\lambda(M_{n-1}^2,M_{n-2}^2,m_{n-1}^2)}}{8 M_{n-1}^2} \int d\Omega_{n-1}^*
\\ \notag &\times& \ldots  
 \int\limits_{(M_{i-1} + m_i)^2}^{(M_n - \sum_{j=i+1}^{n} m_{j})^2} dM_{i-1}^2 
\frac{\sqrt{\lambda(M_{i}^2,M_{i-1}^2,m_{i}^2)}}{8 M_{i}^2} \int d\Omega_{i}^*
\ldots  \\ \notag &\times&
\int\limits_{(m_1 +m_2)^2}^{(M_n - \sum_{j=3}^{n} m_{j})^2} dM_{2}^2 
\frac{\sqrt{\lambda(M_{2}^2,m_1^2,m_2^2)}}{8 M_{2}^2} \int d\Omega_{2}^*,
\end{eqnarray}
where one first samples the mass $\rm M_{2}$, $\rm M_{3}$\ldots
$M_{n-1}$ in the appropriate limits.

In some topologies symmetric cases of mass generation can appear (as
shown in Figure \ref{f:ttbar}) where the integration sequence is
ambivalent (e.g. in Figure \ref{f:ttbar} the ambivalence is which top
quark invariant mass to generate first\ldots) and after a choice is
made (since one of the two cases in the symmetric pair has to take
precedence) the procedure itself remains not entirely
symmetric. Detailed studies have shown that it proves useful to
include all permutations of such ambiguous sequences into the MC
algorithm in order to 'symmetrise' the solution and thus make it
easier to process by further additions (e.g. adaptive algorithms).

\subsubsection{The t-type Branching Algorithms}
\begin{figure}[ht]
\begin{center}
     \epsfig{file=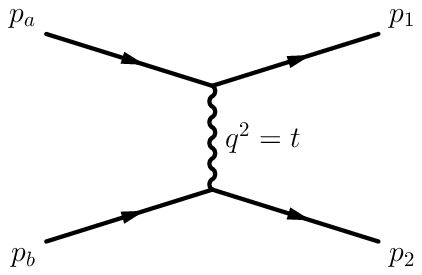,width=3.5cm}
\end{center}
\isucaption{
\small A diagram of a generic $\rm 2 \to 2$ t-channel process.
\label{f:tchan}}
\end{figure}

The t-splits are a specialty of the KB formalism due to the advanced
calculation of the limits on the (massive) t-variable. The formalism can be
introduced by observing that in case of a $\rm p_a + p_b \to p_1 + p_2$
scattering the momentum transfer is characterised by the (Mandelstam) variable
$\rm t = (p_1 - p_a)^2$ (c.f. Fig \ref{f:tchan}). It is thus sensible to replace the $d\Omega_1^* =
d\cos\theta_1^* d\varphi_1^*$ integration in the two body phase space integral of
Eq. \ref{e:phi2s} with integration over the $\rm t$ variable. Writing the definition of
$\rm t$ in the centre-of-mass system one gets:
\begin{eqnarray}
t &=& q^2 = (p_a - p_1)^2 \\ \notag
 &=& m_a^2 + m_1^2 -2E_a^* E_1^* +2p_a^* p_1^* \cos\theta_1^*
\label{e:tdef}
\end{eqnarray}
and hence:
\begin{equation}
 dt = 2 p_a^* p_1^* d\cos\theta^*
\end{equation}
Using the latter substitution together with Eq. \ref{e:cmse},\ref{e:cmsp} and
the analogue for $\rm p_a$:
\begin{equation}
p_a^* = \frac{\sqrt{\lambda(s,m_a^2,m_b^2)}}{2 \sqrt{s}}
\label{e:pa}
\end{equation}
one obtains in place of Eq. \ref{e:phi2s}
\begin{eqnarray}
\Phi_2(s,m_1,m_2)
&=&\frac{p_1^*(s,m_1,m_2)}{4 \sqrt{s}} \int d\Omega_1^* \\  \notag
&=&\frac{1}{8\sqrt{s} p_a^*} \int dt\: d\varphi^*\\ \notag
&=&\frac{1}{4\sqrt{\lambda(s,m_a^2,m_b^2)}} \int\limits_{t^{-}}^{t^{+}} dt \int\limits_0^{2\pi} d\varphi^*
\label{e:phi2t}
\end{eqnarray}
With the integration variable change the integration domain changes from $\rm
[-1,1]$ for $\rm d\cos\theta^*$ to $\rm [t^{-},t^{+}]$ for the $\rm dt$ integration. 
The $\rm t^\pm$ limits are obtained by inserting the $\rm \cos\theta^*$ limits
into Equation \ref{e:tdef}:
\begin{equation}
t^\pm = m_a^2 + m_1^2 -2E_a^* E_1^* \pm 2p_a^* p_1^*
\end{equation}
or in the Lorentz invariant form (c.f. Eq. \ref{e:cmse},\ref{e:cmsp}):
\begin{eqnarray}
t^\pm &=& m_a^2 + m_1^2 - \frac{(s+m_a^2 - m_b^2)(s+m_1^2 -m_2^2)}{2 s} \\ \notag
&\pm & \frac{\sqrt{\lambda(s,m_a^2,m_b^2)\lambda(s,m_1^2,m_2^2)}}{2 s}
\end{eqnarray}

As a step towards generalisation one has to note that the kinematic limits
$t^\pm$ can also be derived from \emph{the basic four-particle kinematic
function G(x,y,z,u,v,w)}\cite{Nyborg65a,KB}, where the function G can be
expressed as a Cayley determinant:
\begin{equation}
G(x,y,z,u,v,w) = -\frac{1}{2} \left|\;
\begin{matrix} 
0 & 1 & 1 & 1 & 1 \\
1 & 0 & v & x & z \\
1 & v & 0 & u & y \\
1 & x & u & 0 & w \\
1 & z & y & w & 0 
\end{matrix}
\; \right|
\end{equation}
The kinematic limits on $\rm t$ are in this case given by the condition 
\begin{equation}
G(s,t,m_2^2,m_a^2,m_b^2,m_1^2) \leq 0,
\end{equation}
it should be noted that the above condition gives either $\rm t^\pm$ limits given a
fixed value of $\rm s$ or equivalently  $\rm s^\pm$ limits given a fixed $\rm t$
value. 

\begin{figure}[ht]
\begin{center}
     \epsfig{file=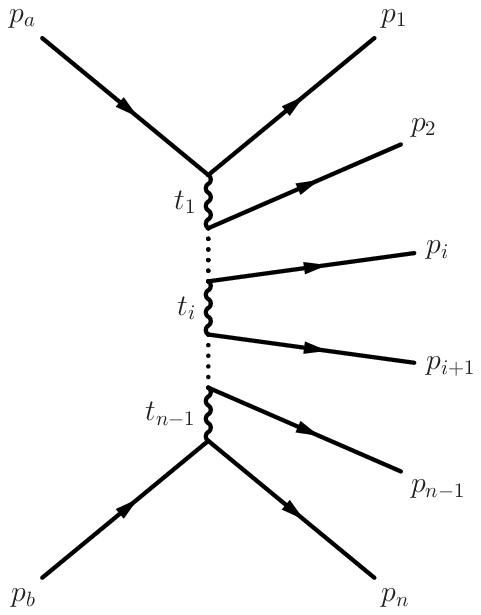,width=6.0cm}\hspace{-1cm}
     \epsfig{file=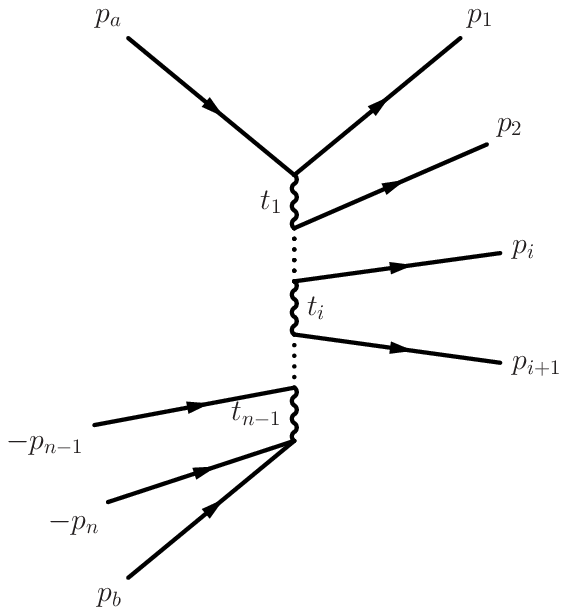,width=6.0cm}\hspace{-1cm}
     \epsfig{file=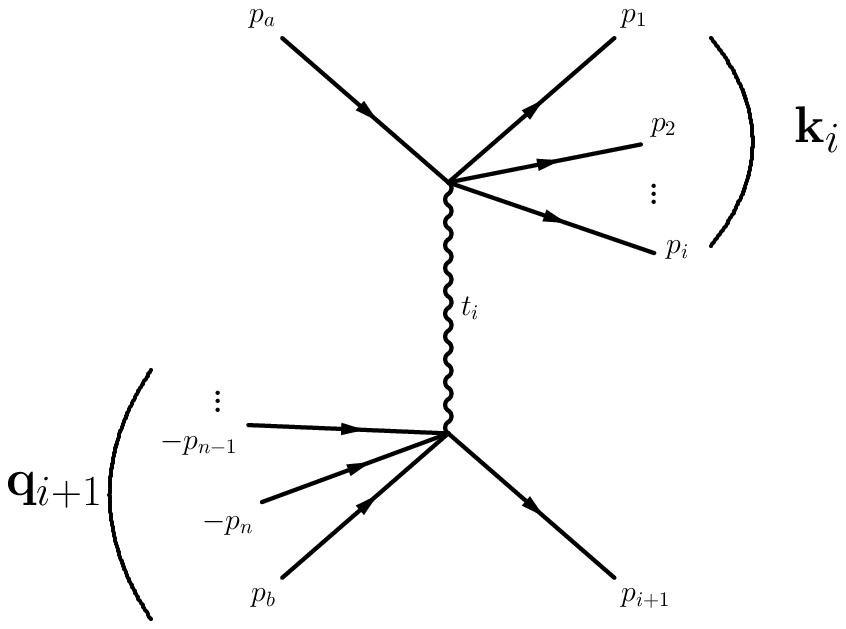,width=6.0cm}
\end{center}
\isucaption{
\small The diagrammatic representation of the method applied in translating the multi-(peri)pheral
splits into a $\rm 2 \to 2$ t-channel configuration.
\label{f:tsplits}} 
\end{figure}

In search of a recursion relation involving t-variables one can note that in
Eq. \ref{e:recur} the angle in $\cos\theta_n^*$ is equivalent to the scattering
angle in the centre-of-mass system of the reaction $\rm p_a + p_b \to k_{n-1} +
p_n$ and thus given by:
\begin{eqnarray}
t_{n-1} &=& (p_a - k_{n-1})^2 \\ \notag
 &=& m_a^2 + M_{n-1}^2 -2E_a^* k_{n-1}^{0*} +2p_a^* k_{n-1}^* \cos\theta_{n-1}^*
\label{e:tndef}
\end{eqnarray}
with the $\rm t_{n-1}^\pm$ limits expressed by:
\begin{equation}
G(M_n^2,t_{n-1},m_n^2,m_a^2,m_b^2,M_{n-1}^2) \leq 0,
\label{e:gtilim}
\end{equation}
and the $p_a^*$ given by Eq. \ref{e:pa}. In order to produce a more general
picture it can further be deduced that the next angle in the recursion
$\theta_{n-1}^*$, is the scattering angle of the subsequent process  
$\rm p_a + (p_b - p_n) \to k_{n-2} + p_{n-1}$ in the centre-of-mass system of
$k_{n-1}$; the $\rm (p_b - p_n)= q_{n-1}$ is in this case considered as a virtual
incoming particle with momentum $\rm q_{n-1}$ (c.f. Figure \ref{f:tsplits}).

\noindent
It immediately follows that for a general process $p_a + q_{i+1} \to k_{i} + p_{i+1}$
with:
\begin{equation}
q_{i} = p_b - \sum_{j=i+1}^{n} p_j = p_a - k_{i}; ~~~~ q_{i}^2 = t_{i}; ~~~
q_n^2 = t_n  = m_b^2
\end{equation}
a general expression for $\rm t_{i}$ becomes in the centre-of-mass frame of $\rm k_{i+1}$:
\begin{eqnarray}
t_{i} &=& (p_a - k_{i})^2 \\ \notag
 &=& m_a^2 + M_{i}^2 -2E_a^{*(i+1)} k_{i}^{0*(i+1)} +2p_a^{*(i+1)} k_{i}^{*(i+1)} \cos\theta_{i}^*
\label{e:tidef}
\end{eqnarray}
where momenta in centre-of-mass frame of $\rm k_{i+1}$, denoted with the 
superscript $\rm *(i+1)$, are given by:
\begin{eqnarray}
k_{i}^{*(i+1)} &=& \frac{\sqrt{\lambda(M_{i+1}^2,M_{i}^2,m_{i+1}^2)}}{2 M_{i+1}}\\
p_a^{*(i+1)}  &=& \frac{\sqrt{\lambda(M_{i+1}^2,m_a^2,t_{i+1})}}{2 M_{i+1}}
\end{eqnarray}
and the corresponding energies  $k_{i}^{0*(i+1)}$  and $\rm E_a^{*(i+1)}$ can
simply be obtained by using  the
analogues of Equations \ref{e:cmse},\ref{e:cmsp} or the usual Einstein
mass-energy relations directly.
The corresponding $\rm t_{i}^\pm$ limits given by:
\begin{equation}
G(M_{i+1}^2,t_{i},m_{i+1}^2,m_a^2,t_{i+1},M_{i}^2) \leq 0,
\end{equation}
and the recursion relation of Eq. \ref{e:recur} becomes:
\begin{eqnarray}
\Phi_n(M_n^2,m_1,m_2,\ldots,m_n) &=& \label{e:recurt} \\
&=& \int\limits_{(\sum_{i=1}^{n-1}m_i)^2}^{(M_n - m_n)^2} 
\frac{dM_{n-1}^2 }{4\sqrt{\lambda(M_n^2,m_{a}^2,t_n)}} \int\limits_0^{2\pi} d\varphi_n^*
\int\limits_{t_{n-1}^{-}}^{t_{n-1}^{+}} dt_{n-1}\: 
 \Phi_{n-1}(M_{n-1}^2,m_1,m_2,\ldots,m_{n-1}), \notag
\end{eqnarray}

As already argued the resulting set of $\rm (s_i=M_i^2,t_i)$ can again be
sampled in any direction with respect to the cascade by applying the appropriate
change in the integration limits (c.f. Eq. \ref{e:recur} and \ref{e:revlim}). 
The recommended approach (i.e. the introduced modification of the
algorithm) is again to first sample the invariant masses in the reverse cascade
direction (i.e. in the sequence $\rm M_2,M_3,\ldots ,M_n$) and then the $t_i$
values within the limits calculated from Eq. \ref{e:gtilim} down the cascade
(i.e. in the order of $\rm t_{n-1},t_{n-2},\ldots,t_1$).

To sum up, it has been shown that using the Kajantie--Byckling formalism 
the phase space for any topology can be split in a set
of s-type and t-type $\rm 2 \to 2$ branching steps (modules) 
given by recursive formulae of Equations \ref{e:revlim} and \ref{e:recurt}.
\clearpage

\subsection{Propagator Sampling\label{s:prop}}

A well known theoretical issue is that one can expect the most prominent peaks
in the differential cross-section of a specific process in the phase space
regions of high propagator values in the corresponding probability
density. Consequently, in the scope of complementing the modular structure of
the derived Kajantie-Byckling based phase space sampling, new approaches were also
developed concerning the numerical sampling methods of the relevant kinematic
quantities.

In order to get small variance in the Monte Carlo procedure one would thus like
to include the appropriate peaking dependence of the relevant momentum transfers
$\rm q^2$ in the importance sampling function. It however turns out that since
the momenta transfers $\rm q$ participate also in the propagator numerators
(typically in $\rm p_\mu q^\mu/q^2$) and since in process of interest one mostly
finds several Feynman diagrams contributing to the final probability density,
thus causing interferences, it is very difficult or even impossible to estimate
the exact power of momenta transfers in the sampling functions for different
propagator peaks. In other words, the probability density dependence on the momentum
transfer $\rm q^2$ can in general be approximated with the dependence $\rm
1/(q^2)^\nu$ where the best value of $\rm \nu$ must be determined separately (on
a process by process basis).

In view of the latter, general formulae have been developed for sampling the
$x^{-\nu}$ shape \cite{fermisv,sampler}: Given a pseudo random number $r \in [0,1]$ and limits 
$x \in [x_{-},x_{+}]$ the value x distributed as $x^{-\nu}$ is obtained from the
formulae in Eq. \ref{e:usampling} as:
\begin{eqnarray}
x &=& \left[ x_{-}^{-\nu+1}\cdot(1-r) + x_{-}^{-\nu+1}\cdot r
\right]^{-\frac{1}{\nu+1}}; ~~~~ \nu \neq 1;\\ x &=&
\frac{x_{+}^r}{x_{-}^{r-1}}; ~~~ \nu = 1.
\label{e:xnuni}
\end{eqnarray}  

\noindent
Using the analogous (unitary) approach a recipe for resonant (Breit-Wigner)
propagator contributions of the type:
\begin{equation}
BW(s)=\frac{1}{(s-M^2)^2+M^2 \Gamma^2}
\end{equation}
with $\rm s \in [s_{-},s_{+}]$ and a pseudo random number $r \in [0,1]$ is
available by the prescription:
\begin{eqnarray}
s &=& M^2 + M\Gamma \cdot \tan \left[ (u_{+} - u_{-}) \cdot r + u_{-} \right]\\
u^\pm &=& {\rm atan}\left( \frac{s^\pm - M^2}{M\Gamma} \right)
\label{e:bwuni}
\end{eqnarray}

Following similar arguments as for the non-resonant propagators one can surmise
that the best sampling function for resonant propagators could in general be a 
Breit-Wigner shape modified by a factor $\rm s^\nu, \nu \in [0,1]$. As shown in the following section
it was found that a shape:
\begin{equation}
BW(s)=\frac{s}{(s-M^2)^2+M^2 \Gamma^2}
\end{equation}
works quite well for a set of processes and a corresponding sampling recipe was developed.
In addition, studies in \cite{lichard} show that a resonant $\rm \sqrt{s} \times$ Breit-Wigner shape:
\begin{equation}
BW(s)=\frac{\sqrt{s}}{(s-M^2)^2+M^2 \Gamma^2}
\end{equation}
should be expected in a range of decay processes. Detailed studies have shown
that it is in general better to introduce a $\rm s^\nu, \nu \in [0,1]$
dependence even if it over-compensates the high mass tails of the corresponding
differential cross-section distribution since this provides an overall reduction
of the maximal weight fluctuations in the Monte--Carlo event generation procedure. 

\subsubsection{Breit-Wigner Function with s-dependent Width}

In some topologies of the processes involving $W^\pm$ or $Z^0$ bosons, a bias of the matrix
element towards large values in the high $s^*_{W/Z}$ region is evident, which in turn
means that a more accurate description of the tails of $s^*_{W/Z}$ distribution is
needed. Consequently, the Breit-Wigner sampling function was replaced by\footnote{To our
knowledge this implementation is original and done for the first time in
{\bf AcerMC}.}:
\begin{equation}
BW_s(s^*_W)=\frac{s^*_W}{(s^*_W - M_W^2)^2 + M_W^2 \Gamma_W^2},
\label{e:bwnew}
\end{equation}
which is proportional to the (more accurate) Breit-Wigner function with an $s^*_W$ dependent
width (W in the above formula denotes either a $W^\pm$ or a $Z^0$ boson).

In order to implement a unitary algorithm (an algorithm that produces a result for every
trial, i.e. there is no rejection) of value sampling on the above function one first has
to calculate the normalisation integral (cumulant) and then its inverse
function. Introducing a new variable $\eta = (s^*_W-M_W^2)/(M_W\;\Gamma_W)$ the integral
of the above function can be expressed as:
\begin{equation}
\int BW_s(s^*_W) \; ds^*_W  = \int \biggl\{ \frac{M_W^2}{M_W \Gamma_W} \cdot
\frac{1}{1+\eta^2} + \frac{\eta}{1+\eta^2} \biggr\} \; d\eta,
\label{e:bws}
\end{equation}
where the upper integral limit is left as a free parameter. The integral thus
gives a function:
\begin{equation}
F(\eta)=  \biggl\{ \frac{M_W^2}{M_W \Gamma_W} \cdot {\rm atan}(\eta) \biggr\}  + 
\biggl\{ \frac{1}{2} \cdot \log(\eta^2+1) \biggr\}, =
F_1(\eta) + F_2(\eta)
\label{e:bwint}
\end{equation}
with $F(\eta_{\rm max})-F(\eta_{\rm min})$ defining the normalisation. One of
the undesirable features is that the function $F(\eta)$ does not have a (simple)
analytical inverse, which is a prerequisite for unitary sampling. Taking a
closer look at the two above expressions one can quickly spot another
undesirable feature, namely that the second term  in the Equation
\ref{e:bws} is an odd function of $\eta$, which after the integration gives an
even term $F_2(\eta)$ in $\eta$ in Equation \ref{e:bwint}. In other words the
second term alone is neither a non-negative function nor does it have an unique
inverse - one has to deal with a {\it negative probability}. 
A reasonably elegant solution to this problem has been developed and implemented here:
\begin{itemize}
\item One samples values of $\eta$ by using only the first term of the above
expressions (the {\it usual} Breit-Wigner function). 
\item One then re-samples the obtained value of $\eta$ using the full expression of
Equation \ref{e:bws}: If $\eta$ is less than zero the value is mapped to $-\eta$
with the probability given by Equation \ref{e:bws}.
\end{itemize}

\begin{figure}[ht]
\vspace{-0.6cm}
\begin{center}
\mbox{
     \epsfxsize=7cm
     \epsffile{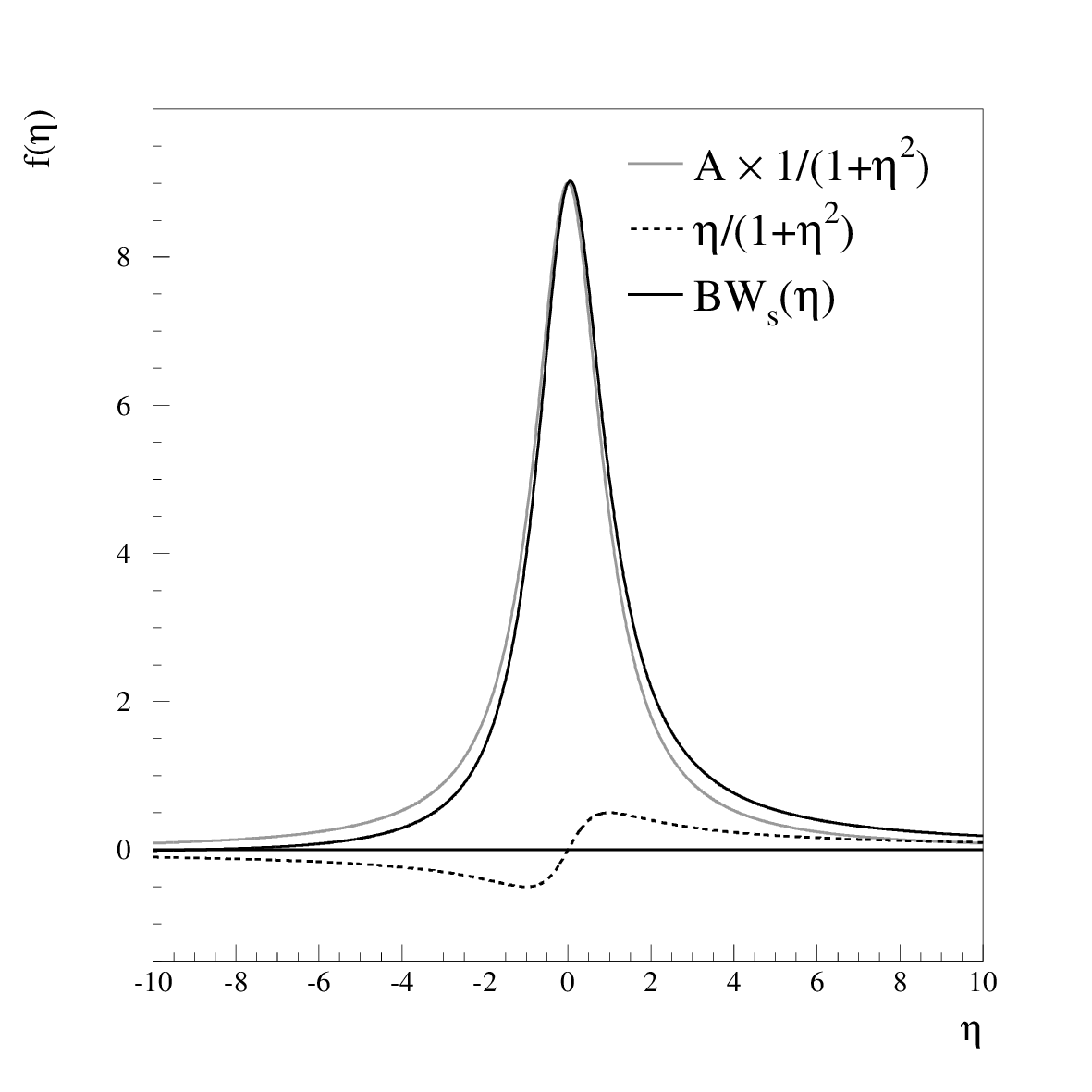} 
     \epsfxsize=7cm
     \epsffile{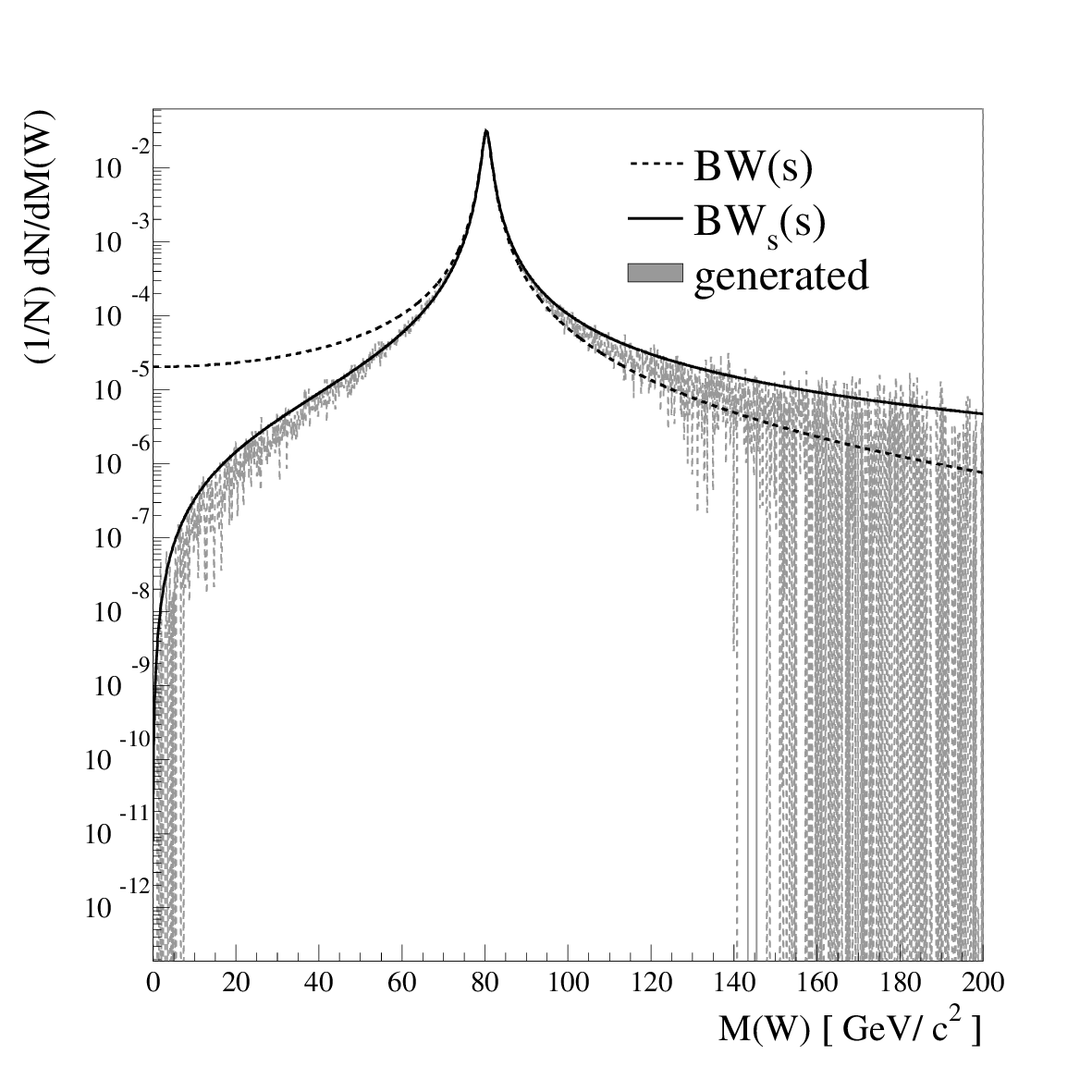} 
}
\end{center}
\vspace{-0.6cm}
\isucaption{\em {\bf Left} Comparisons of the two functional terms of Eq. \ref{e:bws} to 
${\rm BW}_s(\eta)$ given by Equation \ref{e:bwnew}. Note that the scaling factor $A$ is
chosen in view of making the contributions more transparent; it is much too small compared
to the real case of $W^\pm/Z^0$ bosons. \newline {\bf Right} Comparison of the (normalised)
distributions of differential cross-section for the process $q \bar q \to W b \bar b$
(dashed) and sampling functions (solid line) with respect to the variables obtained by
importance sampling, as described in the text.
\label{f:bwcomp}}
\vspace{0.3cm}
\end{figure}

Why this works can quickly be deduced by looking at the Figure \ref{f:bwcomp}:
At negative values of $\eta$ the second term of Equation \ref{e:bws} gives a
{\it negative probability} in the region $\eta<0$, i.e. using a simple
Breit-Wigner (Cauchy) probability function too many events are generated in this
region. Correspondingly, since the  second term of Eq. \ref{e:bws} is an odd
function, exactly the same fraction (distribution) of events is {\it missing} in the
region $\eta>0$. By mapping events with $\eta<0$ over the $\eta=0$ axis one thus
solves both problems at the same time. Using the above re-sampling procedure the
whole approach remains unitary, i.e. no events are rejected when there are no
limits set on the value of $\eta$ or they are symmetric $|\eta_{\rm min}|=\eta_{\rm
max}$. In the contrary case, a small fraction of sampling values is rejected.

After some calculation the whole unitary procedure can thus be listed as follows:
\begin{itemize}
\item Calculate the kinematic limits $\eta_{\rm min}$ and $\eta_{\rm max}$.
\item Calculate the {\it normalisation} factors $\Delta_1=F_1(\eta_{\rm
max})-F_1(\eta_{\rm min})$, $\Delta_2=F_2(\eta_{\rm max})-F_2(\eta_{\rm min})$
and $\Delta_s = \Delta_1 + \Delta_2$; the term $\Delta_2$ can actually be
negative and thus does not represent proper normalisation.

\item Obtain a (pseudo-)random number $\rho_1$. 
\item If $\rho_1 \leq \Delta_2/\Delta_s$ then:
\begin{itemize}
\item Obtain a (pseudo-)random number $\rho_2$;
\item Construct $\eta$ as:
\begin{eqnarray*}
X & = & \Delta_2 \cdot \rho_2 + F_2(\eta_{\rm min}),\\
\eta & = & \sqrt(e^{2X}-1),
\end{eqnarray*}
which is the inverse of the (normalised) cumulant $(F_2(\eta)-F_2(\eta_{\rm
min}))/\Delta_2$.
\item Note that the condition $\rho_1 \leq \Delta_2/\Delta_s$ can be fulfilled
only if $\Delta_2 \geq 0$, which means that $\eta_{\rm max}$ is positive and
greater than $\eta_{\rm min}$.
\end{itemize}
\item Conversely, if $\rho_1 > \Delta_2/\Delta_s$ then:
\begin{itemize}
\item Obtain a (pseudo-)random number $\rho_2$;
\item Construct $\eta$ as:
\begin{eqnarray*}
X & = & \Delta_1 \cdot \rho_2 + F_1(\eta_{\rm min}),\\
\eta & = & \tan(\frac{M_W \Gamma_W}{M_W^2}\cdot X)
\end{eqnarray*}
which is the inverse of the (normalised) cumulant $(F_1(\eta)-F_1(\eta_{\rm
min}))/\Delta_1$.
\item If the obtained $\eta$ is less than zero then calculate the normalised
probability densities:
\begin{eqnarray*}
P_1 & = & \frac{1}{\Delta_1} \cdot \{\frac{M_W^2}{M_W \Gamma_W} \cdot
\frac{1}{1+\eta^2} \} \\
P_s & = &  \frac{1}{\Delta_s} \cdot \{\frac{M_W^2}{M_W \Gamma_W} \cdot
\frac{1}{1+\eta^2} + \frac{\eta}{1+\eta^2} \}
\end{eqnarray*}
\item Obtain a (pseudo-)random number $\rho_3$;
\item If $\rho_3 > P_s/P_1$ map $\eta \to -\eta$.
\item If the new $\eta$ falls outside the kinematic limits $[\eta_{\rm
min},\eta_{\rm max}]$ the event is rejected.
\item Note also that the last mapping can only occur if the original $\eta$ was
negative, since $P_s < P_1$ only in the region $\eta < 0$. 
\end{itemize}
\item Calculate the value of $s^*_W$ using the inverse of $\eta$ definition:
\begin{equation}
s^*_W = (M_W\;\Gamma_W) \cdot \eta + M_W^2
\end{equation}
The weight corresponding to the sampled value $\eta$ is exactly:
\begin{equation}
\Delta_s \cdot \frac{(s^*_W - M_W^2)^2 + M_W^2 \Gamma_W^2}{s^*_W},
\end{equation}
which is the (normalised) inverse of Equation \ref{e:bwnew} as requested.
\end{itemize}
As it turns out in subsequent generator level studies, this generation procedure
gives much better agreement with the  differential distributions than the
{\it usual} (width independent) Breit-Wigner; an example obtained for the  $q \bar q \to W b
\bar b$ process is shown in Figure \ref{f:bwcomp}. The evident consequence is
that the unweighting efficiency is substantially improved due to the reduction
of the event weights in the high $s^*_W$ region.

\subsubsection{The Inclusion of Mass Effects in Propagator Sampling}

Studies have shown that the $x^{-\nu}$ approximation works quite well for
t-channel type propagators since the phase space suppression factor
$\sqrt{\lambda}$ participates in the denominator, as shown in
Eq. \ref{e:recurt}, and thus contributes only to the $x^{-\nu}$
slope. Contrariwise, while the $x^{-\nu}$ approximation still works reasonably
well when describing the s-channel type propagators involving particles with
high virtuality and/or decay products with low masses, it can be shown that this
is not necessarily the case in the LHC environment, where the presence of
massive decay products can significantly affect the invariant mass
distributions.  As it can be seen in Figure \ref{f:acermc} the shape of the
propagator dependence can be strongly suppressed by the phase space $\rm
\sqrt{\lambda}$ (boost) factor at low values; thus the sampling function
approximation for non-resonant propagators could be approximated with something
like:
\begin{equation}
f_{\rm NR}(s) =  \frac{\sqrt{\lambda(s,m_a^2,m_b^2)}}{s} \cdot \frac{1}{s^\nu} = \frac{\sqrt{\lambda(s,m_a^2,m_b^2)}}{s^{\nu+1}}
\label{e:lamxnu}
\end{equation}

and similarly
\begin{equation}
f_{\rm R}(s) = \frac{\sqrt{\lambda(s,m_a^2,m_b^2)}}{s} \cdot \frac{\sqrt{s}}{(s-M^2)^2+M^2 \Gamma^2} = \frac{\sqrt{\lambda(s,m_a^2,m_b^2)}}{\sqrt{s}\cdot((s-M^2)^2+M^2 \Gamma^2)}
\label{e:lamsbw}
\end{equation}

for resonant propagators. 

As it turns out the two functions cannot be sampled by the well known
unitary algorithms (i.e. the biggest collection of recipes \cite{sampler}
yielded no results); already the integral values of the functions yield
complicated expressions which cannot be easily calculated, let alone 
inverted analytically. The solution was to code numerical algorithms
to calculate the integrals (i.e. cumulants) explicitly.

After the integrals are calculated, their inverse and the subsequent
sampling value can again be obtained numerically. Namely, resorting to
the original definition of the unitarity sampling recipe in
Eq. \ref{e:usampling}, by replacing the normalised $\rm g_i(x)$ with:
\begin{equation}
 g_i(x) \to \frac{f(x)}{\int\limits_{x_{-}}^{x_{+}} f(x)\, dx},
\end{equation}
which in turn gives:
\begin{equation}
\int\limits_{x_{-}}^{x} f(x)\; dx = r \cdot \int\limits_{x_{-}}^{x_{+}} f(x)\; dx,
\label{e:sampling}
\end{equation}
where $\rm f(x)$ is the non-negative function one wants to sample from, 
$\rm [x_{-},x_{+}]$ is the range of values of the parameter $\rm x$ we want to sample 
and $\rm r$ a pseudo random number $r \in [0,1]$. As already stated
(c.f. Eq.\ref{e:usampling}), in the case when the integral of the function $\rm f(x)$ 
is an analytic function, $F(x) = \int_{x_{-}}^x f(x)\, dx$, and has a known inverse $\rm F^{-1}(x)$ one 
can construct explicit unitary prescriptions by:
\begin{equation}
x = F^{-1}\left( r \cdot [ F(x_{+})-F(x_{-}) ] + F(x_{-}) \right)
\end{equation}
as given for two particular cases in Eq. \ref{e:xnuni},\ref{e:bwuni}.

In the cases the integral can not be inverted, the prescription of the
Eq. \ref{e:sampling} can directly be transformed into a zero-finding request;
thus, since both the integral and the first derivative (i.e. the sampling
function and its cumulant) are known, the Newton-Rhapson method is chosen as the
optimal one for root finding:

\begin{eqnarray}
g(x) &=& \left\{ \int\limits_{x_{-}}^{x} f(x) dx - r \cdot \int\limits_{x_{-}}^{x_{+}} f(x) dx \right\} = 0\\
g'(x)  &=& \frac{d}{dx} \left\{ \int\limits_{x_{-}}^{x} f(x) dx - r \cdot \int\limits_{x_{-}}^{x_{+}} f(x) dx \right\} =  f(x)
\end{eqnarray}

With a sensible choice of starting points the procedure generally takes
on the order of ten cycles until finding the root with adequate numerical
precision. The overall generation speed is still deemed quite
reasonable. 

\noindent
The integration of the phase-space suppressed resonant propagator of
Eq. \ref{e:lamsbw} yields a rather non-trivial expression:
\begin{eqnarray}
\int\limits_{(m_a+m_b)^2}^{s}f_{\rm R}(s)~ ds &=& \int\limits_{(m_a+m_b)^2}^{s} \frac{\sqrt{\lambda(s,m_a^2,m_b^2)}\, ds}{\sqrt{s}\cdot((s-M^2)^2+M^2 \Gamma^2)}\\  \notag
&=& \int\limits_{a}^{s}\frac{\sqrt{(s-a)(s-b)}~ ds}{\sqrt{s}\cdot((s-M^2)^2+M^2 \Gamma^2)}\\ \notag
&=& \frac{1}{{\sqrt{-b}}\,\Gamma\,M^2} \times  \frac{-2\,i \,a\,b\,\Gamma\,}{( \Gamma^2 + M^2 ) }\\ \notag  &\times& 
\biggl\{  
           \Mfunction{F}\left[ i \,{\rm arcsinh}(\frac{{\sqrt{-b}}}{{\sqrt{a}}}),\frac{a}{b} \right]
-\Mfunction{F}\left[ i \, {\rm arcsinh}(\frac{{\sqrt{-b}}}{{\sqrt{s}}}),\frac{a}{b}\right] \\ \notag
           &+& ( i \,\Gamma + M ) \, 
           ( a + i \,( \Gamma + i \,M ) \,M ) \,
           ( b + i \,( \Gamma + i \,M ) \,M ) \, 
           \Mfunction{\Pi}\left[ \frac{M\,( -i \,\Gamma + M ) }{b}, i \,{\rm arcsinh}(\frac{{\sqrt{-b}}}{{\sqrt{a}}}),
            \frac{a}{b} \right] \\  \notag
           &+& ( \Gamma + i \,M ) \,
           ( b + ( -i \,\Gamma - M ) \,M ) \,
           ( i \,a + ( \Gamma - i \,M ) \,M ) \,
           \Mfunction{\Pi}\left[ \frac{M\,( i \,\Gamma + M ) }{b},
            i \,{\rm arcsinh}(\frac{{\sqrt{-b}}}{{\sqrt{a}}}),
            \frac{a}{b}\right]  \\  \notag
           &-& ( i \,\Gamma + M ) \,
           ( a + i \,( \Gamma + i \,M ) \,M ) \,
           ( b + i \,( \Gamma + i \,M ) \,M ) \,
           \Mfunction{\Pi}\left[ \frac{M\,( -i \,\Gamma + M ) }{b},
            i \,{\rm arcsinh}(\frac{{\sqrt{-b}}}{{\sqrt{s}}}),\frac{a}{b}\right]\\  \notag
           &-& ( \Gamma + i \,M ) \,
           ( b + ( -i \,\Gamma - M ) \,M ) \,
           ( i \,a + ( \Gamma - i \,M ) \,M ) \,
           \Mfunction{\Pi}\left[ \frac{M\,( i \,\Gamma + M ) }{b},
            i \,{\rm arcsinh}(\frac{{\sqrt{-b}}}{{\sqrt{s}}}),
            \frac{a}{b}\right]   \biggr\}
\end{eqnarray}
where the variables $a,b$ stand for $\rm a=(m_a+m_b)^2$ and $\rm b=(m_a-m_b)^2$
and the functions $\rm \Mfunction{F}[\varphi,k]$ and $\rm
\Mfunction{\Pi}[\varphi,k,n]$ are the Legendre's incomplete elliptic integrals
of the second and third kind with complex arguments. In order to perform the
calculations the latter functions had to be coded from scratch since they were
not found in any (publicly available) computer libraries or code repositories.
The prescriptions for calculating them were found in \cite{carlson}; the results
were checked against the values given by {\tt Mathematica$^{\rm TM}$}.

\noindent In the special case $\rm m_a = m_b$ the above expression simplifies into:
\begin{eqnarray}
\int\limits_{(2 m_a)^2}^{s}f_{\rm R}(s)~ ds &=& \int\limits_{(2 m_a)^2}^{s} \frac{\sqrt{\lambda(s,m_a^2,m_a^2)}\, ds}{\sqrt{s}\cdot((s-M^2)^2+M^2 \Gamma^2)}\\ \notag
&=& \int\limits_{a}^{s}\frac{\sqrt{(s-a)}~ ds}{\cdot((s-M^2)^2+M^2 \Gamma^2)}\\ \notag
&=&  \frac{1}{\Gamma\,M\,
     {\sqrt{a + \left( -i \,\Gamma - M \right) \,M}}}\\ \notag
&\times& 
\biggl\{
\left( i \,a + \left( \Gamma - i \,M \right) \,M \right) \,
      \arctan (\frac{{\sqrt{-a + z}}}
        {{\sqrt{a + \left( -i \,\Gamma - M \right) \,M}}}) \\ \notag
&-& 
     i \,{\sqrt{a + \left( -i \,\Gamma - M \right) \,M}}\,
      {\sqrt{a + i \,\left( \Gamma + i \,M \right) \,M}}\,
      \arctan (\frac{{\sqrt{-a + z}}}
        {{\sqrt{a + i \,\left( \Gamma + i \,M \right) \,M}}})  \biggr\}
\end{eqnarray}

\noindent The result of integrating the phase-space suppressed non-resonant
propagator (Eq. \ref{e:lamxnu}) yields a similarly non-trivial result:

\begin{eqnarray}
\int\limits_{(m_a+m_b)^2}^{s} f_{\rm NR}(s)~ds &=& \int\limits_{(m_a+m_b)^2}^{s} \frac{\sqrt{\lambda(s,m_a^2,m_b^2)}~ds}{s^{\nu+1}}\\
&=& \notag
  \frac{1}{2\,{\sqrt{1 - \frac{s}{a}}}\,\nu}  \biggl\{
\frac{-2\,{\sqrt{\left( a - s \right) \,\left( b - s \right) }}\,
         \Mfunction{F_1}\left[ -\nu,-\left( \frac{1}{2} \right) ,
         -\left( \frac{1}{2} \right) ,1 -
         \nu,\frac{s}{a},\frac{s}{b}\right]}{s^\nu\, {\sqrt{1 -
         \frac{s}{b}}}} \\  \notag &+& \frac{{\sqrt{\pi }}\,{\sqrt{\left( -a +
         b \right) \, \left( a - s \right)
         }}\,\Mfunction{\Gamma}\left[1 - \nu\right]\,
         \Mfunction{F}\left[-\nu,-\left( \frac{1}{2} \right) ,
         \frac{3}{2} - \nu,\frac{a}{b}\right]}{a^\nu\,{\sqrt{1 -
         \frac{a}{b}}}\, \Mfunction{\Gamma}\left[\frac{3}{2} -
         \nu \right]} \biggr\} 
\end{eqnarray}

where the function $\rm F[\alpha,\beta,\gamma,x]$ is the Gauss Hypergeometric
function and the $\rm F_1[\alpha,\beta,\beta',\gamma,x,y]$ is the two-parameter
(Appell) Hypergeometric function \cite{appell}. Both functions can be calculated
by using the prescriptions in \cite{appell}; it however turns out that the
calculation of the $\rm F_1[\alpha,\beta,\beta',\gamma,x,y]$ to a certain (high)
precision is almost two times slower than the explicit numerical calculation of
the integral to the same precision. Subsequently the numerical evaluation of the
Gauss Hypergeometric function $\rm F[\alpha,\beta,\gamma,x]$ was retained since
it participates in the $\rm m_a = m_b$ simplification and the calculation of the
integral was done by using a 50-point Gauss-Legendre quadrature with $\sqrt{s}$
weight function; the weights were calculated by \cite{quad}. The implementation of the
(Appell) Hypergeometric function calculation was used as a cross-check to confirm the
correct implementation and precision of the numerical method.

\noindent As already mentioned, the above integral again simplifies for $\rm m_a = m_b$:
\begin{eqnarray}
\int\limits_{(2 m_a)^2}^{s} f_{\rm NR}(s)~ds &=& \int\limits_{(2 m_a)^2}^{s} \frac{\sqrt{\lambda(s,m_a^2,m_b^2)}~ds}{s^{\nu+1}} \\ \notag
&=& \int\limits_{a}^{s} \frac{\sqrt{(s-a)}~ds}{s^{\nu+\frac{1}{2}}} \\ \notag
&=& \frac{2}{3}\,a^{1-\nu}\,s^{\frac{3}{2}}\,\Mfunction{F}\left[ \frac{3}{2},\nu+\frac{1}{2},
      \frac{5}{2},-s \right], 
\end{eqnarray}
and the Gauss Hypergeometric function $\rm F[\alpha,\beta,\gamma,x]$ is in this case
calculated by the methods described in  \cite{appell} with some improvements analogous to
the ones described e.g. in \cite{numrec}. 
\clearpage

\subsection{Application of the Phase Space Generation Algorithms \label{s:num2}}

The {\tt AcerMC 2.0} Monte-Carlo generator uses the multi-channel phase space
generation where each channel corresponds to an expected phase space topology as
derived from the participating Feynman diagrams. In {\tt AcerMC 2.0} this
information was obtained from the modified {\tt MadGraph}\cite{Madgraph} program
which also supplied the probability amplitudes for the implemented
processes. Each channel topology was in turn constructed from the t-type and
s-type modules and sampling functions described in this paper together with some
additional importance sampling techniques for space angles and rapidity
distributions described in detail elsewhere
\cite{Pythia62,fermisv,excal}. The unknown slope parameters (denoted $\nu$
in the text, c.f. Eq. \ref{e:lamxnu}) of the invariant mass sampling functions
for non-resonant propagators were obtained by short training runs of the program
on a process by process basis.

As a further step the multi-channel self-optimisation procedure was implemented
in order to minimise the variance of the event weights further
\cite{Kleiss1994}. Eventually, additional smoothing of the phase space was
obtained by using a modified {\tt VEGAS} routine to improve the generation
efficiency (see next Section).

The procedure of multi-channel importance sampling used in the event generation
can briefly be outlined as follows. The analytically integrable function
$g(\vec{\Phi})$ (c.f. Eq. \ref{e:gsampling}, which aims to approximate the
peaking behaviour of the differential cross-section dependence on various
kinematic quantities is introduced into the differential cross-section equation
as a weighted sum of several channels $g_i(\vec{\Phi})$, each adapted to a
certain event topology:
\begin{equation}
g(\vec{\Phi})  = \sum_i \alpha_i \cdot g_i(\vec{\Phi}). 
\end{equation}
The values of relative weights $\alpha_i$ are determined from  multi-channel
self-optimisation procedure  in order to minimise the variance of the 
weights $w(\vec{\Phi})$ \cite{Kleiss1994}. The phase space points are than sampled from
the function $g(\vec{\Phi})$, first by randomly choosing a channel $i$ according to the
relative frequencies $\alpha_i$ and then deriving the required four momenta from the
chosen $g_i(\vec{\Phi})$ using unitary\footnote{Unitary in this context meaning that 
there is no event rejection in the algorithm.} algorithms \cite{fermisv}. 

\begin{figure}[ht]
\begin{center}
     \epsfig{file=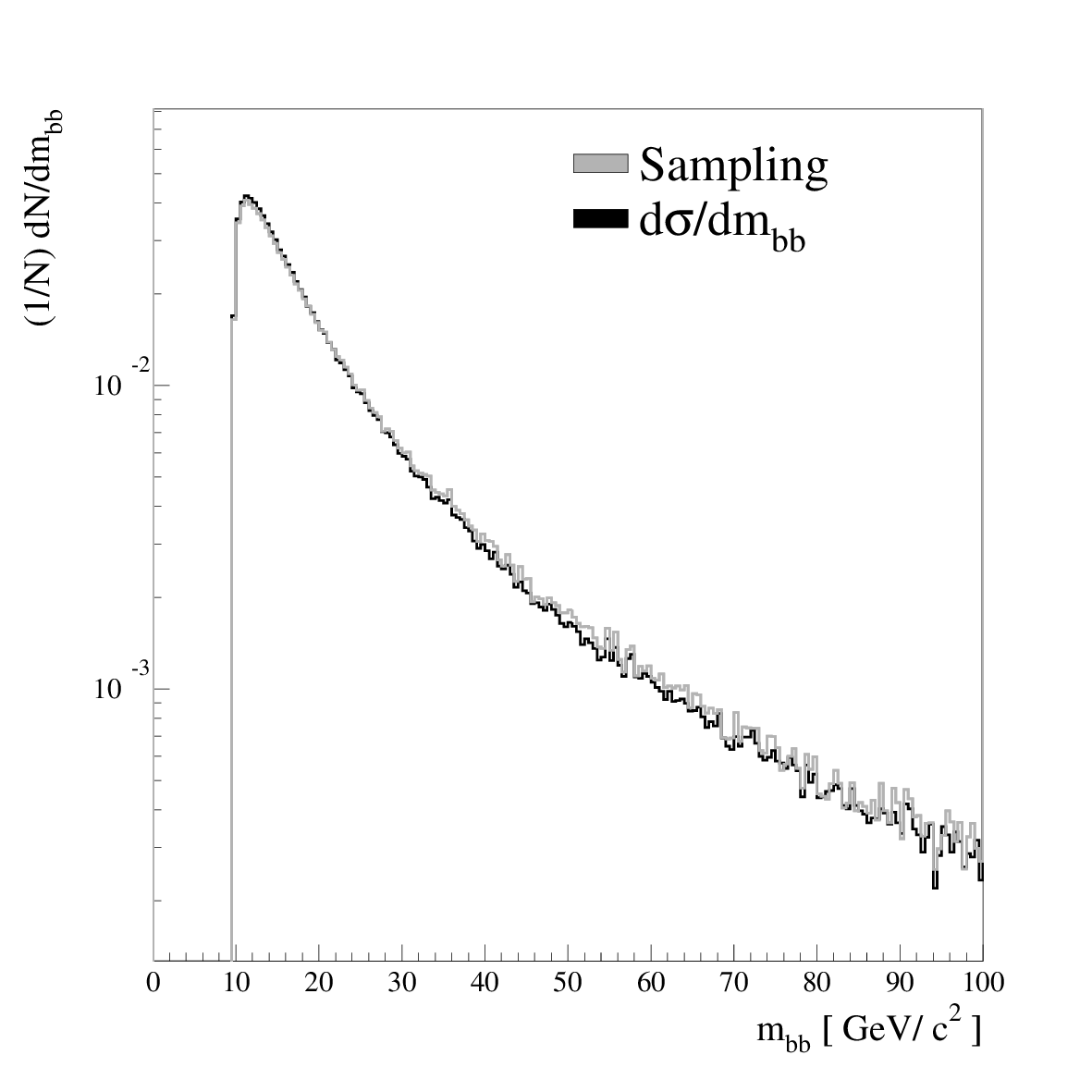,width=5.3cm}\hspace{-0.3cm}
     \epsfig{file=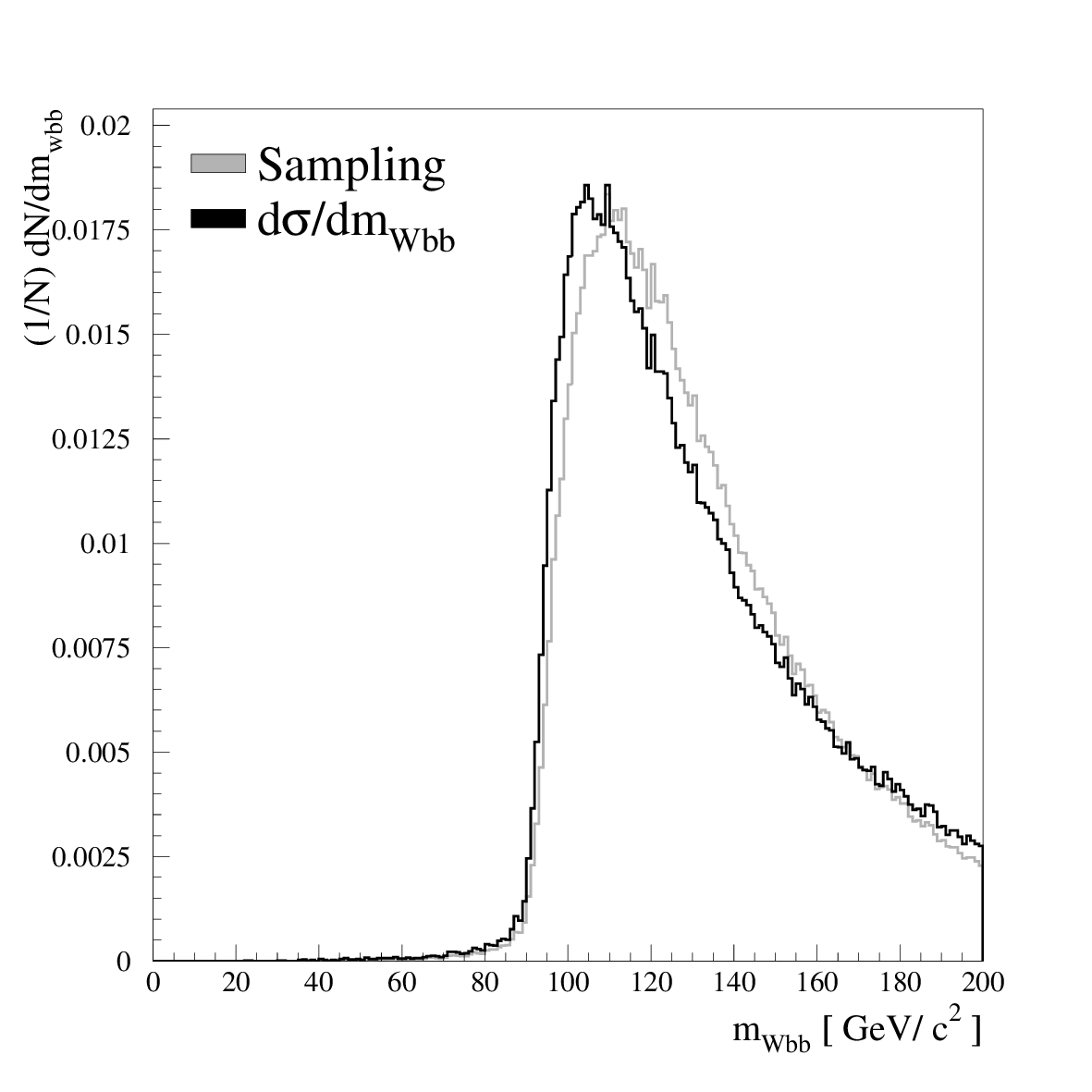,width=5.3cm}\hspace{-0.3cm}
     \epsfig{file=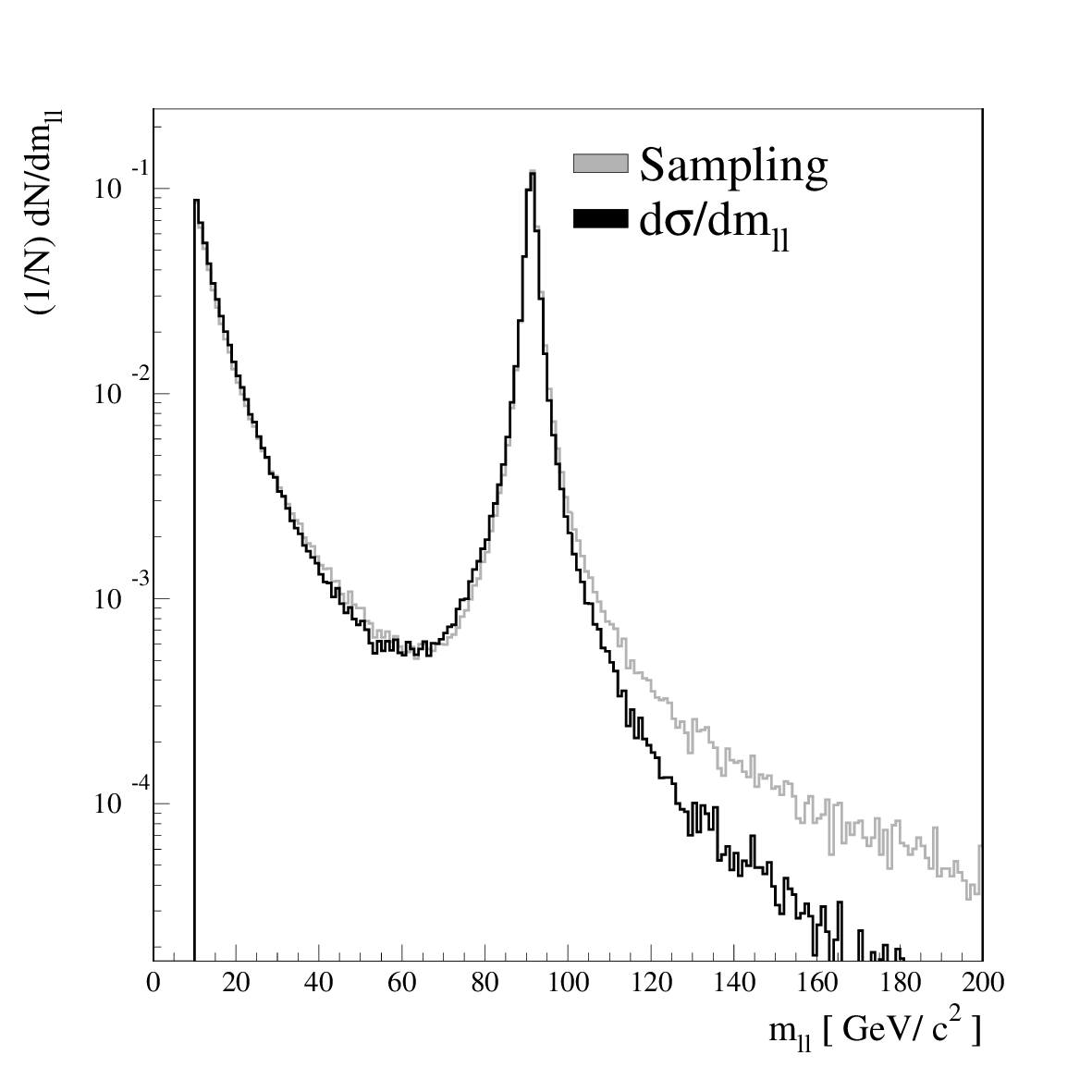,width=5.3cm}\vspace{-0.5cm}
\end{center}
\isucaption{
\small A few representative invariant mass distribution comparisons between the
(normalised) sampling functions and the normalised differential cross-section as obtained
with {\tt AcerMC 2.0} Monte--Carlo generator.  Left: The invariant mass of the
$\rm b \bar{b}$ pair in the process $\rm u \bar{d} \to W^+ g^* \to l^+ \nu_l b
\bar{b}$.  Center: The invariant mass of the $\rm W b \bar{b}$ system
(equivalently the hard centre-of-mass energy $\rm \sqrt{\hat{s}}$) for the same
process.  Right: The invariant mass of the $\rm \ell \bar{\ell}$ pair in the
process $g g \to Z^0/\gamma^* b \bar{b} \to \ell \bar{\ell} b \bar{b}$. All the
distributions were obtained using the prescriptions of this paper without the
adaptive algorithms also used in the {\tt AcerMC 2.0} Monte--Carlo generator. As
one can see the approximations used seem to work quite well.
\label{f:acermc}} 
\end{figure}

A few representative invariant mass distribution comparisons
between the implemented sampling functions and the actual differential
distributions are shown in Figure \ref{f:acermc}.

\subsubsection{Modified {\tt VEGAS} Algorithm \label{s:ac-veg}}

Using the described multi-channel approach, the total generation (unweighting)
efficiency amounts to about $3-10\%$ depending on the complexity of the chosen process.
 In order to further improve the efficiency, a set of modified {\tt VEGAS}
\cite{vegas} routines was used as a (pseudo-)random number generator for
sampling the peaking quantities in each kinematic channel. The conversion into a
{\it (pseudo-)random number generator} consisted of re-writing the calling routines
so that instead of passing the analysed function to {\tt VEGAS} for sampling and
integration, {\tt VEGAS} calls produce only (weighted) random numbers in the region
$[0,1]$ and the corresponding sampling weight, while the {\tt VEGAS} grid training is
done using a separate set of calls.

\begin{figure}[ht]
\begin{center}
\mbox{
     \epsfxsize=7cm
     \epsffile{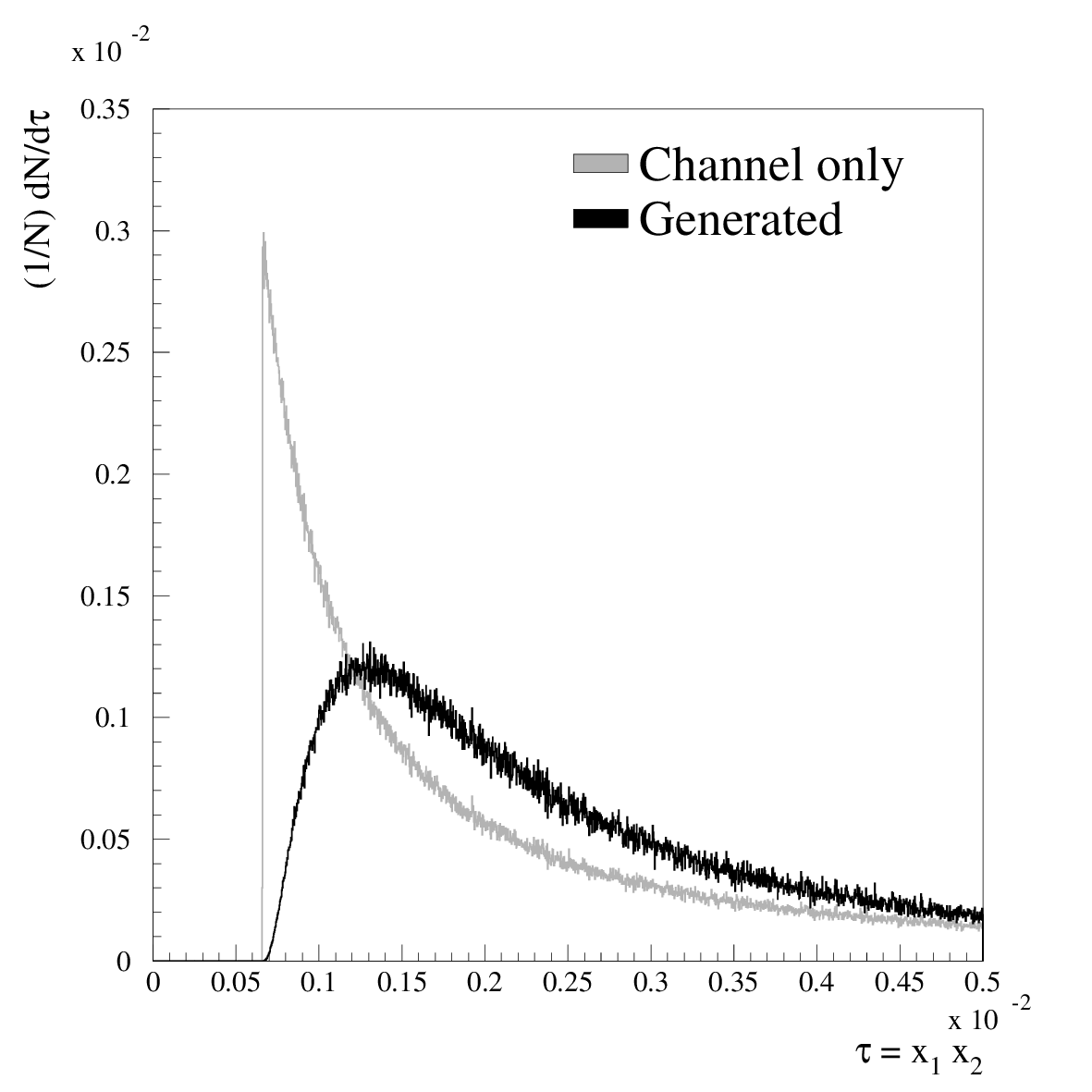}
     \epsfxsize=7cm
     \epsffile{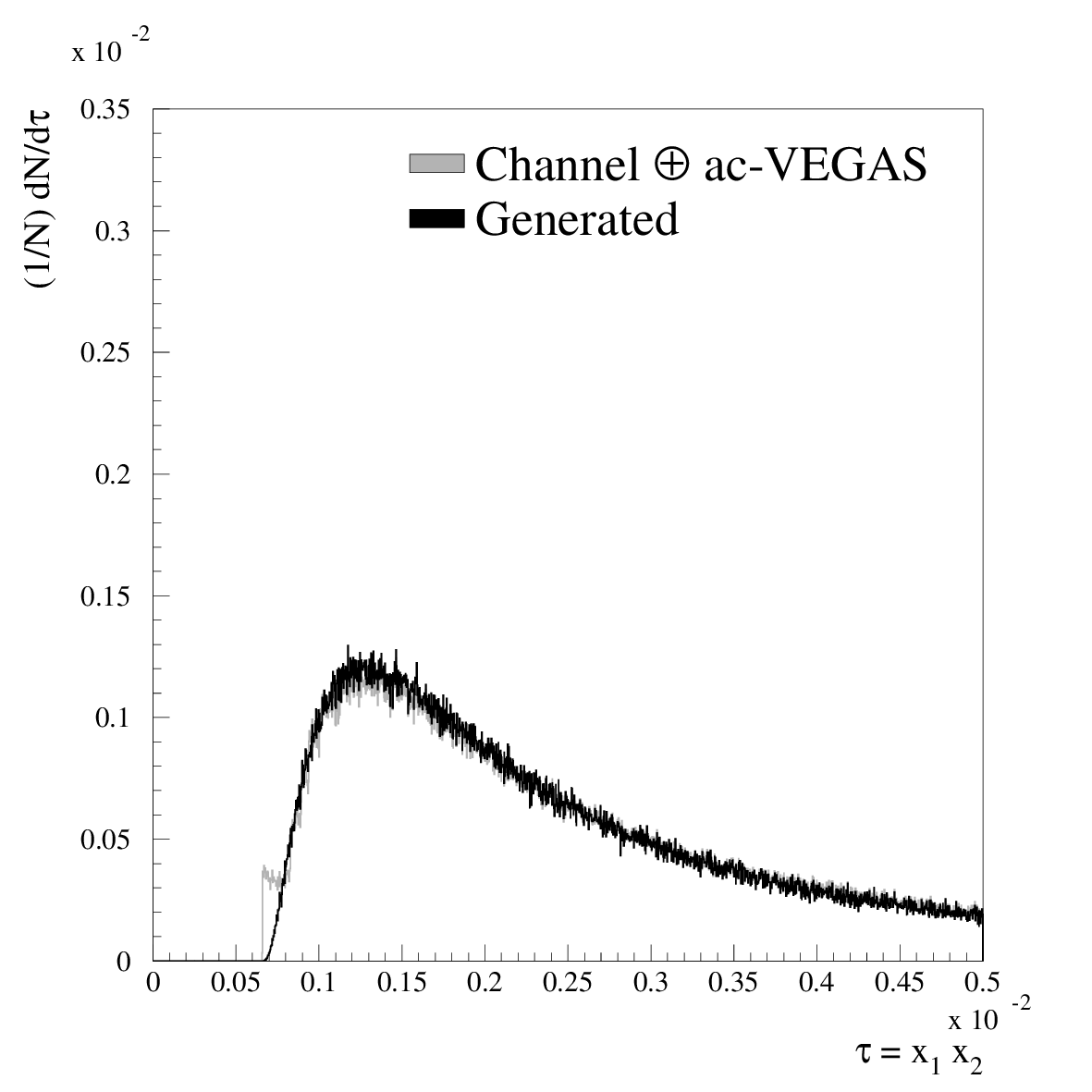}
}
\end{center}
\vspace{-0.4cm}
\isucaption{\em
Comparison between the sampling distribution for the $\tau = \hat{s}/s
\in [\tau_{\rm min},1]$ variable in $g g -> t \bar{t} b \bar{b}$
process before and after the application of modified {\tt ac-VEGAS}
\cite{vegas} smoothing procedure (light gray histogram). The generated (normalised) 
differential cross-section is also drawn (black histogram, labelled {\sf Generated}).
\label{f:tauveg}}
\end{figure}

After training all the sampling grids (of dimensions 4-7, depending on the
kinematic channel), the generation efficiency increased to the order of
$6-14\%$. The motivation for this approach was that in unitary algorithms
only a very finite set of simple sampling functions is available, since the
functions have to have simple analytic integrals for which an inverse function
also exists. Consequently, the non-trivial kinematic distributions can not be
adequately described by simple functions at hand in the whole sampling domain
(e.g. the $\tau$ distribution, c.f. Figure \ref{f:tauveg}) and some additional
smoothing might be welcome. In addition, the random numbers distributions
should, due to the applied importance sampling, have a reasonably flat behaviour
to be approached by an adaptive algorithm such as {\tt VEGAS}\footnote{At this
point also a disadvantage of using the adaptive algorithms of the {\tt VEGAS}
type should be stressed, namely that these are burdened with the need of
training them on usually very large samples of events before committing them to
event generation.}. 

The further modification of {\tt VEGAS}, beside adapting it to function as a
(pseudo-)random number generator instead of the usual {\it integrator}, was based on
the discussions \cite{ohl,foam} that in case of event generation,
i.e. unweighting of events to the weight one, reducing the maximal value of
event weights is in principle of higher importance than achieving the minimal
weight variance. Since the {\tt VEGAS} algorithm was developed with the latter
scope, some modification of the algorithm was necessary. As it turned out, the
modification was fairly easy to implement: Instead of the usual cumulants:
\begin{equation}
<I>_{\rm cell} ~= \sum_{\rm cell} {\rm wt}_i,
\end{equation}
according to the size of which {\tt VEGAS} decides to split its cells, the values:
\begin{equation}
<F>_{\rm cell} ~= \Delta_{\rm cell} \cdot {\rm wt}^{\rm max}_{\rm cell} \;- 
\sum_{\rm cell} {\rm wt}_i,
\end{equation}
were collected and used as the splitting criterion. The above value
(called {\it loss integral} in \cite{foam}) is basically a measure of the deviation
between the maximal weight sampled in the given cell ${\rm wt}^{\rm max}_{\rm
cell}$ and the average weight in the cell $<\rm wt_{\rm cell}> = (\sum_{\rm cell} {\rm
wt}_i)/\Delta_{\rm cell}$ (the quantity $\Delta_{\rm cell}$ denoting the cell
width, i.e. the integration range). Re-writing the above expression as:
\begin{equation}
<F>_{\rm cell} ~= \bigl( \Delta_{\rm cell} \cdot {\rm wt}^{\rm max}_{\rm cell}
\bigr)
\cdot \biggl\{ 1 - \frac{<\rm wt_{\rm cell}>}{{\rm wt}^{\rm max}_{\rm cell}} \biggr\}
\end{equation}
clearly indicates that the value $<F>_{\rm cell}$ is actually a measure of the
generation {\it inefficiency} in the cell, since the term in the curly brackets is
equivalent to one minus the generation efficiency $<\rm wt_{\rm cell}>/{\rm
wt}^{\rm max}_{\rm cell}$. In addition, the inefficiency is weighted with the
{\it crude}/maximal estimation of the function integral over the cell $\Delta_{\rm
cell} \cdot {\rm wt}^{\rm max}_{\rm cell}$ and cells with the highest $ <F>_{\rm
cell}$ are split. 

This method is of relevance because the {\tt VEGAS} cells are actually
projections of the whole phase space on the (chosen) side axes, i.e. {\tt VEGAS}
cannot isolate a maximal weight in a certain point in phase-space and build a
cell around it, which in principle would be an ideal solution.
An implementation with this scope in view has been made in {\tt FOAM}
\cite{foam}, nevertheless we have not found it competitive with respect to
the modified {\tt VEGAS} for the given application.                                                          

The thus modified {\tt ac-VEGAS} algorithm further increased the unweighting
efficiency for almost a factor of two.

One of the sampling distributions is shown in Fig.~\ref{f:tauveg} as a gray
histogram (marked {\it channel}) and the actual ({\it generated})
differential cross-section dependence is drawn in black. In the first figure,
the random variable used for sampling values from $1/\tau^\mu$ distribution was
drawn from a flat probability in the interval $[0,1]$; in the second plot the
{\tt ac-VEGAS} algorithm was used to give an optimal grid for sampling the random
variables needed for parameter generation (the grid is trained for each
kinematic channel separately, the sum of all channels is shown in the plot). The
improvement is evident; one has to stress that the use of {\tt ac-VEGAS} algorithm
to generate the values of $\tau$ directly would be much less efficient since
{\tt VEGAS} gives a grid of 50 bins/dimension, which would give a very crude
description of the $\tau$ distribution compared to the one at hand.
\begin{figure}[ht]
\begin{center}
\mbox{
     \epsfxsize=8cm
     \epsffile{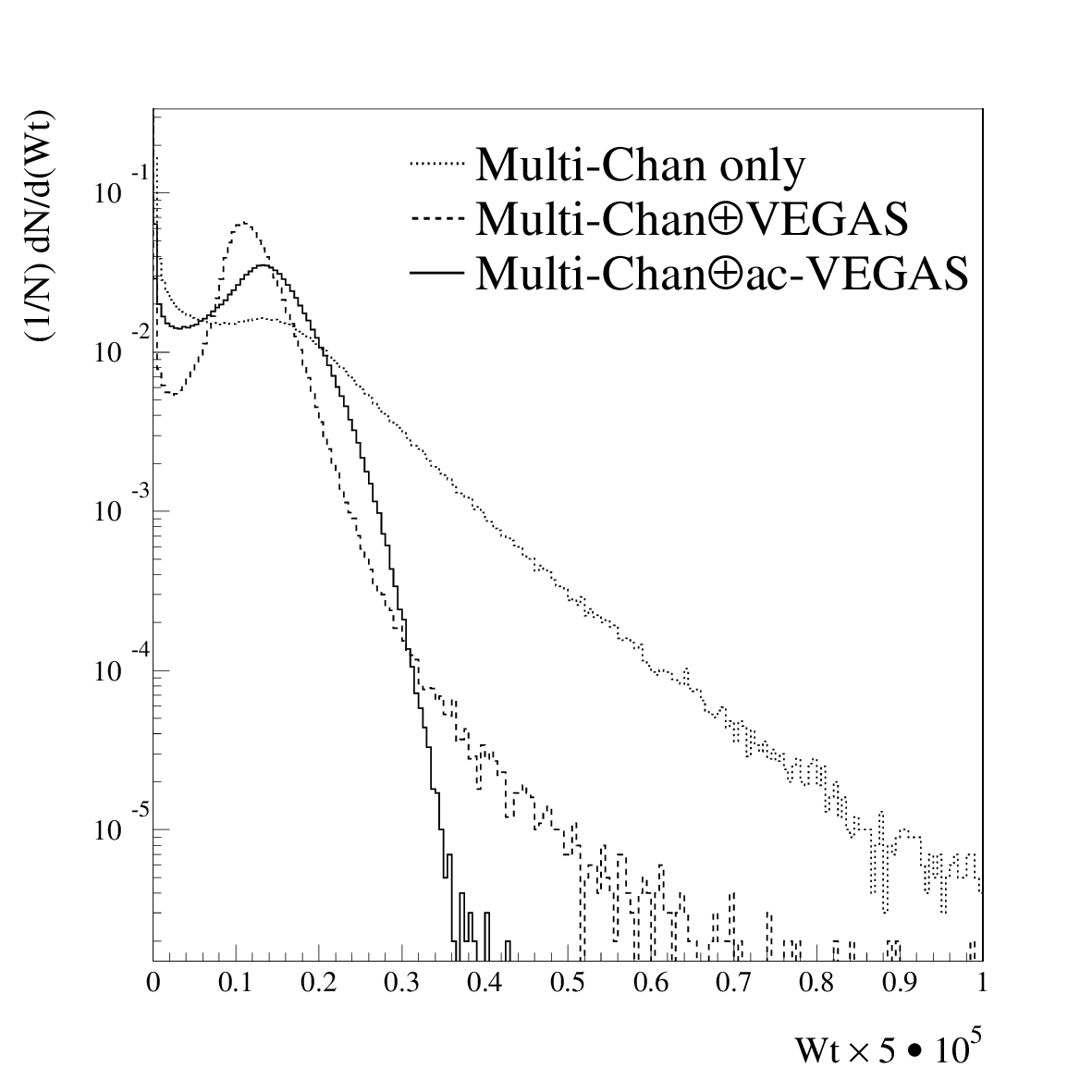}
}
\end{center}
\vspace{-0.5cm}
\isucaption{\em
The distribution of event weights using only the Multi-Channel approach (dotted histogram)
 and after application of {\tt VEGAS} (dashed histogram) and {\tt ac-VEGAS} (full histogram) 
algorithms in the $ g g \to (Z^0 \to) l \bar{l} b \bar{b}$ process. 
\label{f:weights}}
\end{figure}

Observing the distributions of the event weights before and after the inclusion
of the modified {\tt ac-VEGAS} algorithm (Fig.~\ref{f:weights}) it is evident
that {\tt ac-VEGAS} quite efficiently clusters the weights at lower values.
Note that the principal effect of original {\tt VEGAS} is indeed to cluster event
weights in a narrow region, nevertheless a tail towards the high-weight region
remains. On the other hand, the {\tt ac-VEGAS} efficiently reduces the tail in
the high weight region; only a few of the event weights still retain their large
values, thus reducing the generation efficiency.  Given the difference in
distributions, the observed increase of the generation efficiency seems
relatively modest. To better understand this result one should consider that the
formula for the MC generation efficiency is given by:
\begin{equation}
\epsilon = \frac{<{\rm wt}>}{\rm wt_{\rm max}},
\label{e:cleff}
\end{equation}
where $<{\rm wt}>$ is the average weight of the sample and equals the total
event cross-section, while $\rm wt_{\rm max}$ represents the maximum event weight in
the applied generation procedure and is determined through a pre-sampling run
with a high statistic.  Since the average weight $<{\rm wt}>$ equals the total
cross-section of the process, it remains (necessarily) unchanged after the
application of the {\tt VEGAS} refining; consequently the change of efficiency
results in the reduction of the maximum weight $\rm wt_{\rm max}$ by approximately a
factor two, which is from technical point of view quite an achievement.

A further step to profit from the clustering of weights induced by {\tt ac-VEGAS}
is to adopt a re-definition of the MC generation efficiency as proposed by
\cite{bhlumi,foam}. In this approach, the alternative definition of $\rm wt_{\rm
max}$ is: For a given precision level $\alpha << 1$, the $\rm wt_{\rm max}$ is
determined from the total weight distribution in such a way that the
contribution of the events exceeding this value to the total weight sum
(i.e. cross-section integral) equals $\alpha$. Such a quantity is referred to as
$\rm wt^\alpha_{\rm max}$ and the efficiency expression becomes:
\begin{equation}
\epsilon = \frac{<{\rm wt}>}{\rm wt^\alpha_{\rm max}}.
\label{e:neweff}
\end{equation}

The argument presented in \cite{bhlumi,foam} seems to be quite reasonable since
the {\it true} event weight is in any case only estimated from a finite sample of
events and the new definition simply takes into account a certain level of
accuracy in the maximum weight determination. In addition, certain very weak
singularities that might exist in the simulated process and might occasionally
result in a very high event weight are automatically taken into account.  The
use of new $\rm wt^\alpha_{\rm max}$ consequently results in a generation efficiency
of about $\epsilon \geq 20\%$ for all the implemented processes, which is a
significant improvement in terms of time needed for MC generation.

\boldmath
\subsubsection{\label{s:cols}Colour Flow Information}
\unboldmath

Before the generated events are passed to {\tt PYTHIA/HERWIG} to complete the
event generation, additional information on the colour flow/connection of the
event has to be defined.  Below we discuss the implemented method of the colour
flow determination on the example of two processes, $ g g \to t \bar{t} b \bar{b}$ 
and $ q \bar{q} \to t \bar{t} b \bar{b}$.

\begin{figure}[ht]
\vspace{-0.3cm}
\begin{center}
\mbox{
     \epsfxsize=10cm
     \epsffile{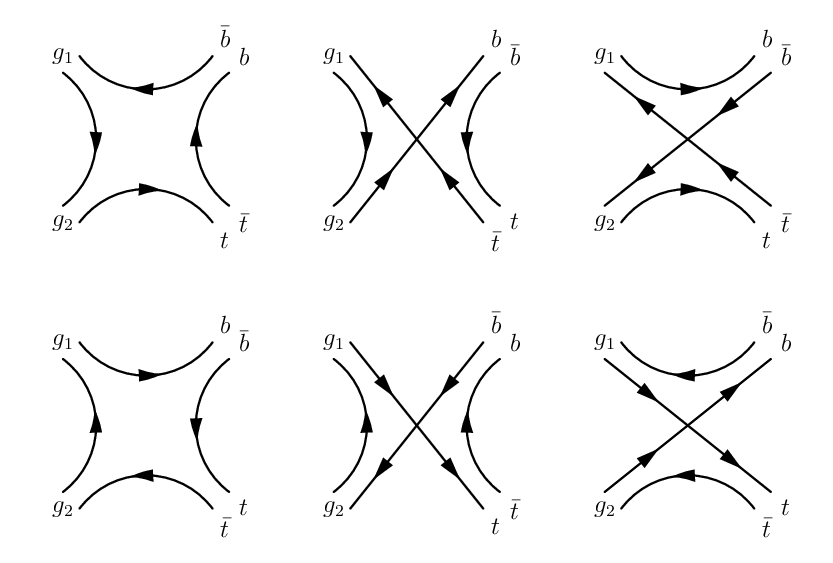}
}
\end{center}
\vspace{-0.3cm}
\isucaption{\em 
A diagrammatic representation of the six colour flow configurations in the process
$g g \to t \bar{t} b \bar{b}$. Certain colour combinations, leading for example to 
colourless (intermediate) gluons, are not allowed.
\label{f:cols}}
\vspace{-0.5cm}
\end{figure}

For the process $ g g \to t \bar{t} b \bar{b}$ six colour flow configurations are
possible, as shown in Figure \ref{f:cols}. With 36 Feynman diagrams contributing to the
process and at least half of them participating in two or more colour flow configurations,
calculations by hand would prove to be very tedious. Consequently, a slightly modified
colour matrix summation procedure from {\tt MADGRAPH} \cite{Madgraph} was used to
determine the colour flow combinations of the diagrams and the corresponding colour
factors. The thus derived squared matrix elements for separate colour flow combinations
$|{\mathcal{M}_{\rm flow}}|^2$ were used as sampling weights on an event-by-event basis to
decide on a colour flow configuration of the event before passing it on to {\tt
PYTHIA/HERWIG} for showering and fragmentation. The procedure was verified to give
identical results regarding the colour flow combinations and corresponding colour factors
when applied to the processes published in \cite{cflows}. As one can see this approach
neglects the interference terms between the distinct colour-ordered amplitudes and is
indeed exact only in the $N_C \to \infty \;$ limit\footnote{The matrix elements used in
the cross-section calculation and event generation are of course complete and do not
employ any approximation.} \cite{mangano_cflow,alpha}. Since there is no {\it a priori}
rule of how to split the interference terms between the colour ordered amplitudes this
approach is generally deemed to be the best one can do; recent developments in this field
\cite{odagiri} however suggest additional improvements to the method that indeed might be
incorporated into later versions of {\bf AcerMC}.

\begin{figure}[ht]
\vspace{-0.4cm}
\begin{center}
\mbox{
     \epsfxsize=8cm
     \epsffile{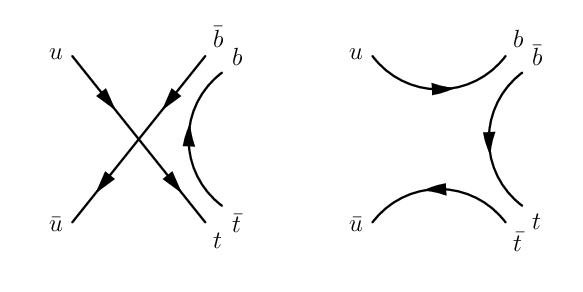}
}
\end{center}
\vspace{-0.3cm}
\isucaption{\em 
A diagrammatic representation of the two colour flow configurations in the process
$q \bar{q} \to t \bar{t} b \bar{b}$.
\label{f:colsqq}}
\end{figure}

The colour flow configuration in the $ q \bar{q} \to t \bar{t} b \bar{b}$ channel is much
simpler since only two colour flow topologies exist (Fig. \ref{f:colsqq}); the choice
between the two has been solved in a manner identical to the one for the $ g g \to t
\bar{t} b \bar{b}$ process, as described above.

Some specifications of the implemented matrix-element-based processes: number of
Feynman diagrams, channels used in the phase-space generation and colour flow
configurations are collected in Table~\ref{t:acmeps}. The {\it control} processes {\sf ID=91-94}
are omitted from the table due to their simplicity.

\begin{table}[ht]
\vspace{0.25cm}
\newcommand{\lstrut}{{$\strut\atop\strut$}}
  \isucaption {\em Some details on matrix-element-based process implementation in {\bf
AcerMC} library. In case of $ q \bar {q}$ initial state the number of Feynman diagrams
corresponds to one flavour combination. The $f=e,\mu,\tau,b$. 
\label{t:acmeps}}
\vspace{-0.3cm} 
\begin{center}
\begin{tabular}{|c|c|c|c|c|} \hline \hline
Process id & Process specification & Feyn. diagrams & Channels & Colour flows \\ 
\hline\hline
 1      &  $gg  \to   t \bar t b \bar b $ & 36 & 12 & 6 \\
\hline
 2      &  $q \bar q  \to   t \bar t b \bar b $ & 7 & 5 & 2 \\
\hline
 3      &  $q \bar q  \to W(\to \ell \nu)  b \bar b $ & 2 & 2 & 1 \\
\hline
 4      &  $q \bar q  \to W(\to \ell \nu)  t \bar t $ & 2 & 2 & 1 \\
\hline
 5      &  $gg \to Z/\gamma^*(\to \ell \ell)  b \bar b $ & 16 & 6 & 2 \\
\hline
 6      &  $q\bar q  \to Z/\gamma^*(\to \ell \ell)  b \bar b $ & 8 & 6 & 1 \\
\hline
 7      &  $gg \to Z/\gamma^*(\to f f, \nu \nu)  t \bar t $ & 16 & 6 & 2 \\
\hline
 8      &  $q \bar q  \to Z/\gamma^*(\to f f, \nu \nu)  t \bar t $ & 8 & 6 & 1 \\
\hline
 9      &  $gg  \to (Z/W/\gamma^* \to) t \bar t b \bar b $ & 72 & 20 & 12 \\
\hline
 10      &  $q \bar q  \to (Z/W/\gamma^* \to) t \bar t b \bar b $ & 28 & 12 & 5 \\
\hline
 11      &  $gg   \to (t \bar t \to) f \bar f b f \bar f \bar b $ & 3 & 2 & 2 \\
\hline
 12      &  $q \bar q   \to (t \bar t \to) f \bar f b f \bar f \bar b $ & 1 & 1 & 1 \\
\hline
 13      &  $gg   \to (WW b \bar b) f \bar f  f \bar f b \bar b $ & 31 & 13 & 2 \\
\hline
 14      &  $q \bar q   \to (WW b \bar b \to) f \bar f  f \bar f b \bar b $ & 14 & 7 & 1 \\
\hline
 15      &  $gg   \to t \bar t t \bar t $ & 72 & 10 & 6 \\
\hline
 16      &  $q \bar q   \to t \bar t t \bar t $ & 14 & 4 & 2 \\
\hline \hline
\end{tabular}
\end{center}
\end{table}

\boldmath
\subsection{The $\alpha_{\rm QED}$ and $\alpha_{s}$ calculations \label{s:alphas}} 
\unboldmath

Native functions of running $\alpha_{\rm QED}(Q^2)$ and $\alpha_{s}(Q^2)$ have been
implemented inside {\bf AcerMC} with the main objective of providing a means to keep the
(total) cross-sections of the processes unchanged when interfacing with the two
supervising generators, since the implementations of the two functions in {\tt
PYTHIA} and {\tt HERWIG} differ to some extent. Especially the  $\alpha_{s}(Q^2)$ is
subject to  experimental and theoretical uncertainties, however obtaining a
different cross-sections for the same {\bf AcerMC} process due to different interface,
could be regarded (at least to some extent) as an inconsistency\footnote{The values  will 
still differ by a small amount in processes containing W bosons (processes 3,4) 
due to different values of the CKM matrix in the two supervising generators.}. 
\begin{figure}[hbt]
\vspace{-0.5cm}
\begin{center}
\mbox{
     \hspace{-0.3cm}
     \epsfxsize=7.4cm
     \epsffile{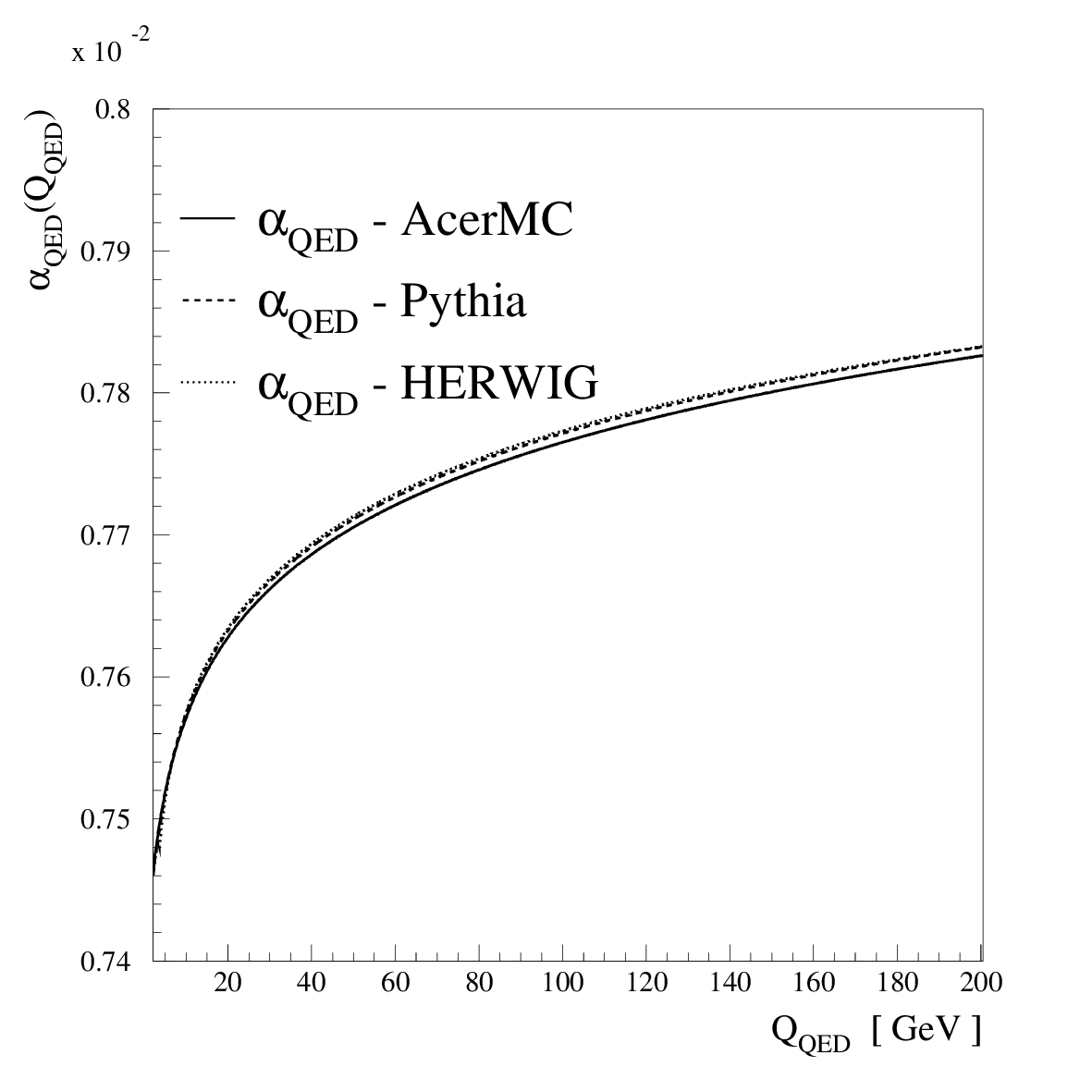}
     \hspace{-0.7cm}
     \epsfxsize=7.4cm
     \epsffile{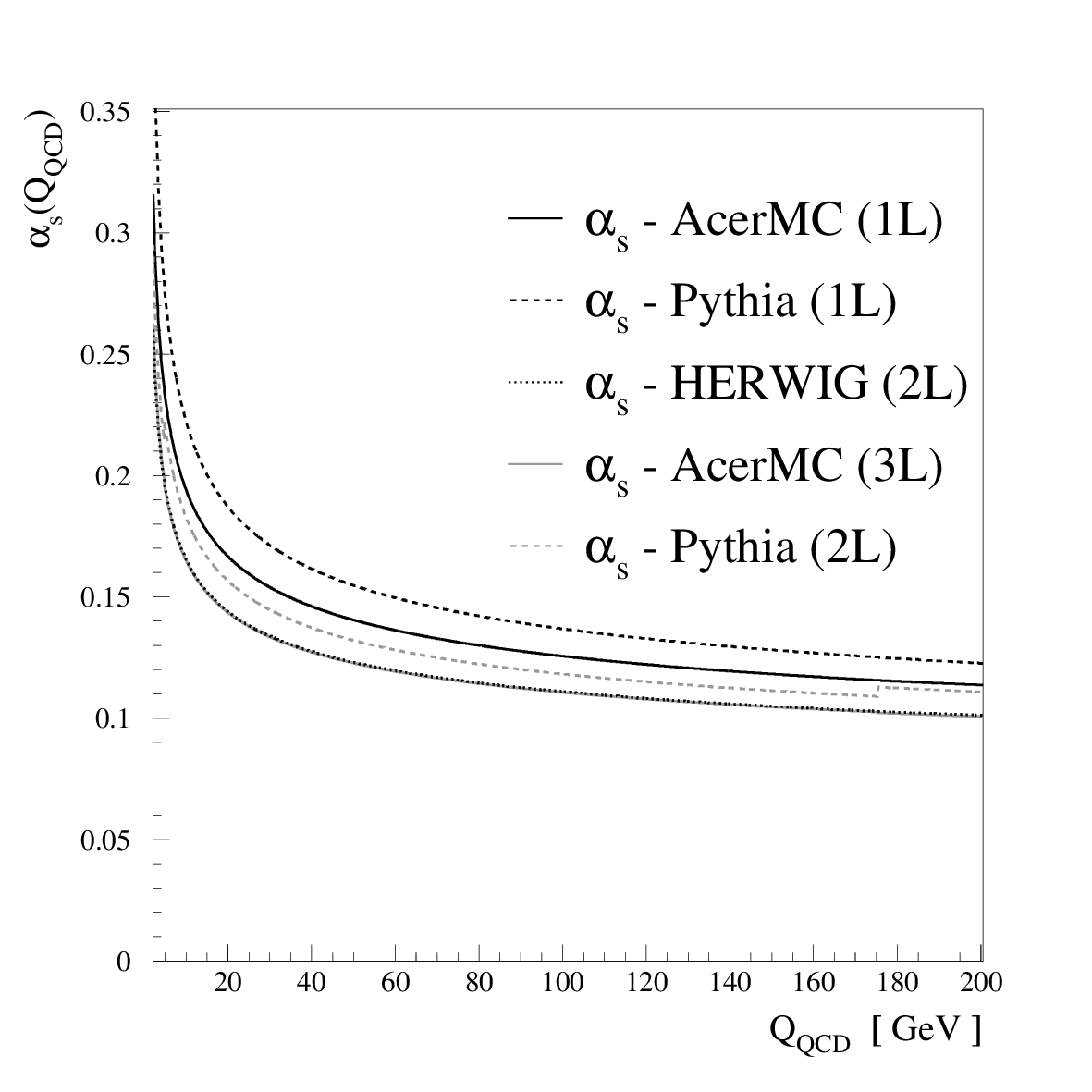}
}
\end{center}
\vspace{-0.9cm}
\isucaption{\em
Comparison between the $\alpha_{\rm QED}(Q^2)$ ({\bf Left}) and $\alpha_{s}(Q^2)$ ({\bf
Right}) implementations in {\bf AcerMC}, {\tt PYTHIA} and {\tt HERWIG}. For
$\alpha_{s}(Q^2)$ calculations with different loop orders ({\sf L}) are given where
applicable.
\label{f:alps}}
\vspace{-0.2cm}
\end{figure}

\begin{itemize}
\item {\boldmath $\alpha_{\rm QED}$} is implemented in {\bf AcerMC} using the formulae given
in \cite{field} and is in complete accordance with the implementations in {\tt
PYTHIA} and {\tt HERWIG} apart from the updated hadronic component published recently by
Burkhardt {\it et. al.} \cite{burk2001}. As one can see in Figure \ref{f:alps}, the latter 
minimally lowers the $\alpha_{\rm QED}$ values.
\item{\boldmath $\alpha_{s}$} has one and three loop implementations in {\bf AcerMC}
following the calculations of W. J. Marciano \cite{marciano} and using
$\Lambda^{(nf)}_{\bar{MS}}$ transformations for flavour threshold matches. The
three loop version gives good agreement with the {\tt HERWIG} implementation
(both functions have been set to the same $\Lambda^{(nf=5)}_{\bar{MS}}$ value)
as one can see in Figure \ref{f:alps}. The {\tt PYTHIA} two loop
implementation deviates somewhat from the latter two; the kinks observed in the plot 
are due to approximate $\Lambda^{(nf)}_{\bar{MS}}$ transformations at flavour thresholds, 
which are exact to one loop only.
\end{itemize}

Although the {\bf AcerMC} and {\tt PYTHIA} one loop implementations are
identical in form the resulting values differ by a small amount because the
default {\tt PYTHIA} implementation reads the $\Lambda^{(nf=4)}_{\bar{MS}}$
value from {\tt LHAPDF} instead of the $\Lambda^{(nf=5)}_{\bar{MS}}$ one used by
{\bf AcerMC} and {\tt HERWIG}; the difference thus occurs due to
$\Lambda^{(nf)}_{\bar{MS}}$ propagation at flavour thresholds.


\boldmath
\section{Structure of the package} 
\unboldmath

The {\bf AcerMC} package consists of a library of the matrix-element-based generators
for selected processes, interfaces to the {\tt PYTHIA 6.4} , {\tt ARIADNE 4.1}  and {\tt HERWIG 6.5}
generators, sets of data files and three main programs: 
{\tt demo\_hw.f}, {\tt demo\_py.f} and {\tt demo\_ar.f}. Provided makefiles
allow to build the executables with either of these generators as the 
{\it supervising generator}:
{\tt demo\_hw.e}, {\tt demo\_py.e} and {\tt demo\_ar.e}.
\enlargethispage{2cm}

\boldmath
\subsection{Main event loop and  interface to {\tt  PYTHIA/HERWIG, TAUOLA and PHOTOS}} 
\unboldmath

\begin{figure}[ht]
\begin{center}
\mbox{
     \epsfxsize=\linewidth
     \epsffile{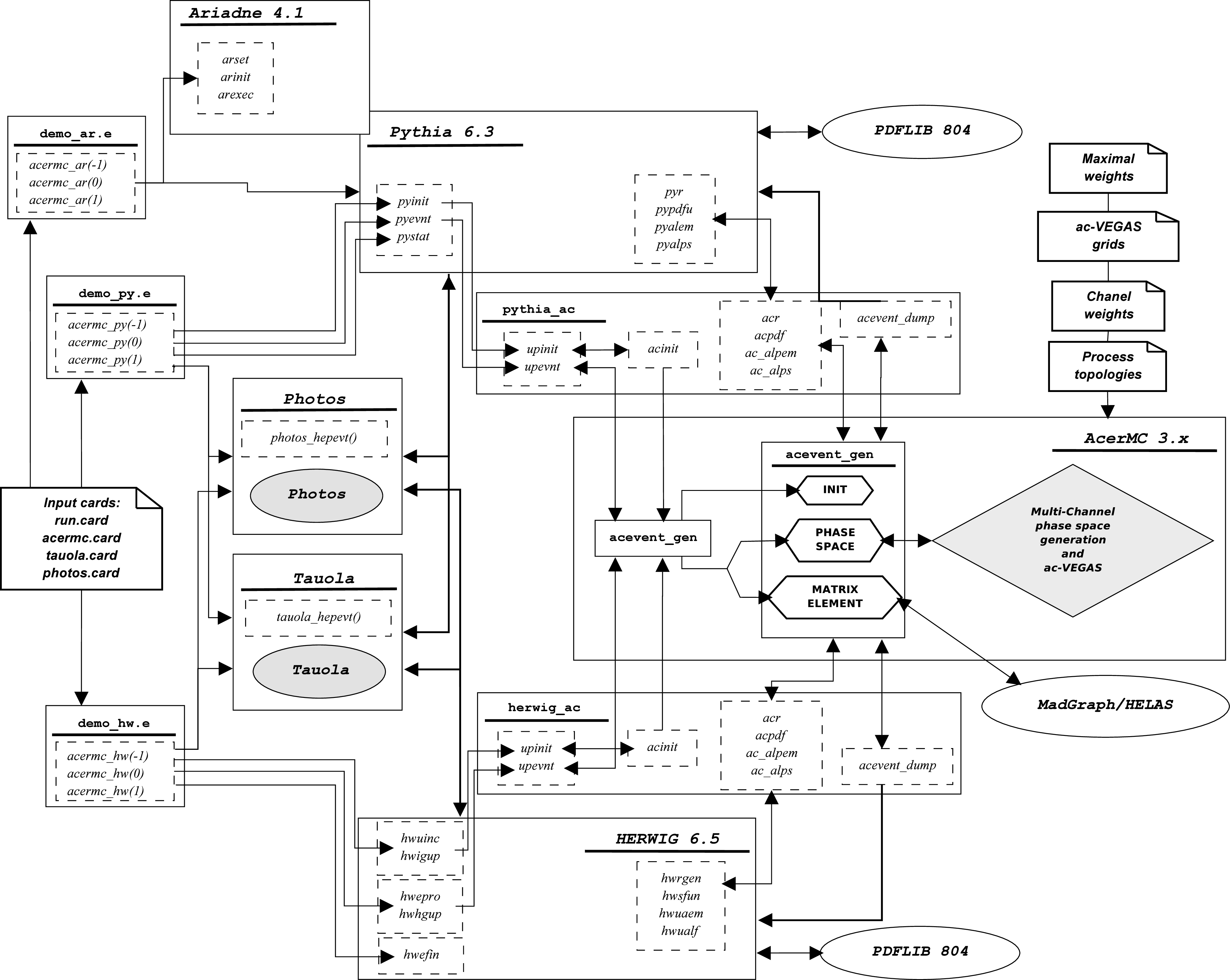}
}
\end{center}
\caption{\em
The calling sequence of the main event generation routine {\tt
acevent\_xx}.  The routine is called either through {\tt demo\_py
$\to$ acermc\_py} sequence when interfacing the {\tt PYTHIA~6.4}
generator or {\tt demo\_hw $\to$ acermc\_hw} sequence when the {\tt
HERWIG~6.5} is linked. When the {\tt ARIADNE 4.1} setup is called via
{acermc\_ar} calls it still relies on {\tt PYTHIA~6.4} for
hadronisation and particle(resonance) decays.  The structure of the
interface subroutines and relations with the corresponding ones from
supervising generators and/or external libraries is also evident.
\label{f:callstruc}}
\vspace{0.4cm}
\end{figure}

The main event loop is coded in the {\tt demo\_hw.f}, {\tt demo\_py.f} and {\tt demo\_ar.f} files,
where the opening/closing of the input/output files, reading of the data-cards
and event-loop execution is performed.  Main event loop consists only
of calls to the {\tt acermc\_py},  {\tt acermc\_ar} or {\tt acermc\_hw} subroutines, with parameter
{\tt MODE = -1,~0,~1} respectively set for initialisation, generation and finalisation of
the event loop.  The call to {\tt acermc\_xx} activates respective procedures of
the supervising generator, which in  turn activates the {\tt acevtgen}
procedure steering the native {\bf AcerMC} generation of the matrix element
event. Fig.~\ref{f:callstruc} illustrates this calling sequence in some details.

As one can deduce from the diagram in Fig.~\ref{f:callstruc}, certain functions
called by {\bf AcerMC}, as e.g. pseudo-random number generator {\tt acr} are
re-routed through the interfaces to the linked supervising generator, depending
on the choice at compilation time (e.g. {\tt acr} function giving
(pseudo-)random numbers is linked to either {\tt pyr} or {\tt hwrgen} as shown
in the plot), providing the internal consistency of the package. The generated
event is rewritten to the format required by the supervising generator by means
of the {\tt acdump\_xx} routines. While {\tt ARIADNE} provides an alternative (colour-dipole)
based implementation of initial and final state radiation it relies on {\tt PYTHIA} 
for hadronisation and particle decays.

The {\tt pythia\_ac.f} and  {\tt herwig\_ac.f} files contain sets of re-routing/interface
functions, specialised for the respective supervising generator. The main library 
of {\bf AcerMC} is well screened from dependencies on the supervising generator,
all dependencies are hidden in {\tt herwig\_ac.f} and {\tt pythia\_ac.f} respectively.

The {\tt PYTHIA}, {\tt ARIADNE} and {\tt HERWIG} libraries remain essentially 
untouched\footnote{For specification of exceptions see Section 7.4, 7.5. and 7.6},
without introducing any dependencies on the {\bf AcerMC} code.
The input cards are common for all interfaces.

The calling sequence is further enhanced by the optional calls to the external 
{\tt TAUOLA} \cite{TAUOLA} library, which handles the $\tau$-lepton decays using the
spin information of the hard process, and the {\tt PHOTOS} \cite{PHOTOS} library which 
adds the final state QED radiation to final state leptons and hadrons. The two interfaces
are controlled by two additional input cards. The details of the interface are given
in Section \ref{s:tauphot}.

\newpage

\boldmath
\subsection{Structure of the {\bf AcerMC} matrix-element and phase-space code} 
\unboldmath

\begin{figure}[ht]
\begin{center}
\mbox{
     \epsfxsize=\linewidth
      \epsffile{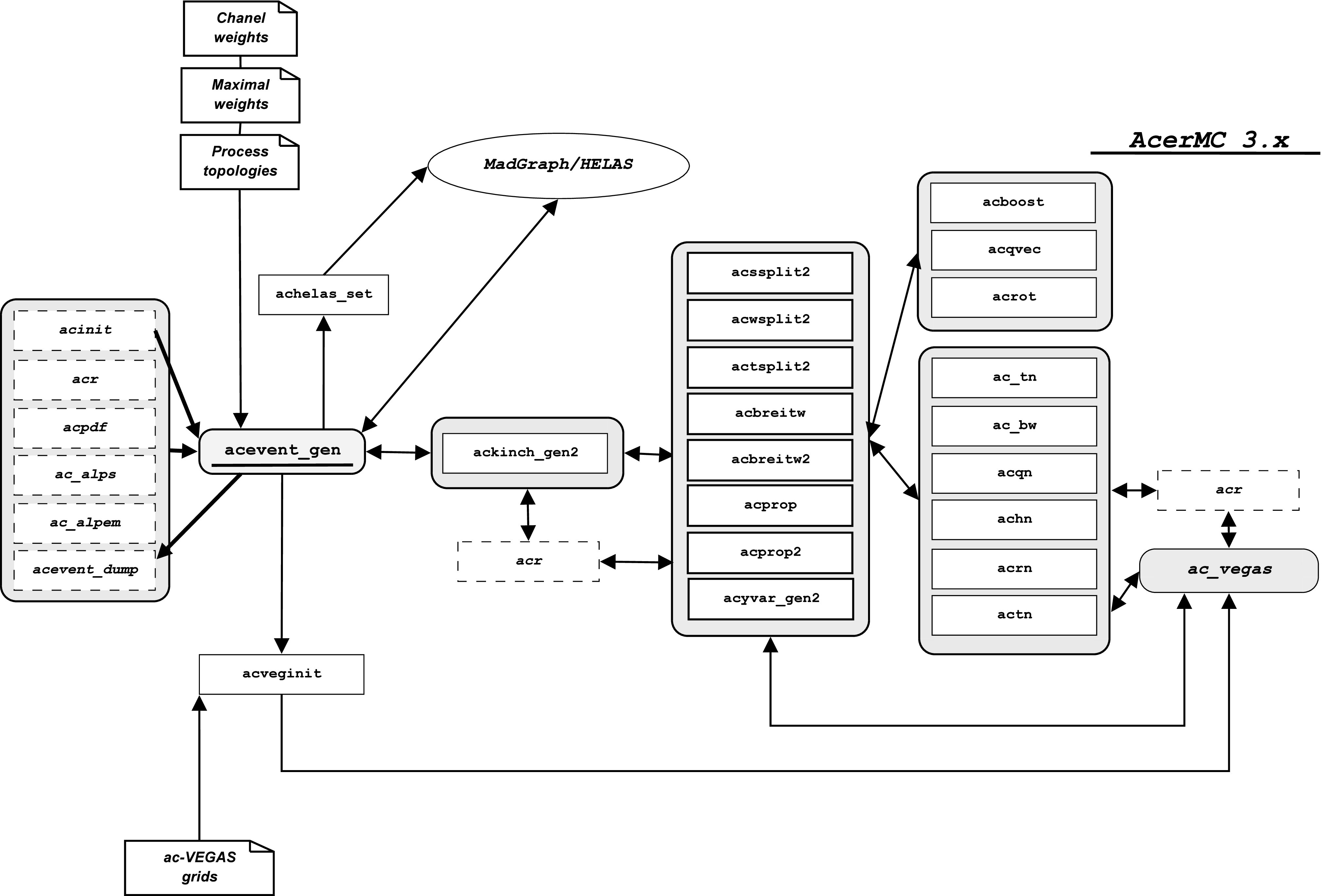}
}
\end{center}
\caption{\em
The event generation sequence controlled by {\tt acevent\_gen} subroutine. Phase
space generation is sequenced by calling the {\tt ackinch\_gen2} routine
to obtain the incoming gluon and the outgoing four-momenta of the participating
particles.  The latter routine handles the possible momenta permutations and
calls the explicit four-momenta generation (and PS weight calculation) in the
sequence prescribed by the event topology. These (channel-specific) routines are
constructed from common building blocks listed in the next two columns.  The
{\tt acevent\_gen} routine also initialises {\tt MADGRAPH/HELAS} package and
retrieves the matrix element values. All the generated four-momenta, as well as
the event weight are finally passed back to the supervising generator via the
{\tt acevent\_dump} call.
\label{f:ttbbflow}}

\vspace{0.25cm}
\end{figure}

The {\bf AcerMC} core code performs the generation of a matrix-element-based event.
Fig.~\ref{f:ttbbflow} illustrates the calling sequence for
generating an unweighted event. The steering subroutine is called
{\tt acevent\_gen} which subsequently constructs the weight by calling the Madgraph/HELAS subroutines
for the matrix element evaluation.  To stress again, this subroutine calls only a
sequence of the native {\bf AcerMC} subroutines, any call to the supervising
generator goes via the respective interface function/subroutine. A more detailed
representation of calling sequence is shown in the Figure \ref{f:ttbbflow}.

\begin{figure}[ht]
\begin{center}
\mbox{
     \epsfxsize=0.6\linewidth
      \epsffile{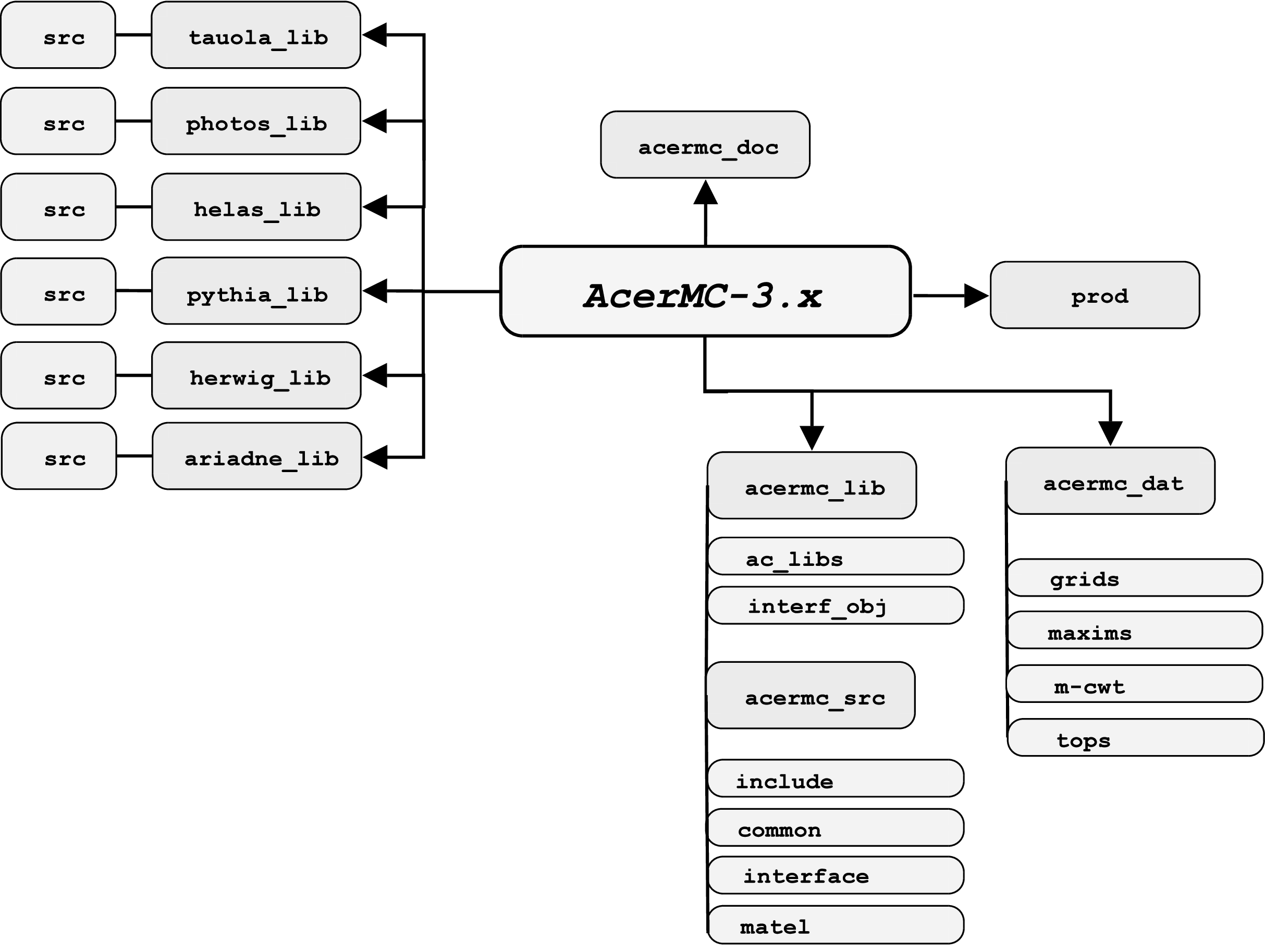}
}
\end{center}
\caption{\em The structure of the {\bf AcerMC} directories.
\label{f:acstruct}}
\end{figure}

Code for the phase space generation is since the version {\tt 2.0} greatly
simplified compared to the earlier versions. Code for matrix element calculations is
grouped together for all processes in subdirectory {\tt matel}. Code with
different utility subroutines, e.g. kinematic transformations used by all
subprocesses, is in the subdirectory {\tt common}.  Subdirectory {\tt interface}
contains code with interfaces to supervising generators, finally
subdirectory {\tt include} contains all include files. The overall view on 
the structure of the {\bf AcerMC} directories is shown in Fig.~\ref{f:acstruct}.

\vspace{0.25cm}

The core code builds one library  {\tt libacermc.a}. 

\boldmath
\subsection{Data files for the phase-space optimisation} 
\unboldmath

The {\bf AcerMC}  matrix-element-based generators are very highly optimised, using
multi-channel optimisation and additional improvement with the {\tt ac-VEGAS} grid. 
The generation modules require three kinds of the input data
to perform the generation of unweighted events:
\begin{itemize}
\item A file descrbing the construction of the topologies relevant for the
chosen process, the implemented sets are stored in the directory  {\tt acermc\_dat/tops}.
\item A file containing the list of the values of relative channel weights obtained by
the multi-channel optimisation, defaults being stored in {\tt acermc\_dat/m-cwt}.
\item A file containing the pre-trained {\tt ac-VEGAS} grid, the pre-trained
(default) ones located in {\tt acermc\_dat/grids}.
\item A file containing the maximum weight $\rm wt_{\rm max}$,  $\alpha$-cutoff
maximum weight $\rm wt_{\rm max}^\alpha$ and the 100 events with the highest
weights, the default ones being provided in {\tt acermc\_dat/maxims}.
\end{itemize}
In case of changing the default running conditions, like parton density functions or
centre-of-mass energy, the user should repeat the process of preparation of the listed
data files containing the inputs for the phase-space generator modules in order to
preserve the initial event generation efficiency.

\begin{figure}[hb]
\vspace{0.2cm}
\begin{center}
\mbox{
     \epsfxsize=0.55\linewidth
      \epsffile{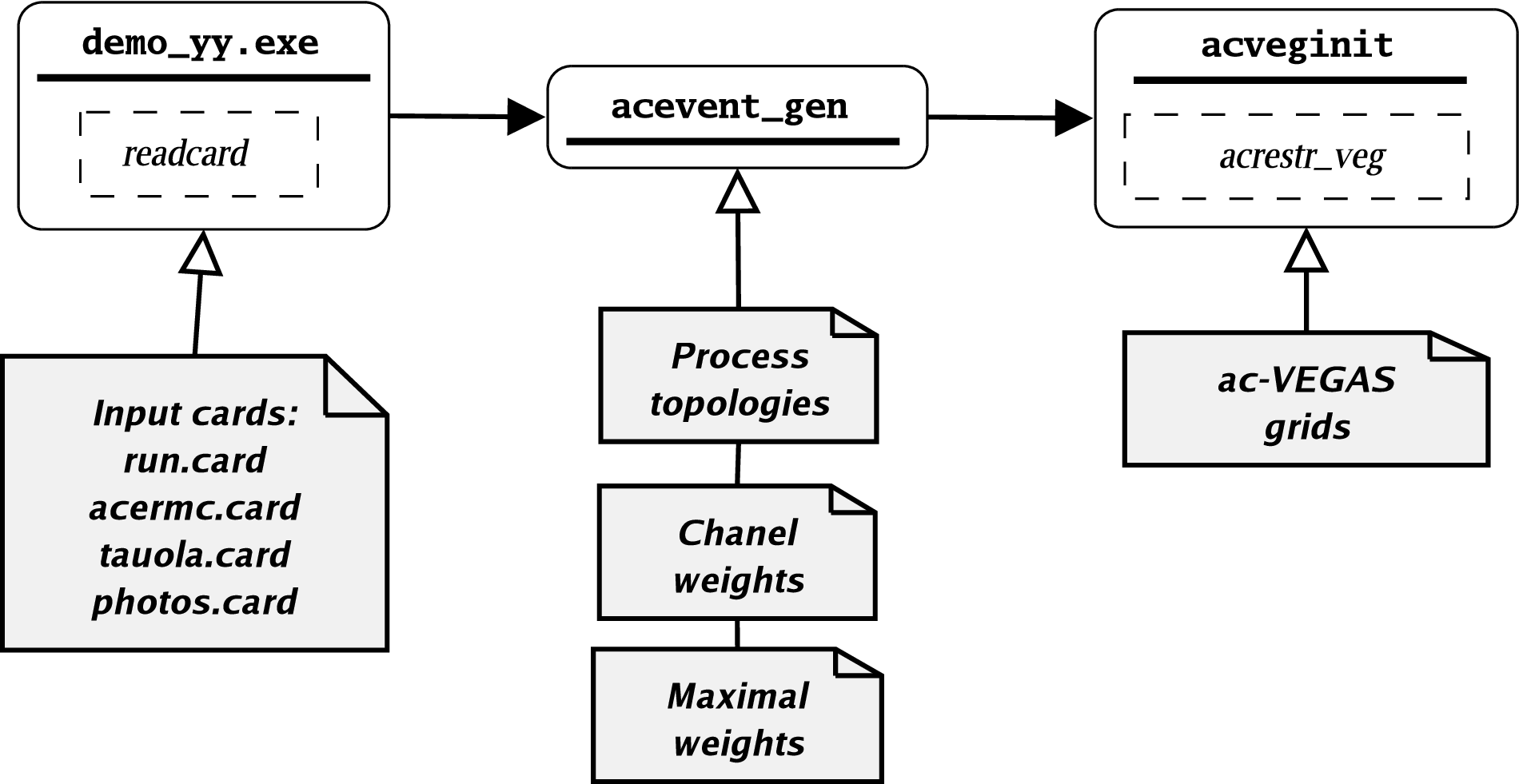}
}
\end{center}
\vspace{-0.2cm}
\caption{\em The reading sequence of the input files and the performing subroutines 
in the {\bf AcerMC} code.
\label{f:acread}}
\end{figure}

The reading sequence of data files inside {\bf AcerMC} is shown in Figure \ref{f:acread}.

Pre-trained data sets are obtained using $\sqrt{s}=14\;$ TeV, {\tt PYTHIA} default
$\alpha_s(Q^2)$ and $\alpha_{\rm QED}(Q^2)$ and CTEQ5L (parametrised) parton density function
set and are provided for each implemented process\footnote{These can also be used for a series of 
other settings, see Section \ref{s:training} for details}.  For these, the relative channel
weights are stored in the {\tt INCLUDE} files in {\tt acermc\_src/include/chanwt\_xx.inc}
where {\tt xx} denotes the process id (c.f. Table \ref{T3:1}); the
default/pre-trained {\tt ac-VEGAS} grids are listed in the directory {\tt
acermc\_dat/grids/vscalA\_xxYYY.veg}, where {\tt A} denotes the scale choice of the
process {\tt xx} and {\tt YYY} denotes the cutoff value of the $m_{Z^0/\gamma^*}$ for the
{\bf AcerMC} processes {\tt xx = 05 $\to$ 08}. The files containing the maximal weights
$\rm wt_{\rm max}$ and $\rm wt_{\rm max}^\alpha$ as well as the 100 events with the highest
weights are stored in the directory {\tt acermc\_dat/grids/vtmaxA\_xxYYY.dat}, following
the same labelling convention. Both the trained {\tt ac-VEGAS} grids and the weight files
were obtained from test runs with at least $2 \cdot 10^6$ weighted events being generated.

The number of required input files might at first look seem large, considering that many
event generators do not require any input files for operation; the difference is not in so
much in the complexity of the phase space generation as in the fact that many event
generators require a {\it warming run} instead, i.e. before the generation of unweighted
events is performed a certain number of weighted events (typically of the order of $10^4$)
is generated in order to obtain the relative multi-channel weights (in case multi-channel
phase space generation is used) and/or the optimised {\tt VEGAS} grid and/or an estimate
of the maximal weight. Such an approach can have an advantage when event generation is
very fast and the phase space regions with the highest weights are well known (as done for
the $2 \to 2$ processes in {\tt PYTHIA}); on the other hand, when the phase space topology
of the process is more complex and the event generation is comparatively slow, generating
a relatively small number of e.g. $10^4$ weighted events {\it every time} a generator is
started can become CPU wasteful and/or inaccurate in terms of maximum weight estimation.

Reasonably accurate estimation of the latter is namely crucial for correct event
unweighting; event generators using {\it warming-up} method for maximal weight search often
find still higher weights during the production run and reset the maximal weight
accordingly. In this case however, statistically correct approach would be to reject all
events generated beforehand and start the event generation anew, which is almost never
implemented due to the CPU consumption and the possibility of hitting a weak singularity
(the same argument leads to the definition of the $\rm wt_{\rm max}^\alpha$, c.f. Section
\ref{s:ac-veg}). With a small pre-sampled set the generator can however badly under-estimate the
maximum weight and a large number of events can be accepted with a too-high
probability. The only hope of obtaining correct results is in such cases that the weight
{\it plateau} will be hit sufficiently early in the event generation process. Consequently,
such approach can be very dangerous when generating small numbers of
events\footnote{{\it Small} being a somewhat relative quantifier, since the size of an
representative sample should depend on the phase space dimension, i.e. the number of
particles in the final state; with e.g. 4 particles in the final state, $10^5$ events can
still be considered a relatively small statistics.}.

In contrast to the {\it warming-up} approach for the {\bf AcerMC} we decided that
using separate {\it training} runs with large numbers of weighted events to obtain the
optimised grids and maximum weight estimates are preferable. In case a user wants
to produce data sets for non-defaults setting, this can easily be done
by configuring the switches in the {\tt acermc.card} (see Section 6.2).


\boldmath
\section{How to use the package} 
\unboldmath

There are two main steering input files: {\bf run.card} and {\bf acermc.card}
which share a common format for both executables.  The {\bf run.card} (see
App. \ref{app:run}) provides switches for modifying: generated process, number
of events, parton density functions, predefined option for
hadronisation/fragmentation in the {\it supervising generator}, random number,
etc..  The {\bf acermc.card} (see App. \ref{app:acmc}) provides switches for
modifying more specialised settings for the {\bf AcerMC} library itself. Once
the user decides on a setup for the generated process, only {\bf run.card} is
very likely to be modified for the job submission. In case the user decides to
use the external {\tt TAUOLA} or {\tt PHOTOS} libraries by selecting the
appropriate switches in the {\bf run.card} there are two additional files {\bf
tauola.card}(see App. \ref{app:tauola}) and {\bf photos.card} (see
App. \ref{app:photos}) which steer the performance of these two
libraries.  All input files are read by AcerMC executables through the {\tt
CERNLIB FFREAD} routines, some commands given in the input files (e.g. LIST
entry, see Appendix B) are internal {\tt FFREAD} commands which should be
disregarded by the user.

The same executables can also be used for running standard {\tt PYTHIA 6.4} and
{\tt HERWIG 6.5} processes. The example how to require such process is provided as
well, in {\tt demo\_hw.f} and {\tt demo\_py.f} respectively.  If the user
requires that the {\bf AcerMC} library is not used, the $ttH$ production will be
generated with {\tt demo\_py.exe} and {\tt HERWIG 6.5} implementation of the
$Zbb$ production will be generated with {\tt demo\_hw.exe}. In this case only
the {\tt run.card} file will be read, so in case the user requires different
processes and/or settings of the supervising generators the user has to
implement her/his steering there or create another xxx.card file, together with
the corresponding code added to the {\tt demo\_xx.f}.
  
\boldmath
\subsection{Steering switches of the overall run} 
\unboldmath

The overall run is controlled by the switches read from the {\tt
run.card} file (see also App. \ref{app:run}).
Some of these general switches are  also passed to the {\bf AcerMC} library.
\begin{scriptsize}
\begin{itemize}

\item{\tt CMS }\ : \ Sets the centre-of-mass energy in GeV.\\
 
\item{\tt ACER}\ : \ Specifies if the internal {\bf AcerMC} process will be used\\
{\tt ACER=0} - use process from {\tt PYTHIA/HERWIG} \\
{\tt ACER=1} - use internal {\bf AcerMC} process\\
{\tt ACER=2} - use internal {\bf AcerMC} process \& dump events into a record\\
{\tt ACER=3} - use internal {\bf AcerMC} process \& read events from a record\\

\item{\tt PROCESS}\ : \ Sets process id\\

\item{\tt HAD}\ : \ Sets predefined option for QCD ISR/FSR and hadronisation\\
{\tt HAD=0} - only hard process \\
{\tt HAD=1} - only ISR (works for {\tt PYTHIA} interface only)\\
{\tt HAD=2} - only ISR and FSR\\
{\tt HAD=3} - full treatment\\
{\tt HAD=4} - only FSR \\
{\tt HAD=5} - only FSR and hadronisation\\


\item{\tt LHAPDF} {\tt NSET}\\
Sets the value of the {\tt LHAPDF/LHAGLUE} {\tt NSET} parton density function choice.

\item{\tt RSEED}\ : \ Choose the random seed for (pseudo-)random generator initialisation\\

\item{\tt TAUOLA}\ : \ Specifies if the {\tt TAUOLA} library will be used for tau decays\\
{\tt TAUOLA=0} - use internal {\tt PYTHIA/HERWIG} mechanisms for $\tau$-lepton decays \\ 
{\tt TAUOLA=1} - use the {\tt TAUOLA}  library for $\tau$-lepton decays.\\

\item{\tt PHOTOS}\ : \ Specifies if the {\tt PHOTOS} routines will be used for final
state photon radiation\\
{\tt PHOTOS=0} - use internal {\tt PYTHIA/HERWIG} mechanisms for FSR photon radiation \\ 
{\tt PHOTOS=1} - use the {\tt PHOTOS} mechanisms for FSR photon radiation.\\

\item{\tt NEVENT}\ : \ Required number of generated events \\

\end{itemize}
\vspace*{-0.5cm}
\end{scriptsize}

\boldmath  
\subsection{Steering switches of the  {\bf AcerMC} processes} 
\unboldmath

The {\bf AcerMC} processes are controlled by values set in a simple arrays specified
in {\tt acermc\_src/include/AcerMC.inc}:
\begin{scriptsize}
{\tt  
\\
C CROSS-TALK PARAMETERS \\
      DOUBLE PRECISION ACSET \\
      INTEGER IACPROC\\
      COMMON/ACPAR1/ACSET(200),IACPROC(200)\\
C PARTICLE PROPERTIES\\
      DOUBLE PRECISION ACCHG,ACMAS,ACCKM \\
      COMMON/ACPAR2/ACCHG(50,4),ACMAS(50,4),ACCKM(4,4) \\ 
C ROUTINE I/O\\	
      INTEGER LACSTD,LACIO \\
      COMMON/ACPAR3/LACSTD,LACIO 
}
\end{scriptsize}

The {\tt IACPROC} array activates the process {\tt IPROC=PROCESS}
(read from {\tt run.card} file) by setting {\tt IACPROC(IPROC)=1}.

The list of currently implemented processes in {\bf AcerMC}
can be found in Table \ref{T3:1}. When running in the generation mode
with {\tt ACER=0} full list of processes implemented in either {\tt PYTHIA 6.4}
or {\tt HERWIG 6.5} can be activated, however the mechanism for passing information 
about process id to either of these generators has to be coded by user individually
in {\tt demo\_xx.f}.

The main control switches reside in the array {\tt ACSET}.
The {\tt COMMON} block {\tt ACPAR2} contains the particle charges, masses and
decay widths as well as the CKM matrix using the {\tt PYTHIA} convention. The
values are filled by the interface routines to be equal to the {\tt PYTHIA/HERWIG}
internal values in order to preserve consistency within the generation
stream. In case the user wants to change some of the particle properties this
should be done through the native {\tt PYTHIA/HERWIG} switches; {\bf AcerMC} will copy
them and use the new values.

The {\tt COMMON} block {\tt ACPAR3} contains the two logical I/O unit numbers
used by {\bf AcerMC}. The {\tt LACSTD} value determines the output unit of the {\bf AcerMC}
messages and the {\tt LACIO} unit is used for reading/writing the {\bf AcerMC} data files. 

The main control switches which reside in the array {\tt ACSET} (see also App. \ref{app:acmc}):
\begin{scriptsize}
\begin{itemize}

\item{\tt ACSET(1)}\ : \ Sets the centre-of-mass energy in GeV. \\

\item{\tt ACSET(2)}\ : \ Scale of the hard process\\
Choose the $Q^2$ scale for the active {\bf AcerMC} process.
The implemented values differ for various processes, the currently implemented
settings are specified in Section~\ref{s:scdef}.\\

\item{\tt ACSET(3)}\ : \ Fermion code\\
The flavour of the final state fermions produced in $W^{\pm}, Z^0/\gamma^* \to f \bar f$ decays of {\bf AcerMC} processes $3 \to 8$. The {\tt PYTHIA/PDG} naming convention is used:\\
{\tt ACSET(3)=1} - $W \to q \bar q$; $Z^0/\gamma^* \to q \bar q$ \\
{\tt ACSET(3)=4} -  $W \to e \nu_e,\mu \nu_{\mu}$; $Z^0/\gamma^* \to e^+ e^-,\mu^+ \mu^-$\\
{\tt ACSET(3)=5} -  $Z^0/\gamma^* \to b \bar{b}$\\
{\tt ACSET(3)=10} -  $W \to e \nu_e,\mu \nu_{\mu},\tau \nu_{\tau}$; $Z^0/\gamma^* \to e^+ e^-,\mu^+ \mu^-,\tau^+ \tau^-$\\
{\tt ACSET(3)=11} - $W \to e \nu_e$; $Z^0/\gamma^* \to e^+ e^-$ \\
{\tt ACSET(3)=12} - $Z^0 \to \nu_e \nu_e,\nu_{\mu} \nu_{\mu}, \nu_{\tau} \nu_{\tau}$,\\\
{\tt ACSET(3)=13} - $W \to \mu \nu_{\mu}$; $Z^0/\gamma^* \to \mu^+ \mu^-$  \\
{\tt ACSET(3)=15} - $W \to \tau \nu_{\tau}$; $Z^0/\gamma^* \to \tau^+ \tau^-$  \\
{\tt ACSET(3)=5} -  $Z^0/\gamma^* \to b \bar b$ \\
At present the {\tt ACSET(3)=5} is implemented only for processes $7 \to 8$.\\

\item{\tt ACSET(4)}\ : \ $Z^0/\gamma^*$ propagator\\
Use full $Z^0/\gamma^*$ propagator instead of the pure $Z^0$ propagator in
matrix element calculation for the {\bf AcerMC} processes $5 \to 8$ and 91.
 The switch is provided
since in some of the analyses the $\gamma^*$ contribution is of relevance in the
selected mass windows; for the analyses selecting the mass window around the
$Z^0$ peak this contribution can safely be neglected.\\
{\tt ACSET(4)=0} - only $Z^0$ propagator. \\
{\tt ACSET(4)=1} - full $Z^0/\gamma^*$ propagator. \\

\item{\tt ACSET(5)}\ : \ $m_{Z^0/\gamma^*}$ mass cut\\
Cutoff value on the invariant mass $m_{Z^0/\gamma^*}$ in GeV when {\tt ACSET(4)=1}. 
Note that the provided data files exist only for values of
{\tt ACSET(5)}=2,5,10,15,30,60, 120,270,300 and 500 GeV which should satisfy most 
user requirements for the analyses foreseen at LHC.
 In case a different value is set the user has also to provide 
the user data files for the run.\\


\item{\tt ACSET(7)}\ : \ Sets the value of the {\tt LHAPDF/LHAGLUE} {\tt NSET}
 parton density function choice.\\

\item{\tt ACSET(8)}: The implementation of $\alpha_s(Q^2)$\\
Selects the implementation of $\alpha_s(Q^2)$ to be used in the matrix element
calculation:\\
{\tt ACSET(8)=0} - Use the $\alpha_s(Q^2)$ as provided by the supervising generator\\
{\tt ACSET(8)=1} - Use the  $\alpha_s(Q^2)$ (one loop) provided by the {\bf AcerMC}; this
option gives $\alpha_s(Q^2)$ values equal to the default {\tt PYTHIA} implementation.\\
{\tt ACSET(8)=2} - Use the  $\alpha_s(Q^2)$ (three loop) provided by the {\bf AcerMC}.\\

\item{\tt ACSET(9)}: $\alpha_s(M_Z^2)$ value\\
Sets the $\alpha_s(M_Z^2)$ value to be used in the $\alpha_s(Q^2)$
calculations in case the {\bf AcerMC} native implementation ({\tt ACSET(8)=1}) is
used.\\
{\tt ACSET(9)=-1} - The $\Lambda^{(nf=5)}_{\bar{MS}}$ value is taken from the
{\tt LHAPDF} for the selected parton density function set.\\
{\tt ACSET(9)$>$0} - The provided value is taken.\\

\item{\tt ACSET(10)}: The implementation of $\alpha_{\rm QED}(Q^2)$\\
Selects the implementation of $\alpha_{\rm QED}(Q^2)$ to be used in the matrix
element calculation:\\ 
{\tt ACSET(10)=0} - Use the $\alpha_{\rm QED}(Q^2)$ as provided by
the  supervising  generator.\\ 
{\tt ACSET(10)=1} - Use the $\alpha_{QED}(Q^2)$ implemented in the {\bf AcerMC}.\\

\item{\tt ACSET(11)}: $\alpha_{\rm QED}(0)$ value\\
Specifies the value of $\alpha_{\rm QED}(0)$ for {\bf AcerMC} $\alpha_{\rm QED}(Q^2)$
calculation.\\
{\tt ACSET(11)=-1} - The $\alpha_{\rm QED}(0)$ value is set to 
$\alpha_{\rm QED}(0) = 0.0072993$. \\
{\tt ACSET(11)$>$0} - The provided value is taken.\\

\item{\tt ACSET(12)}: Decay mode of the produced $t \bar t$ pair\\
Sets the decay mode of the W boson pair from the $t \bar t$ final
state in the {\bf AcerMC} processes 1,2,4,7,8,9,92 and 93. 
For {\tt ACSET(12)$>$0} the combinatoric  value of the $\sigma \times BR$ is
recalculated and printed in the output. This switch was implemented
since the supervising generators 
({\tt PYTHIA/HERWIG}) do not allow for forcing specific decays of the
top quark pairs generated by external processes. This switch imposes a  modification 
of the decay tables of the supervising generators on an event by event
basis.\\ 
{\tt ACSET(12)=0} - both W bosons decay according to {\tt
PYTHIA/HERWIG} switches.\\ 
{\tt ACSET(12)=1} - $W_1 \to e \nu_e$ and $W_2 \to q \bar q$.\\ 
{\tt ACSET(12)=2} - $W_1 \to \mu \nu_{\mu}$ and $W_2 \to q \bar q$.\\ 
{\tt ACSET(12)=3} - $W_1 \to \tau \nu_{\tau}$ and $W_2 \to q \bar q$.\\ 
{\tt ACSET(12)=4} - $W_1 \to e \nu_e, \mu \nu_{\mu}$ and $W_2 \to q \bar q$.\\ 
{\tt ACSET(12)=5} - $W_1 \to e \nu_e, \mu \nu_{\mu}$ and $W_2 \to q \bar q$. \\ 
The setting {\tt ACSET(12)=5} works for {\tt PROCESS=4} only and implies
leptonic decay of the $W$-boson with the same charge as the one of the primary 
$W$ boson produced in the hard process. Folowing configurations are posible:\\
$ q \bar q \to W^+ t \bar t 
\to (W^+ \to) L^+ \nu_L ~~~ (W^+_1 \to) l^+ \nu_l\; b ~~~ (W^-_2 \to) q' \bar{q}' \; \bar b,$\\
or:\\
$ q \bar q \to W^- t \bar t 
\to  (W^- \to) L^- \bar{\nu}_L ~~~ (W^+_1 \to) q' \bar{q}' \;  b  ~~~ (W^-_2 \to) l^- \bar{\nu}_l \; \bar b ,$\\
where $L^\pm$ is the lepton from the primary $W$ decay (controlled by {\tt ACSET(3)} switch) 
and $l^\pm$ is either an $e^\pm$  or $\mu^\pm$ as for {\tt ACSET(12)=4}. Since the charge of the 
semi-leptonic decaying $W$ is correlated
with the charge of the primary W boson, the $\sigma \times BR$ is consequently a factor 
two smaller than the one for {\tt ACSET(12)=4}.\\

\item{\tt ACSET(13)}: Decay mode of the produced $WW$ pair, works for
PROC=11-14 and PROC=20,21,23 only\\
{\tt ACSET(13)=0} - both W bosons decay according to {\tt
PYTHIA/HERWIG} switches.\\ 
{\tt ACSET(13)=1} - $W_1 \to e \nu_e$ and $W_2 \to q \bar q$.\\ 
{\tt ACSET(13)=2} - $W_1 \to \mu \nu_{\mu}$ and $W_2 \to q \bar q$.\\ 
{\tt ACSET(13)=3} - $W_1 \to \tau \nu_{\tau}$ and $W_2 \to q \bar q$.\\ 
{\tt ACSET(13)=4} - $W_1 \to e \nu_e, \mu \nu_{\mu}$ and $W_2 \to q \bar q$.\\ 
{\tt ACSET(13)=5} - $W_1 \to e \nu_e, \mu \nu_{\mu}, \tau \nu_{\tau}$ and $W_2 \to q \bar q$.\\ 
{\tt ACSET(13)=6} - one or both $W \to e \nu_{e} \text{or} \mu \nu_{\mu} \text{or} \tau \nu_{\tau}$, the remaining one hadronically.\\ 
{\tt ACSET(13)=11} -both $W \to e \nu_e$.\\ 
{\tt ACSET(13)=13} -both $W \to \mu \nu_{\mu}$.\\ 
{\tt ACSET(13)=15} -both $W \to \tau \nu_{\tau}$.\\ 
{\tt ACSET(13)=17} -both $W \to e \nu_{e} \text{or} \mu \nu_{\mu}$.\\ 
{\tt ACSET(13)=19} -both $W \to e \nu_{e} \text{or} \mu \nu_{\mu} \text{or} \tau \nu_{\tau}$.\\ 
{\tt ACSET(13)=20} -both $W  \to q \bar q$.\\ 

\item{\tt ACSET(50)}: {\bf AcerMC} training mode\\
The switch controls the mode in which {\bf AcerMC} is run:\\
{\tt ACSET(50)=0} - production run, generate unweighted events.\\
{\tt ACSET(50)=1} - perform multi-channel optimisation and output the user file
with channel weights.\\
{\tt ACSET(50)=2} - perform {\tt ac-VEGAS} grid training and output the user file
with trained {\tt ac-VEGAS} grid.\\
{\tt ACSET(50)=3} - perform {\tt ac-VEGAS} grid training as in {\tt ACSET(50)=2}
but do this by updating a provided grid.\\

\item{\tt ACSET(51)}: Required number of generated events {\tt NEVENT}\\
In case the switch {\tt ACSET(50)} is set to the non-zero value (i.e. in one of
the training modes) the  {\tt ACSET(51)} entry is used and defines the number of
(weighted) events that will be generated; this information is necessary for the
learning algorithms to decide on steps in the learning sequence. \\

\item{\tt ACSET(52)}: User data files\\
Use the data files provided by user:\\
{\tt ACSET(52)=0} - no, use native (default) {\bf AcerMC} data files.\\
{\tt ACSET(52)=1} - use the user's multi-channel optimisation and {\tt VEGAS} grid
files.\\ 
{\tt ACSET(52)=2} - use the default multi-channel optimisation
 and user's {\tt VEGAS} grid files.\\
{\tt ACSET(52)=3} - use the default multi-channel optimisation and {\tt VEGAS}
                    grid files; read the user maximal weight file.\\

\item{\tt ACSET(53)}: Maximum weight search\\
Mode for the maximum weight search needed for unweighting procedure:\\
{\tt ACSET(53)=0} - no, use the provided files containing maximal weights.\\
{\tt ACSET(53)=1} - use the provided files for max. weights and re-calculate 
the max. weights using the stored 100 events with the highest weight.\\
{\tt ACSET(53)=2} - perform the search and give the new {\tt wtmax\_xx\_new.dat}
file; the switch is  equivalent to generation of weighted events.\\

\item{\tt ACSET(54)}: Maximum weight choice\\
Use the $\alpha$-cutoff maximal weight $\rm wt_{\rm max}^\alpha$ or the overall maximal
weight $\rm wt_{\rm max}$ found in training (see Section \ref{s:ac-veg} for the explanation on
these two options).\\
{\tt ACSET(54)=0} - use the $\rm wt_{\rm max}^\alpha$ weight.\\
{\tt ACSET(54)=1} - use the $\rm wt_{\rm max}$ weight.\\

\item{\tt ACSET(56)}: Naive QCD correction for width calculations\\
Use the naive QCD multiplicative corrections in the resonant width calculations of 
the weak boson and top quark decay widths. It is recommended to use them except for
some specific cases or studies.\\
{\tt ACSET(56)=1} - use the naive QCD corrections\\
{\tt ACSET(56)=0} - don't use naive QCD corrections.

\item{\tt ACSET(57)}: Use the Collins derived PDF-s for showering\\
applicable to processes with the merging of processes as described in \cite{acot}.\\
At present these are the AcerMC processes 17, 18 and 20. \\
{\tt ACSET(57)=1} - use the Collins PDF-s\\
{\tt ACSET(57)=0} -  don't use  the Collins PDF-s.

\item{\tt ACSET(58)}: Value of Z-prime mass in TeV/c$^2$\\ 
Note that the provided data files exist only for values of\\
Z-prime mas=1 TeV and 0.5 TeV  which should satisfy most \\
users. In case a different value is set the user has also to provide \\
the user data files for the run.\\
The corresponding width of the boson is calculated internally.\\

\item{\tt ACSET(59)}: Values of Z-prime coupling sets\\
Note that the provided data files exist only for values of\\
{\tt ACSET(59)=0} - Standard Model values of $Z^{0\prime}$ couplings \\
{\tt ACSET(59)=1} - $Z^{0\prime}_R$ values of $Z^{0\prime}$ couplings as described in (hep-ph/0307020). \\
{\tt ACSET(59)=2} - Pure V-A values of $Z^{0\prime}$ couplings (Weinberg angle set to zero for $Z^{0\prime}$)\\
In case a different value is needed the user should contact the AcerMC authors.\\

\item{\tt ACSET(60)}: Leptonic top coupling\\
Vertex coupling of top, b-quark and leptonic decaying W\\
The coupling is given by:\\
{\tt GTF = ACSET(60)*$\rm G_L$+(1-ACSET(60)*$\rm G_R$)}\\
meaning that {\tt  ACSET(60)=1} is the Standard model value

\item{\tt ACSET(61)}: Hadronic top coupling\\
Vertex coupling of top, b-quark and hadronic decaying W\\
The coupling is given by:\\
{\tt GTF = ACSET(61)*$\rm G_L$+(1-ACSET(61)*$\rm G_R$)}\\
meaning that {\tt  ACSET(61)=1} is the Standard model value.
 
\end{itemize}
\end{scriptsize}

\boldmath
\subsection{Steering {\tt TAUOLA} and {\tt PHOTOS} \label{s:tauphot}} 
\unboldmath

It is highly recommended that in case {\tt TAUOLA} is used for $\tau$ decays
that {\tt PHOTOS} is also activated and the setting {\tt PMODE} is set at least
to {\tt 2} in order to keep the $\tau$-leptons radiating (see below), since it
gives the necessary contribution to the expected $\tau$ branching ratios. 

It should also be stressed that the actions of the interfaced {\tt TAUOLA} and
{\tt PHOTOS} are meaningful only when the full hadronisation procedure of the
event is selected (by setting {\tt HAD 3} in {\sf run.card}).

The switches and settings specific to the {\tt TAUOLA} library are set in the 
{\sf tauola.card}:
\begin{scriptsize}
\begin{itemize}

\item{\tt  POLAR}\ : \ Polarisation switch for tau decays \\
{\tt POLAR=0} - switch polarisation off \\
{\tt POLAR=1} - switch polarisation on \\

\item{\tt  RADCOR}\ : \ Order(alpha) radiative corrections for tau decays\\
{\tt RADCOR=0} - switch corrections off\\
{\tt RADCOR=1} - switch corrections on\\

\item{\tt  PHOX}\ : \ Radiative cutoff used in tau decays\\ 
{\tt PHOX=0.01} - default value by TAUOLA authors\\

\item{\tt  DMODE}\ : \ Tau and tau pair decay mode\\
{\tt DMODE=0} - all decay modes allowed\\
{\tt DMODE=1} - (LEPTON-LEPTON): only leptonic  decay modes\\
{\tt DMODE=2} - (HADRON-HADRON): only hadronic  decay mode\\
{\tt DMODE=3} - (LEPTON-HADRON): one tau decays into leptons and the other one into hadrons\\
{\tt DMODE=4} - ($\tau \to \pi \nu$)  : taus are restricted to decay to a pion and neutrino\\

\item{\tt  JAK1/JAK2}\ : \ Decay modes of taus according to charge, the list is
taken from the {\tt TAUOLA} output. The listing gives only $\tau^-$ modes, the
$\tau^+$ are charge conjugate, neutrinos are omitted. {\tt JAK1}\ : \  decay
mode of $\tau^+$, {\tt JAK2}\ : \  decay mode of $\tau^-$:\\
{\tt JAK1/2 = 1}   - $\tau^-  \to  e- $\\
{\tt JAK1/2 = 2}   - $\tau^-  \to  \mu-$  \\
{\tt JAK1/2 = 3}   - $\tau^-  \to  \pi^-$ \\
{\tt JAK1/2 = 4}   - $\tau^-  \to  \pi^-, \pi^0 $\\
{\tt JAK1/2 = 5}   - $\tau^-  \to  A_1^-$ (two subch)\\
{\tt JAK1/2 = 6}   - $\tau^-  \to  K^- $\\
{\tt JAK1/2 = 7}   - $\tau^-  \to  K^{*-}$ (two subch)\\
{\tt JAK1/2 = 8}   - $\tau^-  \to  2\pi^-,  \pi^0,  \pi^+$\\
{\tt JAK1/2 = 9}   - $\tau^-  \to  3\pi^0,        \pi^-$\\
{\tt JAK1/2 = 10}  - $\tau^-  \to  2\pi^-,  \pi^+, 2\pi^0$ \\
{\tt JAK1/2 = 11}  - $\tau^-  \to  3\pi^-, 2\pi^+$\\
{\tt JAK1/2 = 12}  - $\tau^-  \to  3\pi^-, 2\pi^+,  \pi^0 $\\
{\tt JAK1/2 = 13}  - $\tau^-  \to  2\pi^-,  \pi^+, 3\pi^0$\\
{\tt JAK1/2 = 14}  - $\tau^-  \to  K^-, \pi^-,  K^+$\\
{\tt JAK1/2 = 15}  - $\tau^-  \to  K^0, \pi^-, \bar{K^0} $\\
{\tt JAK1/2 = 16}  - $\tau^-  \to  K^-,  K^0, \pi^0 $\\
{\tt JAK1/2 = 17}  - $\tau^-  \to  \pi^0  \pi^0   K^-$\\
{\tt JAK1/2 = 18}  - $\tau^-  \to  K^-  \pi^-  \pi^+ $\\
{\tt JAK1/2 = 19}  - $\tau^-  \to  \pi^- \bar{K^0}  \pi^0  $\\
{\tt JAK1/2 = 20}  - $\tau^-  \to  \eta  \pi^-  \pi^0$\\
{\tt JAK1/2 = 21}  - $\tau^-  \to  \pi^-  \pi^0  \gamma$\\
{\tt JAK1/2 = 22}  - $\tau^-  \to  K^-  K^0$
\end{itemize}

The switches and settings specific to the {\tt PHOTOS} routines are set in the 
{\sf photos.card}:
\begin{itemize}

\item{\tt  PMODE}\ : \ Radiation mode of photos:\\
{\tt PMODE=1} - enable radiation of photons for leptons and hadrons\\
{\tt PMODE=2} - enable radiation of photons for $\tau$-leptons only\\
{\tt PMODE=3} - enable radiation of photons for leptons only\\

\item{\tt  XPHCUT}\ : \ Infrared cutoff for photon radiation:\\
{\tt XPHCUT=0.01} - default value by PHOTOS authors\\

\item{\tt  ALPHA}\ : \ Alpha(QED) value used in {\tt PHOTOS}:\\
{\tt ALPHA < 0} - leave default (0.00729735039)\\

\item{\tt  INTERF}\ : \ Photon interference weight switch\\ 
{\tt INTERF = 1} - interference is switched on\\
{\tt INTERF = 0} - interference is switched off\\

\item{\tt  ISEC}\ : \ Double bremsstrahlung switch:\\
{\tt ISEC=1} - double bremsstrahlung is switched on\\
{\tt ISEC=0} - double bremsstrahlung is switched off\\

\item{\tt  IFTOP}\ : \ Switch for $gg(qq) \to t\bar{t}$ process radiation:\\ 
{\tt IFTOP=1} - the procedure is is switched on\\
{\tt IFTOP=0} - the procedure is is switched off\\

\end{itemize}
\end{scriptsize}
Detailed information about {\tt TAUOLA} and {\tt PHOTOS} implemetations in the
{\bf AcerMC} setup can be found in Sections \ref{s:tauola} and \ref{s:photos}.
\clearpage 

\boldmath 
\subsection{How to prepare data-files for the non-default setup \label{s:training}} 
\unboldmath

The following actions are possible, to recover better efficiency of the generator
modules with the non-default settings:

\begin{itemize}
\item{\it The user wants to generate events using different parton density function
sets and/or different coupling values (e.g. {\bf AcerMC} third order $\alpha_s(Q^2)$
instead of the first order one)}:\\
 
It should suffice to set the the switch {\tt ACSET(53)=1}, which signals {\bf AcerMC} to
re-calculate the $\rm wt_{\rm max}$ and $\rm wt_{\rm max}^\alpha$ using the 100 events stored in
the file \\ {\tt acermc\_dat/grids/vtmaxA\_xxYYY.dat}. The coupling and
parton density functions values should not change significantly the process topology but affect
foremost the overall scale of the event weights; thus, the stored hundred events should
still remain the ones with the highest weights and the re-calculated approximate estimates
of the highest weight should be accurate enough.

In case the user is not confident in the obtained result, the new maximal weight estimation
can be initiated by setting the switch {\tt ACSET(53)=2}, which will result in generation
of weighted events. The number of generated events is determined by the usual {\tt NEVENT}
in {\tt run.card}.  At the end of the run {\bf AcerMC} will produce a file called {\tt
wtmax\_xx\_new.dat}, with {\tt xx} specifying the process number. The user should then
start the generation of unweighted events with the setting {\tt ACSET(52)=3} and
linking(renaming) the new file to {\tt wtmax\_xx\_usr.dat}, with {\tt xx} denoting the
process number (e.g. {\tt wtmax\_01\_usr.dat}).\\

\item{\it The user wants to generate events using different values of particle/boson
masses or other significant changes of the parameters apart from the
centre-of-mass energy and/or $m_{Z^0/\gamma^*}$ cutoff value for processes 5-8}:\\

In this case the user should re-train the {\tt VEGAS} grid since the process
topology is assumed to undergo minor changes. This is done by setting the switch
{\tt ACSET(50)=2} or {\tt ACSET(50)=3}; in the first case {\bf AcerMC} starts with an
untrained grid and in the second one it starts modifying the existing grid
provided for the process at the selected hard process scale. In general the
second option should be preferable since the topology should still be close to
the pre-trained one. {\bf AcerMC} again produces weighted events and at the end of the
run outputs a file {\tt grid\_xx\_new.veg}. The number of generated events is determined
by the usual {\tt NEVENT} in {\tt run.card}.  As
in the previous case, the user should re-name the file to {\tt grid\_xx\_usr.veg},
re-set the switch to {\tt ACSET(50)=0} and repeat the maximal weight search
procedure described above, by setting the switch {\tt ACSET(53)=2} etc..
When the maximum weight search is completed the user switch {\tt ACSET(52)=2},
which will cause {\bf AcerMC} to read the {\tt wtmax\_xx\_usr.dat} as well as {\tt
grid\_xx\_usr.veg} files and produce unweighted events with the new setup.\\ 

\item{\it The user wants to generate events at a different $m_{Z^0/\gamma^*}$
cutoff value and/or different centre-of-mass energy $\sqrt{s}$}:\\

When the user changes at least one of these two parameters the event topology
 is significantly changed
as well as the contributions from different kinematic channels. The user should thus start
with a new multi-channel optimisation by setting the mode switch {\tt ACSET(50)=1} and
start an {\bf AcerMC} run.  The number of generated events is determined by the usual {\tt
NEVENT} in {\tt run.card}.  At the end of the run {\bf AcerMC} will produce a file {\tt
chanwt\_xx\_new.dat} which should be renamed/linked to {\tt chanwt\_xx\_usr.dat}. The
user should then set the switch {\tt ACSET(52)=1} and first put {\tt ACSET(50)=2} and
and repeat the {\tt VEGAS} grid training as described above and consequently {\tt
ACSET(50)=0} and {\tt ACSET(53)=2} to perform the maximum weight search. After obtaining
all three user files the {\tt ACSET(53)} should again be put back to {\tt ACSET(53)=0} and
a normal run should be started; the switch {\tt ACSET(52)=1} will in this case
force {\bf AcerMC} to read all three user files and produce unweighted events.
\end{itemize}
At the first look procedure for listed action scenaria might seem a bit complex
but should after a few trials and errors become a straightforward routine; it is
expected that the vast majority of users would have to deal with at most the
first scenario.

\newpage
\boldmath 
\subsection{Details on the interface to {\tt PYTHIA 6.4}} 
\unboldmath

The {\bf AcerMC} interface to {\tt Pythia 6.4} is implemented close to the new
standard specified at the Les Houches workshop 2001 \cite{Boo01}. The full
description of the standard can be found in the {\tt PYTHIA 6.4} manual
(\cite{Pythia62}). In addition to the {\tt UPINIT} and {\tt UPEVNT} routines the file
{\tt acermc\_src/interface/pythia\_ac.f} provides links between a list of {\bf AcerMC}
routines and the corresponding {\tt PYTHIA} ones, as e.g. the (pseudo-)random
number generator, $\alpha_s$ and $\alpha_{\rm QED}$ calculations as well as a
series of routines that re-write the {\bf AcerMC} event output to the required {\tt
PYTHIA} format. Using this strategy, the native {\bf AcerMC} code is completely
de-coupled from the linked hadronisation library (at the moment {\tt
\tt PYTHIA/HERWIG}) and new interfaces can thus easily be added. 
The special {\bf AcerMC}
requirement is the call to the {\tt ACFINAL} subroutine at the end of the run
which signals the {\bf AcerMC} to close the various I/O files and produce the final
output. An example of the implementation of the {\tt PYTHIA}/{\bf AcerMC} interface
can be found in the provided {\tt demo\_py.f}. The {\tt PYTHIA} code is
unmodified apart from making a small modification in {\tt PYINIT} routine:

\noindent {\tt CALL UPINIT(1)}\\
\noindent {\sf ..parameter initalisation..}\\
\noindent {\tt CALL UPINIT(2)}\\
\noindent {\sf ..process initialisation..}

since the user-supplied processes in this new interface are not
allowed to (re-)estimate
maximal weights (as e.g. the native {\tt PYTHIA} processes do). 
In the original code
the call to {\tt UPINIT} is set before the {\tt PYTHIA}  parameters and functions 
(e.g. {\tt PYALPS}
for $\alpha_s(Q^2)$ calculation) are initialised with the user settings\footnote{This
was however possible in the old {\tt PYTHIA 6.1} interface}. 

An additional modification was added in the {\tt PYEVNT} routine so that undecayed
resonances from AcerMC (e.g. top quarks) are decayed by {\tt PYTHIA} before the {\tt ARIADNE}
routines are called (when requested by the user).


The user can thus add the most recent {\tt PYTHIA} library without
other neccessary modifications but for the two lines of code in {\tt PYINIT}
routine as described above (the dummy routines {\tt UPINIT, UPEVNT,
STRUCTM, STRUCTP} and {\tt PDFSET} however have to be removed from the
code for an external process to work and to activate the {\tt LHAPDF/LHAGLUE}
interface).

By setting {\tt ACER=1} user decides to generate hard process from
{\bf AcerMC} library.  Modeling of ISR/FSR shower, hadronisation and
decays are generated by {\tt PYTHIA} generator. All steering
parameters, relevant for these steps of full event generation remain
the same as in standard {\tt PYTHIA} execution.

By setting {\tt ACER=0} user decides to generate standard {\tt PYTHIA}
process. The simple example how to generate $t \bar t$ and $t \bar tH$ production process
within {\bf AcerMC} framework is provided in {\tt demo\_py.f}.

The additional settings {\tt ACER=2} and {\tt ACER=3}, which allow the user to
produce event records according to the Les Houches standard are given in the Section 7.8.

\boldmath 
\subsection{Details on the interface to {\tt ARIADNE 4.1}} 
\unboldmath

The {\bf AcerMC} interface to {\tt ARIADNE 4.1} \cite{ariadne41} is
done via the {\tt ARIADNE $\leftrightarrow$ PYTHIA} interface provided
in the {\tt ARIADNE} distribution and the further {\tt PYTHIA} $\leftrightarrow$ {\bf AcerMC}
interface as described in the previous section. This is necessary since {\tt ARIADNE}, 
while providing the advanced colour-dipole model of initial/final state radiation, is
still relying on {\tt PYTHIA} for particle/resonance decays and hadronisation. Some modifications
were made to the  {\tt ARIADNE $\leftrightarrow$ PYTHIA} interface in order to accomodate
resonance (e.g. top-quark) decays in Pythia before the event is passed to ARIADNE for shower
addition and to properly search for Drell-Yan type processes for hard processes not implemented
in {\tt PYTHIA} (but e.g. in {\bf AcerMC}). 
  
\newpage

\boldmath
\subsection{Details on the interface to {\tt HERWIG 6.5} \label{s:hwintf}} 
\unboldmath

Interfacing the {\bf AcerMC} to {\tt HERWIG 6.5} is almost identical with the
{\tt PYTHIA} implementation since the interfaced version now also complies with
the Les Houches standard \cite{Boo01}. The interface routines are written in
accordance with the Les Houches description; in the {\tt HERWIG 6.5} interface
the {\tt UPINIT} routine also has to be called in two steps in order to enable
the user to change the {\tt HERWIG 6.5} default settings and get the correct
re-evaluation of the maximal weight (as in the {\tt PYTHIA} interface):\\
 
\noindent {\tt CALL UPINIT(1)}\\
\noindent {\sf ..user values..}\\
\noindent {\tt CALL UPINIT(2)}\\

The first {\tt UPINIT} call is made from the original location in the {\tt
HWIGUP} routine and the second call to {\tt UPINIT} is placed at the end of the
{\tt HWUINC} routine, after all the internal {\tt HERWIG} settings have been
modified.  As in the {\tt PYTHIA} implementation all the interface subroutines
needed for communication between {\tt HERWIG} and {\bf AcerMC} are stored in
{\tt acermc\_src/interface/herwig\_ac.f}.  

One minor additional modification of the original {\tt HERWIG 6.5} code was albeit
necessary, namely the {\tt IMPLICIT NONE} was commented out in the {\tt
herwig6500.inc} file; this was needed since the {\bf AcerMC} code is written
with the implicit {\tt IMPLICIT DOUBLE PRECISION(A-H,O-Z)}.

In principle the implemented changes should be very easy and transparent for the transfer
into new {\tt HERWIG} releases;  an example of the use of {\bf AcerMC}/{\tt HERWIG}
interface is provided in the file {\tt demo\_hw.f}.

By setting {\tt  ACER=1} the user decides to generate a hard process from {\bf AcerMC} library.
Modeling of ISR/FSR shower, hadronisation and decays are generated by {\tt HERWIG} 
generator. All steering parameters, relevant for these steps of full event generation
remain valid as for the standard {\tt HERWIG} execution. 

By setting {\tt  ACER=0} the user decides to generate standard {\tt HERWIG} processes. The simple
example how to generate $Zb \bar b$ production process within the {\bf AcerMC} framework
 is provided in {\tt demo\_hw.f}.

The additional settings {\tt ACER=2} and {\tt ACER=3}, which allow the user to
produce event records according to the Les Houches standard are given in the Section 7.8.
\newpage

The output logs of the run are produced in the directory {\tt prod}, the {\tt acermc.out}
file containing the {\bf AcerMC} specific information and the outputs {\tt pythia.out}
and/or {\tt herwig.out} listing the outputs of the respective supervising generators. The
information about the input values of the steering files is stored in {\tt run.out} in
order to facilitate the event generation 'bookkeeping'. The sample outputs are given in
Appendix C.

Specific information produced by {\tt TAUOLA} and {\tt PHOTOS} is stored in respective
files {\sf tauola.out} and {\sf photos.out}. A point to stress is that in case
the tau decay was restricted the hard process cross-section given in 
{\sf pythia.out},{\sf herwig.out} or {\sf acermc.out} should be multiplied by a
branching ratio as detailed at the end of {\sf tauola.out}, for example:
\begin{verbatim}
         ------< TAUOLA BRANCHING RATIO FOR TAU DECAYS >------
 
          THE TAU DECAYS ARE RESTRICTED TO A:
 
                    LEPTON-HADRON  DECAY MODE

          THE PROCESS CROSS-SECTION MUST BE MULTIPLIED BY:

           -> A BRANCHING RATIO =   0.459303E+00 FOR TWO TAUS
 
          IN THE HARD PROCESS DECAY PRODUCTS!!!
 
         ------> TAUOLA BRANCHING RATIO FOR TAU DECAYS <------
\end{verbatim}

\boldmath
\subsection{Definition of the energy scale \label{s:scdef}} 
\unboldmath

A few different values of scale $Q^2$ used in the evolution of parton density functions 
as well as the running couplings $\alpha_s(Q^2)$ and $\alpha_{\rm QED}(Q^2)$ can be set by
the switch {\tt ACSET(2)} (remember that the factorisation and renormalisation scales are
assumed to be equal in {\bf AcerMC}). Note that the {\it correct} value of the scale to be used for
certain processes is in principle not known; what was implemented in {\bf AcerMC} are the
most probable/usual choices on the market; in measurements the {\it best}
value will have to be determined by data analysis.
\begin{scriptsize}
\begin{itemize}
\item Processes $1,2,9,10$:\\
{\sf 
  ACSET(2): (D=1)\\
\hspace{1cm} 1 - $Q^2 = \hat{s}$ \\       
\hspace{1cm} 2 - $Q^2 = \sum{({p^i_T}^2 + m_i^2)}/4~=\;<m_T^2>$\\
\hspace{1cm} 3 - $Q^2 = \sum{({p^i_T}^2)}/4 ~=\;<p_T^2>$\\
\hspace{1cm} 4 - $Q^2 = (m_t + m_H/2)^2,~~~m_H = 120$~GeV/$c^2$ 
}
\item Processes $3 \to 4$:\\
{\sf 
  ACSET(2): (D=1)\\
\hspace{1cm} 1 - $Q^2 = M_W^2$ \\       
\hspace{1cm} 2 - $Q^2 = s^*_{q \bar q}$, where $q=b,t$ \\       
\hspace{1cm} 3 - $Q^2 = M_W^2+pT_W^2$ \\       
\hspace{1cm} 4 - $Q^2 = 0.5 \cdot (s^*_W + s^*_{q \bar q}) +(p_T^W)^2$, where $q=b,t$        
}
\item Processes $5 \to 8$:\\
{\sf 
  ACSET(2): (D=1)\\
\hspace{1cm} 1 - $Q^2 = M_Z^2$ \\
\hspace{1cm} 2 - $Q^2 = s^*_{q \bar q}$, where $q=b,t$ \\
\hspace{1cm} 3 - $Q^2 = s^*_Z$ \\
\hspace{1cm} 4 - $Q^2 = pT_Z^2 + s^*_{q \bar q})/2$, where $q=b,t$
}
\item Processes $11 \to 14$:\\
{\sf 
  ACSET(2): (D=1)\\
\hspace{1cm} 1 - $Q^2 =(2 \cdot m_t^2) $ \\
\hspace{1cm} 2 - $Q^2 =\sum{({p^i_T}^2 + m_i^2)}/4 $ \\
\hspace{1cm} 3 - $Q^2 =\sum{({p^i_T}^2)}/2 $ \\
\hspace{1cm} 4 - $Q^2 =\hat{s} $
}
\item Processes $15 \to 16$:\\
{\sf 
  ACSET(2): (D=1)\\
\hspace{1cm} 1 - $Q^2 = \hat{s}$ \\       
\hspace{1cm} 2 - $Q^2 = \sum{({p^i_T}^2 + m_i^2)}/4~=\;<m_T^2>$\\
\hspace{1cm} 3 - $Q^2 = \sum{({p^i_T}^2)}/4 ~=\;<p_T^2>$\\
\hspace{1cm} 4 - $Q^2 = (m_t + m_H/2)^2,~~~m_H = 120$~GeV/$c^2$ 
}
\item Processes $17,100,101$:\\
{\sf 
  ACSET(2): (D=1)\\
\hspace{1cm} 1 - $Q^2 = \hat s_{\rm top}$ \\       
\hspace{1cm} 2 - $Q^2 = (\sum{({p^i_T}^2)} + m_t^2)/2$ \\       
\hspace{1cm} 3 - $Q^2 = (60~\mathrm{GeV})^2$ \\       
\hspace{1cm} 4 - $Q^2 = m_t^2$        
}

\item Processes $18,97,98$:\\
{\sf 
  ACSET(2): (D=1)\\
\hspace{1cm} 1 - $Q^2 = \hat s_{Z}$ \\       
\hspace{1cm} 2 - $Q^2 = (\sum{({p^i_T}^2)} + M_Z^2)/2$ \\       
\hspace{1cm} 3 - $Q^2 = \sum{({p^i_T}^2)}/2$ \\  
\hspace{1cm} 4 - $Q^2 = M_Z^2$        
}

\item Processes $19$:\\
{\sf 
  ACSET(2): (D=1)\\
\hspace{1cm} 1 - $Q^2 = \hat s_{\rm top}$ \\       
\hspace{1cm} 2 - $Q^2 = (\sum{({p^i_T}^2)} + m_t^2)/2$ \\       
\hspace{1cm} 3 - $Q^2 = (60~\mathrm{GeV})^2$ \\       
\hspace{1cm} 4 - $Q^2 = m_t^2$        
}

\item Processes $20,21,105,107$:\\
{\sf 
  ACSET(2): (D=1)\\
\hspace{1cm} 1 - $Q^2 =(2 \cdot m_t^2) $ \\
\hspace{1cm} 2 - $Q^2 =\sum{({p^i_T}^2 + m_i^2)}/4 $ \\
\hspace{1cm} 3 - $Q^2 =m_t^2$ \\
\hspace{1cm} 4 - $Q^2 =\hat{s} $
}

\item Processes $22$:\\
{\sf 
  ACSET(2): (D=1)\\
\hspace{1cm} 1 - $Q^2 = \hat s_{\rm Z^{0\prime}}$ \\       
\hspace{1cm} 2 - $Q^2 = (\sum{({p^i_T}^2)} + M(Z^{0\prime})^2)/2$ \\       
\hspace{1cm} 3 - $Q^2 = m_t^2$    \\    
\hspace{1cm} 4 - $Q^2 = M(Z^{0\prime})^2$        
}

\item Processes $26,27$:\\
{\sf 
  ACSET(2): (D=1)\\
\hspace{1cm} 1 - $Q^2 = M_Z^2$ \\
\hspace{1cm} 2 - not applicable (same as in proc 5-6) \\
\hspace{1cm} 3 - $Q^2 = s^*_Z$ \\
\hspace{1cm} 4 - not applicable (same as in proc 5-6)
}

\item Processes $91$:\\
{\sf 
  ACSET(2): (D=1)\\
\hspace{1cm} 1 - $Q^2 = \hat s$ \\       
\hspace{1cm} 2 - $Q^2 = (\sum{({p^i_T}^2)} + M_Z^2)/2$ \\       
\hspace{1cm} 3 - $Q^2 = \sum{({p^i_T}^2)}/2$ \\       
\hspace{1cm} 4 - $Q^2 = M_Z^2$        
}
\item Processes $92 \to 93$:\\
{\sf 
  ACSET(2): (D=1)\\
\hspace{1cm} 1 - $Q^2 = (2 m_t)^2$ \\
\hspace{1cm} 2 - $Q^2 = \sum{({p^i_T}^2 + m_i^2)}/2 $ \\       
\hspace{1cm} 3 - $Q^2 = \sum{({p^i_T}^2)}/2$ \\       
\hspace{1cm} 4 - $Q^2 = \hat s$ \\       
}
\item Processes $94$:\\
{\sf 
  ACSET(2): (D=1)\\
\hspace{1cm} 1 - $Q^2 = \hat s$ \\       
\hspace{1cm} 2 - $Q^2 = (\sum{({p^i_T}^2)} + M_W^2)/2 $ \\       
\hspace{1cm} 3 - $Q^2 = \sum{({p^i_T}^2)}/2$ \\       
\hspace{1cm} 4 - $Q^2 = M_W^2$        
}
\item Processes $95$:\\
{\sf 
  ACSET(2): (D=1)\\
\hspace{1cm} 1 - $Q^2 =(2 \cdot m_t^2) $ \\
\hspace{1cm} 2 - $Q^2 =\sum{({p^i_T}^2 + m_i^2)}/2 $ \\
\hspace{1cm} 3 - $Q^2 =\sum{({p^i_T}^2)}/2 $ \\
\hspace{1cm} 4 - $Q^2 =\hat{s} $
}
\end{itemize}
\end{scriptsize}
there are the settings implemented so far.

\clearpage

\boldmath
\subsection{Installation procedure \label{s:insprod}} 
\unboldmath

The installation requires availability of the {\tt CERNLIB} fortran library as well as the {\tt LHAPDF} library.
\begin{itemize}
\item
Ungzip and untar distribution file.
\item
Modify the main Makefile to specify the {\tt LHAPDF} (LHPATH variable) and {\tt CERNLIB} paths if needed.
\item
In the main directory type {\tt make demo\_py}, {\tt make demo\_hw} or  {\tt make demo\_ar}.
It will compile {\tt demo\_py.f}, {\tt demo\_hw.f} or  {\tt demo\_ar.f}and produce 
the executables {\tt demo\_py.e}, {\tt demo\_hw.e} or  {\tt demo\_ar.e} depending 
on the selected option. The first-time call will also build and install all the
required libraries; this will not be repeated when the {\tt demo\_xx.f} are subsequently
changed. 

\item
To execute the programs type {\tt make run\_py},{\tt make run\_hw} or {\tt make run\_ar}.\newline
The scripts will change directory to {\tt prod} and create respective links to data
directories there. The execution will also be performed there. All input files
should be accessible/routed from directory {\tt prod}, the output files will
also be produced in that directory.
\end{itemize}

\boldmath
\subsection{Storing and reading events using the Les Houches accord \label{s:leshdump}}
\unboldmath

When setting the {\tt ACER=2} switch the {\bf AcerMC 3.8} is instructed to dump
the events generated by the {\bf AcerMC} library in combination with {\tt
PYTHIA} or {\tt HERWIG} into a pair of output files using the Les Houches format
\cite{Boo01}. The files are {\sf AcerMC\_pXXX\_rYYYYYYY.inparm} and {\sf
AcerMC\_pXXX\_rYYYYYYY.events}, with {\sf XXX} denoting the AcerMC process ID
while {\sf YY..} is currently the random seed used in {\bf AcerMC}, to provide a
unique identifier in the absence of other choices (run number \& similar).

The {\sf *.inparm} file is basically an information header, containing the
relevant run parameters in accordance with the Les Houches standard
\cite{Boo01}, e.g. the cross-section of the process and the number of
dumped/stored events; it is written in a 'human-readable' format in order to
provide the user with the necessary information. The {\sf *.events} file on the
other hand contains the actual record of the hard process events produced by the
{\bf AcerMC}. The event dump is obtained by compiling and running the provided
programs {\tt demo\_xx.e} as detailed in Section
\ref{s:insprod}.

Using the {\tt ACER=3} switch the thus produced event record can subsequently be
read back either into the {\bf AcerMC 3.8} generator to be processed further by
{\tt PYTHIA} or {\tt HERWIG}, i.e. the ISR/FSR and hadronisation can be added.

The same event records can also be read into the stand-alone {\tt PYTHIA} or
{\tt HERWIG} generators, by re-directing the original(dummy) {\tt UPINIT}
routine to the {\tt INITACERMC} routine and the original {\tt UPEVNT} routine to
{\tt USEACERMC} routine (In the {\bf AcerMC 3.8} setup this is done
automatically).

The files containing the respective routines are named {\sf initacermc.f} and
{\sf useacermc.f} and are provided in the {\sf leshouches} directory of the {\bf
AcerMC 3.8} distribution. The reading of a specified set of generated events is
achieved by copying the {\sf *.inparm} and {\sf *.events} files into the running
directory and linking/copying the desired {\sf *.inparm} file to a file named
{\sf inparmAcerMC.dat}; there is no need to modify the name of the {\sf *.events}
file since its name is already stored in the corresponding {\sf *.inparm} file.

\newcommand{\acermc}{{\tt AcerMC }}
\newcommand{\pythia}{{\tt PYTHIA }}
\newcommand{\herwig}{{\tt HERWIG }}
\newcommand{\tauola}{{\tt TAUOLA }}
\newcommand{\photos}{{\tt PHOTOS }}

\boldmath 
\subsection{Interface of \tauola to \pythia and \herwig \label{s:tauola}} 
\unboldmath

\begin{itemize}
\item{\tt Interface to \pythia}:
\end{itemize}

The choice of \tauola library in \pythia sets the $\tau$-s to be stable by setting
{\tt MDCJ(15,1)=0}, thus leaving them to be treated by \tauola procedures. The
\tauola library is called via the {\tt TAUOLA\_HEPEVT(IMODE)} routine, where {\tt
IMODE=-1,0,1} represent the initialisation, operation and finalisation
respectively. In the operation mode the call to\\ {\tt TAUOLA\_HEPEVT(0)} is made
after the {\tt PYEVNT} and {\tt PYHEPC(1)} calls which generate the full event and
translate and fill it into the {\tt HEPEVT} common block. After \tauola has
finished a subsequent call {\tt PYHEPC(2)} is made to translate the new event
back into the \pythia internal structure and a subsequent call to {\tt PYEXEC}
is made so that any undecayed particles (e.g. the $\pi^0$) are decayed. It was
decided against using the provided \pythia interface via the {\tt PYTAUD}
routine since this routine is called for each occurrence of a $\tau$-lepton
separately so no (complex) polarisation options can be included without
substantial changes to \pythia routines.

\begin{itemize}
\item{\tt Interface to \herwig}:
\end{itemize}
The present version 6.5 of \herwig contrary to \pythia does internal tracking of
the polarisations of the $\tau$-leptons for the more complex built-in processes
(e.g. Higgs decays); contrariwise nothing can be done for the external hard
processes passed to \herwig for hadronisation. In order to preserve this feature
the {\tt HWDTAU} routine has been modified to perform the \tauola calls only for
internal \herwig processes which provide the $\tau$ polarisation information; other
$\tau$-lepton decays are again executed via a call to {\tt TAUOLA\_HEPEVT}. Since
\herwig is using the {\tt HEPEVT} common block already as the internal event
record no translation of events is needed. In order to also keep other \herwig
parameters current (e.g. {\tt IDHW} of the particle) a call to a new routine 
{\tt HWHEPC} is provided; this new routine also corrects the vertex positions of
the $\tau$ decay products and sets the status codes of new particles to the
\herwig recognisable values. Subsequently a call to {\tt HWDHAD} is made in
order to decay any undecayed particles.

\boldmath 
\subsection{Interface of \photos to \pythia and \herwig \label{s:photos}} 
\unboldmath

\begin{itemize}
\item{\tt Interface to \pythia}:
\end{itemize}

Choosing \photos to provide the QED final state radiation sets the parameter
{\tt PARJ(90)= $2 \cdot 10^{4}$} order to prevent \pythia to radiate photons off leptons,
thus inducing double counting. The parameter is representing the threshold in
GeV below which leptons do not radiate. Pythia does not contain the routines to
handle radiation of hadrons however some caution is necessary due to some
exceptions/specific decays (e.g. $\pi^0 \to e^+ e^- \gamma$) which are 
generally recognised by \photos itself. The initialisation, execution and finalisation
are done by calls to the new {\tt PHOTOS\_HEPEVT(IMODE), IMODE=-1,0,1} subroutine, constructed
for this purpose by the \acermc authors. Since \photos operates on the {\tt HEPEVT}
record the calls to {\tt PYHEPC} routine are again necessary.

\begin{itemize}
\item{\tt Interface to \herwig}:
\end{itemize}
The present version 6.5 of \herwig does itself not provide the final state QED
radiation of any kind, thus the inclusion of \photos is simple and possibly also
rather necessary. The call sequence is the same as in \pythia, with no
need for event record conversion.

\boldmath 
\subsection{Details of the \tauola implementation}
\unboldmath

The \tauola library is built from the latest distribution source with the 
{\it Cleo} setup option. The native random generator was replaced with a link to
the random generator of the linked event generator (\pythia or \herwig) in the
same manner as done for \acermc, which decreases the number of random seeds
which need to be initialised. 

The {\tt TAUOLA\_HEPEVT} routine is a modification
of the original {\tt TAUOLA} routine provided in the file {\sf tauface\_jetset.f}
in the \tauola distribution. The modifications were restricted to allow for the
'overloaded' use of the HEPEVT record by \pythia and \herwig, where respective mother
and daughter pointers (which should match) sometimes point to different
particles (e.g. hard process copies and similar). The original {\tt TAUOLA}
routine already worked when interfaced to \pythia, albeit requiring the
non-default setting of {\tt MSTP(128)=1}, whereby all the (sometimes useful)
links to the hard record were lost. The modified version also works with the
default \pythia {\tt MSTP(128)=0} setting. 

Additional modifications were made in
order to use the parameters set in {\sf tauola.card} and the call to {\tt
INIMAS} routine was replaced by a call to {\tt TAU\_INIMAS}, which is a copy of
the former but sets the particle masses to the values of \pythia or \herwig
defaults. 

A routine {\tt TAUBRS} that defines the special decay modes (e.g. LEPTON-HADRON,
where one tau decays hadronically and the other one leptonically) was written;
it operates by modifying the internal {\tt GAMPRT} array on an event by event
basis. A related routine {\tt TAUBR\_PRINT} prints the value of the branching
ratio into the {\sf tauola.out} file.No changes to the native tauola code was 
necessary (and therefore not made). 

\boldmath 
\subsection{Details of the \photos implementation}
\unboldmath

The \photos is also built from the latest distribution source. As in \tauola The
native random generator was replaced with a link to the random generator of the
linked event generator (\pythia or \herwig) in the same manner as done for
\acermc.

The {\tt PHOTOS\_HEPEVT} routine and the subsequently called {\tt
PHOTOS\_HEPEVT\_MAKE} are newly written routines that inspects the {\tt HEPEVT}
record for charged particles and finds their highest 'mother' particle, which is
consequently passed to photos routines as the starting point. Photos itself then
walks down the branches and performs radiation where possible. A bookkeeping of
the starting points is made in {\tt PHOTOS\_HEPEVT} in order to prevent multiple
invocations of radiation on the same particle.

Original photos code had to be modified due to the 'overloaded' {\tt HEPEVT}
record, since its requirements for matching mother-daughter pointers were too
strict for either \pythia (with external processes and/or {\tt MSTP(128)=0}
setting) or \herwig. The modification was limited to {\tt PHOTOS\_MAKE} and {\tt
PHOBOS} routines. In addition the tracking of {\tt IDHW} array was added to {\tt
PHOTOS\_MAKE} to accommodate the \herwig event record. A further modification was
however necessary in the {\tt PHOIN} routine since in \herwig the entry {\tt
JMOHEP(2,I)} is not empty but filled with colour flow information, which in turn
inhibited \photos radiation off participating particles\footnote{\photos expects
the non-zero second 'mother' {\tt JMOHEP(2,I)} entry only for $gg(qq) \to
t\bar{t}$ process, which is treated by a set of dedicated routines; this in turn
clashes with \herwig 'overloaded' {\tt JMOHEP(2,I)} entry.}.  No modifications
of the core (physics) \photos code was made.

\boldmath
\section{Outlook and conclusions} 
\unboldmath

In this paper we presented the {\bf AcerMC} Monte Carlo Event Generator, based on the library
of the matrix-element-based generators and interfaces to the universal event generators
{\tt PYTHIA~6.4} and {\tt HERWIG 6.5}. The interfaces are based on the standard proposed
in \cite{Boo01}. 

The presented library fulfills the following goals:
\begin{itemize}
\item
It gives a possibility to generate the few Standard Model background processes which
were recognised as very dangerous for the searches for the {\it New Physics} at LHC,
and generation of which was either unavailable or not straightforward so far.
\item
Although the hard process event is generated with matrix-element-based generator, the
provided interface allows to complete event generation with initial and final
state radiation, multiple interaction, hadronisation, fragmentation and decays,
using implementation of either {\tt PYTHIA~6.4} or {\tt HERWIG 6.5}.
\item
These interfaces can be also used for studying systematic differences between 
 {\tt PYTHIA~6.4} or {\tt HERWIG~6.5} predictions for the underlying QCD processes.
\end{itemize}

The complete list of the native {\bf AcerMC} processes implemented so far is:
$gg, q \bar q \to t \bar t b \bar b$; $q \bar q \to W(\to \ell \nu) b \bar b$;
$gg, q \bar q \to Z/\gamma^*(\to \ell \ell) b \bar b$; $q \bar q \to W(\to \ell
\nu) t \bar t$; $gg, q \bar q \to Z/\gamma^*(\to \ell \ell, \nu \nu, b \bar b) t
\bar t$; $g g, q \bar q \to (Z/W/\gamma^* \to)t \bar t b \bar b$;
$g g, q \bar q \to t \bar t t \bar t$;  $gg, q \bar q  \to (  t \bar t \to) f \bar f b f \bar f \bar b$ ; 
$ gg, q \bar q  \to  (W W b b~ \to) f \bar f f \bar f b \bar b$, single top, $Z^0 b$ and $Z^{0\prime} \to t \bar t$ processes.
  We plan to extend this not too exhaustive, but very much demanded list of
processes, in the near future.

Several improvements of the existing Monte Carlo algorithms/programs
have been developed in the process of this work. Let us make short
list of the most interesting ones: (1) The use of the adapted
Kajantie-Byckling enables one to automatise and modularise the phase
space generation of $\rm n$-particle final states.  (2) The additions
and extensions to the available (multi-channel) phase space algorithms
(e.g. Breit-Wigner function with s-dependent width, mass-threshold
effects) lead to substantial improvement of the unweighting
efficiency; Figs.~\ref{f:bwcomp},~\ref{f:tauveg} and~\ref{f:weights}
illustrate the improvements achieved in the generation efficiency.
(3) The power of the multi-channel optimisation was
enhanced by using the modified {\tt ac-VEGAS} package.  We believe that the
modification in the {\tt VEGAS} code represents a very powerful extension of
this package; (4)the colour flow information has been obtained after some modification of
{\tt MADGRAPH} package. 

Having all these different production processes implemented in the consistent
framework, which can be also directly used for generating standard processes
available in either {\tt PYTHIA~6.4} or {\tt HERWIG 6.5} Monte Carlo, represents
very convenient environment for several phenomenological studies dedicated to
the LHC physics.  Such frame was not available to our knowledge so far. We hope
that it can serve as an interesting example or even a framework.  This way some
tools for discussing the ambiguities due to QCD effects are collected, however
the necessary discussion for the appropriate uncertainties is still not
exhausted.  Nevertheless some discussions using this tool can be already found
in \cite{ATLCOMP013}, \cite{ATLCOMP014}, \cite{ATLCOMP025},
\cite{ATLCOMP032}.

\section*{Acknowledgments}

We would like to thank Ian Hinchliffe, Daniel Froidevaux, Torbjorn Sjostrand, Bryan
Webber and Alessandro Ballestrero, Zbigniew Was and Svjetlana Fajfer
for several very valuable discussions. In particular, we thank and ackgnowledge the cooperation
of Ian Hinchiffe and Liza Mijovic on the work of developing the matrix element and parton shower matching
prescriptions. We would like also to thank all
our colleagues from ATLAS Collaboration who were the first and very
enthusiastic users of the preliminary versions of this package.  We
both very warmly acknowledge the support from the CERN PH division.

\appendix


\boldmath
\section{Feynman Diagrams}
\unboldmath

The 38+7 Feynman diagrams contributing to the 
$g g,q \bar{q} \to t \bar t  b \bar b$ production. Only four flavours are included 
for incoming quarks. Contribution of the incoming b-quarks could be excluded 
from the calculations thanks to very high suppression induced by either 
the parton density functions and/or CKM matrix elements.
\begin{figure}[hb]
\begin{center}
\hspace{-2cm}
\mbox{
     \epsfxsize=6.5cm
     \epsffile{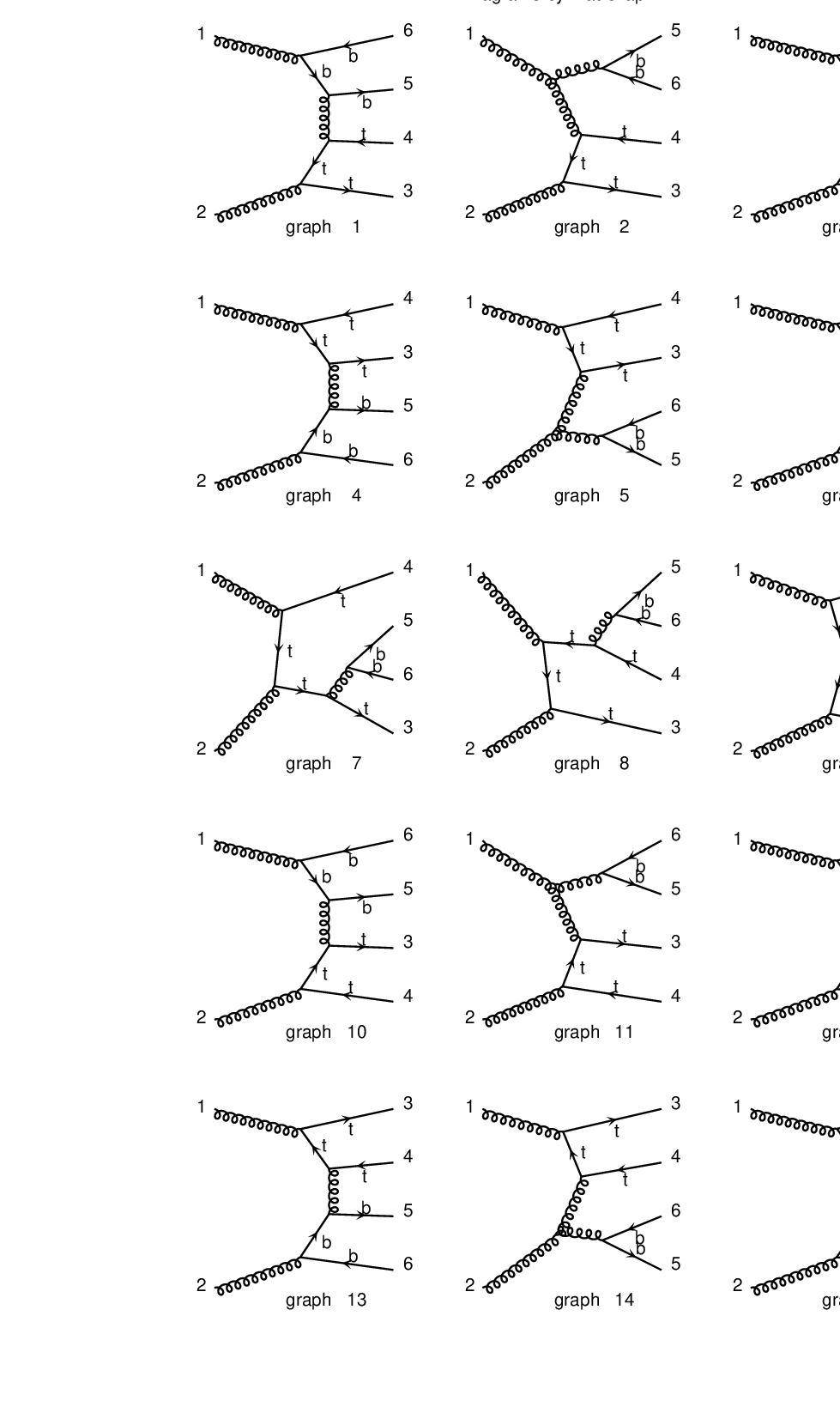}
}
\mbox{
     \epsfxsize=6.5cm
     \epsffile{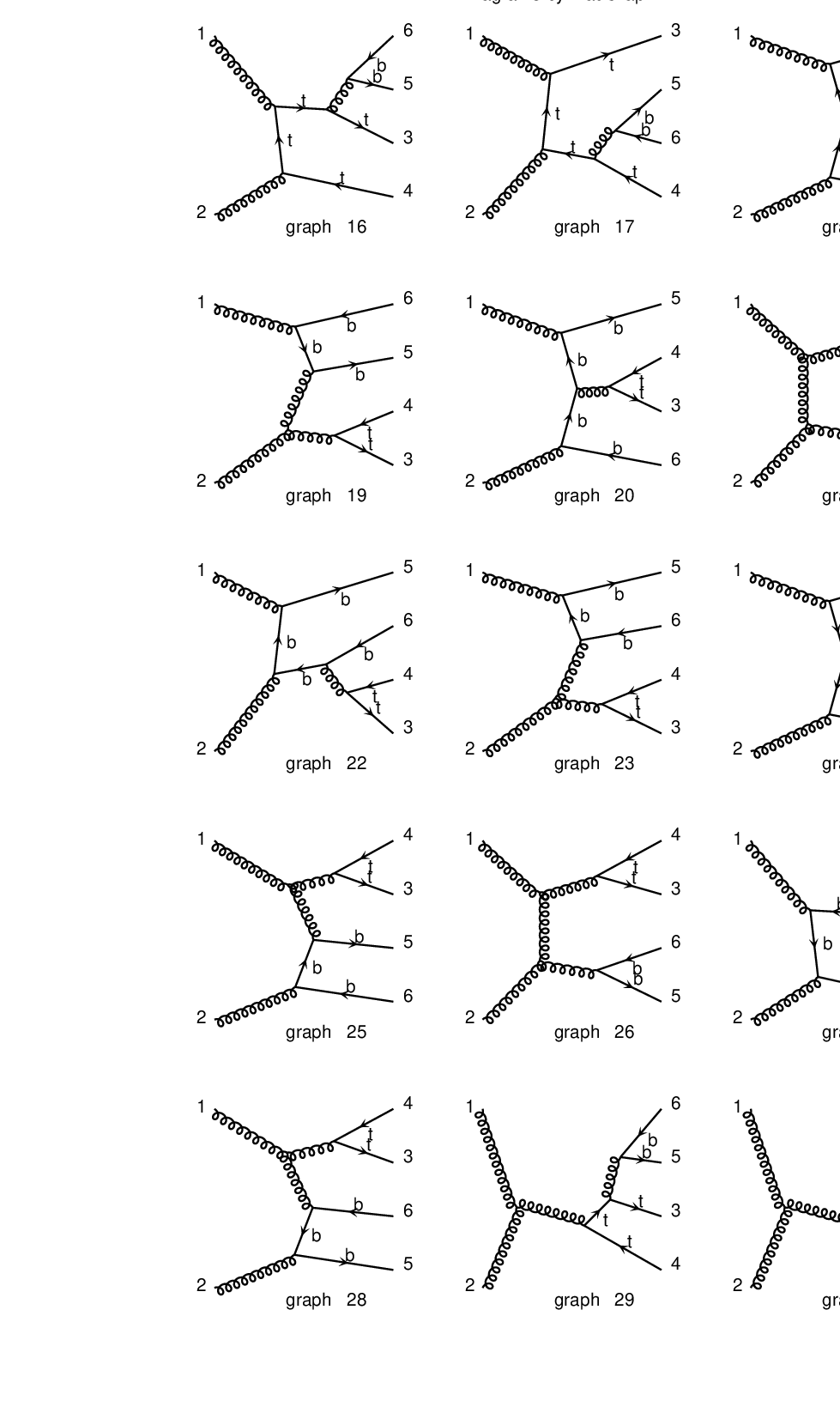}
}
\end{center}
\begin{center}
\hspace{-2cm}
\mbox{
     \epsfxsize=6.5cm
     \epsffile{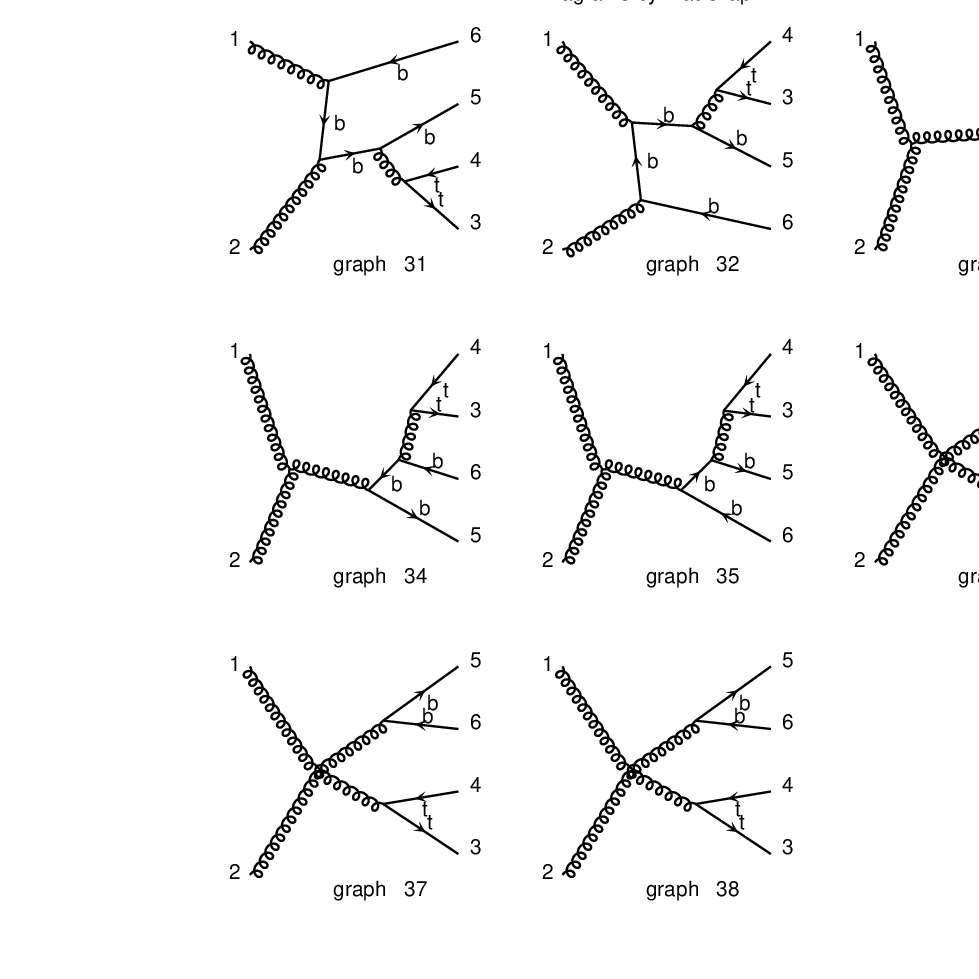}
}
\mbox{
     \epsfxsize=6.5cm
     \epsffile{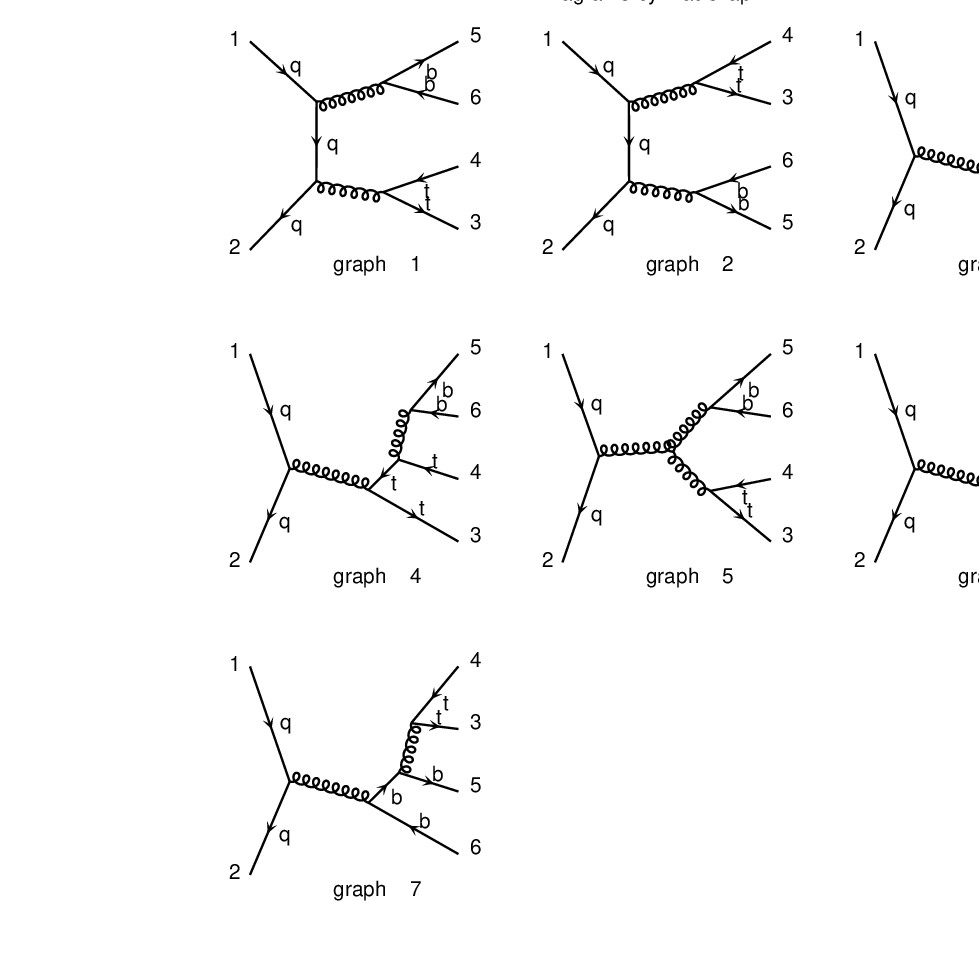}
}
\end{center}

\vglue -0.5cm
\isucaption{\em
The Feynman diagrams for the processes
$gg,q \bar{q} \to t \bar t b \bar b$.
\label{f:ttbbdia}}
\end{figure}

\begin{figure}[hb]
\vspace{1cm}
\begin{center}
\mbox{
     \epsfxsize=12.5cm
     \epsffile{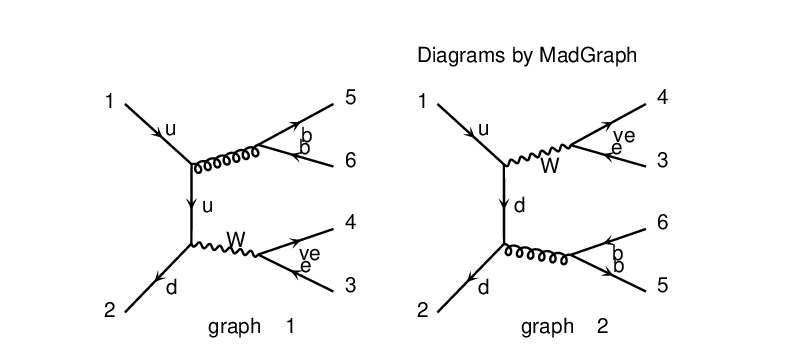}
}
\end{center}
\isucaption{\em
The Feynman diagrams for the process
 $q \bar q \to W b \bar b \to e^+ \nu_e\; b \bar b $. The same set is used for
$q \bar q \to W b \bar b \to e^+ \nu_e\; t \bar t $ process, with b-quarks 
replaced by top-quarks. 
\label{PS:1}}
\end{figure}

\newpage
The Feynman diagrams contributing to the $q \bar q \to W (\to \ell \nu) b \bar b$ and
 $q \bar q \to W (\to \ell \nu) t \bar t$ matrix
element are just two t-channel diagrams with fermion exchange and double conversion into
an off-shell W boson and a virtual gluon; the W boson subsequently decays leptonically
into $\ell \nu$ and the gluon splits into a $b \bar b$ pair or $t \bar t$ pair
 respectively.
(c.f. Figure \ref{PS:1}).

\begin{figure}[hb]
\vspace{1cm}
\begin{center}
\hspace{-2cm}
\mbox{
     \epsfxsize=6.5cm
     \epsffile{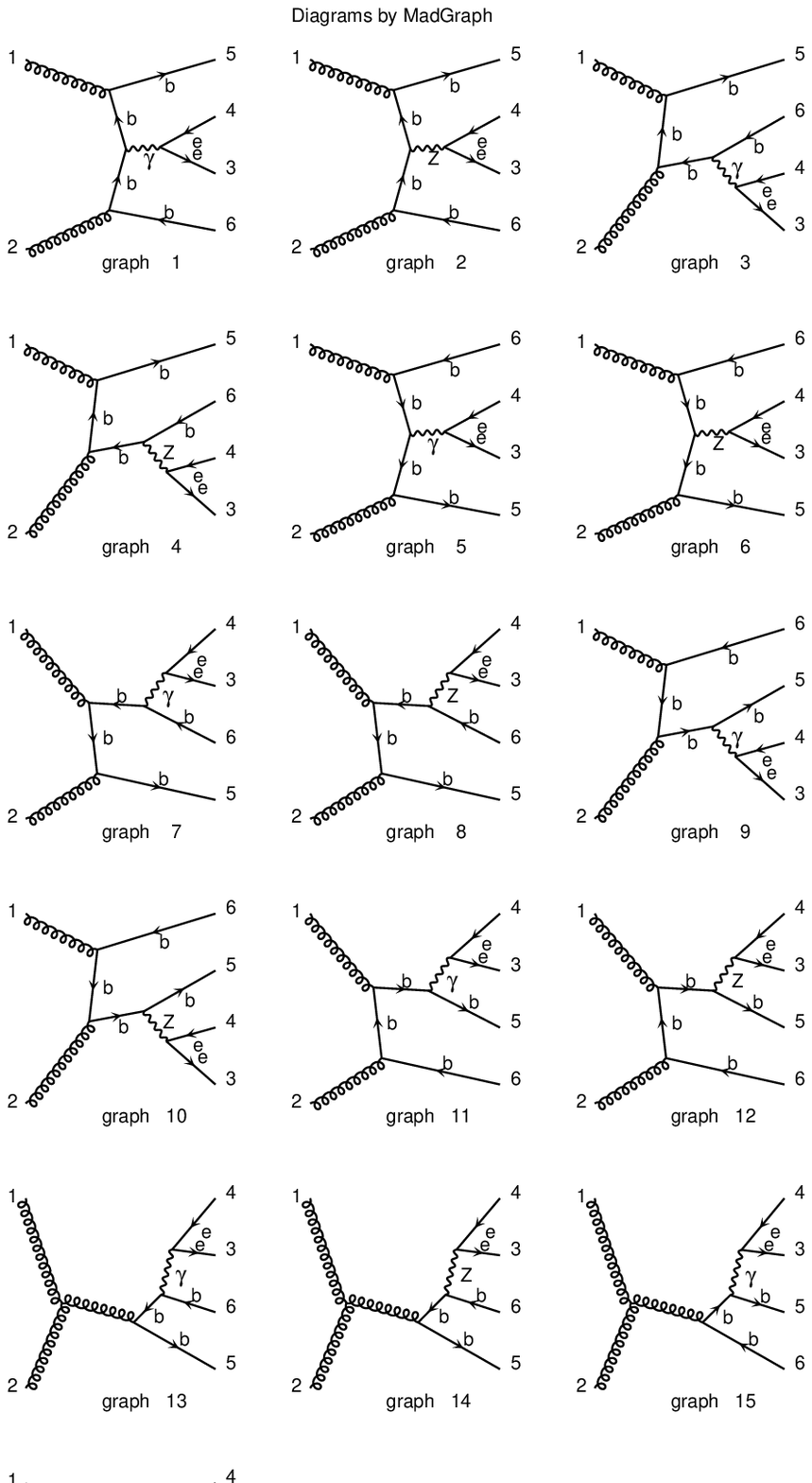}
}
\mbox{
     \epsfxsize=6.5cm
     \epsffile{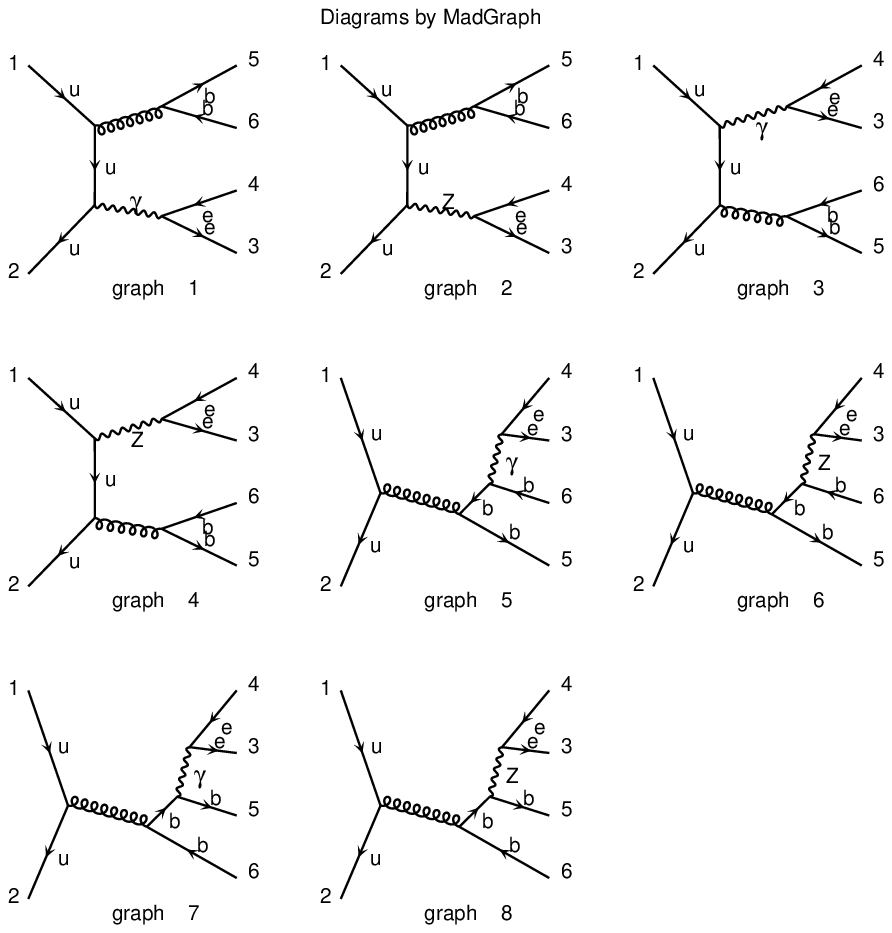}
}
\end{center}
\vspace{1cm}
\isucaption{\em
The Feynman diagrams for the processes
 $gg, q \bar q \to Z/\gamma^* b \bar b \to e e b \bar b $. 
\label{FA3:1}}
\end{figure}

\newpage
The Feynman diagrams contributing to the
 $g g, q \bar q \to Z/\gamma^*(f \bar f, \nu \nu) b \bar b$
production are shown in Figure ~\ref{FA3:1}.  The dominant contribution comes from the (2)
and (6) configurations for the processes with $gg$ initial state and the double conversion
configuration (2),(4) for the ones with $q \bar q$ initial state.
The same set of Feynman diagrams is used for the 
 $g g, q \bar q \to Z/\gamma^*(f \bar f, \nu \nu ) t \bar t$ process, with b-quarks being replaced by the top-quarks. 
If the  $Z/\gamma^*(\to b \bar b)$ decay mode is simulated, 
it represents only subset of the
EW production of $t \bar t b \bar b$ final state.
\clearpage

\begin{figure}[htb]
\begin{center}
\vspace{-0.6cm}
\hspace{-1cm}
\mbox{
     \epsfxsize=6.45cm
     \epsffile{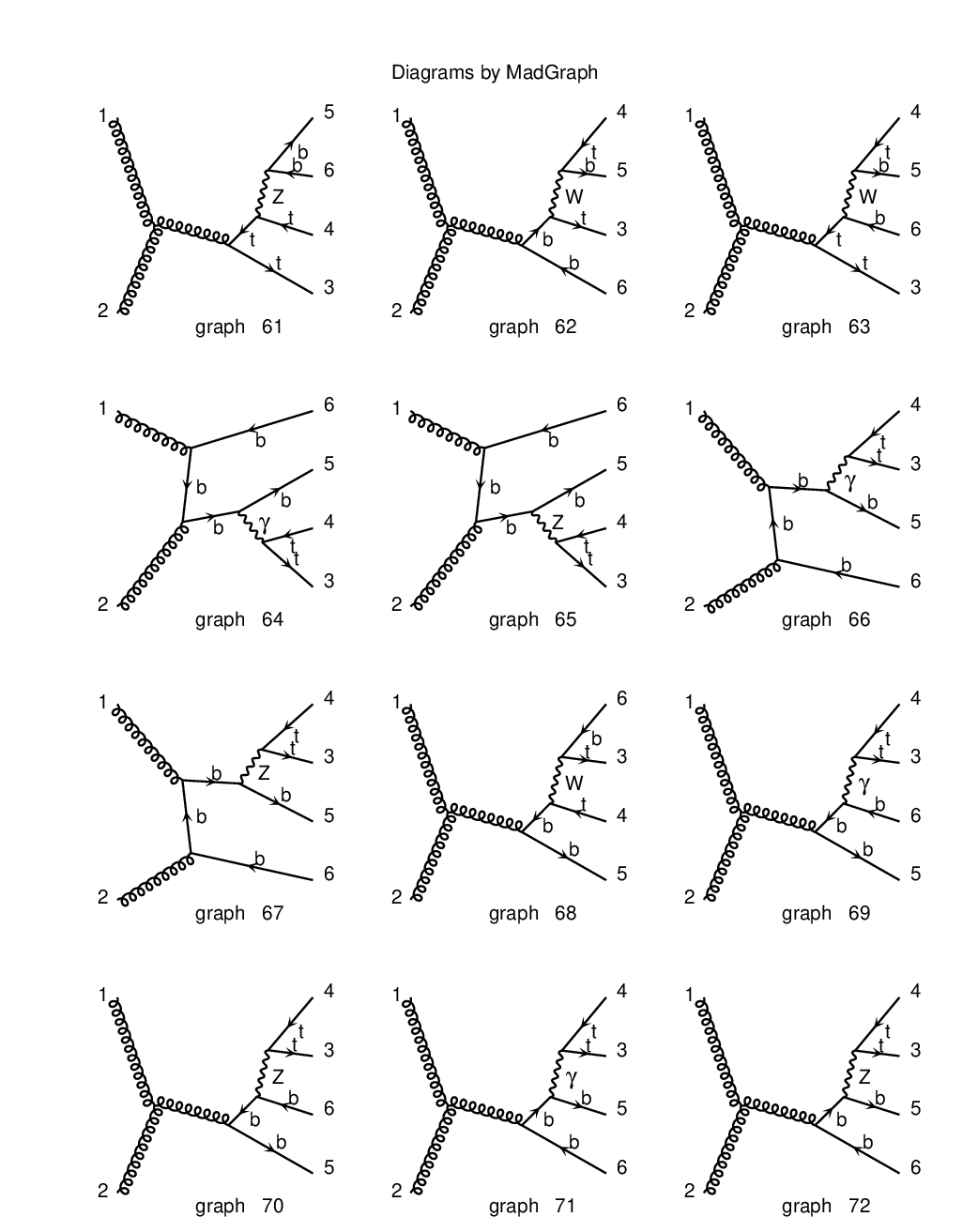}
}
\end{center}
\vspace{-.5cm}
\isucaption{\em
The Feynman diagrams for the processes
 $gg \to (Z/W/\gamma^* \to) b \bar b t \bar t $. 
\label{FA3:2}}
\vspace{-0.8cm}
\end{figure}

The complete set of th the Feynman diagrams contributing to the full electro-weak
 $gg,q\bar{q} \to (Z/W/\gamma^* \to) b \bar b t \bar t$ production mediated by exchange of the
$Z/W/\gamma^*$ bosons is shown in Fig.~\ref{FA3:2}
\begin{figure*}[htb]
\vspace{-0.5cm}
\begin{center}
\hspace{-1cm}
\mbox{
     \epsfxsize=6.45cm
     \epsffile{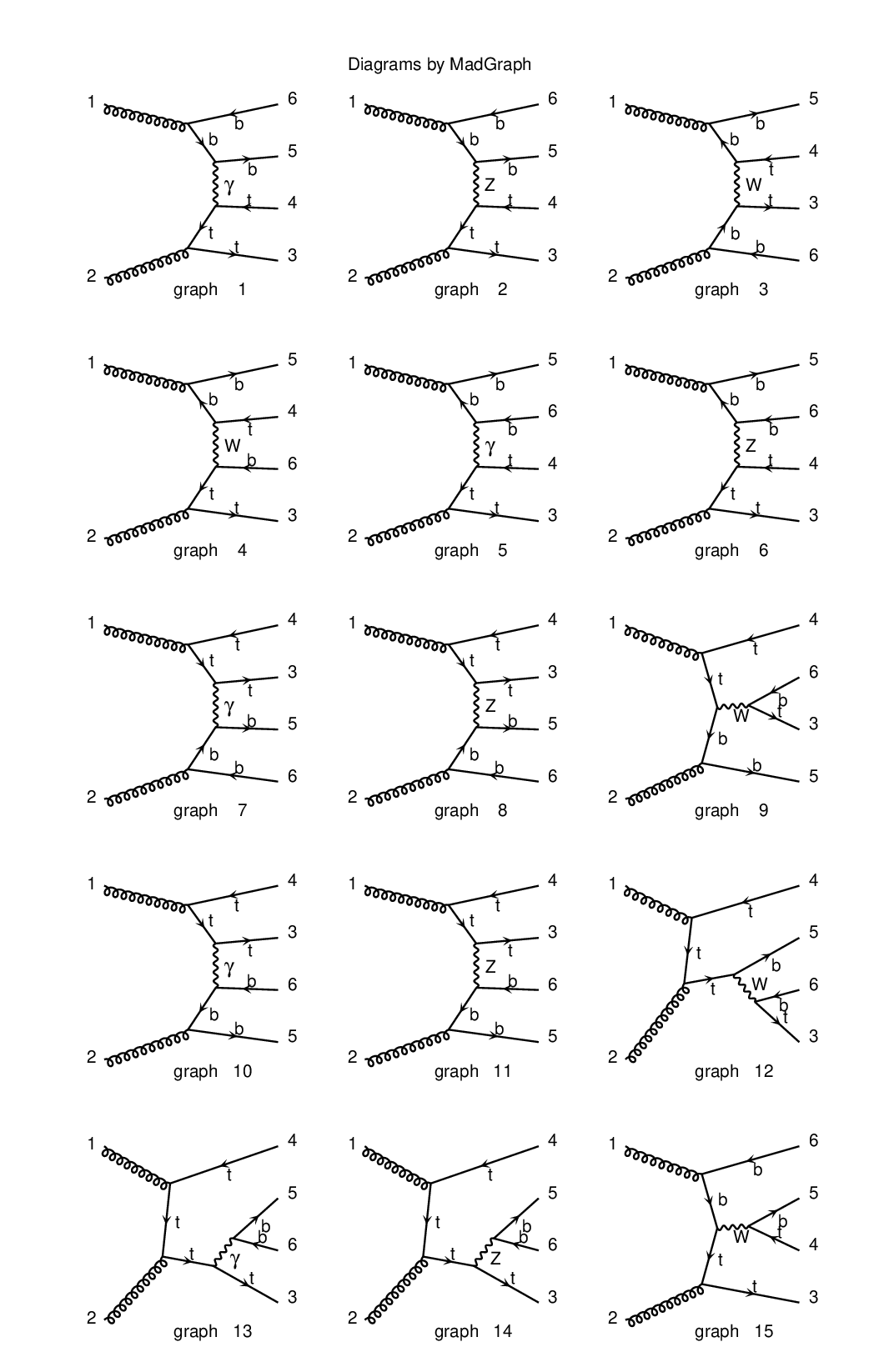}
}
\mbox{
     \epsfxsize=6.45cm
     \epsffile{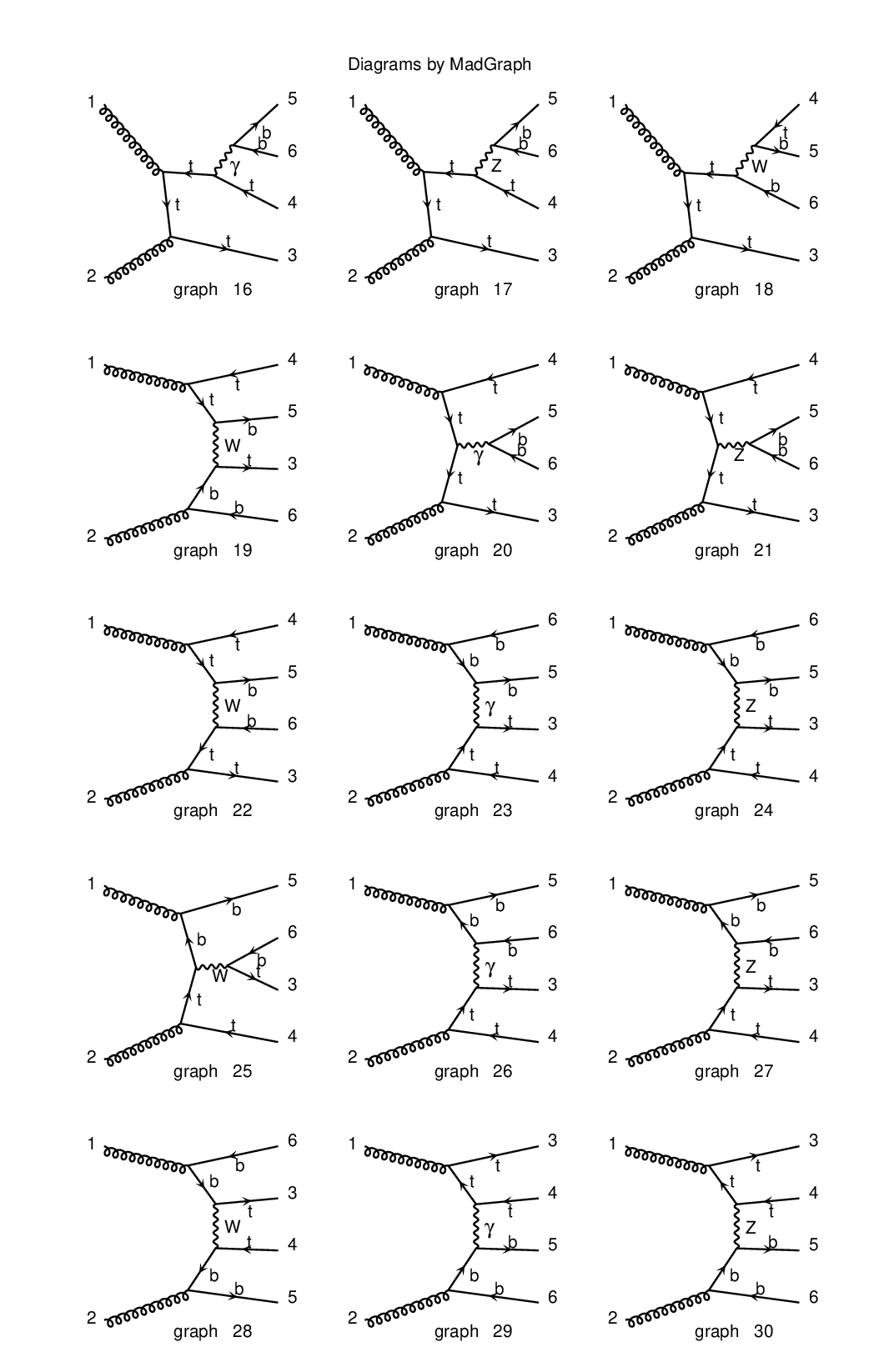}
}
\end{center}
\vspace{-0.8cm}
\begin{center}
\hspace{-1cm}
\mbox{
     \epsfxsize=6.45cm
     \epsffile{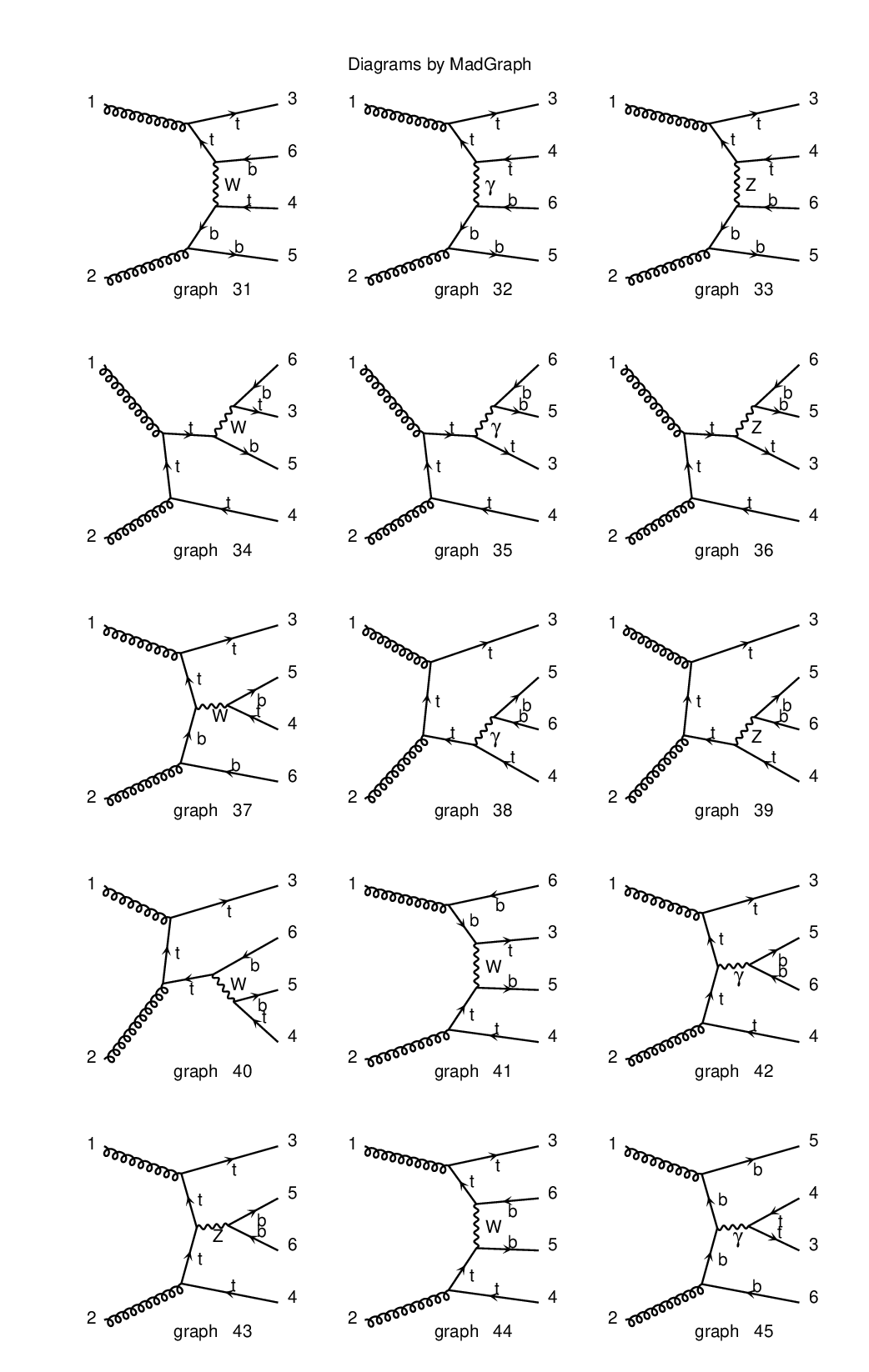}
}
\mbox{
     \epsfxsize=6.45cm
     \epsffile{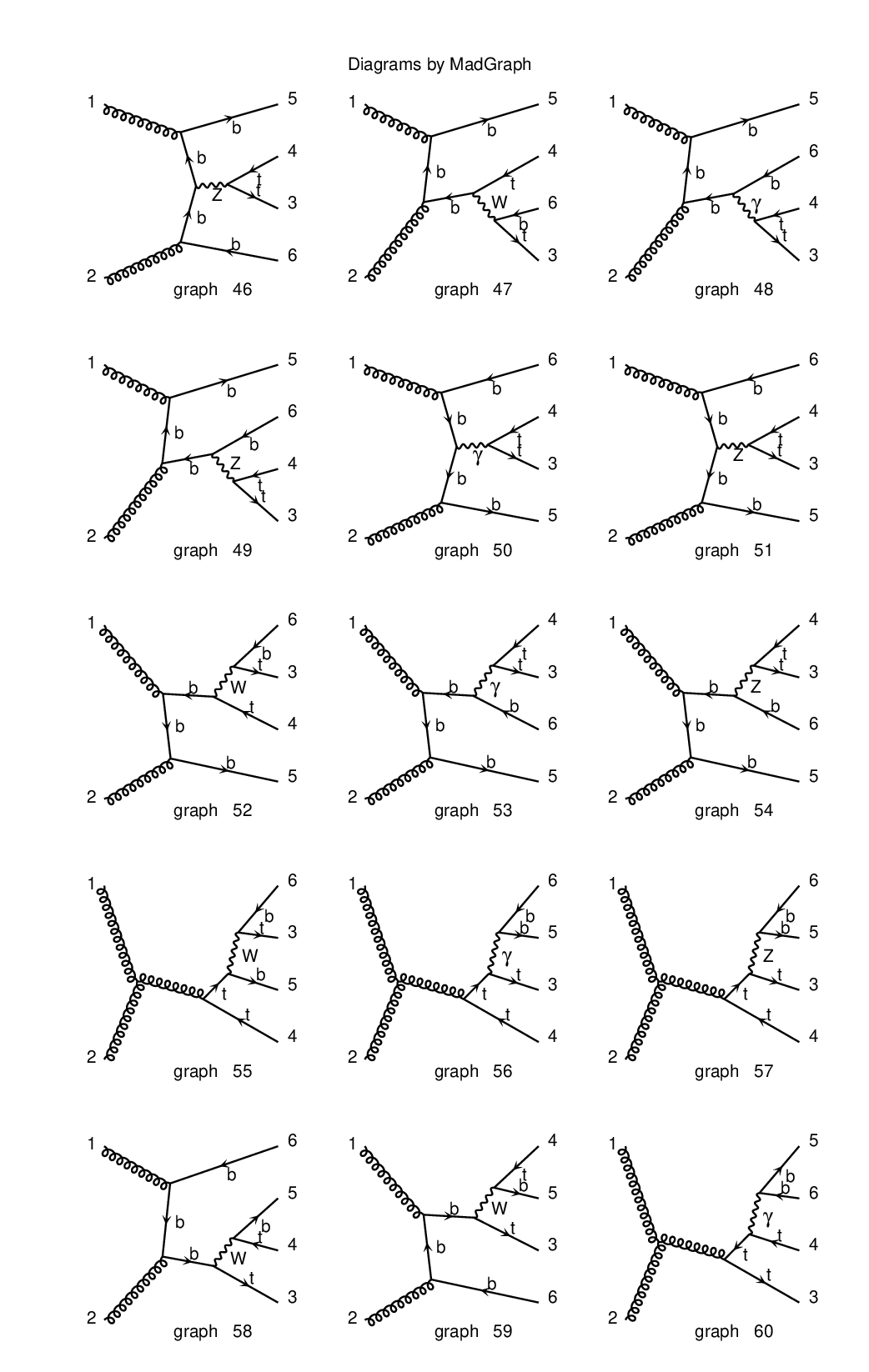}
}
\end{center}
\end{figure*}
\clearpage

\begin{figure}[htb]
\vspace{-0.4cm}
\begin{center}
\hspace{-0.7cm}
\mbox{
     \epsfxsize=7.2cm
     \epsffile{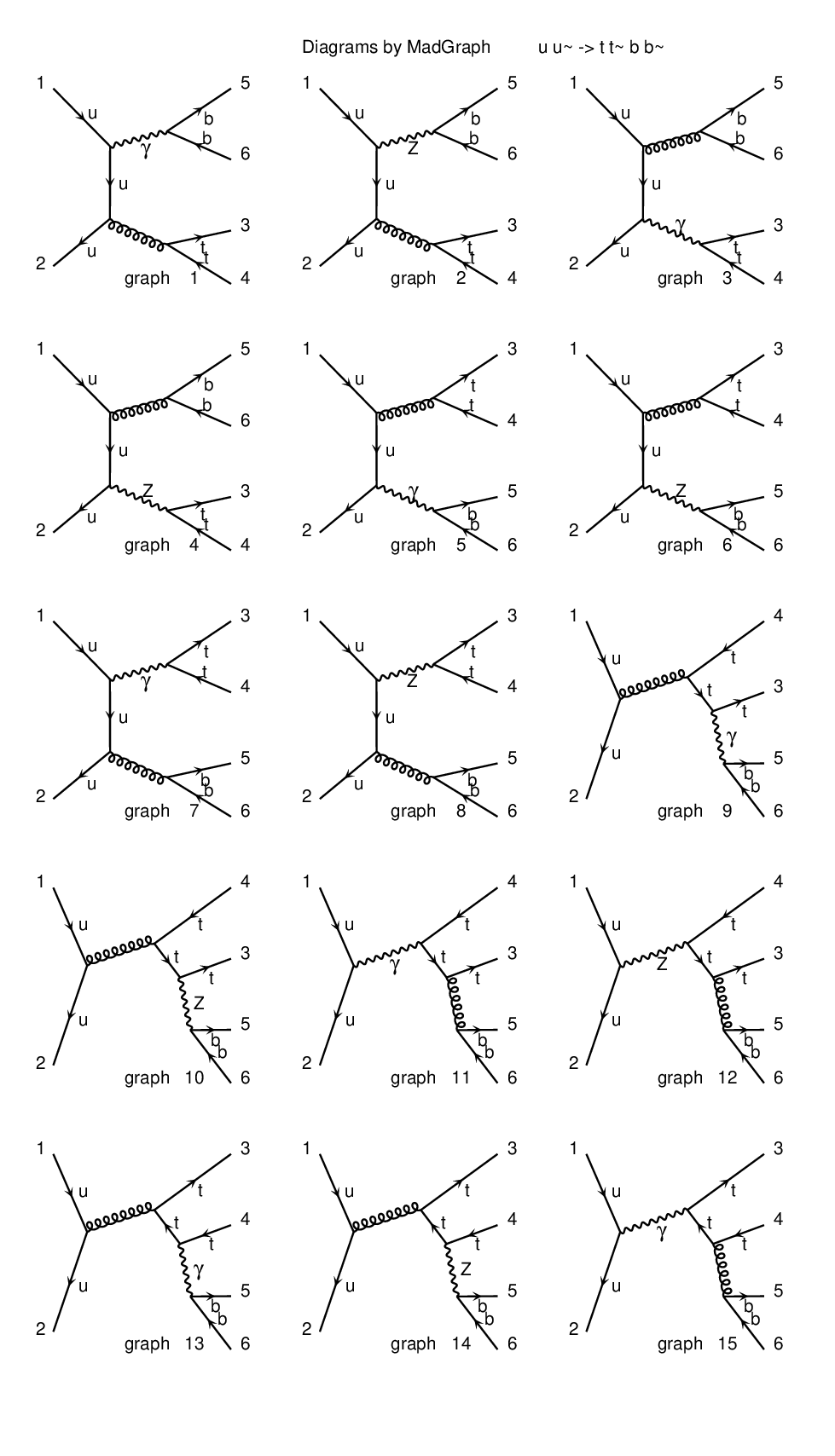}
}\hspace{-0.6cm}
\mbox{
     \epsfxsize=7.2cm
     \epsffile{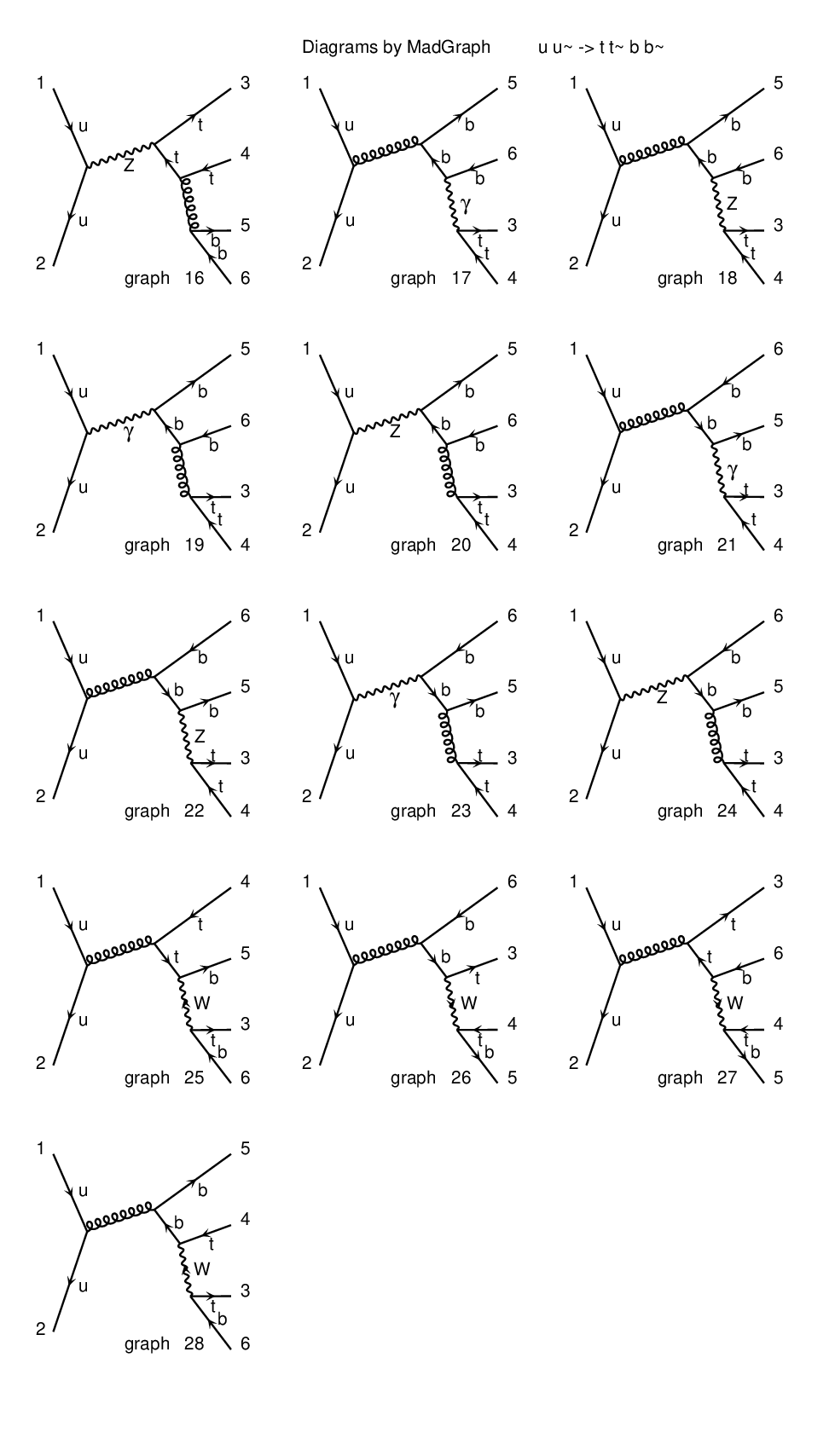}
}
\end{center}
\vspace{-1.4cm}
\isucaption{\em
The Feynman diagrams for the processes
 $q\bar{q} \to (Z/W/\gamma^* \to) b \bar b t \bar t $. 
\label{FA3:3}}
\end{figure}
\clearpage

\begin{figure}[hb]
\vspace{-0.2cm}
\begin{center}
\hspace{-1cm}
\mbox{
     \epsfxsize=6.45cm
     \epsffile{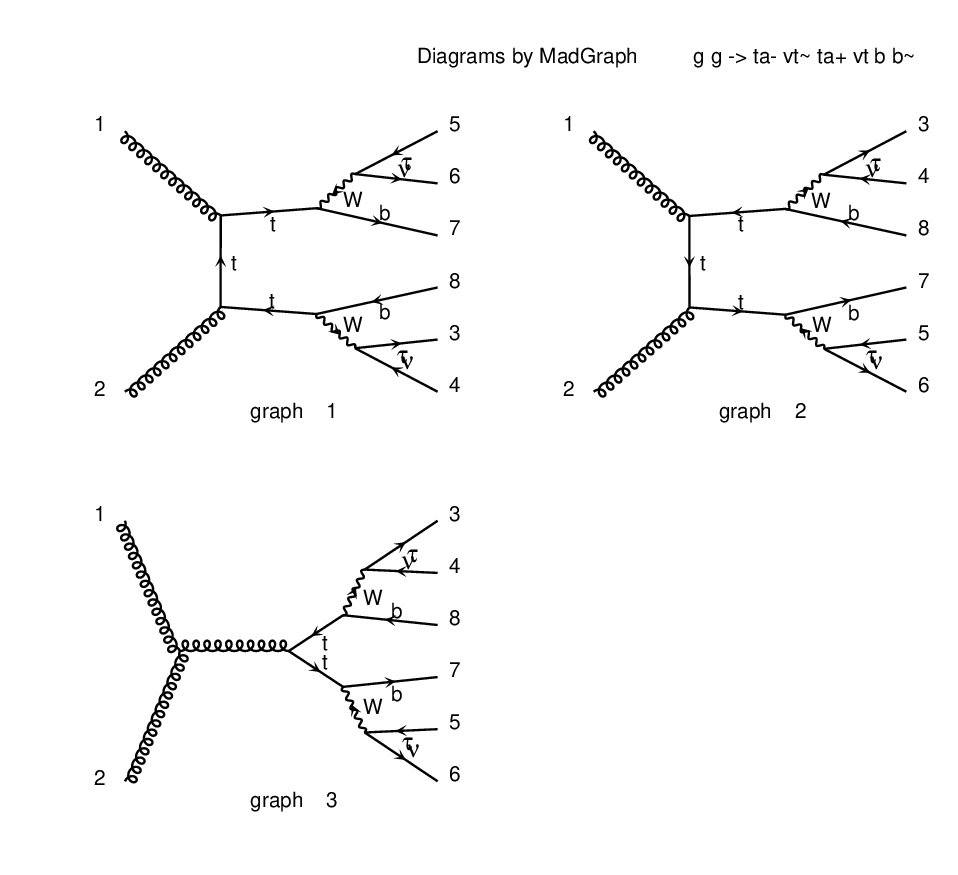}
}
\vspace{-3.5cm}

\mbox{
\hspace{5.4cm}
     \epsfxsize=6.85cm
     \epsffile{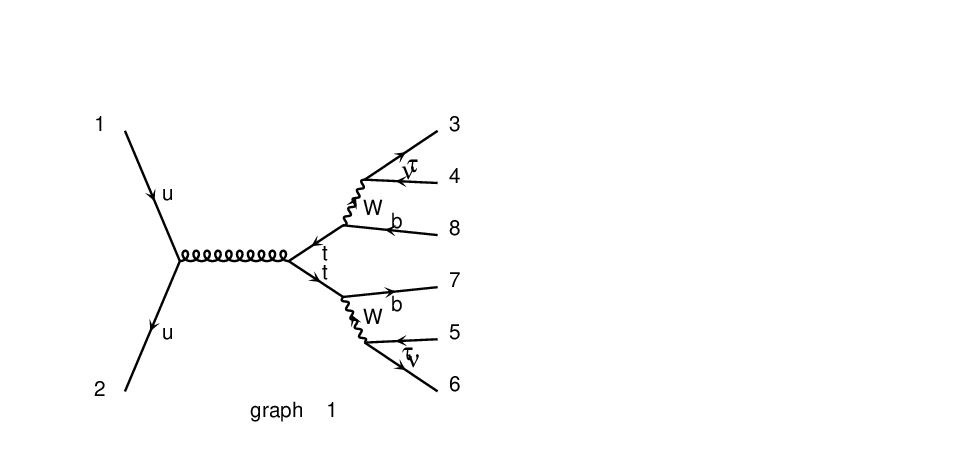}
}

\end{center}
\vspace{-0.8cm}
\isucaption{\em
The Feynman diagrams for the processes
 $gg,q\bar{q} \to (t \bar t \to) f \bar f b f \bar f \bar b $. 
\label{FA3:4}}
\end{figure}

\begin{figure}[hb]
\vspace{-1.2cm}
\begin{center}
\hspace{-0.7cm}
\mbox{
     \epsfxsize=7.2cm
     \epsffile{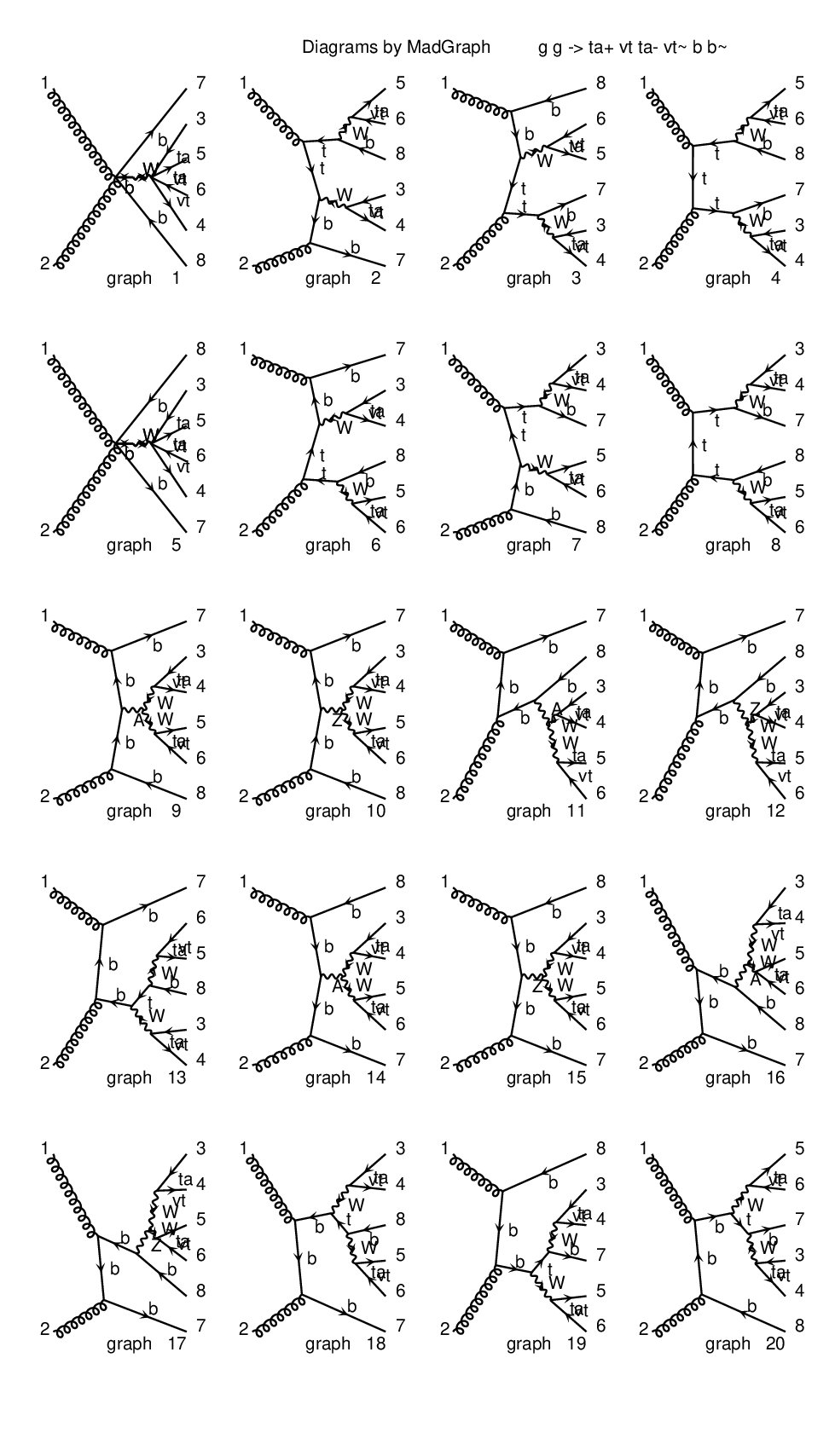}
}\hspace{-0.6cm}
\mbox{
     \epsfxsize=7.2cm
     \epsffile{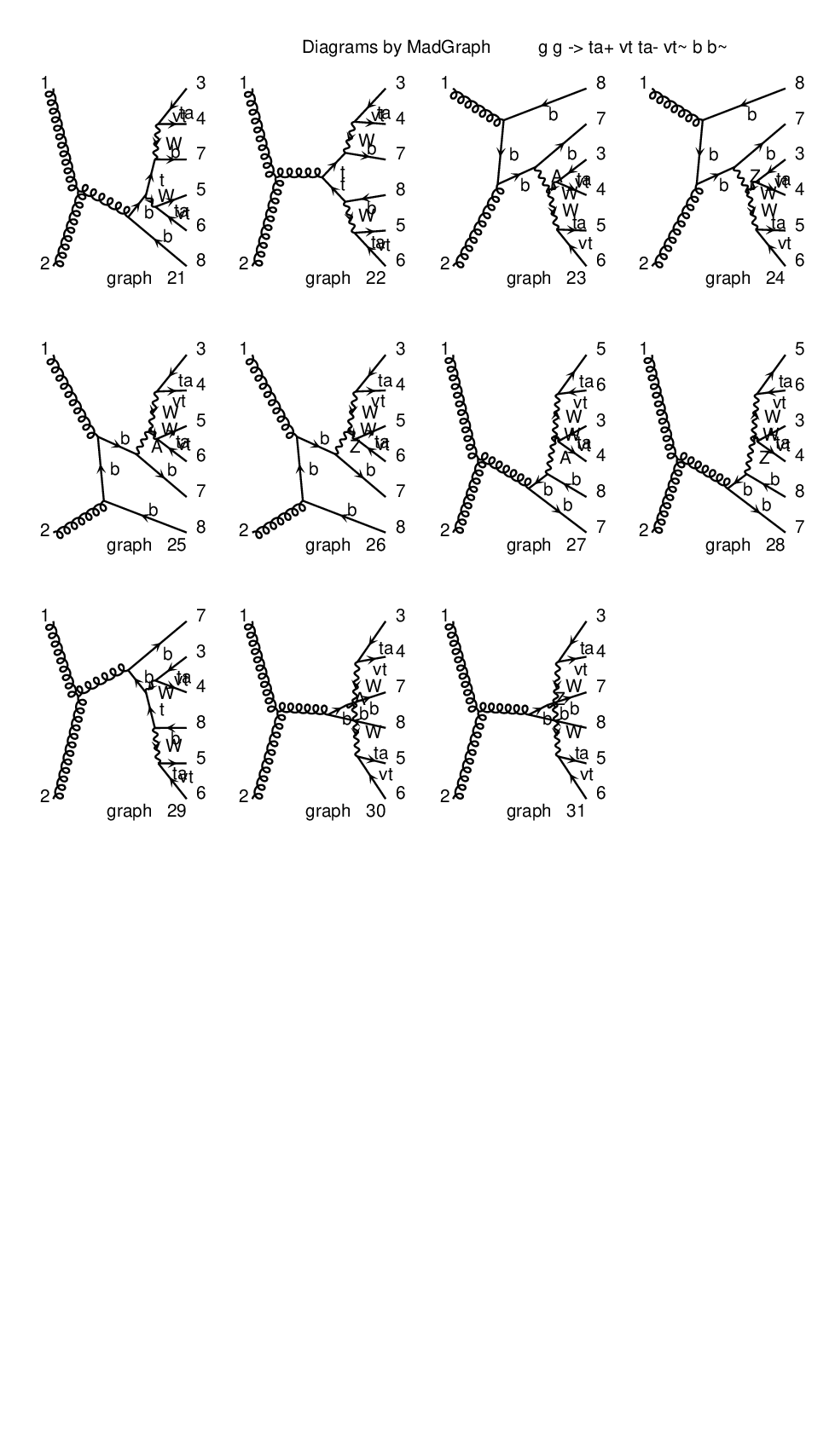}
}
\end{center}
\vspace{-1.4cm}
\isucaption{\em
The Feynman diagrams for the processes
 $gg \to (Wb Wb  \to) f \bar f b f \bar f \bar b $. 
\label{FA3:5}}
\end{figure}

\begin{figure}[hb]
\vspace{-0.8cm}
\begin{center}
\hspace{-1cm}
\mbox{
     \epsfxsize=7.2cm
     \epsffile{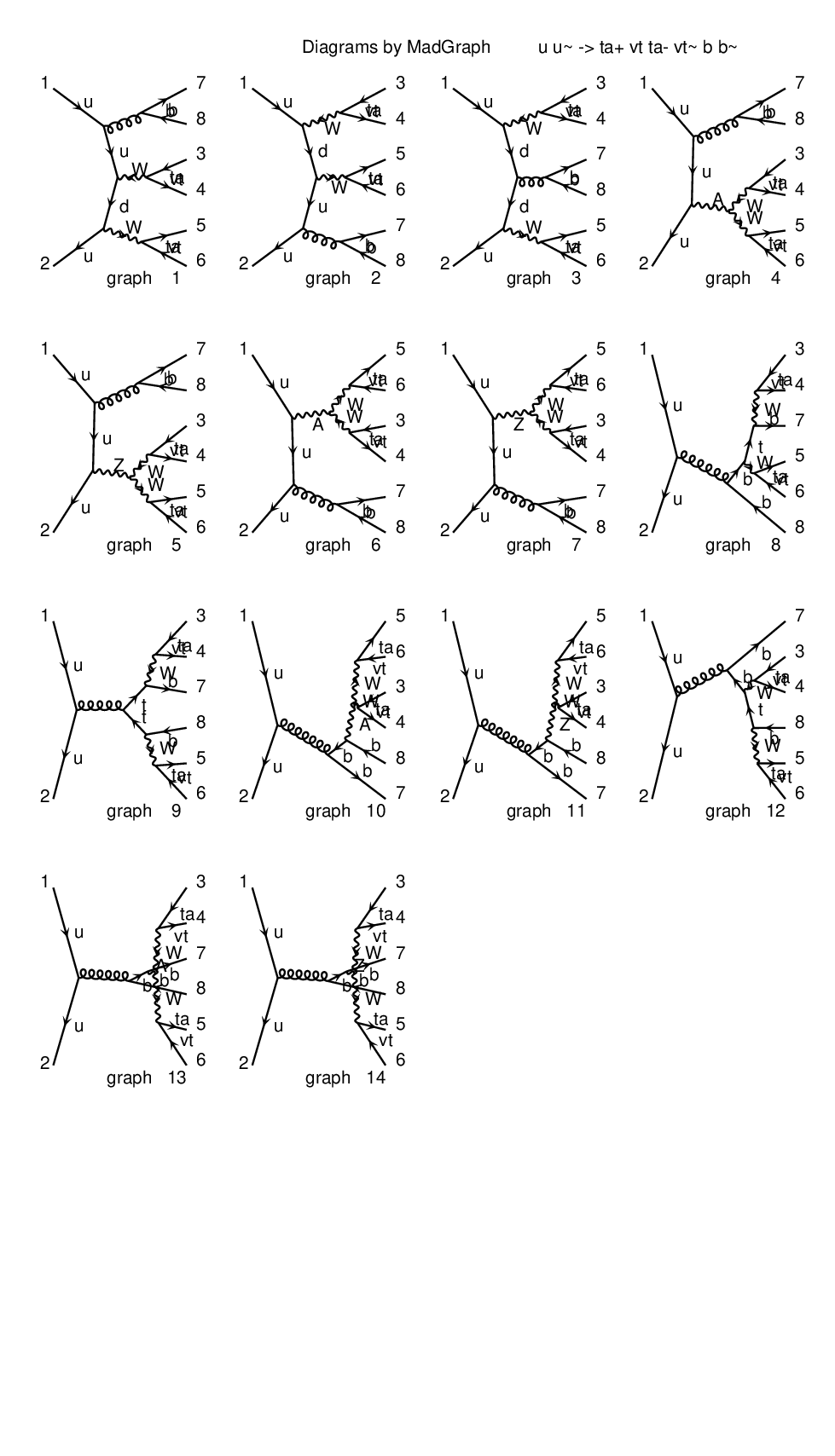}
}
\end{center}
\vspace{-3cm}
\isucaption{\em
The Feynman diagrams for the processes
 $q\bar{q} \to (Wb Wb \to) f \bar f b f \bar f \bar b $. 
\label{FA3:6}}
\end{figure}

\begin{figure}[hb]
\vspace{-0.2cm}
\begin{center}
\hspace{-1cm}
\mbox{
     \epsfxsize=6.45cm
     \epsffile{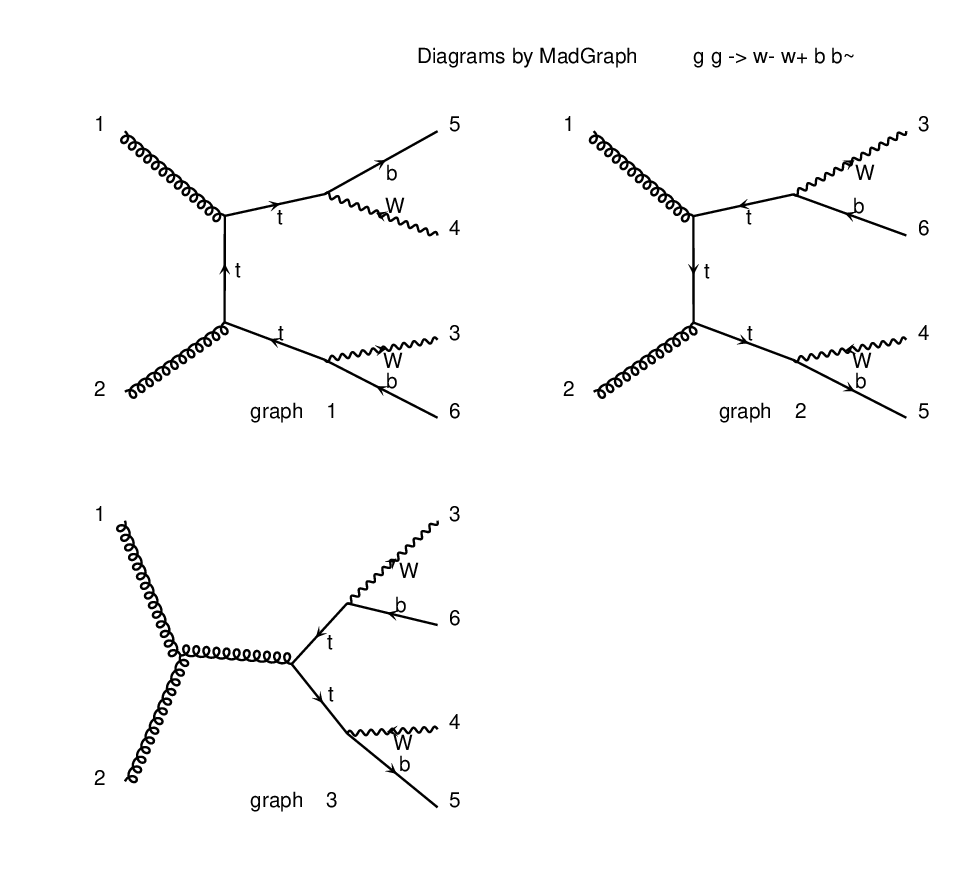}
}
\end{center}
\isucaption{\em
The Feynman diagrams for the processes
 $gg \to Wb Wb $. 
\label{FA3:7}}
\end{figure}

The set of the Feynman diagrams contributing to the $gg,q\bar{q} \to t \bar t$ production implemented
with different approaches: resonant only $2 \to 6$ process  $gg,q\bar{q} \to (t \bar t \to) f \bar f b f \bar f \bar b$,
complete  $2 \to 6$ process  $gg,q\bar{q} \to (W b W \bar b \to ) f \bar f b f \bar f \bar b$ and 
 $2 \to 4$ process  $gg \to W b W \bar b \to $ are shown in Figs.~\ref{FA3:4}-~\ref{FA3:7} .

\clearpage

\begin{figure}[hb]
\begin{center}
\hspace{-2cm}
\mbox{
     \epsfxsize=6.5cm
     \epsffile{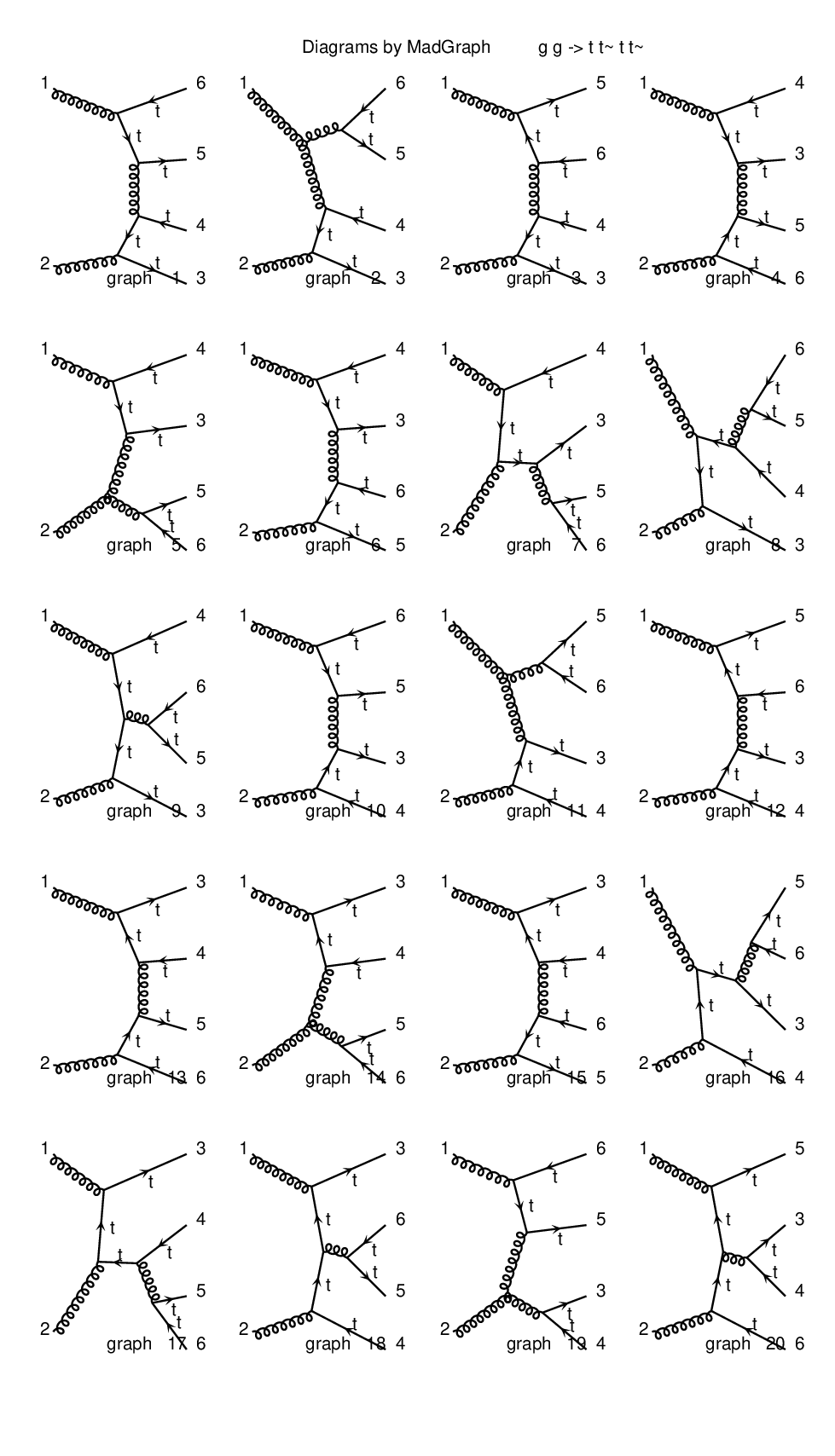}
}
\mbox{
     \epsfxsize=6.5cm
     \epsffile{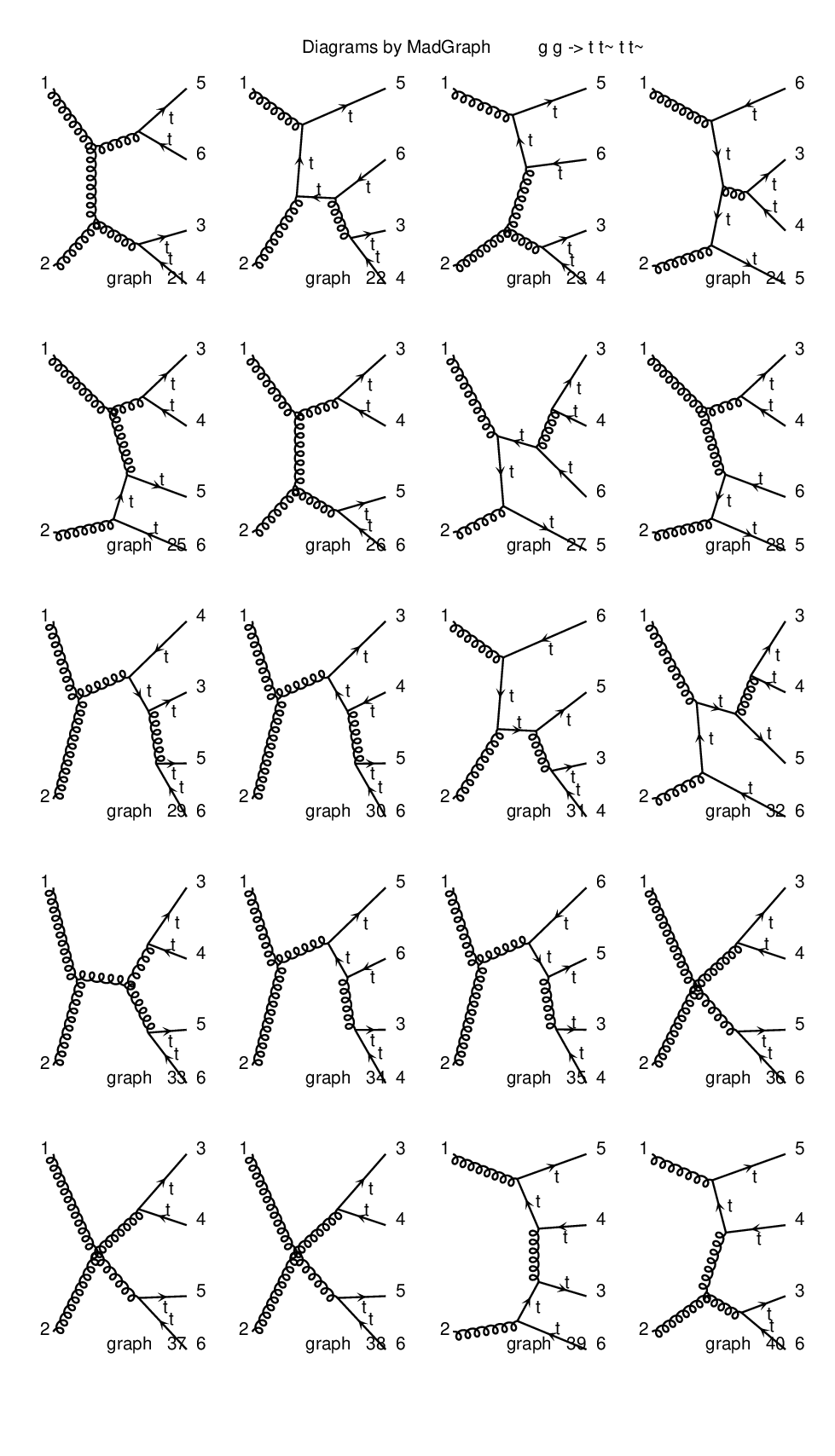}
}
\end{center}
\vspace{-1cm}
\begin{center}
\hspace{-2cm}
\mbox{
     \epsfxsize=6.5cm
     \epsffile{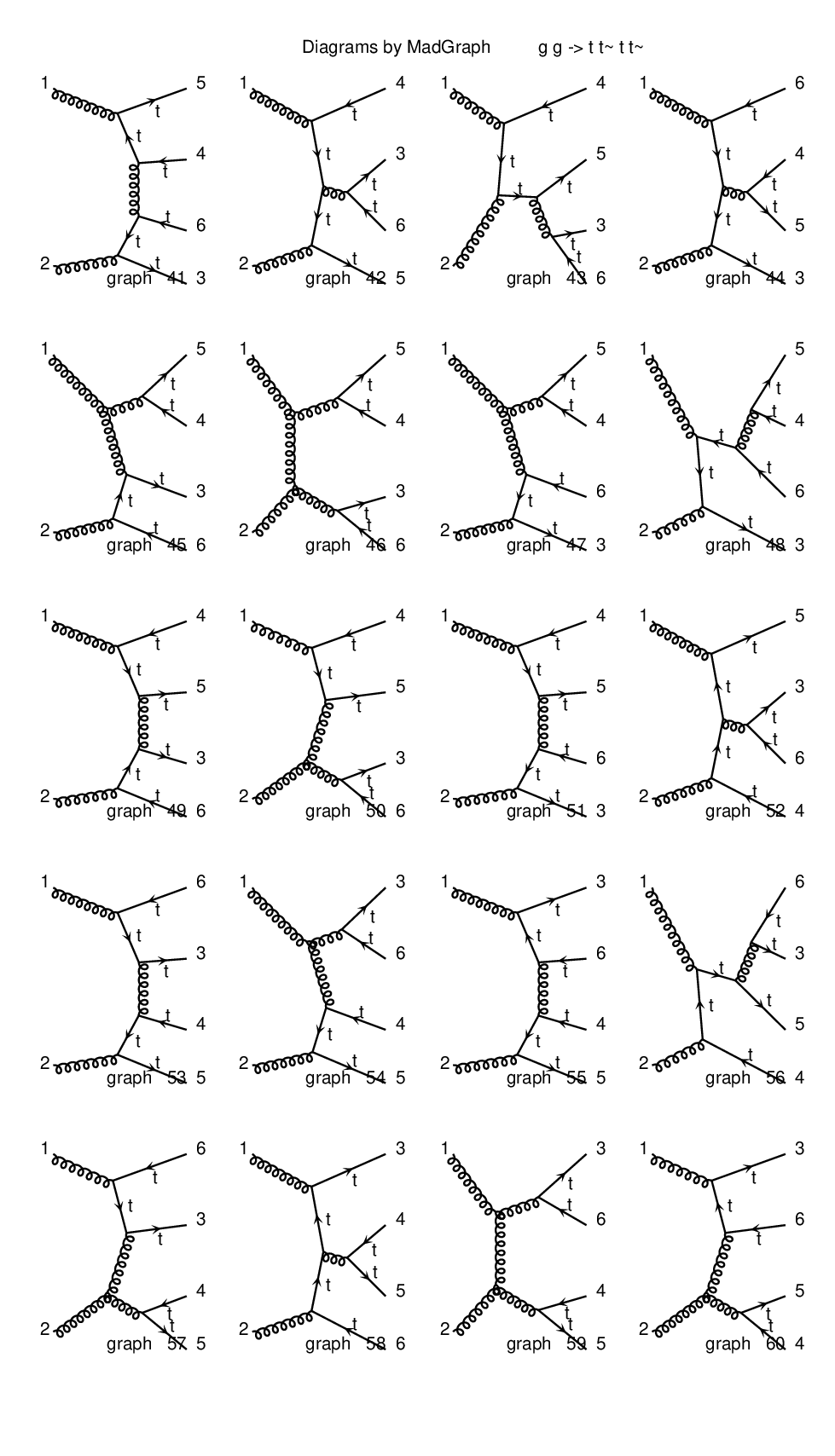}
}
\mbox{
     \epsfxsize=6.5cm
     \epsffile{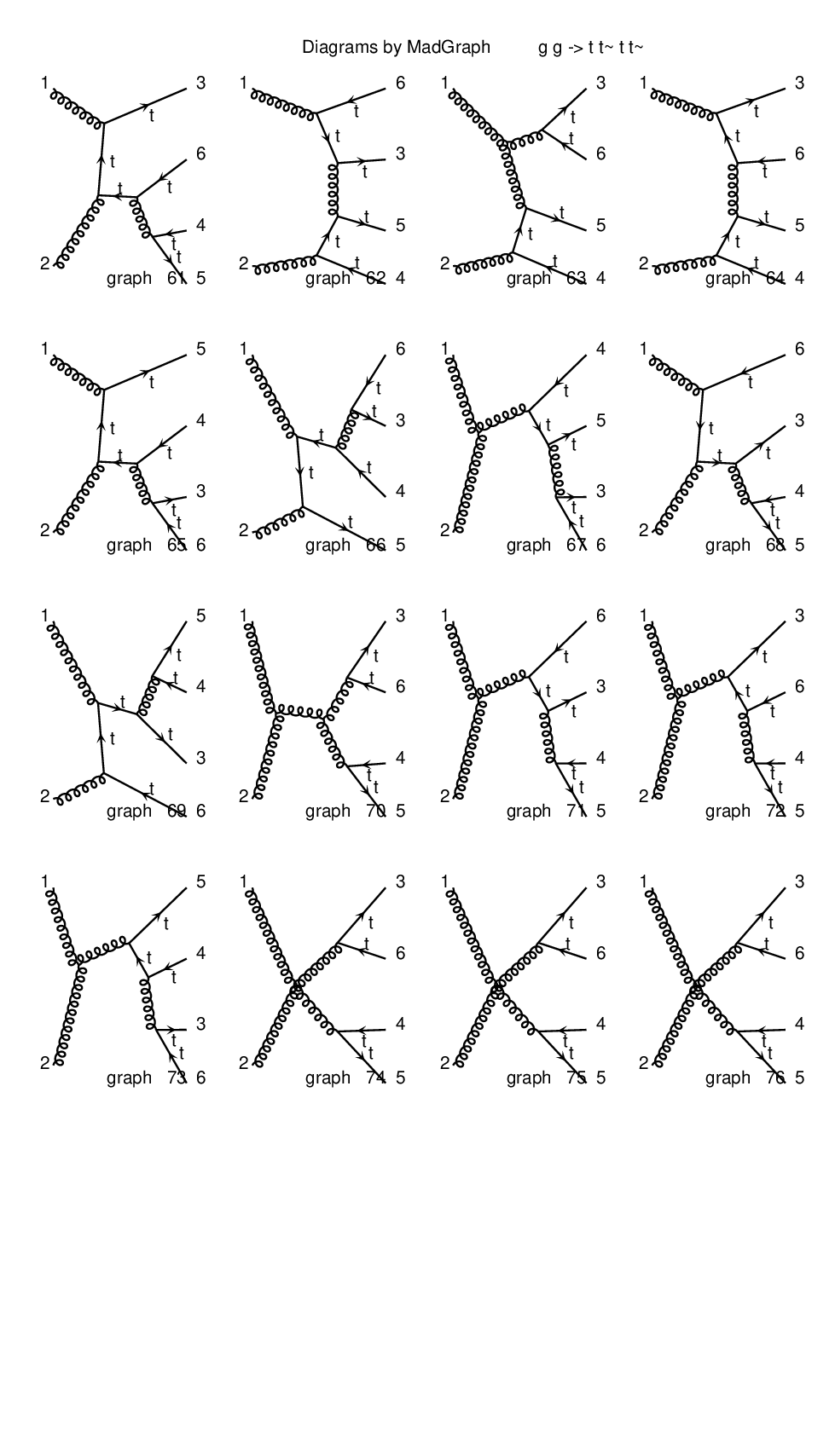}
}
\end{center}

\vspace{-1cm}
\isucaption{\em
The Feynman diagrams for the processes
$gg \to t \bar t t \bar t$.
\label{f:ttttdia}}
\end{figure}

\begin{figure}[hb]
\begin{center}
\hspace{-2cm}
\mbox{
     \epsfxsize=6.5cm
     \epsffile{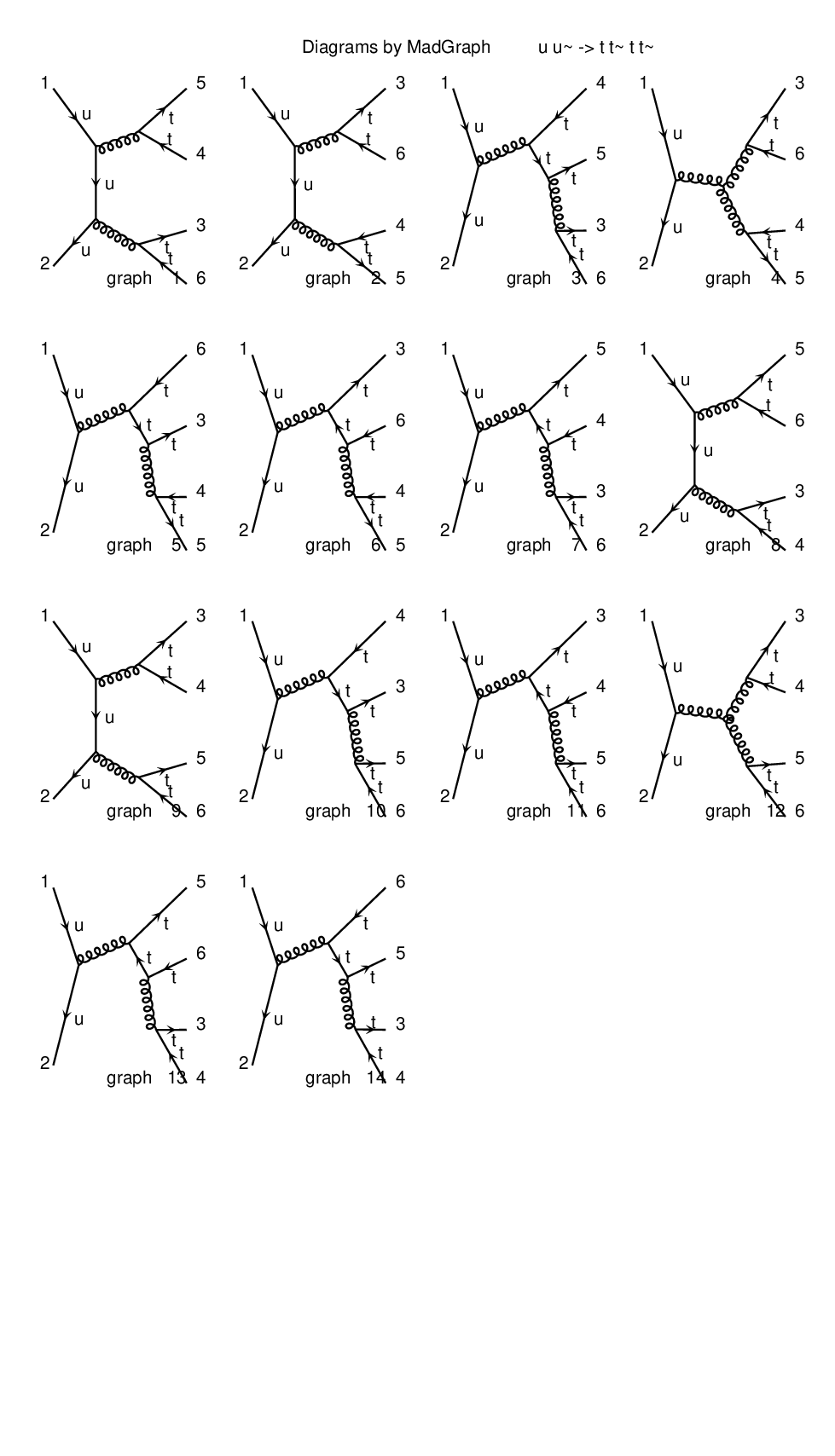}
}
\end{center}
\vspace{-1cm}
\isucaption{\em
The Feynman diagrams for the processes
$q \bar q \to t \bar t t \bar t$.
\label{f:ttttdia1}}
\end{figure}

\vspace{1.3cm}
The 76+14 Feynman diagrams contributing to the $g g,q \bar{q} \to t
\bar t t \bar t$ production are shown in Figures \ref{f:ttttdia} and \ref{f:ttttdia1}. Only
four flavours are included for incoming quarks. Contribution of the
incoming b-quarks could be excluded from the calculations thanks to
very high suppression induced by either the parton density functions
and/or CKM matrix elements.

\clearpage

\vspace{-1cm}

\begin{figure}[hb]
\begin{center}
\mbox{
     \epsfxsize=10.5cm
     \epsffile{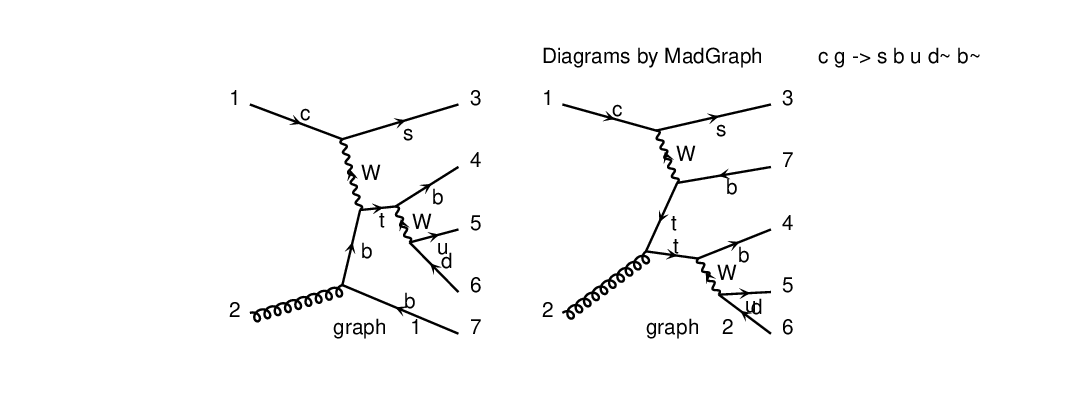}
}
\hspace{-4cm}
\mbox{
     \epsfxsize=6.5cm
     \epsffile{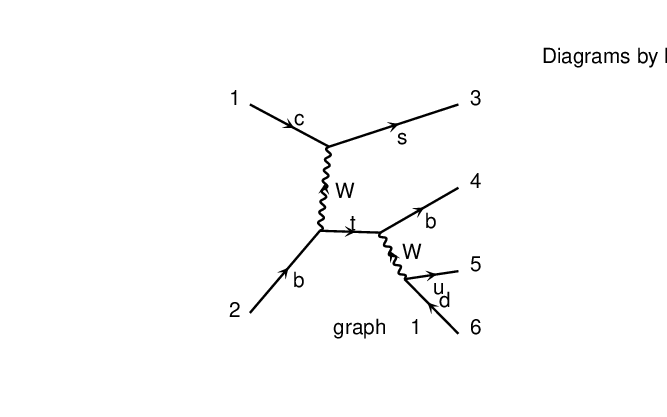}
}
\end{center}

\vglue -0.5cm
\isucaption{\em
The Feynman diagrams for the processes
$ q b \oplus q g \to q t \oplus b \to  q  b f \bar{f} \oplus b $}
\end{figure}

\begin{figure}[hb]
\begin{center}
\mbox{
     \epsfxsize=10.5cm
     \epsffile{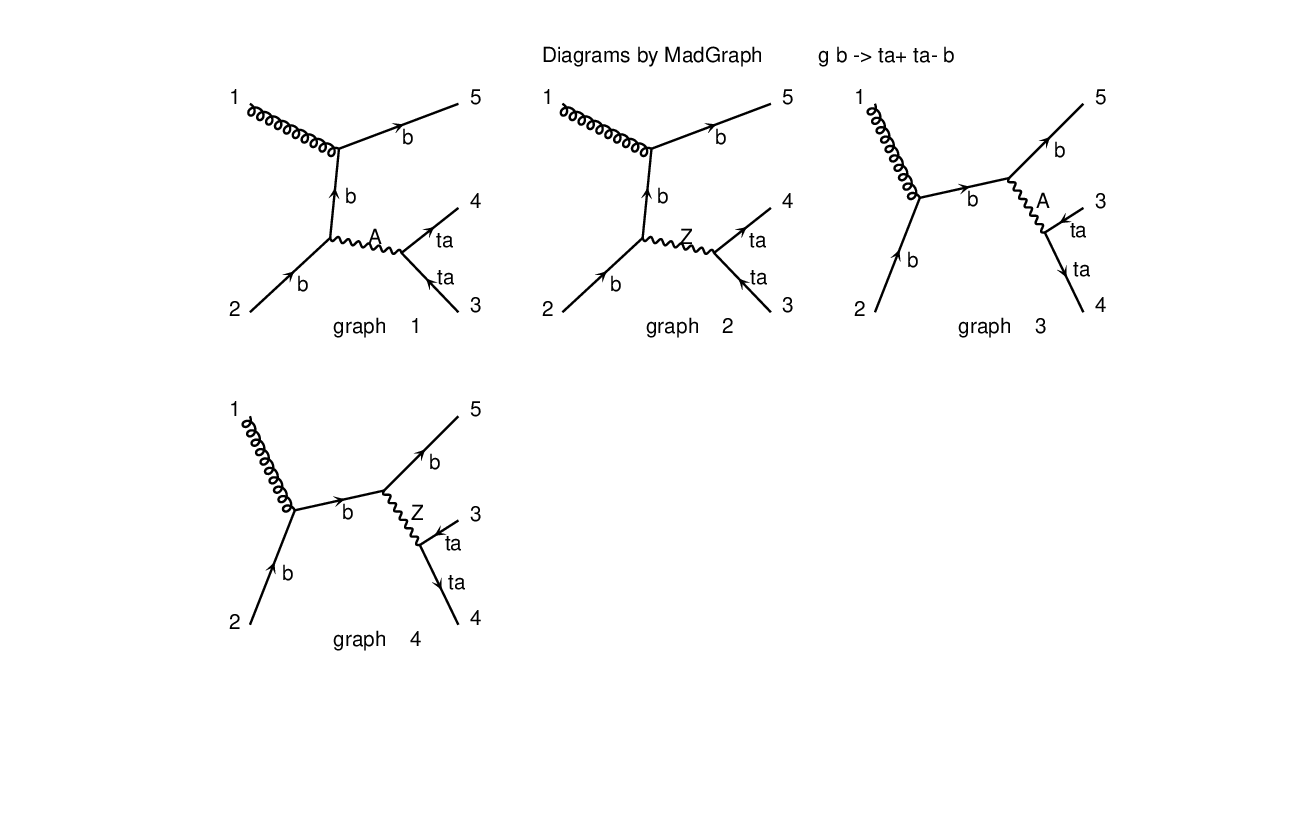}
}
\\
\vspace{-1cm}
\mbox{
     \epsfxsize=10.5cm
     \epsffile{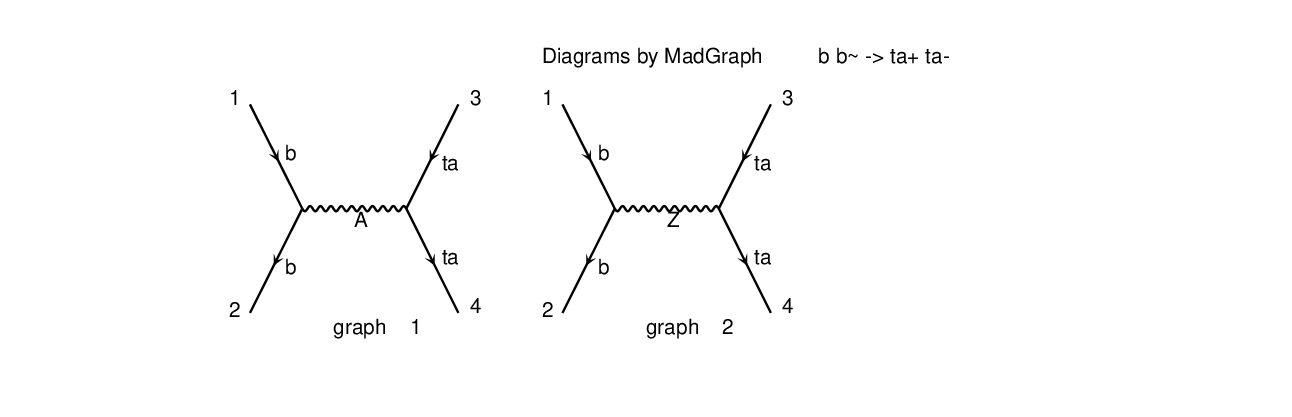}
}
\end{center}

\vglue -0.5cm
\isucaption{\em
The Feynman diagrams for the processes
$ b b \oplus b g \to Z^0 \oplus b \to f \bar f \oplus b$}
\end{figure}

\begin{figure}[hb]
\begin{center}
\mbox{
     \epsfxsize=10.5cm
     \epsffile{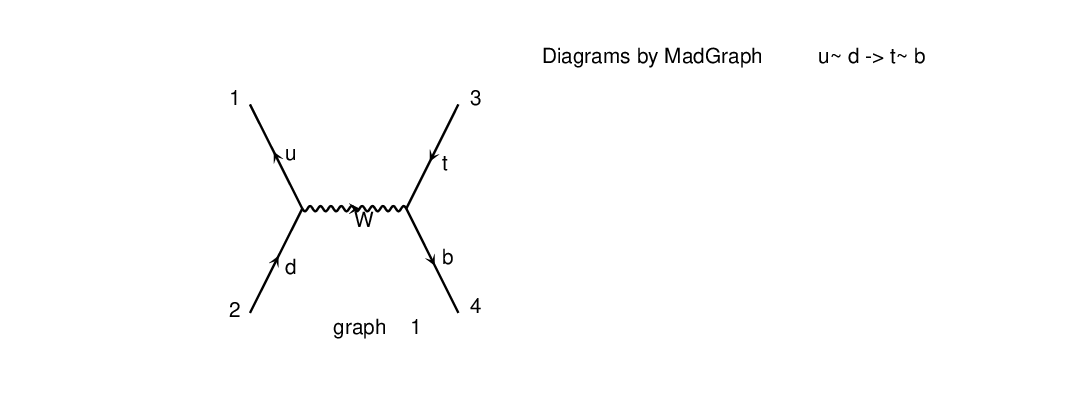}
}
\end{center}

\vglue -0.5cm
\isucaption{\em
The Feynman diagrams for the processes
$ q q \to t b  \to b f \bar{f} b $ (s-channel single top)
}
\end{figure}

\begin{figure}[hb]
\begin{center}
\mbox{
     \epsfxsize=10.5cm
     \epsffile{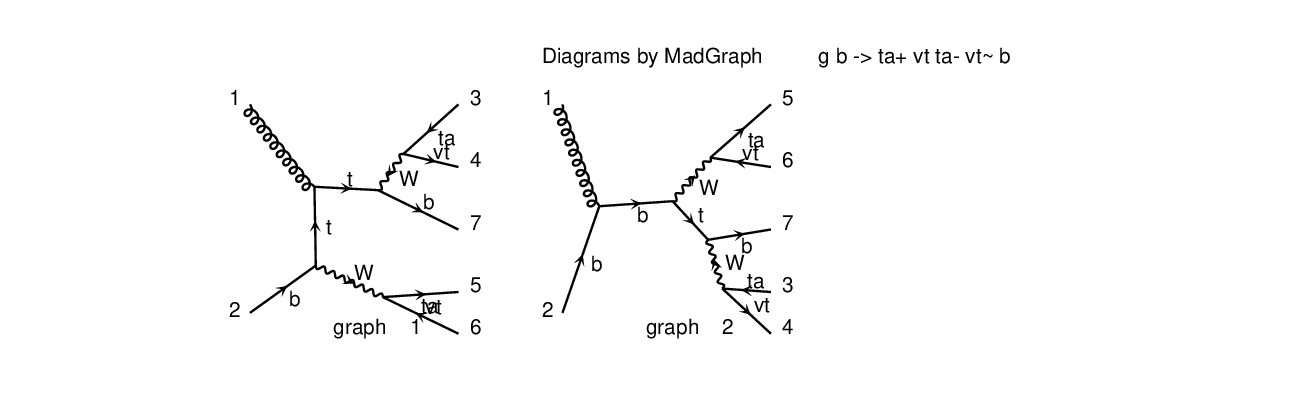}
}
\end{center}

\vglue -0.5cm
\isucaption{\em
The Feynman diagrams for the processes
$ g b \to t W \to b f \bar{f} f \bar f$ (tW-channel single top)
}
\end{figure}


\boldmath
\section{Example input files}
\unboldmath

\boldmath
\subsection{File run.card \label{app:run}}
\unboldmath

{\scriptsize
\begin{verbatim}
C------------------------------------------------------------------------------
C                 STEERING FILE FOR ACERMC (3.8) - BASIC SETTINGS
C------------------------------------------------------------------------------

C==== TURN ON FFKEY STEERING FILE (DEBUG)
LIST 

C==== CMS/ACSET(1)
C Specify the centre-of-mass energy in GeV 

CMS 14000.0
C====

C==== ACER
C Use AcerMC code
C ACER=0 - no
C ACER=1 - yes
C ACER=2 - yes & store events to file; see the manual for details
C ACER=3 - yes & read events from file; see the manual for details

ACER 1

C====

C==== PROCESS/IACSET ARRAY
C Specify the process to generate. The available AcerMC processes are:
C
C  1)  g + g  ->  t t~ b b~ (MG)    
C  2)  q + q~ ->  t t~ b b~ (MG)    
C  3)  q + q~ ->  (W->) l nu_l b b~ (MG)
C  4)  q + q~ ->  (W->) l nu_l t t~ (MG)
C  5)  g + g  ->  (Z0->) f f~ b b~ (MG) 
C  6)  q + q~ ->  (Z0->) f f~ b b~ (MG) 
C  7)  g + g  ->  (Z0->) f f~ t t~ (MG) 
C  8)  q + q~ ->  (Z0->) f f~ t t~ (MG) 
C  9)  g + g  ->  (Z0/W/gamma->)t t~ b b~ (MG)    
C 10)  q + q  ->  (Z0/W/gamma->)t t~ b b~ (MG)    
C 11)  g + g  ->  t t~ off-shell (MG)    
C 12)  q + q~ ->  t t~ off-shell (MG)    
C 13)  g + g  ->  (W W b b~ ->) ~ 2f_1 2f_2 b b~ (MG)      
C 14)  q + q~ ->  (W W b b~ ->) ~ 2f_1 2f_2 b b~ (MG) 
C 15)  g + g  ->  t t~ t t~ (MG)    
C 16)  q + q~ ->  t t~ t t~ (MG)    

C 17)  ACOT q + g ->  q' t(~) b -> q' b(~) f_1 f_2 b (t-chan) (MG) (100+101)  
C 18)  ACOT b + g ->  (Z0/gamma->) l l~ b  (MG)   (96+97)    
C 19)  q + q~ ->  t b~ ->  f f b b~ (s-chan) (MG)
C 20)  ACOT g + g  ->  (W W b b~ ->) ~ 2f_1 2f_2 b b~ (MG) (13+105)   
C 21)  g + b ->  (t W ->) b 2f_1 2f_2 (W-chan) (MG)    
C 22)  q + q~ -> (Z'/Z0/gamma) ->  t t~ off-shell (MG)    
C 23)  g + g, q + q~  ->  t t~ off-shell (MG) (11+12)    
C 24)  g + g, q + q~  ->  (Z0->) f f~ b b~ (MG)  (5+6)
C 25)  g + g, q + q~  ->  (Z0->) f f~ t t~ (MG) (7+8)
C 26)  ACOT g + g ->  (Z0->) f f~ b b~ (MG) 
C 27)  ACOT g + g, q + q~ ->  (Z0->) f f~ b b~ (MG) (26+6)
C
C 'Control processes'
C
C 91)  q + q~   ->  (Z0/gamma->) l l~ (MG)
C 92)  g + g    ->  t t~ (MG)    
C 93)  q + q~   ->  t t~ (MG)    
C 94)  q + q~   ->  (W->) l nu_l (MG)    
C 95)  g + g    ->  b b~ W W (MG)
C 96)  b + b    ->  (Z0/gamma->) l l~ (MG)     
C 97)  g + b(~) ->  (Z0/gamma->) l l~ b (MG)   
C 98)  q + b(~) ->  q' t(~) (MG)               
C 99)  q + g    ->  q' t(~) b (MG)              
C 100) q + b(~) ->  q' t(~) -> q' b(~) f_1 f_2  (MG)                
C 101) q + g    ->  q' t(~) b -> q' b(~) f_1 f_2 b (MG)            
C 102) q + b(~)    ->  q' t(~) b -> q' b(~) W (MG)              
C 103) q + b(g) ->  q' t(~) b  (MG)  (98+99)          
C 104) g + b -> t l nu_l (W-chan) (MG)         
C 105) g + b ->  (t W ->) b 2f_1 2f_2 (W-chan) (MG) (equal to 21)  
C 106) g + g -> t l nu_l b~ (W-chan) (MG)      
C 107) g + g  ->  (t W b  ->)  2f_1 2f_2 b b~ (MG)    

C
C In case ACER=0 the native Pythia/Herwig conventions should be used 
C e.g. for Herwig PROCESS 1453 1355 3000

PROCESS 11
C====

 C==== HAD
C Control of hadronization/fragmentation/ISR/FSR switches:
C HAD=0 - switch off radiation in initial and final state, multinteraction and
C        hadronization
C HAD=1 - switch off radiation in final state and hadronization
C HAD=2 - switch off hadronization
C HAD=3 - full treatment 
C HAD=4 - switch off radiation in initial state, multiinteraction and hadronization 
C HAD=5 - switch off radiation in initial state and multiinteraction

HAD 3 
C====

C==== PDF-SET/ACSET(7)
C Choose a PDF set according to LHAPDF naming scheme
C PDFSET=19070 represents the CTEQ5L parametrised set
C PDFSET=10042 represents the CTEQ6L parametrised set

PDFSET 10042
C====

C==== RSEED
C Choose the random seed for random generator initialisation

RSEED 945169

C====

C==== TAUOLA
C Tau decays handled by TAUOLA library
C TAUOLA=0 - use internal PYTHIA/HERWIG mechanisms for Tau decays
C TAUOLA=1 - use the TAUOLA library  for Tau decays

TAUOLA 0
C====

C==== PHOTOS
C QED FSR handled by PHOTOS library
C PHOTOS=0 - use internal PYTHIA/HERWIG mechanisms for FSR photon radiation
C PHOTOS=1 - use the PHOTOS routines for FSR photon radiation

PHOTOS 0
C====

C==== NEVENT/ACSET(51)
C Specify the number of events to generate

NEVENT 10

C====

C ------------------------------------------------------------------------------
END
\end{verbatim}      
}

\boldmath
\subsection{File acermc.card \label{app:acmc}}
\unboldmath

{\scriptsize
\begin{verbatim}      
C------------------------------------------------------------------------------
C                 STEERING FILE FOR ACERMC (3.8) - ACERMC SETTINGS
C------------------------------------------------------------------------------

C==== TURN ON FFKEY STEERING FILE (DEBUG)
LIST 

C THE AcerMC EVENT SETTINGS ---------------------------------------------------

C==== SCALE/ACSET(2)
C Choose the Q^2 scale for the active AcerMC process.
C The implemented values differ for various processes, please look into the manual
C for details

ACSET2 1
C====

C==== FERMION/ACSET(3)
C The flavour of the final state leptons produced in W or Z decays of AcerMC
C processes.The Pythia/PDG naming convention is used:
C FERMION=0  - all decays
C FERMION=1  - only hadronic decays
C FERMION=4  - electron and muon
C FERMION=5  - b-quark decay
C FERMION=10 - leptons (el,mu,tau) only
C FERMION=11 - electron only
C FERMION=13 - muon only
C FERMION=15 - tau only
C FERMION=12 - neutrinos, all three flavours are generated and 
C the cross-section is calculated accordingly.
C The setting FERMION=5 works only for processes 7 and 8
ACSET3 10
C====

C==== Z/GAMMA/ACSET(4)
C Use the full Z/gamma* propagator in AcerMC processes 5-8. 
C ZGAMMA=0 - only Z propagator
C ZGAMMA=1 - full Z/gamma* propagator

ACSET4 1
C====

C==== Z/G CUT/ACSET(5)
C Cutoff value on the invariant mass m_Z/gamma* in GeV when ZGAMMA=1. 
C Note that the provided data files exist only for values of
C ZGCUT=10,30,60,120,300 GeV which should satisfy most 
C users. In case a different value is set the user has also to provide 
C the user data files for the run.
ACSET5 60.0
C====

C==== Z-PRIME MASS ACSET(58)
C Value of Z-prime mass in TeV/c^2 
C Note that the provided data files exist only for values of
C Z-PRIME MASS=1 TeV and 0.5 TeV  which should satisfy most 
C users. In case a different value is set the user has also to provide 
C the user data files for the run.
C The corresponding width of the boson is calculated internally.
ACSET58 0.5
C====

C==== Z-PRIME COUPLING SETS  ACSET(59)
C Note that the provided data files exist only for values of
C Z-PRIME COUPLINGS=0 - Standard Model values of Z-prime couplings 
C Z-PRIME COUPLINGS=1 - Z_R couplings as described in (hep-ph/0307020).
C Z-PRIME COUPLINGS=2 - pure V-A: sin(theta_W)=0 for Z-prime.
C In case a different value is needed the user should contact the AcerMC authors. 
ACSET59 2
C====

C==== LEPTONIC TOP COUPLING ACSET(60)
C Vertex coupling of top, b-quark and leptonic decaying W
C The coupling is given by:
C GTF = ACSET(60)*G_L+(1-ACSET(60)*G_R
C meaning that ACSET(60)=1 is the Standard model value
ACSET60 1.
C====

C==== HADRONIC TOP COUPLING ACSET(61)
C Vertex coupling of top, b-quark and hadronic decaying W
C The coupling is given by:
C GTF = ACSET(61)*G_L+(1-ACSET(61)*G_R
C meaning that ACSET(61)=1 is the Standard model value
ACSET61 1.
C====


C THE AcerMC ADVANCED SWITCHES ------------------------------------------------

C==== ALPHA_S/ACSET(8)
C Use the alpha_s provided by the linked generator (Pythia/Herwig) or the
C one provided by AcerMC
C ALPHAS=0 - use the linked generator's alpha_s
C ALPHAS=1 - use the AcerMC's alpha_s (one loop calculation)
C ALPHAS=2 - use the AcerMC's alpha_s (three loop calculation)
ACSET8 0
C====

C==== ALPHA_S(M_Z^2)/ACSET(9)
C Specify the value of alpha_QCD(M_Z^2) for AcerMC alpha_s calculation
C ALPHASMZ=-1 - the value is taken from PDFLIB 
C ALPHASMZ>0 - the provided value is taken
ACSET9 -1.
C====

C==== ALPHA_EM/ACSET(10)
C Use the alpha_QED provided by the linked generator (Pythia/Herwig) or the
C one provided by AcerMC
C ALPHAEM=0 - use the linked generator's alpha_QED
C ALPHAEM=1 - use the AcerMC's alpha_QED
ACSET10 0
C====

C==== ALPHA_EM(0)/ACSET(11)
C Specify the value of alpha_QED(0) for AcerMC alpha_QED calculation
C ALPHAEM0=-1 - the default AcerMC value is used 
C ALPHAEM0>0 - the provided value is taken

ACSET11 -1.
C====

C==== TOP S-L/ACSET(12)
C Specify the decay mode of WW pair produced by external top decays in AcerMC
C processes 1,2,4,7,8,9,92 and 93:
C TOPDEC=0 - both W bosons decay according to Pythia/Herwig switches
C TOPDEC=1 - one W decays into electron + nu and the other one hadronically
C TOPDEC=2 - one W decays into muon + nu and the other one hadronically
C TOPDEC=3 - one W decays into tau + nu and the other one hadronically
C TOPDEC=4 - one W decays into el or mu + nu and the other one hadronically
C TOPDEC=5 - one W decays into el or mu + nu and the other one hadronically, the 
C            W decaying leptonically has the same charge as the primary W; the decay
C            mode makes sense only for AcerMC processes 4!
C When TOPDEC>0 the output cross-section is ALREADY MULTIPLIED by the corresponding 
C branching ratio(s)! (Courtesy of AcerMC authors) 

ACSET12 0
C====

C==== BOSON PAIR DECAYS/ACSET(13)
C Specify the decay mode of boson pairs inside AcerMC processes:
C
C BOSDEC=0 - both bosons decay in all possible modes
C BOSDEC=1 - one boson decays into electron + nu and the other one hadronically
C BOSDEC=2 - one boson decays into muon + nu and the other one hadronically
C BOSDEC=3 - one boson decays into tau + nu and the other one hadronically
C BOSDEC=4 - one boson decays into el or mu + nu and the other one hadronically
C BOSDEC=5 - one boson decays into leptons (el or mu or tau) and the other one hadronically
C BOSDEC=6 - one or both bosons decay into leptons (el or mu or tau) and the remaining one hadronically
C BOSDEC=11 - both bosons decay into el + nu 
C BOSDEC=13 - both bosons decay into muon + nu 
C BOSDEC=15 - both bosons decay into tau + nu 
C BOSDEC=17 - both bosons decay into el or muon + 2 nu 
C BOSDEC=19 - both bosons decay into el or muon or tau + 2 nu 
C BOSDEC=20 - both bosons decay hadronically
C The output cross-section is ALREADY MULTIPLIED by the corresponding 
C branching ratio(s)!

ACSET13 6
C====

C THE AcerMC TRAINING SETUP AND UNWEIGHTING TREATMENT -------------------------

C==== MODE/ACSET(50)
C Specify the AcerMC training mode:
C MODE=0 - normal run, generate unweighted events
C MODE=1 - perform multi-channel optimisation.
C MODE=2 - perform VEGAS grid training.
C MODE=3 - perform VEGAS grid training as MODE=2 but does this by updating a provided grid

C MODE=-1 - all flat
C MODE=-2 - only VEGAS flat

ACSET50 0
C====

C==== USER/ACSET(52)
C Use the data files provided by user
C USER=0 - no, use internal files
C USER=1 - use the user's multi-channel optimisation and VEGAS grid files 
C USER=2 - use the default multi-channel optimisation and user's VEGAS grid files 
C USER=3 - use the default multi-channel optimisation and VEGAS grid files, read the user
C           maximal weight file. 

ACSET52 0
C====

C==== MAXFIND/ACSET(53)
C Search for the maximum weight needed for event unweighting
C MAXFIND=0 - no, use the provided file for max. weights
C MAXFIND=1 - use the provided file for max. weights, re-calculate the max. weights using
C             the stored 100 highest events
C MAXFIND=2 - perform the search and give the wtmax file, equivalent to generation of
C             weighted events

ACSET53 0
C====

C==== EPSILON/ACSET(54)
C Use the epsilon maximal weight or the overall maximal weight found in training (see the 
C manual for the difference)
C EPSILON=0 - use the epsilon max. weight
C EPSILON=1 - use the overall maximal weight

ACSET54 0
C====

C==== NQCD/ACSET(56)
C Use the naive qcd correction for width calculations
C (see the manual for details)
C NQCD=1 - use the naive QCD corrections
C NQCD=0 - don't use naive QCD corrections

ACSET56 1
C====

C==== JCCPDF/ACSET(57)
C Use the Collins derived PDF-s for showering
C (see the manual for details)
C JCCPDF=1 - use the Collins PDF-s
C JCCPDF=0 - don't use  the Collins PDF-s

ACSET57 0
C====

C ------------------------------------------------------------------------------
END
\end{verbatim}      
}
\boldmath
\subsection{File tauola.card \label{app:tauola}}
\unboldmath
\vspace{-0.3cm}

{\scriptsize
\begin{verbatim}
C------------------------------------------------------------------------------
C                 STEERING FILE FOR TAUOLA & ACERMC (3.5) 
C------------------------------------------------------------------------------

C==== TURN ON FFKEY STEERING FILE (DEBUG)
LIST 

C==== POLAR
C Polarisation switch for tau decays
C POLAR=0 - switch polarisation off
C POLAR=1 - switch polarisation on
POLAR 1	
C====

C==== RADCOR
C Order(alpha) radiative corrections for tau decays
C RADCOR=0 - switch corrections off
C RADCOR=1 - switch corrections on

RADCOR 1
C====

C==== PHOX
C Radiative cutoff used in tau decays 
C PHOX=0.01 - default value by TAUOLA authors

PHOX 0.01
C====

C==== DMODE
C Tau and tau pair decay mode
C DMODE=0 - all decay modes allowed
C DMODE=1 - (LEPTON-LEPTON): only leptonic  decay modes
C DMODE=2 - (HADRON-HADRON): only hadronic  decay modes
C DMODE=3 - (LEPTON-HADRON): one tau decays into leptons and the other one into hadrons
C DMODE=4 - (TAU->PI NU)   : taus are restricted to decay to a pion and neutrino

DMODE 2
C====

C==== JAK1/JAK2
C Decay modes of taus according to charge, the list is taken from TAUOLA output
C The listing gives only Tau- modes, the Tau+ are charge conjugate, neutrinos 
C are omitted.
C JAK1/2 = 1   - TAU-  -->  E- 
C JAK1/2 = 2   - TAU-  -->  MU-  
C JAK1/2 = 3   - TAU-  -->  PI- 
C JAK1/2 = 4   - TAU-  -->  PI-, PI0 
C JAK1/2 = 5   - TAU-  -->  A1- (two subch)
C JAK1/2 = 6   - TAU-  -->  K- 
C JAK1/2 = 7   - TAU-  -->  K*- (two subch)
C JAK1/2 = 8   - TAU-  -->  2PI-,  PI0,  PI+
C JAK1/2 = 9   - TAU-  -->  3PI0,        PI-
C JAK1/2 = 10  - TAU-  -->  2PI-,  PI+, 2PI0 
C JAK1/2 = 11  - TAU-  -->  3PI-, 2PI+
C JAK1/2 = 12  - TAU-  -->  3PI-, 2PI+,  PI0 
C JAK1/2 = 13  - TAU-  -->  2PI-,  PI+, 3PI0
C JAK1/2 = 14  - TAU-  -->  K-, PI-,  K+
C JAK1/2 = 15  - TAU-  -->  K0, PI-, K0B 
C JAK1/2 = 16  - TAU-  -->  K-,  K0, PI0 
C JAK1/2 = 17  - TAU-  -->  PI0  PI0   K-
C JAK1/2 = 18  - TAU-  -->  K-  PI-  PI+ 
C JAK1/2 = 19  - TAU-  -->  PI-  K0B  PI0  
C JAK1/2 = 20  - TAU-  -->  ETA  PI-  PI0
C JAK1/2 = 21  - TAU-  -->  PI-  PI0  GAM
C JAK1/2 = 22  - TAU-  -->  K-  K0

C==== DECAY MODE OF TAU+
JAK1 13 
C====

C==== DECAY MODE OF TAU-
JAK2 13
C====

C ------------------------------------------------------------------------------
END
\end{verbatim}      
}
\boldmath
\subsection{File photos.card \label{app:photos}}
\unboldmath

\vspace{-0.3cm}
{\scriptsize
\begin{verbatim}
C------------------------------------------------------------------------------
C                 STEERING FILE FOR PHOTOS & ACERMC (3.5) 
C------------------------------------------------------------------------------

C==== TURN ON FFKEY STEERING FILE (DEBUG)
LIST 

C==== PMODE
C Radiation mode of photos
C PMODE=1 - enable radiation of photons for leptons and hadrons
C PMODE=2 - enable radiation of photons for taus only
C PMODE=3 - enable radiation of photons for leptons only
PMODE 1

C====

C==== XPHCUT
c Infrared cutoff for photon radiation
C XPHCUT=0.01 - default value by PHOTOS authors
XPHCUT 0.01     

C====

C==== ALPHA
C Alpha(QED) value 
C ALPHA < 0 - leave default (0.00729735039)
ALPHA -1.
C====

C==== INTERF
C Photon interference weight switch 
C INTERF = 1 - interference is switched on
C INTERF = 0 - interference is switched off
INTERF 1
C====

C==== ISEC
C Double bremsstrahlung switch
C ISEC=1 - double bremsstrahlung is switched on
C ISEC=0 - double bremsstrahlung is switched off
ISEC 1
C====

C==== IFTOP
C Switch for gg(qq)->tt~ process radiation 
C IFTOP=1 - the procedure is is switched on
C IFTOP=0 - the procedure is is switched off

IFTOP 0
C====
\end{verbatim}      
}
\newpage
\boldmath
\section{Example output files}
\unboldmath

\boldmath
\subsection{File acermc.out}
\unboldmath

{\scriptsize
\begin{verbatim}            
  -----------------------------------------------------------------------------
                                   ._                           
                                 .j%3]:,                        
                               ~!%%%%%%% ,._.                   
                            _|xx%xxxx%%%%%+`                    
                             :~]%xxxx]xx%x_,+_x_%`              
                      -__||x||xx]+]]]+]]]]x|xxx]`               
                       -+%%xxxx]]]]+]]++]+]x]|>- .;..;.:_/`     
                          -+x]]]]|+]+]+]++]++]]+|+|]+|]]+-      
              ,.   .  ., |x]+]+||=++]+]=++++++=|++]=+]|~-       
              -|%x]]];x]]||=++++++++|];|++++++++++=+]=];   ..   ..  :..,; 
                -/]|]+]|]++++++++||||==|:;:=|==++==+;|,   :;;. :;=;;===|` 
                 _|]|+>+++]+]]|+|||=|=;|;;:|;=:===|==;::.;;:;,;:;;;;;;:-  
                     -++]+++:+|x+::||=||:=:::::;;::::::::;:;:;:;:;;;=::   
                   .|x+|+|]++:,-..:::|=;=:-.:.:-::.:.:.::-::::::::;:-     
         .,  .:||_   --||;:|:::.-.:.-:|;||::.:...-.-.-....::::-:::;::.    
     __._;++;;;|=|;. -:::::::---:.--;;|==|:;:.:-:-:.-.-.:::.--:::;:;--    
     -+++=+=======;==:::::-:.::...:.|+=;;===:-..:.:.:::::-...--:  -       
     :|:-:|===;:;:;::::::--::.:-::::.|||===:::.:.:::::.-......--:: .      
        -;|===;;:;;;;::;::.:-:: -:=;;|+||=;==|=::::::::...-.:..-.::..     
         ---|;:,::::.::::=;=:=::  -++===|======:--:   ...-...-:-:.-       
             ---;:::::;:--:===;;;:|-|+=+|+|==;:`    ...-....-..           
             .::;;:::;:;;:--- -- --.|=||=+|=:=,      ..... .              
            .;;=;;:-:::;:::.        -+;,|]| ;;=`                          
              - -:- --:;;:- -         - :|; -                             
                       -                 :`                               
                                          :                               
                                          :                               
                                          .                               
                                                                  
   40000L,                                           |0000i   j000&  .a00000L#0
   --?##0L        .aaaa aa     .aaaa;     aaaa, _aaa, -000A  _0001- _d0!`  -400
     d0 40,     _W0#V9N0#&    d0#V*N#0,   0##0LW0@4#@' 00j#; J0|01  d0'      40
    J0l -#W     #0'    ?#W   ##~    -#0;    j##9       00 4#|01|01  00     
   _00yyyW0L   :0f      ^-  :000###00001    j#1        00 ?#0@`|01  #0      
   ##!!!!!#0;  -0A       _  -0A             j#1        00  HH< |01  j0L       _
 ad0La,  aj0Aa  4#Aaa_aj#0`  ?0Laa_aaa0L  aaJ0Laaa,  _a00aa  _aj0La  *0Aaa_aad
 HHHRHl  HHHHH   `9##009!     `9NW00@!!`  HHHHRHHHl  :HHHHH  ?HHHRH   ?!##00P!`
      
  -----------------------------------------------------------------------------
      
            AcerMC 3.8 (May 2011),  B. P. Kersevan, E. Richter-Was
     
  -----------------------------------------------------------------------------
      
  ---------------------------< ACTIVATED PROCESSES >---------------------------
      
                      1)  g + g  ->  t t~ b b~ (MG)               OFF 
                      2)  q + q~ ->  t t~ b b~ (MG)               OFF 
                      3)  q + q~ ->  (W->) l nu_l b b~ (MG)       OFF 
                      4)  q + q~ ->  (W->) l nu_l t t~ (MG)       OFF 
                      5)  g + g  ->  (Z0->) f f~ b b~ (MG)        OFF 
                      6)  q + q~ ->  (Z0->) f f~ b b~ (MG)        OFF 
                      7)  g + g  ->  (Z0->) f f~ t t~ (MG)        OFF 
                      8)  q + q~ ->  (Z0->) f f~ t t~ (MG)        OFF 
                      9)  g + g  -> (Z0/W/gamma->) t t~ b b~ (MG) OFF 
                     10)  q + q~ -> (Z0/W/gamma->) t t~ b b~ (MG) OFF 
                     11)  g + g  ->  t t~ off-shell (MG)          ON  
                     12)  q + q~ ->  t t~ off-shell (MG)          OFF 
                     13)  g + g  ->  (W W b b~ ->)  (MG)          OFF 
                     14)  q + q~ ->  (W W b b~ ->)  (MG)          OFF 
                     15)  g + g  ->  t t~ t t~ (MG)               OFF 
                     16)  q + q~ ->  t t~ t t~ (MG)               OFF 
                     17)  ACOT q + g  -> q t -> q b(~) f f b (t-chOFF 
                     18)  ACOT b + g  -> (Z0/gamma->) f f~ b (MG) OFF 
                     19)  q + q~ ->  t b~ ->  f f b b~ (s-chan) (MOFF 
                     20)  ACOT q + g  ->  (W W b b~ + t W b ->) (WOFF 
                     21)  g + b -> b f f  l nu_l (W-chan) (MG)    OFF 
                     22)  q + q~ -> Z-prime -> t t~ off-shell (MG)OFF 
                     23)  g + g, q + q~  ->  t t~ off-shell (MG)  OFF 
                     24)  g + g, q + q~  -> (Z0->) f f~ b b~(MG)  OFF 
                     25)  g + g, q + q~  -> (Z0->) f f~ t t~  (MG)OFF 
                     26)  ACOT g + g  -> (Z0->) f f~ b b~(MG)     OFF 
                     27)  ACOT g + g, q + q~  -> (Z0->) f f~ b b~(OFF 
                     91)  q + q~ -> (Z0/gamma->) l l~ (MG)        OFF 
                     92)  g + g  ->  t t~ (MG)                    OFF 
                     93)  q + q~ ->  t t~ (MG)                    OFF 
                     94)  q + q~ ->  (W->) l nu_l (MG)            OFF 
                     95)  g + g  ->  W+ W- b b~ (MG)              OFF 
                     96)  b + b~ -> (Z0/gamma->) f f~ (MG)        OFF 
                     97)  g + b  -> (Z0/gamma->) f f~ b(MG)       OFF 
                     98)  q + b  ->  q t (MG)                     OFF 
                     99)  q + g  ->  q t b (MG)                   OFF 
                     100)  q + b  -> q t -> q b(~) f f (MG)        OFF 
                     101)  q + g  -> q t b -> q b(~) f f b (MG)    OFF 
                     102)  q + b  -> q t -> q b(~) W (MG)          OFF 
                     103)  q + b(g)  -> q t -> q b(~)  (MG)        OFF 
                     104)  g + b -> t l nu_l (W-chan) (MG)         OFF 
                     105)  g + b -> b f f  l nu_l (W-chan) (MG)    OFF 
                     106)  g + g -> t l nu_l b~ (W-chan) (MG)      OFF 
                     107)  g + g -> (t W b-> )  (W-chan) (MG)      OFF 
      
  -----------------------------< ACERMC SETTINGS >-----------------------------
      
                      C.M.S ENERGY  =       14000.00       [ACSET(1)]
                      SCALE CHOICE  =              1       [ACSET(2)]
                      ACERMC ALPHA_QCD =           0       [ACSET(8)]
                      ALPHA_QCD(M_Z)   =   -1.000000       [ACSET(9)]
                      ACERMC ALPHA_QED =           0       [ACSET(10)]
                      ALPHA_QED(0)     =   -1.000000       [ACSET(11)]
                      TOP->W S-L DECAY =           0       [ACSET(12)]
                      BOSON PAIR DECAY =           6       [ACSET(13)]
       
                      OPTIMIZATION   =             0       [ACSET(50)]
                      OPTIM. STEPS   =             1       [ACSET(51)]
                      USER FILES     =             0       [ACSET(52)]
                      MAX. SEARCH    =             0       [ACSET(53)]
                      EPSILON CUTOFF =             0       [ACSET(54)]
                      NAIVE QCD      =             1       [ACSET(56)]
       
  -----------------------------------------------------------------------------
  
  
           READ MAXIMUM WEIGHT(MB)  =   0.196952E-05
           READ EPSILON WEIGHT(MB)  =   0.630858E-06


  
         SET MAXIMUM WEIGHT(MB)  =   0.139683E-05
         SET WEIGHT CORRECTION   =   0.221417E+01


  
    ---------< FINALIZATION FOR PROCESS: 11 >---------
      
      --------------< WEIGHT SURVEY >--------------
      
      ------------< TOTAL STATISTICS >-------------
  
       CROSS-SECTION ESTIMATE =   0.224783E+03 PB
                            +/-   0.711578E+02 PB
            VARIANCE ESTIMATE =   0.506343E+04 PB^2
                            +/-   0.292734E+04 PB^2
  
       MAXIMUM WEIGHT  =   0.567410E-06
       NO.WEIGHTS NE 0 =              7
       NO.WEIGHTS EQ 0 =              0
       NO.WEIGHTS LT 0 =              0
       MAX. (-)WEIGHT  =   0.000000E+00
       MAX. (+)WEIGHT  =   0.567410E-06
       EFFICIENCY FOR ALL WEIGHTS     =  39.616 %
       EFFICIENCY FOR NONZERO WEIGHTS =  39.616 %

       NO.WEIGHTS ABOVE EPSILON-CUT  =           0

      --------------> WEIGHT SURVEY <--------------
\end{verbatim}
}
\newpage

\boldmath
\subsection{File pythia.out}
\unboldmath
\vspace{-0.3cm}

{\scriptsize
\begin{verbatim}                                                                              
 ******************************************************************************
 ******************************************************************************
 **                                                                          **
 **                                                                          **
 **              *......*                  Welcome to the Lund Monte Carlo!  **
 **         *:::!!:::::::::::*                                               **
 **      *::::::!!::::::::::::::*          PPP  Y   Y TTTTT H   H III   A    **
 **    *::::::::!!::::::::::::::::*        P  P  Y Y    T   H   H  I   A A   **
 **   *:::::::::!!:::::::::::::::::*       PPP    Y     T   HHHHH  I  AAAAA  **
 **   *:::::::::!!:::::::::::::::::*       P      Y     T   H   H  I  A   A  **
 **    *::::::::!!::::::::::::::::*!       P      Y     T   H   H III A   A  **
 **      *::::::!!::::::::::::::* !!                                         **
 **      !! *:::!!:::::::::::*    !!       This is PYTHIA version 6.416      **
 **      !!     !* -><- *         !!       Last date of change:  7 Mar 2008  **
 **      !!     !!                !!                                         **
 **      !!     !!                !!       Now is  0 Jan 2000 at  0:00:00    **
 **      !!                       !!                                         **
 **      !!        lh             !!       Disclaimer: this program comes    **
 **      !!                       !!       without any guarantees. Beware    **
 **      !!                 hh    !!       of errors and use common sense    **
 **      !!    ll                 !!       when interpreting results.        **
 **      !!                       !!                                         **
 **      !!                                Copyright T. Sjostrand (2008)     **
 **                                                                          **
 ** An archive of program versions and documentation is found on the web:    **
 ** http://www.thep.lu.se/~torbjorn/Pythia.html                              **
 **                                                                          **
 ** When you cite this program, the official reference is to the 6.4 manual: **
 ** T. Sjostrand, S. Mrenna and P. Skands, JHEP05 (2006) 026                 **
 ** (LU TP 06-13, FERMILAB-PUB-06-052-CD-T) [hep-ph/0603175].                **
 **                                                                          **
 ** Also remember that the program, to a large extent, represents original   **
 ** physics research. Other publications of special relevance to your        **
 ** studies may therefore deserve separate mention.                          **
 **                                                                          **
 ** Main author: Torbjorn Sjostrand; CERN/PH, CH-1211 Geneva, Switzerland,   **
 **   and Department of Theoretical Physics, Lund University, Lund, Sweden;  **
 **   phone: + 41 - 22 - 767 82 27; e-mail: torbjorn@thep.lu.se              **
 ** Author: Stephen Mrenna; Computing Division, GDS Group,                   **
 **   Fermi National Accelerator Laboratory, MS 234, Batavia, IL 60510, USA; **
 **   phone: + 1 - 630 - 840 - 2556; e-mail: mrenna@fnal.gov                 **
 ** Author: Peter Skands; Theoretical Physics Department,                    **
 **   Fermi National Accelerator Laboratory, MS 106, Batavia, IL 60510, USA; **
 **   and CERN/PH, CH-1211 Geneva, Switzerland;                              **
 **   phone: + 41 - 22 - 767 24 59; e-mail: skands@fnal.gov                  **
 **                                                                          **
 **                                                                          **
 ******************************************************************************
 ******************************************************************************
1****************** PYINIT: initialization of PYTHIA routines *****************
 ==============================================
 PDFset name ../lhapdf-5.2.3/../prod/lhapdf/PDFsets/cteq6ll.LHpdf                            
 with          1 members
 ====  initialized. ===========================
 Strong coupling at Mz for PDF is:  0.12978

 ==============================================================================
 I                                                                            I
 I         PYTHIA will be initialized for p+ on p+ user configuration         I
 I            with   7000.000 GeV on   7000.000 GeV beam energies             I
 I                                                                            I
 I           corresponding to  14000.000 GeV center-of-mass energy            I
 I                                                                            I
 ==============================================================================

 ******** PYMAXI: summary of differential cross-section maximum search ********

           ==========================================================
           I                                      I                 I
           I  ISUB  Subprocess name               I  Maximum value  I
           I                                      I                 I
           ==========================================================
           I                                      I                 I
           I    4   User process 611              I    1.3933E-06   I
           I                                      I                 I
           ==========================================================

 ********************** PYINIT: initialization completed **********************

          Event listing of user process at input (simplified)

   I IST     ID Mothers   Colours    p_x      p_y      p_z       E        m
   1 -1      21   0   0  503  504    0.000    0.000  226.463  226.463    0.000
   2 -1      21   0   0  504  505    0.000    0.000 -207.598  207.598    0.000
   3  2      -6   1   2    0  505  -52.411   59.433  111.701  220.818  173.218
   4  2       6   1   2  503    0   52.411  -59.433  -92.835  213.243  174.857
   5  1      -5   3   3    0  505  -49.693   86.081   33.256  104.920    4.800
   6  2     -24   3   3    0    0   -2.718  -26.648   78.445  115.898   81.001
   7  1       5   4   4  503    0    8.066  -24.988   39.743   47.875    4.800
   8  2      24   4   4    0    0   44.345  -34.445 -132.578  165.368   81.343
   9  1       3   6   6  501    0   -0.532  -19.396  -13.491   23.638    0.500
  10  1      -4   6   6    0  501   -2.186   -7.252   91.937   92.260    1.500
  11  1     -15   8   8    0  502   -5.290  -43.817  -38.536   58.619    1.777
  12  1      16   8   8  502    0   49.635    9.373  -94.042  106.749    0.000



                            Event listing (summary)

    I particle/jet KS     KF  orig    p_x      p_y      p_z       E        m

    1 !p+!         21    2212    0    0.000    0.000 7000.000 7000.000    0.938
    2 !p+!         21    2212    0    0.000    0.000-7000.000 7000.000    0.938
 ==============================================================================
    3 !g!          21      21    1    3.297   -0.541  226.461  226.486    0.000
    4 !g!          21      21    2    1.429    0.698 -207.595  207.601    0.000
    5 !g!          21      21    3    3.297   -0.541  226.461  226.486    0.000
    6 !g!          21      21    4    1.429    0.698 -207.595  207.601    0.000
    7 !tbar!       21      -6    0  -49.617   59.219  112.072  220.303  173.218
    8 !t!          21       6    0   54.343  -59.062  -93.206  213.784  174.857
    9 !bbar!       21      -5    7  -48.442   86.036   33.694  104.437    4.800
   10 !W-!         21     -24    7   -1.174  -26.817   78.379  115.866   81.001
   11 !b!          21       5    8    8.732  -25.079   39.640   47.954    4.800
   12 !W+!         21      24    8   45.611  -33.982 -132.845  165.830   81.343
   13 !s!          21       3   10   -0.330  -19.345  -13.545   23.623    0.500
   14 !cbar!       21      -4   10   -0.844   -7.472   91.923   92.243    1.500
   15 !tau+!       21     -15   12   -4.810  -43.678  -38.641   58.542    1.777
   16 !nu_tau!     21      16   12   50.421    9.696  -94.204  107.288    0.000
 ==============================================================================
   17 (W-)         11     -24    3   -1.174  -26.817   78.379  115.866   81.001
   18 (W+)         11      24    3   45.611  -33.982 -132.845  165.830   81.343
   19 tau+          1     -15   18   -4.810  -43.678  -38.641   58.542    1.777
   20 nu_tau        1      16   18   50.421    9.696  -94.204  107.288    0.000
   21 bbar      A   2      -5    3  -48.442   86.036   33.694  104.437    4.800
   22 u         V   1       2    2   -0.682   -0.409 -479.483  479.483    0.000
   23 b         A   2       5    3    8.732  -25.079   39.640   47.954    4.800
   24 uu_1      V   1    2203    1   -1.767    0.501 6672.062 6672.062    0.000
   25 s         A   2       3   17   -0.330  -19.345  -13.545   23.623    0.500
   26 cbar      V   1      -4   17   -0.844   -7.472   91.923   92.243    1.500
   27 d         A   2       1    1   -1.529    0.040  101.455  101.467    0.000
   28 ud_0      V   1    2101    2   -0.748   -0.290-6312.901 6312.901    0.000
 ==============================================================================
                   sum:  2.00          0.00     0.00     0.00 14000.00 14000.00
1********* PYSTAT:  Statistics on Number of Events and Cross-sections *********

 ==============================================================================
 I                                  I                            I            I
 I            Subprocess            I      Number of points      I    Sigma   I
 I                                  I                            I            I
 I----------------------------------I----------------------------I    (mb)    I
 I                                  I                            I            I
 I N:o Type                         I    Generated         Tried I            I
 I                                  I                            I            I
 ==============================================================================
 I                                  I                            I            I
 I   0 All included subprocesses    I            1             9 I  2.197E-07 I
 I   4 User process 611             I            1             9 I  2.197E-07 I
 I                                  I                            I            I
 ==============================================================================

 ********* Total number of errors, excluding junctions =        0 *************
 ********* Total number of errors, including junctions =        0 *************
 ********* Total number of warnings =                           0 *************
 ********* Fraction of events that fail fragmentation cuts =  0.00000 *********
\end{verbatim}
}

\newpage

\boldmath
\subsection{File herwig.out}
\unboldmath

{\scriptsize
\begin{verbatim}
          HERWIG 6.510  31st Oct. 2005

          Please reference:  G. Marchesini, B.R. Webber,
          G.Abbiendi, I.G.Knowles, M.H.Seymour & L.Stanco
          Computer Physics Communications 67 (1992) 465
                             and
          G.Corcella, I.G.Knowles, G.Marchesini, S.Moretti,
          K.Odagiri, P.Richardson, M.H.Seymour & B.R.Webber,
          JHEP 0101 (2001) 010

          INPUT CONDITIONS FOR THIS RUN

          BEAM 1 (P       ) MOM. =   7000.00
          BEAM 2 (P       ) MOM. =   7000.00
          PROCESS CODE (IPROC)   =    -611
          NUMBER OF FLAVOURS     =    6
          STRUCTURE FUNCTION SET =    8
          AZIM SPIN CORRELATIONS =    T
          AZIM SOFT CORRELATIONS =    T
          QCD LAMBDA (GEV)       =    0.1800
          DOWN     QUARK  MASS   =    0.3200
          UP       QUARK  MASS   =    0.3200
          STRANGE  QUARK  MASS   =    0.5000
          CHARMED  QUARK  MASS   =    1.5500
          BOTTOM   QUARK  MASS   =    4.8000
          TOP      QUARK  MASS   =  175.0000
          GLUON EFFECTIVE MASS   =    0.7500
          EXTRA SHOWER CUTOFF (Q)=    0.4800
          EXTRA SHOWER CUTOFF (G)=    0.1000
          PHOTON SHOWER CUTOFF   =    0.4000
          CLUSTER MASS PARAMETER =    3.3500
          SPACELIKE EVOLN CUTOFF =    2.5000
          INTRINSIC P-TRAN (RMS) =    0.0000
          DECAY SPIN CORRELATIONS=    T
          SUSY THREE BODY ME     =    T
          SUSY FOUR  BODY ME     =    F
          MIN MTM FRAC FOR ISR   =1.0000E-04
          1-MAX MTM FRAC FOR ISR =1.0000E-06

          NO EVENTS WILL BE WRITTEN TO DISK

          B_d: Delt-M/Gam =0.7000 Delt-Gam/2*Gam =0.0000
          B_s: Delt-M/Gam = 10.00 Delt-Gam/2*Gam =0.2000

          LHAPDF USED FOR BEAM 1: SET 10042 OF HWLHAPDF            
          LHAPDF USED FOR BEAM 2: SET 10042 OF HWLHAPDF            


          Checking consistency of particle properties


          Checking consistency of decay tables


          CHECKING SUSY DECAY MATRIX ELEMENTS

          INPUT EVT WEIGHT   =  1.0000E+00
          INPUT MAX WEIGHT   =  0.0000E+00

          SUBROUTINE TIMEL CALLED BUT NOT LINKED.
          DUMMY TIMEL WILL BE USED. DELETE DUMMY
          AND LINK CERNLIB FOR CPU TIME REMAINING.


 EVENT       1: 7000.00 GEV/C P        ON  7000.00 GEV/C P        PROCESS:  -611
 SEEDS:      945169 &     1890338   STATUS:   10 ERROR:   0  WEIGHT:  1.0000E+00


                                                   ---INITIAL STATE---    

 IHEP    ID      IDPDG IST MO1 MO2 DA1 DA2   P-X     P-Y     P-Z  ENERGY    MASS
    1 P           2212 101   0   0   0   0    0.00    0.00 7000.0 7000.0    0.94
    2 P           2212 102   0   0   0   0    0.00    0.00-7000.0 7000.0    0.94
    3 CMF            0 103   1   2   0   0    0.00    0.00    0.014000.014000.00

                                                  ---HARD SUBPROCESS---   

 IHEP    ID      IDPDG IST MO1 MO2 DA1 DA2   P-X     P-Y     P-Z  ENERGY    MASS
    4 GLUON         21 121   6   5   9   8    0.00    0.00  123.0  123.0    0.00
    5 GLUON         21 122   6   7  17   4    0.00    0.00 -269.8  269.8    0.00
    6 HARD           0 120   4   5   7   8   47.77   44.58 -146.8  398.2  364.29
    7 TBAR          -6 123   6   8  37   5   32.54   62.26  -18.0  163.6  146.60
    8 TQRK           6 124   6   4  41   7  -32.54  -62.26 -128.8  229.2  176.11

                                                  ---PARTON SHOWERS---    

 IHEP    ID      IDPDG IST MO1 MO2 DA1 DA2   P-X     P-Y     P-Z  ENERGY    MASS
    9 GLUON         94 141   4   6  11  16   14.47   13.15  133.2  130.3  -33.75
   10 CONE           0 100   4   8   0   0   -0.46   -0.89   -0.5    1.1    0.00
   11 GLUON         21 149   9  12   0  44    0.05    1.15    1.8    2.3    0.75
   12 GLUON         21 149   9  13   0  11   -1.71    1.21  201.4  201.4    0.75
   13 GLUON         21 149   9  14   0  12   -3.02   -0.20  442.8  442.8    0.75
   14 UD          2101 147   9  15   0  13    0.00    0.00 5639.1 5639.1    0.37
   15 UQRK           2 149   9  16   0  14   -6.84  -12.99  579.5  579.7    0.32
   16 GLUON         21 149   9  19   0  15   -2.96   -2.32    2.2    4.4    0.75
   17 GLUON         94 142   5   6  19  36   33.30   31.43 -280.0  267.9  -93.53
   18 CONE           0 100   5   7   0   0    0.46    0.89   -0.3    1.0    0.00
   19 GLUON         21 149  17  20   0  16    4.14    0.22   -6.6    7.8    0.75
   20 GLUON         21 149  17  21   0  19    2.88   -1.12   -3.2    4.5    0.75
   21 GLUON         21 149  17  22   0  20   -7.66    8.20  -25.7   28.1    0.75
   22 GLUON         21 149  17  23   0  21   -9.76    7.26  -37.6   39.5    0.75
   23 DBAR          -1 149  17  24   0  22   -3.10    0.62   -7.6    8.2    0.32
   24 DQRK           1 149  17  25   0  23   -0.58    0.41   -1.1    1.3    0.32
   25 GLUON         21 149  17  26   0  24   -1.13    0.90   -9.5    9.6    0.75
   26 GLUON         21 149  17  27   0  25    2.86  -10.41  -90.0   90.7    0.75
   27 GLUON         21 149  17  28   0  26   -1.11   -8.72  -46.6   47.5    0.75
   28 GLUON         21 149  17  29   0  27   -2.32   -3.50  -38.1   38.3    0.75
   29 GLUON         21 149  17  30   0  28  -12.62  -22.40 -171.6  173.5    0.75
   30 GLUON         21 149  17  31   0  29   -7.53   -3.71  -68.2   68.7    0.75
   31 GLUON         21 149  17  32   0  30    0.55    0.88  -10.9   11.0    0.75
   32 GLUON         21 149  17  33   0  31   -1.25   -0.76-1639.8 1639.8    0.75
   33 GLUON         21 149  17  34   0  32   -0.57   -1.22-1331.1 1331.1    0.75
   34 UD          2101 148  17  35   0  33    0.00    0.00-3074.7 3074.7    0.31
   35 UQRK           2 149  17  36   0  34    1.50    1.40  -74.3   74.3    0.32
   36 GLUON         21 149  17  39   0  35    2.41    0.52  -83.4   83.4    0.75
   37 TBAR          94 143   7   6  39  40   48.68   72.74  -25.3  178.2  153.10
   38 CONE           0 100   7   5   0   0   -0.25   -0.38   -0.9    1.0    0.00
   39 GLUON         21 149  37  50   0  36   -3.29    2.66    3.0    5.2    0.75
   40 TBAR          -6   3  37  37  45  45   51.97   70.08  -28.3  172.9  146.60
   41 TQRK          94 144   8   6  43  44   -0.91  -28.16 -121.5  220.0  181.22
   42 CONE           0 100   8   4   0   0   -0.03   -0.99    0.1    1.0    0.00
   43 TQRK           6   3  41  41  51  51   -2.06  -22.90 -120.1  214.4  176.11
   44 GLUON         21 149  41  11   0  59    1.15   -5.26   -1.5    5.6    0.75

                                               ---HEAVY PARTICLE DECAYS---

 IHEP    ID      IDPDG IST MO1 MO2 DA1 DA2   P-X     P-Y     P-Z  ENERGY    MASS
   45 TBAR          -6 155  37  51  46  47   51.97   70.08  -28.3  172.9  146.60
   46 W-           -24 123  45  46  48  46   62.93   35.09   22.8  109.9   79.79
   47 BBAR          -5 124  45  45  49  45  -10.96   34.99  -51.1   63.0    4.80
   48 W-           -24   3  46  46  60  60   62.93   35.09   22.8  109.9   79.79

                                                  ---PARTON SHOWERS---    

 IHEP    ID      IDPDG IST MO1 MO2 DA1 DA2   P-X     P-Y     P-Z  ENERGY    MASS
   49 BBAR          94 144  47  45  50  50  -10.96   34.99  -51.1   63.0    4.80
   50 BBAR          -5 149  49  57   0  39  -10.96   34.99  -51.1   63.0    4.80

                                               ---HEAVY PARTICLE DECAYS---

 IHEP    ID      IDPDG IST MO1 MO2 DA1 DA2   P-X     P-Y     P-Z  ENERGY    MASS
   51 TQRK           6 155  41  44  52  53   -2.06  -22.90 -120.1  214.4  176.11
   52 W+            24 123  51  52  54  52  -32.46  -56.57  -32.3  115.5   89.67
   53 BQRK           5 124  51  51  55  51   30.40   33.67  -87.7   98.9    4.80
   54 W+            24   3  52  52  71  71  -31.54  -55.20  -32.8  114.7   89.67

                                                  ---PARTON SHOWERS---    

 IHEP    ID      IDPDG IST MO1 MO2 DA1 DA2   P-X     P-Y     P-Z  ENERGY    MASS
   55 BQRK          94 144  53  51  57  59   29.48   32.30  -87.2   99.7   20.27
   56 CONE           0 100  53  51   0   0   -0.31   -0.63   -0.8    1.1    0.00
   57 BQRK           5 149  55  58   0  50   29.77   33.42  -78.0   90.0    4.80
   58 GLUON         21 149  55  59   0  57    0.49   -1.26   -5.8    6.0    0.75
   59 GLUON         21 149  55  44   0  58   -0.78    0.14   -3.4    3.6    0.75

                                               ---HEAVY PARTICLE DECAYS---

 IHEP    ID      IDPDG IST MO1 MO2 DA1 DA2   P-X     P-Y     P-Z  ENERGY    MASS
   60 W-           -24 155  46  45  61  62   62.93   35.09   22.8  109.9   79.79

                                                ---H/W/Z BOSON DECAYS---  

 IHEP    ID      IDPDG IST MO1 MO2 DA1 DA2   P-X     P-Y     P-Z  ENERGY    MASS
   61 SQRK           3 123  60  62  63  62   62.95   54.96    5.9   83.8    0.50
   62 CBAR          -4 124  60  61  67  61   -0.02  -19.87   16.9   26.1    1.55

                                                  ---PARTON SHOWERS---    

 IHEP    ID      IDPDG IST MO1 MO2 DA1 DA2   P-X     P-Y     P-Z  ENERGY    MASS
   63 SQRK          94 143  61  60  65  66   61.50   53.24    6.1   82.5   12.02
   64 CONE           0 100  61  62   0   0   -0.03    0.03   -0.6    0.6    0.00
   65 SQRK           3 149  63  66   0  70   56.79   44.14    6.0   72.2    0.50
   66 GLUON         21 149  63  69   0  65    4.72    9.10    0.1   10.3    0.75
   67 CBAR          94 144  62  60  69  70    1.42  -18.15   16.6   27.4   12.03
   68 CONE           0 100  62  61   0   0    0.55    0.72   -0.7    1.2    0.00
   69 GLUON         21 149  67  70   0  66   -2.29   -0.27   -0.3    2.5    0.75
   70 CBAR          -4 149  67  65   0  69    3.71  -17.88   17.0   25.0    1.55

                                               ---HEAVY PARTICLE DECAYS---

 IHEP    ID      IDPDG IST MO1 MO2 DA1 DA2   P-X     P-Y     P-Z  ENERGY    MASS
   71 W+            24 155  52  51  72  73  -31.54  -55.20  -32.8  114.7   89.67

                                                ---H/W/Z BOSON DECAYS---  

 IHEP    ID      IDPDG IST MO1 MO2 DA1 DA2   P-X     P-Y     P-Z  ENERGY    MASS
   72 MU+          -13 123  71  73  74  73  -53.75  -11.21  -35.5   65.4    0.11
   73 NU_MU         14 124  71  72  75  72   22.21  -43.99    2.6   49.4    0.00
   74 MU+          -13 190  72  71   0   0  -53.75  -11.21  -35.5   65.4    0.11
   75 NU_MU         14 190  73  71   0   0   22.21  -43.99    2.6   49.4    0.00

          OUTPUT ON LES HOUCHES EVENTS


      PROC CODE  XSECT(pb)        XERR(pb)      Max wgt(nb) No. of events

        -611    0.22478E+03     0.65879E+02     0.13968E+01        1

          OUTPUT ON ELEMENTARY PROCESS

          N.B. NEGATIVE WEIGHTS NOT ALLOWED

          NUMBER OF EVENTS   =           1
          NUMBER OF WEIGHTS  =           7
          MEAN VALUE OF WGT  =  2.2478E-01
          RMS SPREAD IN WGT  =  0.0000E+00
          ACTUAL MAX WEIGHT  =  1.3968E+00
          ASSUMED MAX WEIGHT =  1.3968E+00

          PROCESS CODE IPROC =        -611
          CROSS SECTION (PB) =   224.8    
          ERROR IN C-S  (PB) =   65.88    
          EFFICIENCY PERCENT =   16.09    

\end{verbatim}
}

\bibliographystyle{ieeetr}
\bibliography{acermc_bib}

\end{document}